\pdfoutput=1
\documentclass[PAPER, atlasdraft=false, cernpreprint, UKenglish, texmf, orcidlogo]{atlasdoc}
\usepackage[subcaption, rotating, longtable, multirow]{atlaspackage}
\usepackage{atlasbiblatex}

\usepackage{mathrsfs}

\usepackage[hepparticle, hepprocess, jetetmiss, snippets, xref=false, unit=false]{atlasphysics}
\usepackage[italic]{hepnicenames}

\graphicspath{{logos/}{figures/}}

\usepackage{ANA-TOPQ-2021-21-PAPER-defs}

\AtlasTitle{Measurement of $tZq$ inclusive and differential cross-sections in $pp$ collisions at $\sqrt{s} = 13~\text{TeV}$ with the ATLAS detector}

\AtlasAbstract{%
The $t$-channel production of a top quark in association with a $Z$ boson, $tZq$, is measured in the trilepton channel.
The measured $pp \rightarrow t\ell^+\ell^-q$ process is defined for $m_{\ell\ell} > 30\,\text{GeV}$ and includes both resonant $Z$-boson production and non-resonant dilepton production.
The data collected by the ATLAS experiment in 2015 to 2018 in proton--proton collisions at a centre-of-mass energy of $\sqrt{s} = 13\,\text{TeV}$ are used, corresponding to an integrated luminosity of $140\,\text{fb}^{-1}$.
Events containing three light leptons ($e$ or $\mu$) and two to four jets, one or two of which are identified as $b$-quark-initiated jets, are selected.
The inclusive $pp \rightarrow t \ell\ell q$ cross-section with $m_{\ell\ell} > 30\,\text{GeV}$ is measured, along with the separate cross-sections for top-quark and top-antiquark production and their ratio.
The inclusive cross-section is measured to be $95.2\;^{+8.5}_{-7.9}\;\text{(stat.)}\;^{+5.5}_{-5.3}\;\text{(syst.)}\,\text{fb}$ and found to be in agreement with the most advanced Standard Model predictions.
Differential cross-section measurements as a function of many kinematic variables are also presented.
Both absolute and normalised differential cross-section measurements are made at particle level and parton level in their respective phase spaces.
The top-quark spin asymmetry is measured from the unfolded measurement of the top-quark polarisation angle as $A = 0.34 \pm 0.14 \;\text{(stat.)} \pm 0.06\;\text{(syst.)}$, in good agreement with the Standard Model prediction.
The differential cross-section measurement results are interpreted in the framework of Standard Model effective field theory and limits are set on dimension-6 operators involving the top quark.
}

\AtlasRefCode{TOPQ-2021-21}

\PreprintIdNumber{CERN-EP-2026-244}

\AtlasJournal{\EPJC}
\AtlasCoverEgroupAnalysisTeam{atlas-topq-2021-21-analysis-team@cern.ch}

\hypersetup{pdftitle={ATLAS document},pdfauthor={The ATLAS Collaboration}}

\begin{document}

\maketitle

\tableofcontents

%
%
%

%
%

%
%
%

%
%
%
%

%

%

%
\section{Introduction}%
\label{sec:intro}

This paper presents measurements of the electroweak production of a single top quark in association with a \PZ boson in the $t$-channel (\tZq) using the full LHC~\cite{Evans:2008zzb} \RunTwo dataset of proton--proton (\pp) collisions recorded by the ATLAS detector at a centre-of-mass energy of $\sqrt{s}=\SI{13}{\TeV}$, corresponding to an integrated luminosity of \lumi.

\begin{figure}[htbp]
\centering
\begin{subfigure}[b]{.30\textwidth}%
\includegraphics[width=\textwidth]{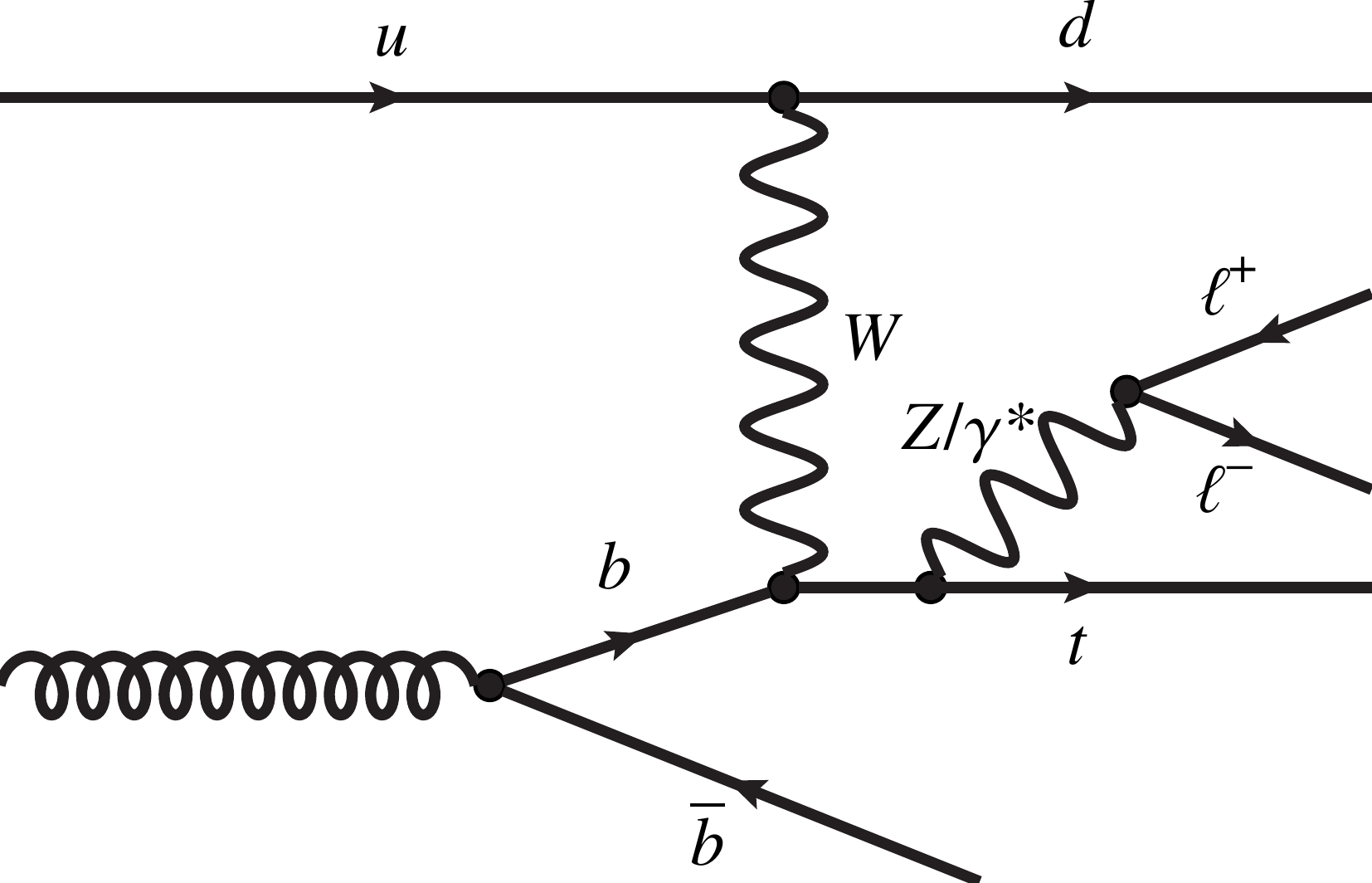}
\caption{}
\label{fig:intro:feynman:a}
\end{subfigure}%
\quad
\quad
\begin{subfigure}[b]{.30\textwidth}%
\includegraphics[width=\textwidth]{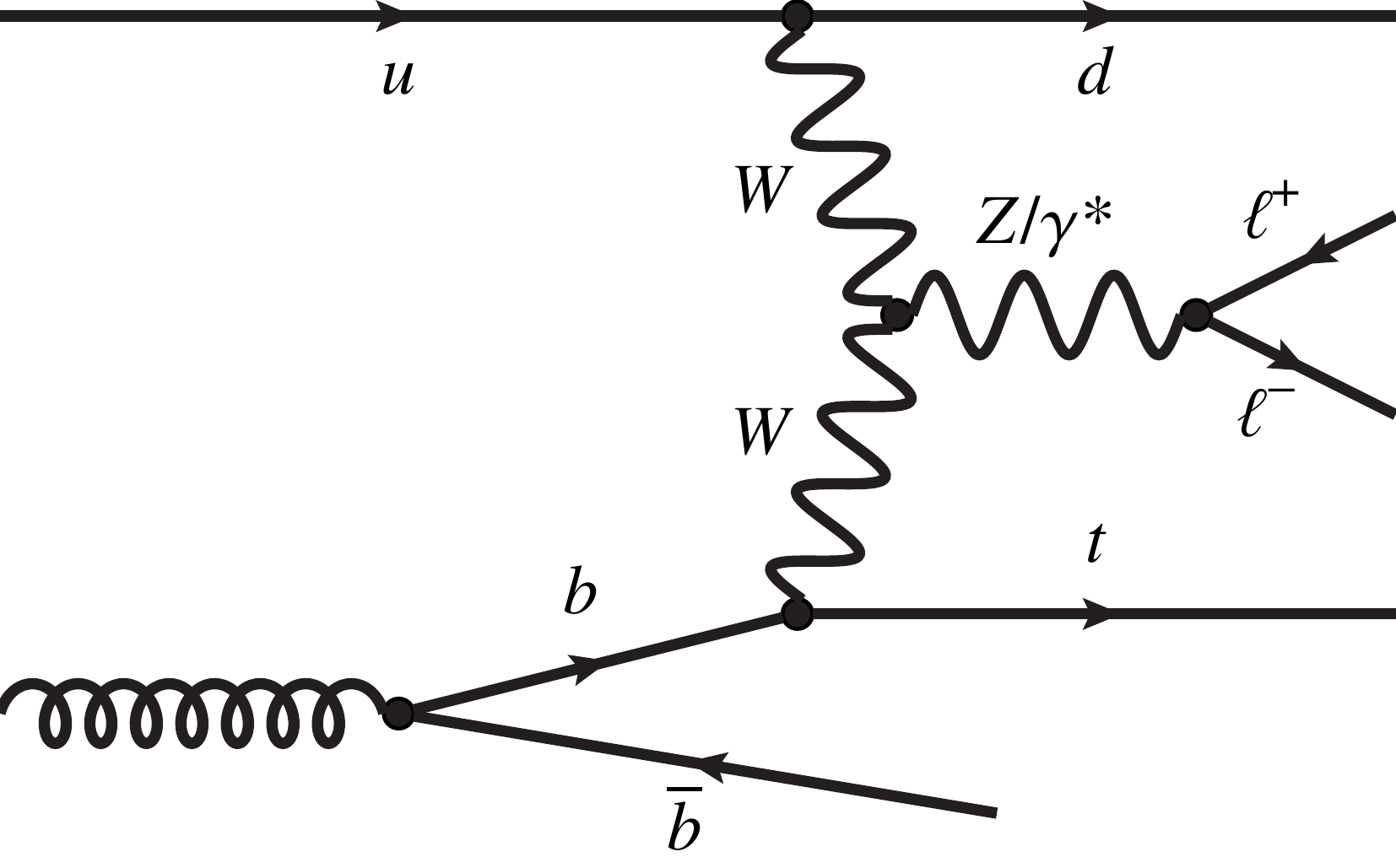}
\caption{}
\label{fig:intro:feynman:b}
\end{subfigure}%
\quad
\quad
\begin{subfigure}[b]{.30\textwidth}%
\includegraphics[width=\textwidth]{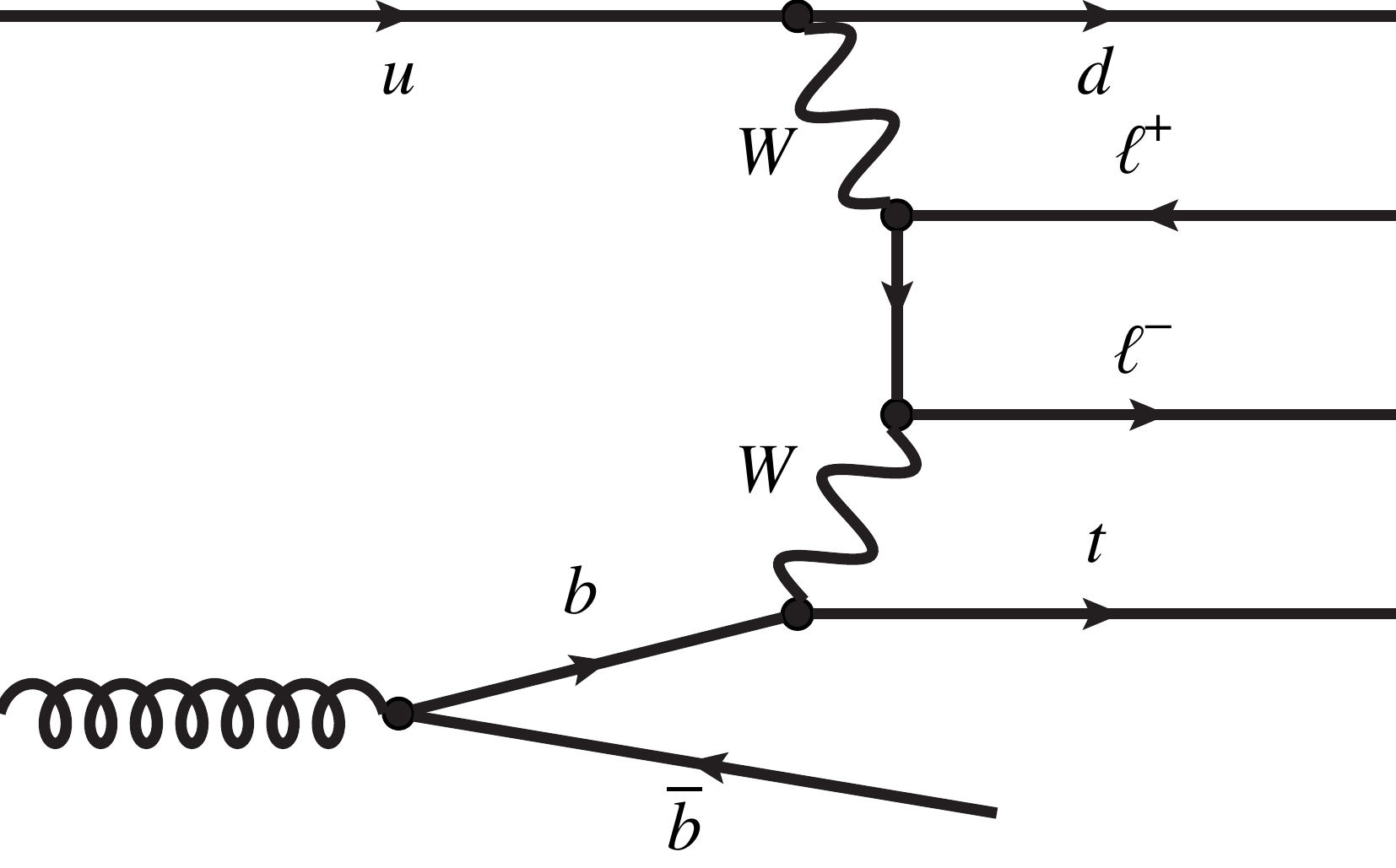}
\caption{}
\label{fig:intro:feynman:c}
\end{subfigure}%
\caption{Example Feynman diagrams of the lowest-order amplitudes for the \tZq process, including (a, b) resonant and (c) non-resonant $\ell^+\ell^-$ production. In the four-flavour scheme, the \Pqb-quark originates from gluon splitting.}
\label{fig:intro:feynman}
\end{figure}

At leading order (LO) in the Standard Model (SM), single top-quark production and decay proceed via the electroweak interaction. The dominant LO diagrams for \tZq production correspond to $t$-channel single top-quark production with an additional \PZ boson radiated either from the $t$-channel \PW-boson propagator or from one of the quark lines, as illustrated in \cref{fig:intro:feynman}. These diagrams provide direct sensitivity to the \Pqt{}\PZ and \PW{}\PZ couplings within a single analysis.
The process is characterised by the presence of a light-flavour spectator quark produced through the $t$-channel exchange, which tends to be emitted at large pseudorapidity and gives rise to a distinctive event topology.
In addition to resonant \PZ-boson production, a non-resonant \Plp{}\Plm contribution, with $\ell=e,\mu,\tau$, to the same \tllq final state is included, as illustrated in \cref{fig:intro:feynman:c}.

The analysis is performed in the trilepton final state, where the \PZ boson decays into a pair of electrons or muons and the \PW boson, originating from the top-quark, decays leptonically. Contributions from $\tau$-lepton decays into electrons or muons are included. The trilepton final state is used due to the best signal-to-background ratio compared to the other \tZq final states.

Throughout this paper, the shorthand \tZq denotes the complete $pp\to t\ell^+\ell^-q$ signal, including both resonant and non-resonant dilepton production.
Unless otherwise stated, the quoted cross-sections correspond to this process in the dilepton phase space considered: they include the leptonic branching fraction of the resonant or non-resonant dilepton system, but are not multiplied by the branching fraction of the top-quark decay.
The inclusive measurement presented in this paper is defined for $m_{\ell\ell}>30\ \si{\giga\electronvolt}$.

Precise measurements of \tZq production provide important tests of the electroweak sector of the SM. In contrast to the \ttZ process, whose dependence on the strong coupling constant $\alpha_S$ is at $O(\alphas^2)$ at LO, \tZq production occurs purely via electroweak interactions at LO and receives relatively small next-to-leading-order (NLO) QCD corrections~\cite{tZq-Pagani,Frixione:2015zaa}. This makes the process particularly well suited for probing possible deviations from the SM predictions within the framework of the Standard Model Effective Field Theory (SMEFT)~\cite{Degrande:2018fog}. In particular, differential measurements of \tZq production probe the high-energy regime of the process and can provide enhanced sensitivity to dimension-six SMEFT operators affecting the electroweak interactions of the top quark~\cite{Maltoni:2019}. Furthermore, \tZq production constitutes an important background to \tHq production and must therefore be precisely characterised.

The \tZq process has previously been observed by the ATLAS Collaboration~\cite{TOPQ-2018-01} using \pp collision data recorded at $\sqrt{s}=\qty{13}{\TeV}$, corresponding to an integrated luminosity of \qty{139}{\per\fb}.
The measured inclusive cross-section for dilepton invariant mass $\mll > \SI{30}{\GeV}$ was \qty[parse-numbers=false]{97~^{+13}_{-12} \stat~\pm 7 \syst}{\fb} = \qty[parse-numbers=false]{97~^{+14}_{-13}}{\fb},
corresponding to a relative uncertainty of \qty{14}{\%}.
The CMS Collaboration has reported inclusive and differential measurements~\cite{CMS-TOP-20-010} using the full Run~2 dataset of \pp collisions at a centre-of-mass energy of $\sqrt{s}=\qty{13}{\TeV}$, corresponding to an integrated luminosity of \qty{138}{\per\fb}.
The inclusive measurement yielded a cross-section of \qty[parse-numbers=false]{87.9~^{+7.5}_{-7.3} \stat~^{+7.3}_{-6.0} \syst}{\fb} for a dilepton invariant mass greater than \qty{30}{\GeV}, reaching a relative uncertainty of \qty{11}{\%}.
Furthermore, the CMS Collaboration measured the ratio of the cross-sections for the top-quark and the top-antiquark production in association with a \PZ boson to be \qty[parse-numbers=false]{2.37~^{+0.56}_{-0.42} \stat~^{+0.27}_{-0.13} \syst}{}.

The \tZq process can be described in different flavour schemes. In the five-flavour scheme (5FS), the bottom quark is treated as a massless parton in the proton, whereas in the four-flavour scheme (4FS) it is treated as massive and is not included in the proton Parton Distribution Functions (PDFs), but appears explicitly in the hard-scattering matrix element.
Calculations in both schemes are used in this analysis: the nominal 4FS sample models the signal acceptance and kinematic distributions, while dedicated 5FS samples are used to determine the overall reference normalisation.
Predictions obtained in both schemes are also considered in the interpretation of the measurements.

Various calculations for the predicted total \tZq SM cross-section are available.
The predictions considered here are defined for the $t$-channel \tZq process; contributions from $s$-channel production, interference with \tWZ production with hadronic \PW-boson decays, and Higgs-boson-mediated diagrams are not included.
These predictions are inclusive with respect to the \PZ-boson decay and are therefore not restricted to leptonic final states.
Ref.~\cite{tZq-Kidonakis} computes a total cross-section of \qty{850}{}$\pm$\qty{14}{\fb} at approximate next-to-next-to-leading-order (aNNLO) in the 5FS using the MSHT2020 parton distribution functions set \cite{Bailey:2020ooq} and renormalisation and factorisation scales set to the top-quark mass $m_\text{top}=172.5\ \si{\giga\electronvolt}$.

In Ref.~\cite{tZq-Pagani}, setting the renormalisation and factorisation scales to $H_\mathrm{T}/6$, with \mtop set to \qty{173.3}{\GeV}, and using the \NNPDF{3.1} PDF set~\cite{Ball:2017nwa}, the predicted total cross-section is \qty{805}{}$^{+45}_{-8}$\qty{}{\fb}, computed at NLO in QCD in the 5FS.\footnote{$H_\mathrm{T}/6 = \frac{\sum_{i}m_{T,i}}{6}$, with $i = \Pqt, \PZ(\ell^+\ell^-), \Pqb$. \Pqt and \Pqb refer to both quarks and antiquarks.} The corresponding prediction for the dilepton final state with $m_{\ell\ell}>\qty{30}{\GeV}$ is \qty{84.0}{}$^{+4.7}_{-1.0}$\qty{}{\fb}.
For comparison, Ref.~\cite{tZq-Pagani} also provides a 5FS NLO QCD+EW prediction including the $t$-channel, $s$-channel, and \tW-associated contributions. The corresponding cross-section is \qty{904}{}$^{+42}_{-19}$\qty{}{\fb} for inclusive \PZ-boson decays, and \qty{89.6}{}$^{+5.1}_{-1.7}$\qty{}{\fb} for the dilepton final state with $m_{\ell\ell}>\qty{30}{\GeV}$. The $k$-factor for the ratio of the NLO(QCD+EW) and NLO(QCD) predictions is \qty{0.93}{}~\cite{tZq-Pagani}, where EW denotes electroweak corrections.
In Ref.~\cite{tZq-Pagani}, predictions are also provided in the 4FS. Using the same scale choice, top-quark mass, and PDF set, the predicted total cross-section is \qty{764}{}$^{+33}_{-48}$\qty{}{\fb}, computed at NLO in QCD. The corresponding prediction for the dilepton final state with $m_{\ell\ell}>\qty{30}{\GeV}$ is \qty{80.2}{}$^{+3.7}_{-5.0}$\qty{}{\fb}.
A complete off-shell calculation of \tZq production at NLO accuracy in QCD and electroweak corrections is also available~\cite{tZq-Denner}. This calculation includes spin correlations and resonant and non-resonant contributions in a realistic trilepton fiducial phase space.

In this paper, an improved inclusive cross-section measurement is presented with respect to Ref.~\cite{TOPQ-2018-01}.
This improvement is driven by a revised analysis based on a looser event selection to increase the signal acceptance and thus reduce the previously dominant statistical uncertainty. The analysis also benefits from improved object reconstruction and identification, including particle-flow jet reconstruction~\cite{PERF-2015-09}, an updated \Pqb-tagging algorithm~\cite{FTAG-2019-07,ATL-PHYS-PUB-2017-013}, and improved lepton-isolation requirements~\cite{EGAM-2021-01,EGAM-2021-02,MUON-2018-03,MUON-2022-01} that reduce the contamination from non-prompt and misidentified leptons.
In addition, separate measurements of the \(\Pqt\PZ\Pq\) and \(\Paqt\PZ\Pq\) cross-sections and their ratio are performed.
This ratio is sensitive to the proton PDFs, since top-quark and top-antiquark production in the $t$-channel are initiated by different quark flavours in the proton.
In the following, these processes are denoted \tZqt and \tZqtbar, respectively.
Moreover, this analysis provides the first differential cross-section measurements of \tZq production at parton and particle level with ATLAS data. The differential measurements are extracted using a profile-likelihood unfolding~(PLU) method~\cite{TOPQ-2020-20}.

The unfolded measurement of the top-quark polarisation angle \cosTheta in the top-quark rest frame is used to extract the top-quark spin asymmetry, expected to be equal to unity at LO~\cite{Je_abek_1994,Aguilar_Saavedra_2010}.\footnote{$\cos\theta^{\star}=\frac{\vec{p}^\star(q^{\prime})\cdot \vec{p}^\star(\ell_t)}{\lvert \vec{p}^\star(q^{\prime})\rvert \lvert \vec{p}^\star(\ell_t)\rvert }$ where $q^{\prime}$ and $\ell_t$ are the spectator quark and lepton from the top-quark decay, measured in the top-quark rest frame.}
The polarisation of top quarks and antiquarks produced in single top-quark $t$-channel events~\cite{Aguilar_Saavedra_2014} was measured by the ATLAS Collaboration~\cite{TOPQ-2018-10}, finding good agreement with the SM prediction. In \tZq production, the top quark is also produced through the single top-quark $t$-channel mechanism.
However, the additional electroweak interactions involved in \tZq production can modify the top-quark polarisation with respect to inclusive single top-quark $t$-channel production, making the spin asymmetry sensitive to the underlying production dynamics.

An interpretation in the SMEFT framework is carried out using detector-level distributions of the reconstructed \PZ-boson transverse momentum. Nine Wilson coefficients corresponding to seven dimension-six operators of the Warsaw basis~\cite{Grzadkowski:2010es}, which affect top-quark-boson vertices through weak dipole and Higgs-current operators and four-quark interactions, are constrained, including both the real and imaginary parts of the weak dipole operators:
\begin{equation}
\text{Re}(\mathcal{C}_{\Pqt \PW}), \text{Im}(\mathcal{C}_{\Pqt \PW}),
\text{Re}(\mathcal{C}_{\Pqt \PB}), \text{Im}(\mathcal{C}_{\Pqt \PB}),
\mathcal{C}_{\PH Q}^{(1)}, \mathcal{C}_{\PH Q}^{(3)},
\mathcal{C}_{\PH \Pqt},
\mathcal{C}_{Q q}^{(3,1)}, \mathcal{C}_{Q q}^{(3,8)}.
\end{equation}
The EFT sensitivities of the \tZq and \ttZ processes are complementary and can help resolve degeneracies or flat directions in the SMEFT parameter space. Therefore, SMEFT effects at detector level are considered for both processes.

This paper is organised as follows.
\Cref{sec:atlas} describes the ATLAS detector.
\Cref{sec:datamc} summarises the data and simulated samples used in the analysis.
\Cref{sec:obj} describes the reconstruction and selection of physics objects, while \cref{sec:sel} defines the event selection in the signal and control regions and outlines the non-prompt-lepton backgrounds estimate.
The multivariate analysis, based on a neural network, is presented in \cref{sec:mva}.
Systematic uncertainties are discussed in \cref{sec:sys}.
The results of the inclusive cross-section measurements are presented in \cref{sec:inclusive}, while the differential cross-section measurements are described in \cref{sec:differential}.
The interpretation in the SMEFT framework is given in \cref{sec:smeft}.


%

%
\newcommand{\AtlasCoordFootnote}{%
ATLAS uses a right-handed coordinate system with its origin at the nominal interaction point~(IP)
in the centre of the detector and the \(z\)-axis along the beam pipe.
The \(x\)-axis points from the IP to the centre of the LHC ring,
and the \(y\)-axis points upwards.
Polar coordinates \((r,\phi)\) are used in the transverse plane,
\(\phi\) being the azimuthal angle around the \(z\)-axis.
The pseudorapidity is defined in terms of the polar angle \(\theta\) as \(\eta = -\ln \tan(\theta/2)\) and is equal to the rapidity
$ y = \frac{1}{2} \ln \left( \frac{E + p_z}{E - p_z} \right) $ in the relativistic limit.
Angular distance is measured in units of \(\Delta R \equiv \sqrt{(\Delta y)^{2} + (\Delta\phi)^{2}}\).}

\section{ATLAS detector}
\label{sec:atlas}

The ATLAS detector~\cite{PERF-2007-01} at the LHC covers nearly the entire solid angle around the collision point.\footnote{\AtlasCoordFootnote}
It consists of an inner tracking detector surrounded by a thin superconducting solenoid, electromagnetic and hadronic calorimeters,
and a muon spectrometer incorporating three large superconducting air-core toroidal magnets.

The inner-detector system~(ID) is immersed in a \qty{2}{\tesla} axial magnetic field
and provides charged-particle tracking in the range \(|\eta| < 2.5\).
The high-granularity silicon pixel detector covers the vertex region and typically provides four measurements per track,
the first hit generally being in the insertable B-layer~(IBL) installed before Run~2~\cite{ATLAS-TDR-19,PIX-2018-001}.
It is followed by the SemiConductor Tracker~(SCT), which usually provides eight measurements per track.
These silicon detectors are complemented by the transition radiation tracker~(TRT),
which enables radially extended track reconstruction up to \(|\eta| = 2.0\).
The TRT also provides electron identification information
based on the fraction of hits (typically 30 in total) above a higher energy-deposit threshold corresponding to transition radiation.

The calorimeter system covers the pseudorapidity range \(|\eta| < 4.9\).
Within the region \(|\eta|< 3.2\), electromagnetic calorimetry is provided by barrel and
endcap high-granularity lead/liquid-argon~(LAr) calorimeters,
with an additional thin LAr presampler covering \(|\eta| < 1.8\)
to correct for energy loss in material upstream of the calorimeters.
Hadronic calorimetry is provided by the steel/scintillator-tile calorimeter,
segmented into three barrel structures within \(|\eta| < 1.7\), and two copper/LAr hadronic endcap calorimeters.
The solid angle coverage is completed with forward copper/LAr and tungsten/LAr calorimeter modules
optimised for electromagnetic and hadronic energy measurements, respectively.

The muon spectrometer~(MS) comprises separate trigger and
high-precision tracking chambers measuring the deflection of muons in a magnetic field generated by the superconducting air-core toroidal magnets.
The field integral of the toroids ranges between \num{2.0} and \qty{6.0}{\tesla\metre}
across most of the detector.
Three layers of precision chambers, each consisting of layers of monitored drift tubes, cover the region \(|\eta| < 2.7\),
complemented by cathode-strip chambers in the forward region, where the background is highest.
The muon trigger system covers the range \(|\eta| < 2.4\) with resistive-plate chambers in the barrel, and thin-gap chambers in the endcap regions.

The luminosity is measured mainly by the LUCID--2~\cite{LUCID2} detector that records Cherenkov light produced in the quartz windows of photomultipliers located close to the beampipe.

Events were selected by the first-level trigger system implemented in custom hardware,
followed by selections made by algorithms implemented in software in the high-level trigger~\cite{TRIG-2016-01}.
The first-level trigger accepted events from the \qty{40}{\MHz} bunch crossings at a rate close to \qty{100}{\kHz},
which the high-level trigger further reduced in order to record complete events to disk at about \qty{1.25}{\kHz}.

A software suite~\cite{ATL-SOFT-PUB-2021-001} is used in data simulation, in the reconstruction
and analysis of real and simulated data, in detector operations, and in the trigger and data acquisition
systems of the experiment.


%

%

%
\section{Data and simulation samples}%
\label{sec:datamc}

The data sample used in this article corresponds to \lumi~\cite{DAPR-2021-01} of \pp collisions at \(\sqrt{s} = \qty{13}{\TeV}\) collected with the ATLAS detector during 2015--2018, after requiring stable LHC beams and that all detector subsystems were operational~\cite{DAPR-2018-01}. The average number of interactions per bunch crossing for the period was 33.7.

Candidate events are required to satisfy one of the single-electron triggers or one of the single-muon triggers~\cite{TRIG-2016-01,TRIG-2018-05,TRIG-2018-01}. Unprescaled single-lepton triggers with the lowest transverse momentum, \pT, thresholds and standard isolation requirements are combined in a logical OR with higher-threshold triggers with looser identification criteria and without isolation requirements, resulting in an efficiency of almost \qty{100}{\%} for events passing the analysis selection. The lowest \pT threshold used for electrons is \qty{24}{\GeV}~(\qty{26}{\GeV}) in 2015~(2016--2018), while for muons the threshold is \qty{20}{\GeV}~(\qty{26}{\GeV}).

To evaluate the effects of the detector resolution and acceptance on the signal and background, and to estimate the backgrounds, simulated event samples were produced using a \GEANT-based Monte Carlo~(MC) detector simulation~\cite{Agostinelli:2002hh, SOFT-2010-01}. Some of the samples used for evaluating systematic uncertainties did not use the full \GEANT simulation but instead relied on a parameterisation of the calorimeter response~\cite{ATL-PHYS-PUB-2010-013}, called \textsc{AtlFast~II}.

The effect of multiple interactions in the same and neighbouring bunch crossings (\pileup) was modelled by overlaying the simulated hard-scattering event with inelastic \pp events generated with \PYTHIA[8.186]~\cite{Sjostrand:2007gs} using the \NNPDF[2.3lo] set of PDFs~\cite{Ball:2012cx} and the A3 set of tuned parameters~\cite{ATL-PHYS-PUB-2016-017}. The simulated samples were reweighted to reproduce the observed distribution of the average number of collisions per bunch crossing in each data-taking period.

The top-quark mass in the event generators described below was set to \SI{172.5}{\GeV}.
The versions of the generator, parton shower and PDF used for each of the nominal MC samples, together with the reference cross-sections used for their normalisation, are given in Table~\ref{tab:datamc:samples}.
Unless otherwise stated, samples interfaced to \PYTHIA[8] used the A14 tune~\cite{ATL-PHYS-PUB-2014-021} with the \NNPDF[2.3lo] PDF set~\cite{Ball:2012cx} for the parton shower.

\begin{table}[tbp]
\tablesetup[1.3]
\caption{Versions of the generator, parton shower and PDF used for the nominal MC samples, together with the reference cross-sections used for their normalisation. The values correspond to the effective cross sections of the generated samples, including decay-mode and generator-filter requirements where applicable. The diboson and triboson values refer to leptonic final states. The \tZq normalisation is described in the text.}
\label{tab:datamc:samples}
\begin{tabular}{l lllc}
\toprule
Process & \multicolumn{1}{c}{Generator} & Parton shower & \multicolumn{1}{c}{PDF} & \multicolumn{1}{c}{Reference cross-section [fb]}\\
\midrule
\tZq   & \MGNLO[3.5.1]  & \PYTHIA[8.309] & \NNPDF[3.0nlo\_4FS] & (see text) \\ %
\ttZ      & \MGNLO[2.8.1]   & \PYTHIA[8.244]  & \NNPDF[3.0nlo]      & \num{876} \\
\tWZ      & \MGNLO[2.6.7]   & \PYTHIA[8.244]  & \NNPDF[3.0nlo]      & \num{16.1} \\
Diboson   & \SHERPA[2.2.12] & \SHERPA[2.2.12] & \NNPDF[3.0nnlo]     & \num{2.17e4} \\
Triboson  & \SHERPA[2.2.2]  & \SHERPA[2.2.2]  & \NNPDF[3.0nnlo]     & \num{13.7} \\
\Zjets    & \SHERPA[2.2.11] & \SHERPA[2.2.11] & \NNPDF[3.0nnlo]     & \num{6.66e6} \\
\ttbar    & \POWHEGBOX[v2]  & \PYTHIA[8.230]  & \NNPDF[3.0nlo]      & \num{8.339e5} \\
\tW       & \POWHEGBOX[v2]  & \PYTHIA[8.230]  & \NNPDF[3.0nlo]      & \num{7.93e4} \\
\ttW      & \SHERPA[2.2.10] & \SHERPA[2.2.10] & \NNPDF[3.0nnlo]     & \num{746} \\
\ttH      & \POWHEGBOX[v2]  & \PYTHIA[8.230]  & \NNPDF[3.0nlo]      & \num{507} \\
\ttt      & \MGNLO[2.2.2]   & \PYTHIA[8.186]  & \NNPDF[2.3lo]       & \num{1.64} \\
\tttt     & \MGNLO[2.3.3]   & \PYTHIA[8.230]  & \NNPDF[3.1nlo]      & \num{12.0} \\
\bottomrule
\end{tabular}
\end{table}

The \tZq signal sample was simulated using the \MGNLO[3.5.1]~\cite{Alwall:2014hca} generator at NLO with the \NNPDF[3.0nlo\_4FS]~\cite{Ball:2014uwa} PDF set. The events were interfaced with \PYTHIA[8.309]~\cite{Bierlich:2022pfr}. Off-resonant events away from the \PZ-boson mass peak were included, for $\mll > \qty{5}{\GeV}$.
The top quark was decayed at LO using \MADSPIN~\cite{Frixione:2007zp,Artoisenet:2012st} to preserve spin correlations.
The 4FS was used, where all the quark masses are set to zero, except for the top and bottom quarks.
Following the discussion in Ref.~\cite{tZq-Pagani}, the renormalisation and factorisation scales (\muR, \muF) were set to $H_\mathrm{T}/6$.
The decays of the bottom and charm hadrons were simulated using the \EVTGEN[1.7.0] programme~\cite{Lange:2001uf}.

The nominal 4FS signal sample is used to model the signal acceptance and kinematic distributions. Its overall nominal normalisation is rescaled to the corresponding 5FS cross-section using dedicated 4FS and 5FS samples generated with otherwise consistent settings and with $m_{\ell\ell}>\qty{5}{\GeV}$. The correction factor is defined as
\begin{equation}
C_{\mathrm{5FS/4FS}}^{5}
=
\frac{\sigma_{\mathrm{5FS}}(m_{\ell\ell}>\qty{5}{\GeV})}{\sigma_{\mathrm{4FS}}(m_{\ell\ell}>\qty{5}{\GeV})} = 1.30
\label{eq:5FS4FSnormalisation}
\end{equation}

The inclusive cross-section is reported in the $m_{\ell\ell}>\qty{30}{\GeV}$ phase space. The conversion from the generated phase space is performed using the fraction
\begin{equation}
f^{30/5}_{\mathrm{4FS}}
=
\frac{\sigma_{\mathrm{4FS}}(m_{\ell\ell}>\qty{30}{\GeV})}{\sigma_{\mathrm{4FS}}(m_{\ell\ell}>\qty{5}{\GeV})}
= 0.75,
\label{eq:mllphasefactor}
\end{equation}
derived from the nominal 4FS sample. Applying the 4FS-to-5FS normalisation correction and the dilepton phase-space fraction to the nominal 4FS generator cross-section gives
\begin{equation}
\sigma_{\mathrm{ref}}
(pp\to t\ell\ell q,\,
m_{\ell\ell}>\qty{30}{\GeV})
=
\sigma_{\mathrm{4FS}}
(m_{\ell\ell}>\qty{5}{\GeV})
C_{\mathrm{5FS/4FS}}^{5}
f^{30/5}_{\mathrm{4FS}}
=
\qty{96.8}{\fb}.
\end{equation}
The corresponding cross-sections for top-quark and top-antiquark production are \tZqthxsect and \tZqthxsectbar, respectively, yielding a ratio of \tZqthxsecr.

To evaluate the modelling uncertainties on the signal prediction, several alternative \tZq MC samples are considered.
These include a sample generated with the same \MGNLO version as the nominal sample, but interfaced to \HERWIG[7.2.3]~\cite{Bahr:2008pv,Bellm:2015jjp} for the simulation of the parton shower and H7.2-Default tune for the underlying event
model.
Two additional samples with the same settings as the nominal \tZq sample, but with an up-and-down variation of the \texttt{Var3c} parameter in the A14 tune are used to evaluate uncertainties associated with the modelling of the initial-state radiation~(ISR), following a similar approach to the one described in Ref.~\cite{ATL-PHYS-PUB-2016-005}.

A further \tZq sample generated in the 5FS, with settings otherwise identical to those of the nominal sample, is used only for comparisons of the predicted differential distributions and is not included in the evaluation of the modelling uncertainties.

The possible impact of the omitted $s$-channel and $tW$-associated contributions is investigated. These contributions are found to be small and to exhibit kinematic properties different from those of the nominal $t$-channel signal. They are therefore expected to preferentially populate the background-dominated region of the Neural Network (NN) discriminant (see~\cref{sec:mva}). Their impact on the result is consequently expected to be limited, and no correction is applied.

The associated production of a top-quark-antiquark pair with a \PZ boson (\ttZ) is modelled using the \MGNLO[2.8.1] generator, which provides matrix elements at NLO in the strong coupling constant \alphas with the \NNPDF[3.0nlo] PDF set. The functional form of the renormalisation and factorisation scales (\muR, \muF) is set to the default scale $0.5 \times \sum_i \sqrt{m^2_i+p^2_{T,i}}$, where the sum runs over all the particles generated from the matrix-element calculation.
Top quarks are decayed at LO using \MADSPIN to preserve all spin correlations.
The events are interfaced with \PYTHIA[8.244]~\cite{Sjostrand:2014zea} for the parton shower and hadronisation.
\EVTGEN[1.7.0] was used to model bottom- and charm-hadron decays.

To evaluate the modelling uncertainties on the \ttZ background prediction, several alternative \ttZ MC samples are considered.
These include a sample simulated with the same \MGNLO version as the nominal sample, but interfaced to \HERWIG[7.2.1] for the simulation of the parton shower and H7.2-default tune for the underlying event model.
Similarly to  the signal sample, two additional samples with the same settings as the nominal \ttZ sample, but with an up-and-down variation of the Var3c parameter in the A14 tune are used to evaluate uncertainties associated with the modelling of ISR.

The associated production of top quarks with a \PW and \PZ boson (\tWZ) was modelled using the \MGNLO[2.6.7] generator at NLO with the \NNPDF[3.0nlo] PDF set.
The events were interfaced with \PYTHIA[8.244].
Heavy-flavour hadron decays were simulated with \EVTGEN[1.6.0].
The nominal \tWZ sample follows the \enquote{diagram removal 1} (DR1) scheme described in Ref.~\cite{Demartin:2016axk} and ignores any Feynman diagrams containing two resonant top quarks.
An alternative sample was generated within the DR2 scheme, which additionally considers the interference terms (at the level of squared amplitudes) between single- and double-resonant top-quark pair production.

Samples of diboson final states ($VV$) were simulated with the \SHERPA[2.2.12]~\cite{Bothmann:2019yzt} generator. Fully leptonic final states were generated using matrix elements at NLO accuracy in QCD for up to one additional parton and at LO accuracy for up to three additional parton emissions. The matrix-element calculations were matched and merged with the \SHERPA parton shower based on Catani--Seymour dipole factorisation~\cite{Gleisberg:2008fv,Schumann:2007mg} using the MEPS@NLO prescription~\cite{Hoeche:2011fd,Hoeche:2012yf,Catani:2001cc,Hoeche:2009rj}. The virtual QCD corrections were provided by the \OPENLOOPS library~\cite{Buccioni:2019sur,Cascioli:2011va,Denner:2016kdg}.
The \NNPDF[3.0nnlo] set of PDFs was used together with a dedicated tune of the \SHERPA parton shower.

Triboson ($VVV$) production was simulated with the \SHERPA[2.2.2] generator using the same PDF set, parton-shower tune, and matching and merging prescription as for diboson production. Matrix elements were calculated at NLO accuracy in QCD for the inclusive process and at LO accuracy for up to two additional parton emissions.

The production of \Zjets was simulated with the \SHERPA[2.2.11] generator using NLO matrix elements for up to two partons, and LO matrix elements for up to five partons calculated with the Comix~\cite{Gleisberg:2008fv} and \OPENLOOPS libraries. They were matched with the \SHERPA parton shower using the \MEPSatNLO prescription together with a dedicated parton-shower tune.
The \NNPDF[3.0nnlo] set of PDFs was used and the samples were normalised to a next-to-next-to-leading-order (NNLO) prediction~\cite{Anastasiou:2003ds}.

The production of \ttbar events was modelled using the \POWHEGBOX[v2]~\cite{Frixione:2007nw,Nason:2004rx,Frixione:2007vw,Alioli:2010xd} generator at NLO with the \NNPDF[3.0nlo] PDF set and the \hdamp parameter set to 1.5\,\mtop~\cite{ATL-PHYS-PUB-2016-020}.\footnote{The \hdamp parameter is a resummation damping factor and one of the parameters that controls the matching of \POWHEG matrix elements to the parton shower and thus effectively regulates the high-\pt radiation against which the \ttbar system recoils.}
The events were interfaced to \PYTHIA[8.230] to model the parton shower, hadronisation, and underlying event.
\EVTGEN[1.6.0] was used to model bottom- and charm-hadron decays.

To evaluate the modelling uncertainties on the \ttbar background prediction, additional samples are used.
The matching uncertainty is evaluated by comparing the nominal sample with an alternative sample obtained by setting the \texttt{pThard} \PYTHIA parameter to 1 (when the default value is 0). This parameter regulates the definition of the vetoed region of the showering, which is important to avoid gaps or overlaps in the phase space filled by \POWHEG and \PYTHIA.
This recommendation follows the description included in Ref.~\cite{Hoche:2021mkv}.
The impact of using a different parton-shower and hadronisation model is evaluated by showering the same \POWHEGBOX[v2] hard-scattering events used for the nominal \ttbar sample with \HERWIG[7.2.1] instead of \PYTHIA[8]. The alternative sample uses the H7.2-default tune for the underlying event model and the \MMHT[lo] PDF set~\cite{Harland-Lang:2014zoa}. Heavy-flavour hadron decays were simulated with \EVTGEN[1.7.0].
An additional sample generated with the same settings as the nominal one but with the \hdamp parameter set to 3.0\,\mtop is also used.

The associated production of top quarks with \PW bosons (\tW) was modelled by the \POWHEGBOX[v2]~\cite{Re:2010bp,Nason:2004rx,Frixione:2007vw,Alioli:2010xd} generator at NLO in QCD using the \NNPDF[3.0nlo] set of PDFs.
The DR1 scheme~\cite{Frixione:2008yi} was used to remove interference and overlap with \ttbar production.
The events were interfaced to \PYTHIA[8.230].

For top-quark pair production with a \PW boson (\ttW), the \SHERPA[2.2.10] generator and default parton shower are used at NLO accuracy in QCD, with multi-leg merging of up to one additional parton at NLO and up to two additional partons at LO. %
Additionally, a sample was generated at LO with \SHERPA[2.2.10] for the $\ttW j$ final state, and is used to account for subleading electroweak contributions to \ttW production.

The production of a top-quark-antiquark pair with a \PH boson (\ttH) was modelled using the \POWHEGBOX[v2]~\cite{Hartanto:2015uka} generator at NLO with the \NNPDF[3.0nlo] PDF set.
The events were interfaced to \PYTHIA[8.230].
\EVTGEN[1.6.0] was used to model bottom- and charm-hadron decays.

The production of \ttt events was modelled using the \MGNLO[2.2.2] generator at LO.
The events were interfaced with \PYTHIA[8.186].
Heavy-flavour hadron decays were simulated with \EVTGEN[1.2.0].

The production of \tttt events was modelled using the \MGNLO[2.3.3] generator at NLO with the \NNPDF[3.1nlo] PDF set.
The events were interfaced with \PYTHIA[8.230].
Bottom- and charm-hadron decays were simulated with \EVTGEN[1.6.0].

Finally, for the SMEFT interpretation, additional MC samples were produced at LO in QCD for the \ttZ and \tZq processes, using the \MADGRAPH[2.9.3] generator and \PYTHIA[8.245] parton shower with the default A14 tune settings.
They rely on the SMEFTsim\,3.0~\cite{Brivio:2020onw} UFO model~\cite{Degrande:2011ua} implemented in \MADGRAPH with FeynRules~\cite{Alloul:2013bka}, in the $m_{\PW}$ electroweak input scheme~\cite{Brivio:2021yjb} with the top-flavour restrictions in the five-flavour scheme. The renormalisation scale is fixed to $\mu=\sum_i m_i$, where $i$ runs over the massive final-state resonances.
Nominal \ttZ and \tZq events were generated according to the SM, and the \MADGRAPH reweighting module was used to compute a large number of alternative event weights
corresponding to the inclusion of dimension-6 EFT vertices and propagators in the Feynman diagrams with values for the Wilson Coefficients in the range of previously best limits. The EFT insertions are restricted to the production process and no EFT effects on the decay amplitudes and the top-quark width are considered.
The EFT samples were passed through the \textsc{AtlFast~II} detector simulation.


%

%

%
\section{Object reconstruction and selection}%
\label{sec:obj}
The primary vertex~\cite{IDTR-2021-01} is selected as the \pp vertex candidate with the highest sum of the squared transverse momenta of all associated tracks with $\pT > \SI{500}{\MeV}$.

Electron candidates are reconstructed from energy clusters in the electromagnetic calorimeter matched to inner-detector tracks~\cite{EGAM-2021-01,EGAM-2021-02}.
They are required to satisfy $\abseta < 2.47$, excluding the calorimeter transition region $1.37 < \abseta < 1.52$, and to have a transverse energy $\ET > \SI{7}{\GeV}$.
The track associated with the electron must satisfy the requirements $|z_0\sin(\theta)| < \SI{0.5}{mm}$ and $|d_{0}| / \sigma(d_{0}) < 5$, where $z_{0}$ describes the longitudinal impact parameter relative to the reconstructed primary vertex, \dzero is the transverse impact parameter relative to the beam axis, and $\sigma(d_{0})$ is the uncertainty in \dzero.
A likelihood-based identification discriminant, built from variables sensitive to shower shapes and tracking information, is applied using an $\eta$- and \ET-dependent working point with an efficiency of about \qty{88}{\%} for electrons from \PZ-boson decays. Electrons are further required to satisfy isolation criteria based on a multivariate discriminant.
For the estimation of backgrounds with non-prompt electrons, a looser identification working point with an efficiency of about \qty{93}{\%} is used, without imposing isolation requirements.
Correction factors are applied to simulated electrons to account for residual differences in trigger, reconstruction, identification, and isolation efficiencies between data and simulation.

Muon candidates are reconstructed by combining tracks from the inner detector and the muon spectrometer~\cite{MUON-2018-03,MUON-2022-01}.
They are required to satisfy $\pt > \SI{7}{\GeV}$ and $\abseta < 2.5$.
In addition, the track associated with the muon candidate must have $|z_0\sin(\theta)| < \SI{0.5}{mm}$ and $|d_{0}| / \sigma(d_{0}) < 3$.
Quality requirements, based on tracking and muon spectrometer information, are applied to suppress misidentified muons originating primarily from pion and kaon decays. An isolation requirement based on a multivariate discriminant is also imposed, resulting in an efficiency of about \qty{96}{\%} for muons from \PW-boson decays in simulated \ttbar events.
For the estimation of backgrounds with non-prompt muons, the same identification criteria are used without imposing isolation requirements.
Correction factors are applied to simulated muons to account for residual differences in trigger, reconstruction, identification, and isolation efficiencies between data and simulation.

Jets are reconstructed from particle flow objects~\cite{PERF-2015-09} using the \antikt algorithm~\cite{Cacciari:2008gp} with the radius parameter $R=0.4$.
These jets are then calibrated to the particle level by the application of a jet energy scale derived from simulation and in situ corrections. Jets are required to have $\pT > \SI{25}{\GeV}$ and $\abseta < 4.5$.
To suppress jets originating from \pileup, the jet vertex tagger (JVT), based on a two-dimensional likelihood discriminant~\cite{JETM-2018-05}, is applied to central jets with $\pT < \SI{60}{\GeV}$ and $\abseta < 2.4$, using a working point with about \qty{92}{\%} efficiency for hard-scatter jets and rejecting approximately \qty{98}{\%} of \pileup and noise jets. For forward jets ($\abseta > 2.5$), the forward jet vertex tagger (fJVT)~\cite{PERF-2016-06} is required, achieving up to \qty{97}{\%} efficiency for hard-scatter jets and a \pileup-jet efficiency of about \qty{53}{\%} at $\pT = \SI{40}{\GeV}$.

To identify jets containing a \Pqb-hadron (\bjets), a multivariate algorithm based on the \texttt{DL1r} discriminant is employed~\cite{FTAG-2019-07,ATL-PHYS-PUB-2017-013}. This algorithm combines information from track impact parameters, reconstructed secondary vertices, and decay topologies of \Pqb- and \Pqc-hadrons within the jet. Due to its reliance on inner detector tracking, \Pqb-tagging is restricted to jets within $\abseta < 2.5$.
A working point corresponding to an efficiency of \qty{85}{\%} for \bjets in simulated \ttbar events is used for the preselection.
At this working point, the misidentification rates are about \qty{30}{\%} for \Pqc-jets and \qty{4}{\%} for light-flavour jets, depending on the jet \pT.
In addition, exclusive intervals of the discriminant corresponding to different efficiency working points are defined and used as input to the multivariate analysis presented in \cref{sec:mva}.
Correction factors are applied to simulated events to account for residual differences in \Pqb-tagging efficiencies and mistag rates between data and simulation~\cite{FTAG-2019-07,PERF-2016-05,ATLAS-CONF-2018-006}.

The missing transverse momentum, with magnitude \MET, is defined as the negative vector sum of the transverse momenta of all selected and calibrated physics objects.
An additional soft term, constructed from tracks associated with the primary vertex but not matched to these objects, is included to account for soft hadronic activity~\cite{PERF-2016-07}.

Ambiguities arising from the independent reconstruction of electrons, muons, and jets are resolved using a sequential overlap removal procedure to avoid double counting of physics objects.
If an electron and a muon share the same track, the electron is removed. Jets within $\DeltaR < 0.2$ of an electron are discarded, while electrons are removed if $0.2 < \DeltaR < 0.4$ to any remaining jet.
For muons, if a jet lies within $\DeltaR < 0.4$, the muon is removed if the jet has more than two associated tracks; otherwise, the jet is discarded.

\paragraph{Parton- and particle-level objects}%
For the differential measurements of the \tZq cross-section, two definitions of generator-level objects are considered: parton level and particle level. The parton-level definition corresponds to a phase space defined by an on-shell \PZ-boson requirement, facilitating comparisons with fixed-order matrix-element calculations, while the particle-level definition is used to define a fiducial phase space closely matching the detector-level selection.

Parton-level objects are constructed from the MC generator history of the \tZq system. The top (anti)quarks and \PZ bosons are identified as the last instances of these particles in the \enquote{truth} record after parton showering, but prior to their decay into \tWb and \Zll, respectively. Leptons originating from the \PW and \PZ bosons are defined as the first instances following the decay of the corresponding parent bosons. The spectator quark is defined as the light-flavour quark produced in association with the top quark through the $t$-channel exchange. It is identified using the MC truth record as a quark that does not originate from the top-quark or \PZ-boson decay.

Particle-level objects are defined using stable particles in the MC simulation, with a lifetime of $\tau \ge \qty{30}{\pico\second}$, and without including any detector simulation or additional \pp interactions. In contrast to the parton-level definition, hadronisation effects are included.
Particle-level leptons are defined as electrons or muons originating from the decay of a \PW or \PZ boson. Their four-momenta are dressed by adding the four-momenta of all photons within a cone of size $\DeltaR = 0.1$ around the lepton direction, excluding photons from hadron decays. The parent particles of the selected leptons are required not to be hadrons or light/heavy quarks (\Pqu, \Pqd, \Pqs, \Pqc, \Pqb).
Particle-level jets are reconstructed using the \antikt algorithm with a radius parameter of $R = 0.4$, using all stable particles except for the selected leptons, their associated photons, and neutrinos originating from \PW- and \PZ-boson decays. A jet is identified as a $b$-jet if it is ghost-associated~\cite{Cacciari:2008gn,FTAG-2018-01} with a \Pqb-hadron with $\pt > \qty{5}{\GeV}$.
To maintain consistency with the reconstruction-level definition, the particle-level spectator jet is defined as the non-\Pqb-identified jet that yields the largest invariant mass when combined with the selected particle-level \Pqb-identified jet, as described in~\cref{sec:mva}.
The particle-level missing transverse momentum is defined as the vector sum of the transverse momenta of all neutrinos in the event record, excluding those originating from hadron decays.


%

%

%
\section{Event selection and background estimation}%
\label{sec:sel}

The \tZq signal targeted in this measurement is characterised by the presence of three charged leptons (electrons or muons), missing transverse momentum, one \bjet from the top-quark decay and an additional jet that does not satisfy the \btag requirement (untagged jet). This additional jet targets the spectator-quark topology of the $t$-channel process, for which the corresponding light-flavour quark is expected to be emitted preferentially at high $\abseta$.
To help separate the \tZq signal from the backgrounds that do not contain a \PZ boson and a top quark (diboson, \Zjets and \ttbar events), both the \PZ-boson and the top-quark invariant masses are reconstructed.
The top-quark candidate is formed by combining the reconstructed leptonically decaying \PW-boson candidate with the highest-\pT \btagged jet.

The data are divided into several non-overlapping regions: one signal region (SR) designed to select \tZq events and six control regions (CRs) designed to enhance the selection of the main sources of background events.
The CRs are used to adjust the normalisation and reduce the associated systematic uncertainties in the main backgrounds, using constraints from the data.
The main source of background is from \ttZ events, which feature a higher jet multiplicity than the signal.
Another important source of background comes from diboson events, mainly \WZ, produced in association with jets. The diboson background is split into \VVBF, \VVCF and \VVLF based on the flavour of the additional jets, using generator-level information: if one of the associated jets is identified as originating from a \Pqb-quark, using particle-level information, then it is considered as \VVBF, otherwise if one of the associated jets is identified as originating from a \Pqc-quark then it is considered \VVCF, otherwise it is considered as \VVLF. %
The last source of background comes from events with non-prompt or \enquote{fake} leptons, which originate from several sources, including photon conversions, kaon or pion decays, and light jets accidentally creating lepton-like detector signatures. The predominant source of non-prompt leptons is the semileptonic decay of heavy-flavour (HF) hadrons. Events from \ttbar and \Zjets productions enter the selection because of the presence of such non-prompt leptons. In the following, fake and non-prompt leptons are referred to as non-prompt leptons, abbreviated as NPL.

The SR requires exactly three leptons, as defined in~\cref{sec:obj}. The three leptons are sorted by their \pT, irrespective of flavour, and required to have transverse momenta of at least 27, 15 and \qty{10}{\GeV}, respectively.   At least one of the leptons needs to match, with $\Delta R < 0.15$, the lepton reconstructed by the trigger.
At least one opposite-sign, same-flavour (\OSSF) lepton pair is required, to reconstruct the \PZ boson.
In the \mee and \emm channels the pair is uniquely identified, whereas in the \eee and \mmm channels both of the possible combinations are considered.
Every OSSF lepton pair must have a minimum $m_{\ell\ell}$ of \qty{20}{\GeV} and the pair with the invariant mass closer to the \PZ-boson mass is required to have $|\mll-\mZ|<\SI{10}{\GeV}$, where $\mZ = \qty{91.2}{\GeV}$~\cite{Zyla:2020zbs}. This selection suppresses backgrounds that do not contain a \PZ boson.
The remaining lepton and the \MET are used to reconstruct the leptonically decaying \PW boson.\footnote{%
The longitudinal component of the neutrino four-momentum is obtained by using the mass constraint of the \PW boson. The twofold ambiguity is resolved by choosing the solution with the smaller $|p_z^{\nu}|$, since the \PW boson is expected to be produced with small pseudorapidity~\cite{TOPQ-2011-18}.}
The transverse mass of the \PW boson is defined as
\begin{equation}
\mtw =
\sqrt{
2\,\pT(\ell_{\PW})\,\MET
\left[1-\cos\Delta\phi\bigl(\ell_{\PW},\vec{p}_{\mathrm{T}}^{\,\mathrm{miss}}\bigr)\right]
},
\end{equation}
where $\ell_{\PW}$ denotes the lepton not assigned to the \PZ-boson candidate.
A requirement of $\mtw > \qty{30}{\GeV}$ is applied to suppress events from processes with non-prompt leptons.
Events are required to contain between two and four jets, of which one or two are required to be \btagged.
Events with exactly two jets that are both \btagged, are not considered.

To control the two main background sources, \ttZ and \VVjets events, two CRs are defined. These two CRs are orthogonal to the SR and are used in the fit to data, as described in \cref{sec:inclusive}.
The \ttZ CR, \enquote{\CRttZ{}} in the following, is defined using the same selection as the SR but requiring at least five jets, of which at least one must be \btagged, since trileptonic \ttZ events generally have a higher jet multiplicity.
The \VVjets CR, \enquote{\CRVV{}} in the following, differs from the SR by vetoing events with at least one jet passing the \qty{85}{\%} \Pqb-tagging working point. The \CRVV region is dominated by the \VVLF component, although it also contains a non-negligible \VVCF contribution.

The contribution of non-prompt-lepton backgrounds in the signal region is estimated with the semi-data-driven \enquote{template fit method}~\cite{TOPQ-2020-20}.
This method relies on using the data to normalise dedicated MC simulation NPL templates built from MC truth-record information about the origin and flavour of the non-prompt leptons.
Four distinct MC simulation templates are defined for the major sources of NPL in this analysis:
electrons from heavy-flavour sources (\enquote{\NPLelHF{}});
electrons from other sources (\enquote{\NPLelOther{}});
muons from heavy-flavour sources (\enquote{\NPLmuHF{}});
and any other non-prompt leptons that do not belong to the other sets or that come from events containing multiple non-prompt leptons (\enquote{\NPLOther{}}).
These MC templates are then normalised to the data in dedicated control regions,
except for the last category, which is subdominant and is instead assigned a \qty{50}{\%} normalisation uncertainty.

The NPL control regions are designed to be kinematically close to the SR by using the same lepton \pT and jet multiplicity requirements.
To ensure orthogonality with the SR, exactly one lepton is required to fail the identification and isolation requirements applied to the signal leptons, referred to as the \enquote{loose} lepton in the following.
The heavy-flavour non-prompt components can be isolated in \ttbar-enriched control regions that select events with similar (\Pqb-)jet multiplicities to the SR but with $\lvert m_{\ell\ell} - m_Z \rvert \geq 10\ \si{\giga\electronvolt}$, where $m_{\ell\ell}$ is the OSOF pair with invariant mass closer to $m_Z$. This contribution is then split by the type of the loose lepton, creating two control regions: \enquote{\CRelHF{}} for the electron HF contribution and \enquote{\CRmuHF{}} for the muon HF contribution.
To target non-prompt-lepton contributions in \Zjets events, events with zero \bjets and two OSSF leptons within the \PZ-boson mass window are selected. The regions are split according to the flavour of the loose lepton, defining \enquote{\CRelOther{}} and \enquote{\CRmuOther{}}, respectively.
The control-region labels refer to the targeted process topology and the flavour of the loose lepton, and do not imply a one-to-one correspondence with the NPL template categories. In particular, no dedicated NPL-$\mu$-Other template is defined: events in \CRmuOther{} containing a non-prompt muon from a heavy-flavour decay are described by the \NPLmuHF{} template, while the remaining non-prompt-lepton contributions are included in the \NPLOther{} template.

\begin{figure}[!tbp]
\centering
\begin{tabular}{cc}
\begin{subfigure}[b]{.48\textwidth}
\includegraphics[width=\textwidth]{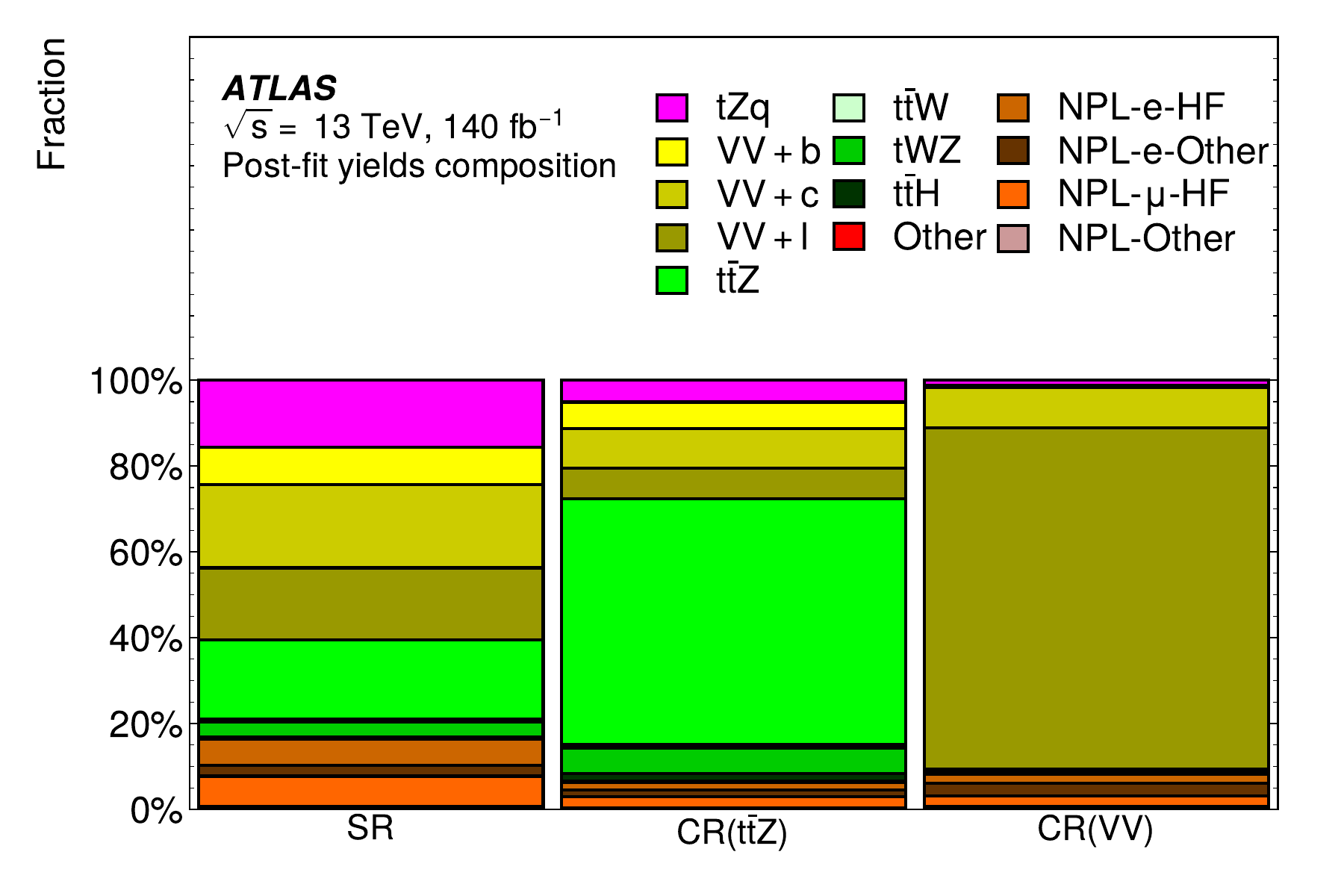}
\caption{}
\end{subfigure}
&
\begin{subfigure}[b]{.48\textwidth}
\includegraphics[width=\textwidth]{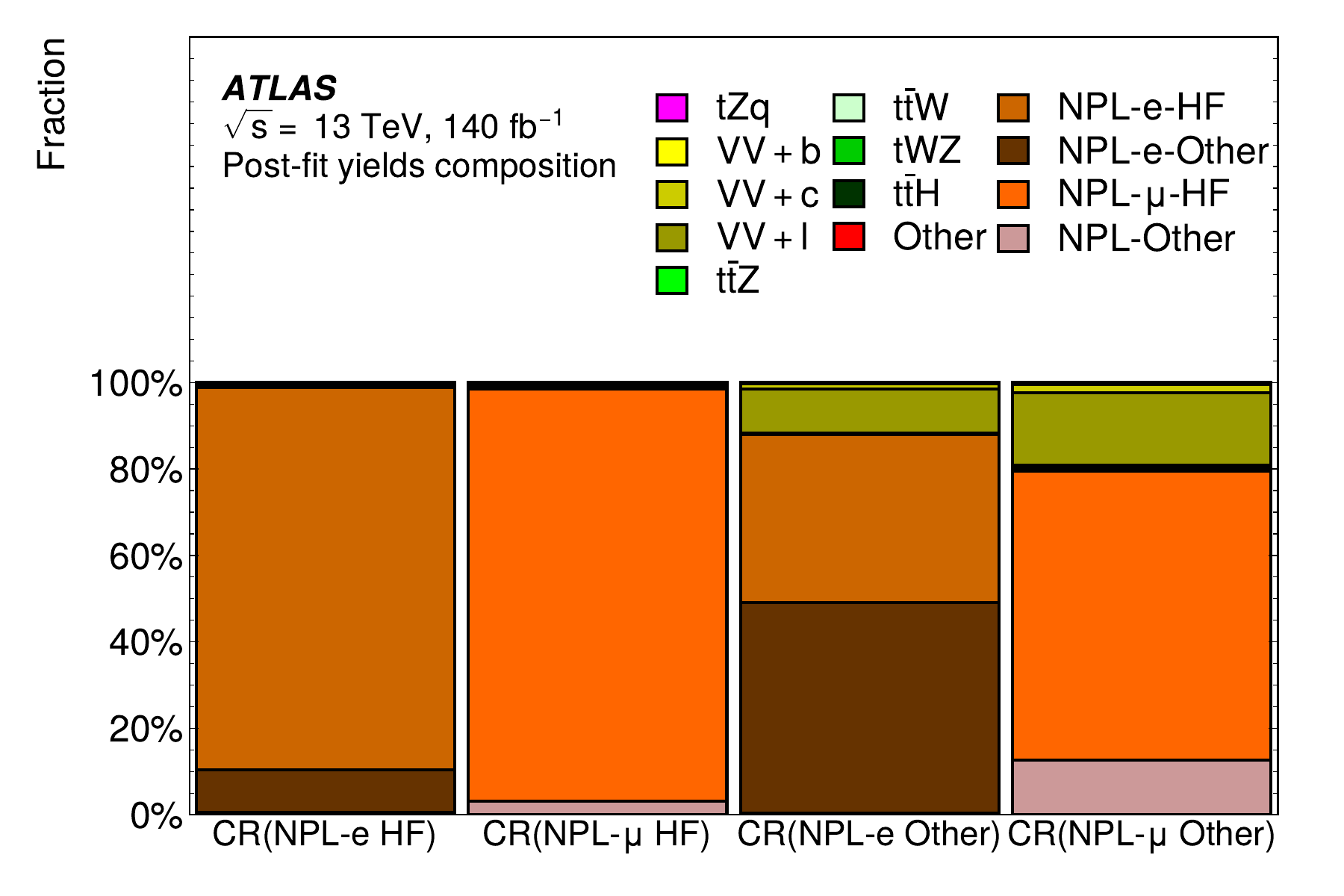}
\caption{}
\end{subfigure}
\end{tabular}
\caption{
Yield composition after the fit to data (\enquote{Post-fit}) in
(a) the SR, \CRttZ and \CRVV, and
(b) the non-prompt-lepton (NPL) CRs.
}
\label{fig:sel:comp}
\end{figure}

\Cref{tab:fullcutsummary} summarises the selection criteria applied in the SR and all CRs, while \cref{fig:sel:comp} shows the composition of the SR and the CRs obtained after the likelihood fit to data described in \cref{sec:inclusive}.

\begin{table}[!tbp]
\tablesetup[1.3]
\caption{Overview of the requirements applied at (a) preselection level and when selecting events in (b) the signal region (SR) and the control regions \CRttZ and \CRVV, and (c) the non-prompt-lepton (NPL) control regions.
OSSF denotes an opposite-sign same-flavour lepton pair.
The $\min(m_{\ell\ell})$ cut applies to any OSSF lepton pair while the $|m_{\ell\ell} - \mZ|$ cut applies to at least one OSSF lepton pair.
The term \enquote{loose lepton} denotes a lepton that does not satisfy the identification and isolation requirements applied to signal leptons.}
\label{tab:fullcutsummary}
\subcaptionbox{Preselection\label{tab:fullcutsummary:presel}}{%
\begin{tabular}{l l}
\toprule
Variable & \mcc{Preselection} \\
\midrule
$N_\ell~(\ell=e,\mu)$ & $=3$ \\
& $\ge 1$ OSSF lepton pair \\
$\pt(\ell_1,\ell_2,\ell_3)$ & $> 27, 15, \qty{10}{\GeV}$ \\
$\min(m_{\ell\ell})$  & $> \qty{20}{\GeV}$ \\
\mtw                  & $> \qty{30}{\GeV}$ \\
\bottomrule
\end{tabular}
}\\[1em]
\subcaptionbox{SR, \CRttZ and \CRVV definitions\label{tab:fullcutsummary:SRCR1}}{%
\begin{tabular}{l lll}
\toprule
Variable & SR & \CRttZ & \CRVV \\
\midrule
$N_{\ell}^{\text{loose}}~(\ell=e,\mu)$  & -- & -- & -- \\
Flavour of loose lepton    & -- & -- & -- \\
$|m_{\ell\ell} - \mZ|$                   & $\qty{10}{\GeV}$ & $\qty{10}{\GeV}$  & $\qty{10}{\GeV}$ \\
$N_{\text{jets}} (\pt > \qty{25}{\GeV})$ & 2--4             & $\ge5$ & 2--4  \\
$N_{\bjets}$ @ 85\%                      & 1--2 (no $2j2b$) & $\ge1$ & 0     \\
\bottomrule
\end{tabular}
}\\[1em]
\subcaptionbox{NPL CR definitions\label{tab:fullcutsummary:SRCR2}}{%
\begin{tabular}{l llll}
\toprule
Variable & \CRelHF & \CRmuHF & \CRelOther & \CRmuOther \\
\midrule
$N_{\ell}^{\text{loose}}~(\ell=e,\mu)$  & $=1$ & $=1$ & $=1$ & $=1$ \\
Flavour of loose lepton    & Electron & Muon & Electron & Muon \\
$|m_{\ell\ell} - \mZ|$                   & $> \qty{10}{\GeV}$ & $> \qty{10}{\GeV}$ & $< \qty{10}{\GeV}$ & $< \qty{10}{\GeV}$ \\
$N_{\text{jets}} (\pt > \qty{25}{\GeV})$ & 2--4 & 2--4 & 2--4 & 2--4 \\
$N_{\bjets}$ @ 85\%                      & 1--2 & 1--2 & 0 & 0 \\
\bottomrule
\end{tabular}
}
\end{table}

\paragraph{Parton- and particle-level phase-space definitions}%
\label{sec:sel:fid}
The parton-level fiducial volume is defined as follows.
The \PZ boson is required to decay into opposite-sign same-flavour lepton pairs of \Pe or \Pgm ($Z\rightarrow e^+e^-,\ \mu^+\mu^-$) and the top quark decaying leptonically ($\ell\nu b$, $\ell=e,\mu$). Events where a \PZ boson is not present are considered if they contain an OSSF lepton pair with invariant mass $\lvert m_{\ell\ell} - m_Z \rvert \leq 15\ \si{\giga\electronvolt}$.
Events containing \Pgt leptons coming from either \PZ-boson decay or \PW-boson decay from the top quark, are removed from the parton-level fiducial volume, irrespective of their successive decay.
The differential variables are reconstructed from the top quark after final-state radiation, immediately before its decays.
The invariant mass of the lepton pair should be within $\pm \qty{15}{\GeV}$ of the mass of the \PZ boson, $\mZ$.
This requirement defines an on-shell \PZ-boson phase space suitable for comparison with theoretical predictions. It is therefore not intended to reproduce exactly the narrower detector-level mass window, which is chosen to suppress backgrounds in the event selection.

The particle-level fiducial region is constructed to closely follow the detector-level regions, using the particle-level objects defined in \cref{sec:obj}, and the selections are summarised in \cref{tab:sel:fid}.
The requirements on the minimum dilepton invariant mass and the transverse mass of the \PW boson are each relaxed by \qty{10}{\GeV} relative to the detector-level selection. This defines a slightly broader particle-level fiducial region and increases its overlap with the detector-level selected event sample.

\begin{table}[htbp]
\tablesetup[1.3]
\centering
\caption{Definition of the fiducial volume at particle level. Leptons refer exclusively to electrons and muons; they are dressed with additional photons at particle level.}
\label{tab:sel:fid}
\begin{tabular}{l l}
\toprule
Variable & Selection \\
\midrule
$N_{\Pl}(e,\mu)$ & 3, $\geq 1$ OSSF lepton pair \\
$p_T(\Pl_1,\Pl_2,\Pl_3)$ & $\geq (27,15,10)$\,\unit{\GeV} \\
$|\eta_e|$ &  $<2.47$, $\notin [1.37, 1.52]$ \\
$|\eta_\mu|$ & $<2.5$ \\
$|\mll-\mZ|$ & $< \qty{15}{\GeV}$ \\
min($\mll$) & $> \qty{10}{\GeV}$ \\
\mtw & $> \qty{20}{\GeV}$ \\
$\pT (\text{jets})$ & $> \qty{25}{\GeV}$ \\
$|\eta (\text{jets})|$ & $<4.5$ \\
$N_{\text{jets}}$ & $2-4$\\
$|\eta (\bjets)|$ & $<2.5$ \\
$N_{\bjets}$ & $\geq 1$ (ghost-matched) \\
\bottomrule
\end{tabular}
\end{table}

%

%

%
%
%
%
%
%

%
%
%
%
%
%
%
%
%
%
%

%
%
%
%
%
%
%
%

%
%
%
%
%
%
%
%
%

%
%
%
%
%

%
%
%
%
%

%
%

%
%
%
%
%

%
%

%

%
%
%
%

%
%
%
%
%
%
%
%
%
%
%
%
%
%
%
%
%
%
%
%
%
%
%
%
%

%
%
%
%
%
%
%
%

%
%
%
%
%
%
%
%
%

%
%
%
%
%
%

%
%
%
%

%

%
%
%
%
%
%
%


%
\FloatBarrier

%

%
\section{Multivariate analysis}
\label{sec:mva}

To improve the separation between signal and background, a NN is used to derive a discriminant that takes correlations among the input variables into account. The framework is based on the \emph{Keras}~\cite{chollet2015keras} functional API within Tensorflow~\cite{tensorflow2015-whitepaper}. A binary NN is used, with signal events assigned to one class and all background processes assigned to the other class. The NN output discriminant, denoted \ONN, is defined as the score assigned by the network to the signal class, with larger values corresponding to more signal-like events. All background processes are considered in the training and are weighted according to their expected event yields in the SR.

The NN consists of three hidden layers with 80 nodes each, using exponential linear unit (ELU) activation functions. Batch normalisation and dropout with a rate of 0.2 are applied. The network is trained using the Nadam optimiser with a learning rate of 0.0005 and a batch size of 5000 events.
The network architecture and training hyperparameters are selected using a standard grid-search optimisation, in which approximately 3000 hyperparameter configurations are evaluated.
A three-fold $k$-fold cross-validation scheme is employed during the training.

Only variables that provide good separation and are well modelled are used in the NN\@.
A total of 19 variables, listed in \cref{tab:mva:vars}, are used for the training.
These are selected through an optimisation procedure based on permutation importance~\cite{Altmann:2010}, in which the importance of each input variable is evaluated from the reduction in the network performance when its values are randomly permuted. Variables providing no significant contribution to the discrimination between signal and background are removed, and the network is later retrained using the retained variables.
Variables related to the reconstructed \PZ boson help to reduce the \ttbar background, while top-quark-related quantities are useful in separating the signal from processes such as \WZ and \Zjets, in which no top quark is produced.

\begin{table}[htbp]
\tablesetup[1.3]
\caption{Variables used as input to the neural network.
Angular distances are calculated using \(\eta\) instead of \(y\). \pTZ and $\abs{\eta(\sjet)}$ are the two most important variables.
}%
\label{tab:mva:vars}
\centering
\begin{tabular}{l p{12cm}}
\toprule
Variable & Definition \\
\midrule
$N_{\text{central-jets}}$ & Number of central jets (with \abseta < 2.5) \\
$N_{\text{forward-jets}}$ & Number of forward jets (with \abseta > 2.5) \\
$N_{\bjets}$ & Number of \bjets \\
\texttt{b1-b2-btagbin} & Variable related to the $b$-tagging score, see text \\
$\abs{\eta(\sjet)}$ & Absolute value of pseudorapidity for the spectator jet \\
$\pT(\sjet)$ & Transverse momentum of the spectator jet \\
$m(b,\sjet)$ & Invariant mass of the highest-\pT $b$-jet+spectator-jet system \\
$\pT(\ell^Z_1)$ & Transverse momentum of the highest-\pT lepton coming from the \PZ boson \\
$\pT(\ell^Z_2)$ & Transverse momentum of the lowest-\pT lepton coming from the \PZ boson \\
\pTZ & Transverse momentum of the reconstructed \PZ boson \\
\mll & Invariant mass of the dilepton system coming from the \PZ boson\\
min(\mll) & Minimum invariant mass between all possible OSSF lepton pairs \\
\mtop & Reconstructed mass of the top quark \\
$Q(\ell^W)\cdot \pT(\ell^W)$ & Product of the charge and \pT of the lepton coming from \PW boson \\
$\abs{\eta(\ell^W)}$ & Absolute value of pseudorapidity of $\ell^W$ \\
$H_\mathrm{T}$ & Scalar sum of transverse momenta of all leptons and jets \\
$\Delta R(\sjet,Z)$ & Angular distance between the spectator jet and reconstructed \PZ boson \\
$\Delta R(\ell^Z_2,b)$ & Angular distance between second highest-\pT lepton associated with the \PZ boson and highest-\pT $b$-jet \\
$\Delta \phi(W,j)$ & Azimuthal angular distance between the reconstructed \PW boson and closest jet \\
\bottomrule
\end{tabular}
\end{table}

The variable $N_{\text{central-jets}}$ denotes the total number of central jets ($\abseta<2.5$), including both \Pqb-tagged and non-\Pqb-tagged jets, while $N_{\text{forward-jets}}$ denotes the number of forward jets ($\abseta>2.5$), for which no \Pqb-tagging information is available. For the signal, the jet that comes from the spectator quark in the hard-scattering process tends to have higher $\abseta$. Information about this jet, called the \enquote{spectator jet} (\sjet), helps the NN provide separation from diboson and \ttZ background events that tend to produce more central jets. The \sjet is non-\Pqb-tagged and when more than one non-\Pqb-tagged jet is in the event, \sjet is the one for which the invariant mass of this non-\Pqb-tagged jet and the highest-\pT \Pqb-tagged jet is the largest.
The output of the \Pqb-tagging discriminant for each \Pqb-tagged jet is grouped into five ordered categories corresponding to exclusive intervals between calibrated \Pqb-tagging working points.
These categories range from the lowest b-tagging purity interval with a score of 1, corresponding to jets passing the 85\% but failing the 77\% efficiency working points, to the highest-purity interval with a score of 5, corresponding to the 60\% efficiency working point.
The resulting per-jet score is denoted by \texttt{bjet-btagbin}.
These per-jet scores are combined into the event-level variable \texttt{b1-b2-btagbin} that incorporates the b-tagging information from both jets. In events containing a single \Pqb-tagged jet, \texttt{b1-b2-btagbin} is equal to the \texttt{bjet-btagbin} score of that jet. In events containing two \Pqb-tagged jets, the jets are ordered in transverse momentum, with $b1$ denoting the leading-\pT jet and $b2$ the subleading-\pT jet.%
\begin{equation}
\texttt{b1-b2-btagbin} =
5\left(\texttt{bjet-btagbin-b2}-1\right)
+\left(\texttt{bjet-btagbin-b1}-\texttt{bjet-btagbin-b2}+1\right).
\label{eq:b1_b2_btagbin}
\end{equation}
This construction incorporates the \Pqb-tagging information from both jets into a single event-level category. Including the score of the second \Pqb-tagged jet improves the discrimination between processes. Background contributions from non-prompt leptons and diboson production predominantly populate the lower-score categories, whereas \ttZ events are more frequently found in the higher-score categories.

\begin{figure}
\centering
\begin{tabular}{cc}
\subcaptionbox{}{%
\includegraphics[width=0.4\textwidth]{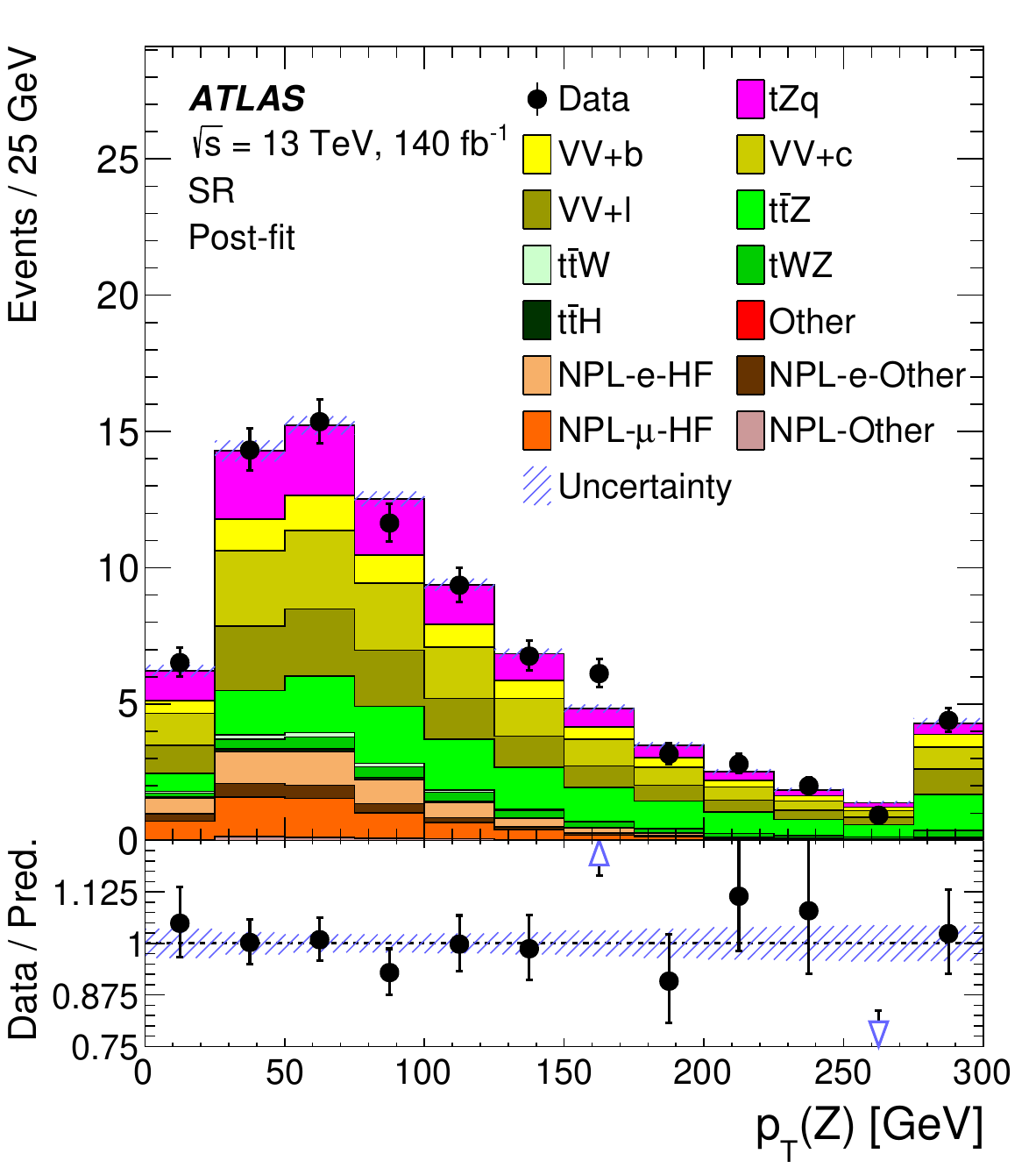}
}
\subcaptionbox{}{%
\includegraphics[width=0.4\textwidth]{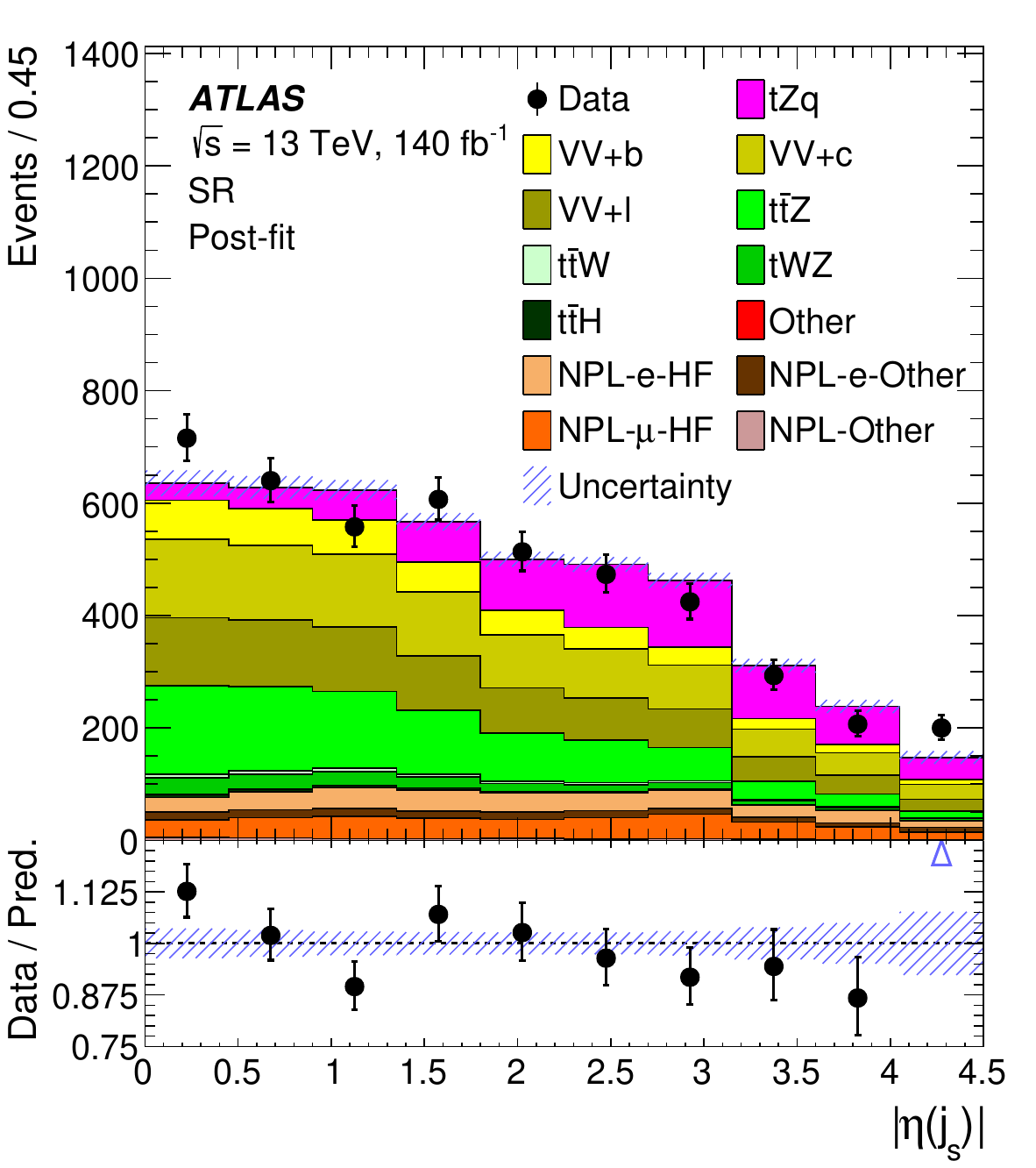}
}\\
\subcaptionbox{}{%
\includegraphics[width=0.4\textwidth]{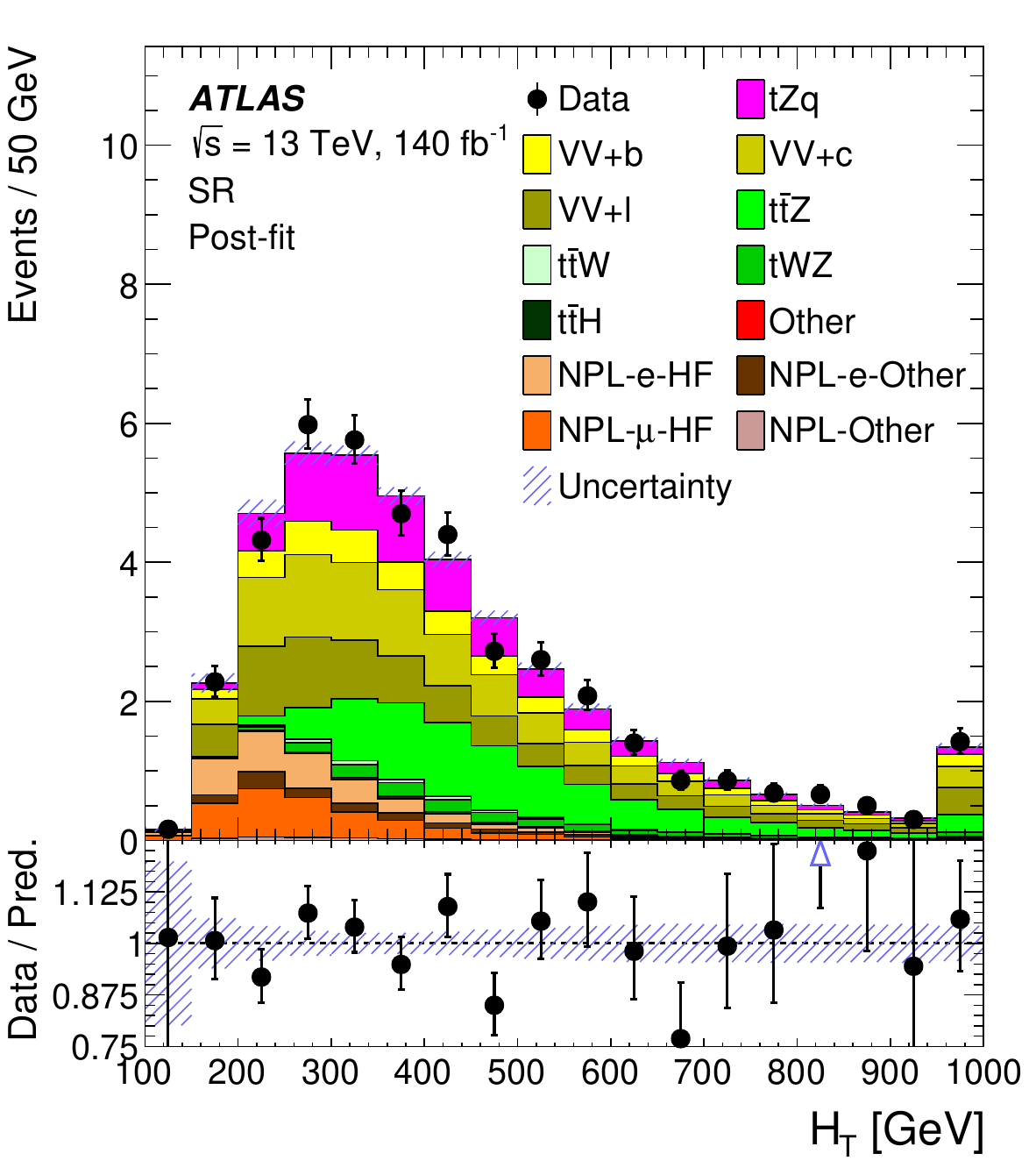}
}
\subcaptionbox{}{%
\includegraphics[width=0.4\textwidth]{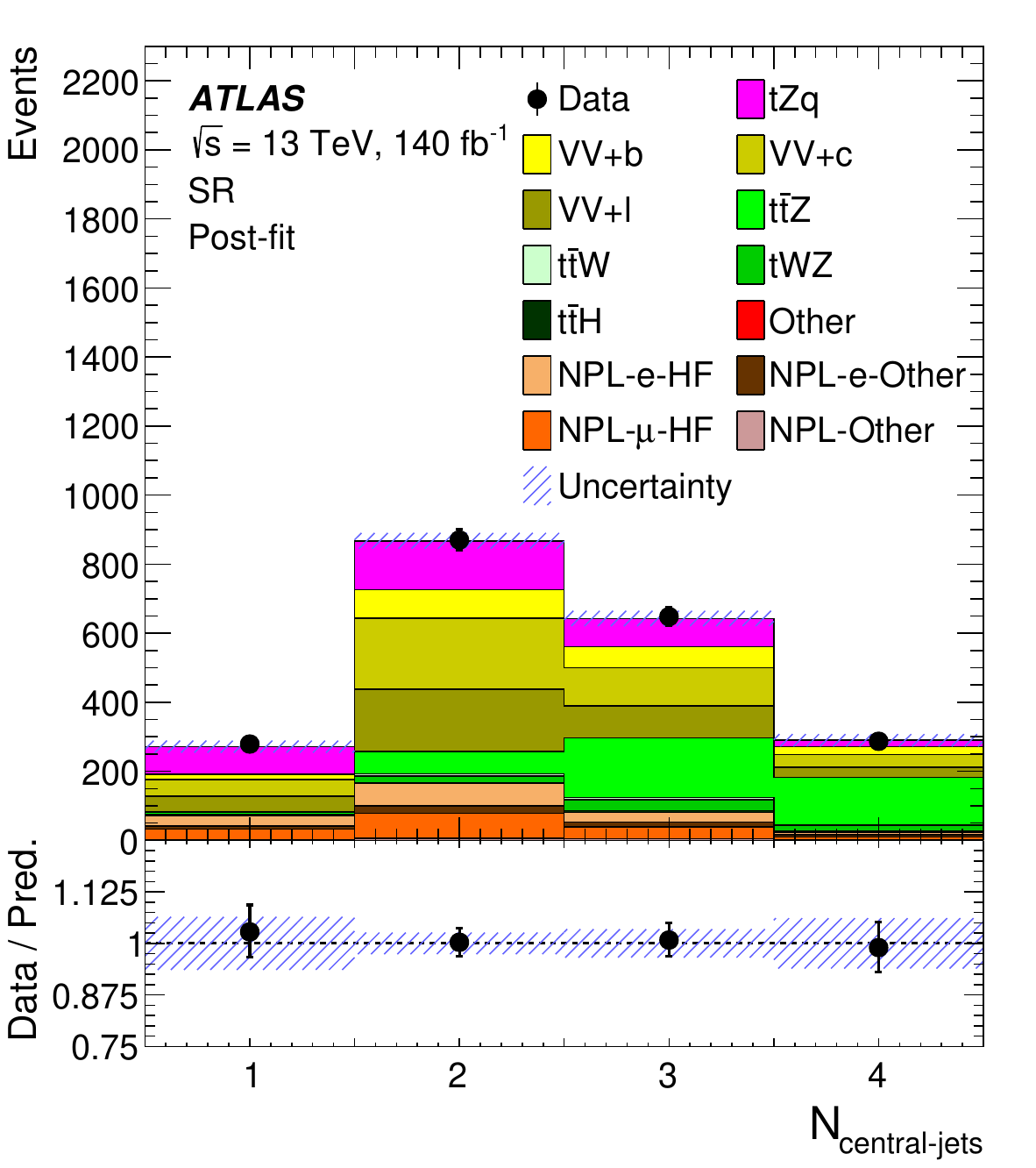}
}
\end{tabular}
\caption{Comparison between data and prediction (\enquote{Pred.}) for four NN input variables after the fit to data (\enquote{Post-fit}). %
The variables displayed are:
(a) the transverse momentum of the reconstructed \PZ boson,
(b) the absolute value of pseudorapidity for the spectator jet,
(c) the scalar sum of transverse momenta of all leptons and jets, and
(d) the number of central jets (with \(\abseta < 2.5\)).
The uncertainty band includes both the statistical and systematic uncertainties as obtained by the fit.
Where applicable, overflow events are included in the rightmost bin.
The lower panels show the ratios of the data to the prediction.
The open triangles indicate points that are outside the vertical range of the figure.
}
\label{fig:mva:vars}
\end{figure}

The post-fit distributions of four of the input variables amongst the ones with the highest discrimination power, obtained after the likelihood fit to data described in \cref{sec:inclusive}, are shown in \cref{fig:mva:vars}.
Good agreement between data and prediction is observed.

%

%

%
\section{Systematic uncertainties}
\label{sec:sys}

The signal and background predictions are affected by several sources of experimental and theoretical systematic uncertainties. These are treated correlated between the different regions.
The uncertainties are considered for both the inclusive and differential measurements presented in \cref{sec:inclusive,sec:differential}.
The uncertainties can be classified into the different categories described in the following.
\paragraph{Detector-related uncertainties}
\label{sec:sys:detector}

The integrated luminosity corresponding to the 2015--2018 period of data taking has an uncertainty of \qty{0.83}{\%} \cite{DAPR-2021-01}. This uncertainty affects all processes that are determined from MC simulations for which the normalisation is not extracted from data.

The uncertainty in \pileup modelling is evaluated by varying the \pileup reweighting of the simulated samples to cover the uncertainty in the ratio of the predicted and measured inelastic $pp$ cross-sections.

The uncertainties related to the lepton selections arise from the trigger, reconstruction, identification, and isolation efficiencies of electrons and muons, and from the calibration of their measured energies and momenta. For electrons, uncertainties associated with the energy scale and resolution are considered, while for muons uncertainties related to the momentum scale and resolution are applied~\cite{EGAM-2021-01,EGAM-2021-02,MUON-2018-03,MUON-2022-01}.

The uncertainties related to the jet selection arise from Jet Energy Scale~(JES), Jet Energy Resolution~(JER), JVT and fJVT selections.
The JES uncertainties are derived using information from test-beam data, collision data and MC simulations~\cite{JETM-2018-05}.
The JER is measured separately for data and MC, and the uncertainty is quoted as the quadratic difference between the resolutions for data and MC.

The uncertainty in \MET due to a possible miscalibration of the soft-track component of the \MET is derived from data--MC comparisons of the \pT balance between the hard and soft \MET components~\cite{PERF-2016-07}.

The efficiency of the \texttt{DL1r} \Pqb-tagging algorithm is calibrated using \ttbar and \Zjets data events, and the efficiencies are corrected by $\pT$ and $\eta$-dependent factors for the jets~\cite{FTAG-2019-07,FTAG-2018-01}. The uncertainties are estimated as variations in the tagging efficiencies.

\paragraph{Signal modelling uncertainties}
\label{sec:sys:signal}

The uncertainty on the signal related to missing higher-order corrections is estimated by variations of the renormalisation and factorisation scales from the nominal value of $\mu_R = \mu_F = H_\mathrm{T}/6$ by factors of 0.5 and 2.
The uncertainties related to the choice of PDF set are evaluated by following the PDF4LHC prescriptions~\cite{Ball:2022oua,Butterworth:2015oua}.
The uncertainty associated with the modelling of the parton shower in the \tZq process is evaluated by comparing the nominal sample generated with \MGNLO interfaced with \PYTHIA[8] to an alternative sample interfaced with \HERWIG[7].
Additionally, the uncertainty on the ISR modelling is evaluated by comparing the nominal sample with alternative samples simulated using the \texttt{Var3cUp} and \texttt{Var3cDown} variations of the \PYTHIA A14 tune~\cite{ATL-PHYS-PUB-2014-021}.

\paragraph{Background modelling uncertainties}
\label{sec:systematics:bkg}

The \ttZ process is assigned the same categories of modelling uncertainties as the signal, namely renormalisation and factorisation scale variations, PDF uncertainties, parton-shower modelling uncertainties, and ISR modelling uncertainties. These uncertainties are evaluated following the same procedure as for the signal, using variations appropriate to the nominal \ttZ simulation setup.

For \tWZ events, renormalisation and factorisation scale uncertainties are evaluated following the same methodology as for the signal, using variations appropriate to the nominal \tWZ simulation setup.
Moreover, to account for the interference between \ttZ and \tWZ, a modelling uncertainty on the \tWZ background is applied considering the difference between samples generated with the DR1 and DR2 diagram removal scheme.
The PDF uncertainties on the prompt backgrounds \ttZ and \tWZ are treated as correlated because both processes use the same PDF set and the same PDF uncertainty prescription.

Several uncertainty sources are considered for the \VVBF, \VVCF and \VVLF backgrounds.
The \muR and \muF scale uncertainties are treated as uncorrelated between the three flavour components, since the relative fractions of \VVBF, \VVCF and \VVLF events are affected differently by the modelling of additional heavy-flavour radiation.
The uncertainties associated with the CKKW matching scale~\cite{Catani:2001cc,Hoeche:2009rj}, the QSF resummation scale~\cite{Bothmann:2019yzt,Hoeche:2011fd}, and the alternative Catani--Seymour shower recoil scheme (CSSKIN)~\cite{Schumann:2007mg,PhysRevD.81.034026} are also treated as uncorrelated between flavour components and are evaluated using alternative truth-level samples.
The difference is parameterised at truth level as a function of the number of truth jets and then applied at reconstruction level based on the number of truth jets of the nominal samples.
Approximate next-to-leading-order electroweak~(NLO EW) corrections are applied as event weights, computed using the electroweak virtual approximation~\cite{Kallweit:2014xda}.
Finally, uncertainties related to the choice of the PDFs are also considered, according to the PDF4LHC prescriptions. These uncertainties are treated as correlated among the different flavours.
The component-specific uncertainties account for uncertainties in the relative \VVBF, \VVCF and \VVLF composition and in the extrapolation of each component from the \CRVV region to the SR.%

For the \ttbar process, the uncertainty on the modelling of the parton shower is evaluated by comparing the nominal \POWHEGBOX interfaced with \PYTHIA[8] sample to an alternative sample interfaced with \HERWIG[7].
The uncertainty associated with the matrix-element modelling (NLO matching) is estimated by using alternative samples with modified \texttt{pThard} and \hdamp parameters. The \texttt{pThard} variation probes the matching procedure, while the \hdamp variation probes the modelling of additional QCD radiation.
These uncertainties are applied to \ttbar events in the \NPLelHF, \NPLmuHF, and \NPLelOther categories.
An additional shape uncertainty is assigned to the \NPLelHF and \NPLmuHF components to account for the residual non-closure observed between data and the templates in the transverse momentum distribution of the third lepton. This uncertainty is implemented through event weights derived from a fit to the ratio of data to prediction in the corresponding NPL CRs as a function of the third-lepton transverse momentum.
For the \NPLelOther component, an additional normalisation uncertainty of \qty{12.5}{\%} is applied to account for the residual non-closure.
For the \NPLOther component, a conservative \qty{50}{\%} uncertainty is assigned to the normalisation. Given the small expected number of events of this background in the SR, the exact uncertainty has no impact on the results.
The \ttH process is assigned a $^{\qty[retain-explicit-plus]{+6.8}{\%}}_{\qty{-9.9}{\%}}$ uncertainty on the cross-section, following Ref.~\cite{deFlorian:2016spz}.
Finally, all remaining, minor backgrounds (\ttt, \tttt~and triboson) are assigned a conservative \qty{50}{\%} normalisation uncertainty.

%
%
%
%
%
%
%
%
%
%

%
%
%
%
%
%
%
%


%

%

%
\section{Inclusive cross-section measurement}
\label{sec:inclusive}

The measurement is performed using a binned profile-likelihood fit to the data in the signal and control regions with the \textsc{TRExFitter} framework~\cite{TRExFitter}.
The likelihood function is constructed as the product of Poisson probability terms over all bins of the fitted distributions, with the expected yield in each bin given by the sum of the signal and background contributions.
Systematic uncertainties are incorporated through nuisance parameters in the likelihood. Gaussian constraint terms are used for most of systematic uncertainties, while MC statistical uncertainties are described by Poisson-constrained parameters.

For the inclusive measurement, the \tZq normalisation is measured together with normalisation factors for the dominant background contributions. In particular, the normalisations of the \ttZ, and non-prompt-lepton background components (\NPLelHF, \NPLmuHF, and \NPLelOther) are left free-floating in the fit.
A single free-floating normalisation factor is assigned to the diboson background and is applied coherently to the \VVBF, \VVCF and \VVLF components. Nevertheless, these components are modelled as separate templates, with their relative yields in the different regions taken from simulation.
The fit includes the SR and the dedicated CRs enriched in \ttZ, diboson, and non-prompt-lepton backgrounds.
The fitted observables and binning used in each region are summarised in \cref{tab:inclusive:regions}.
In the SR and \CRttZ, the full binned \ONN distribution is used in the fit, and no selection is applied on \ONN.
For the \CRttZ templates, the NN trained in the SR is evaluated on the events selected in the corresponding CR\@.
The binning of the fitted \ONN distributions is determined using an automatic procedure based on the expected signal and background yields. Starting from a finely binned distribution, neighbouring \ONN intervals are combined: each final bin is required to contain no more than one quarter (one third) of the total expected signal yield and no more than one sixth (all) of the total expected background yield, for the SR (\CRttZ)

\begin{table}[htbp]
\tablesetup[1.3]
\centering
\caption{Overview of the regions and fitted observables used in the profile-likelihood fit.}
\label{tab:inclusive:regions}
\begin{tabular}{l l c}
\toprule
Region & Fitted observable & Number of bins \\
\midrule
SR(\tZq)     & \ONN & 10 \\
\CRttZ       & \ONN & 4 \\
\CRVV        & $\pT(\ell_3)$ & 5 \\
\CRelHF      & $N_{\text{jets}}$ & 3 \\
\CRmuHF      & $N_{\text{jets}}$ & 3 \\
\CRelOther   & $\pT(\ell_3)$ & 4 \\
\CRmuOther   & \mtw & 4 \\
\bottomrule
\end{tabular}
\end{table}

\begin{figure}[!htbp]
\centering
\begin{tabular}{cc}
\begin{subfigure}[b]{.45\textwidth}
\includegraphics[width=\textwidth]{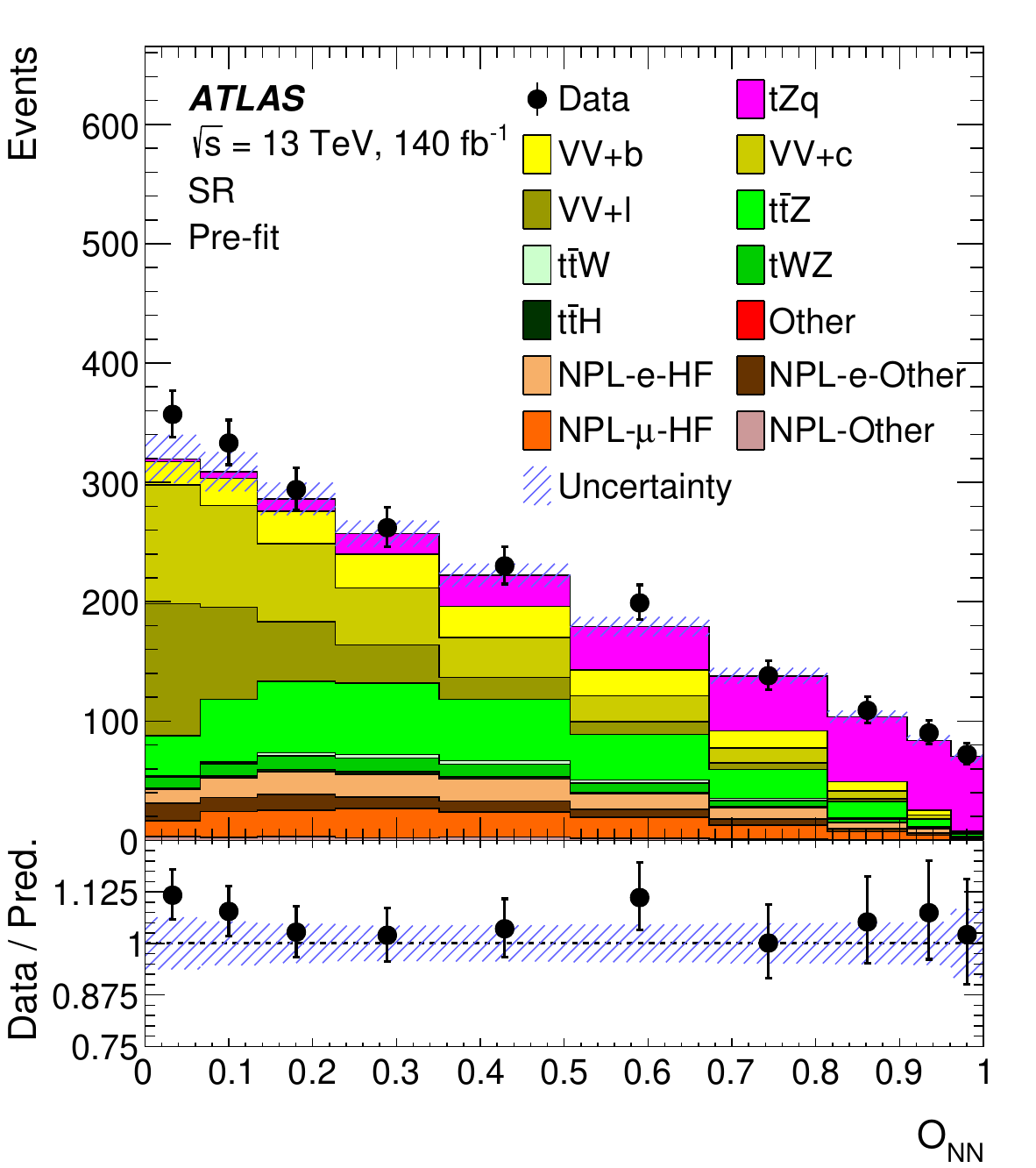}
\caption{}
\end{subfigure}
&
\begin{subfigure}[b]{.45\textwidth}
\includegraphics[width=\textwidth]{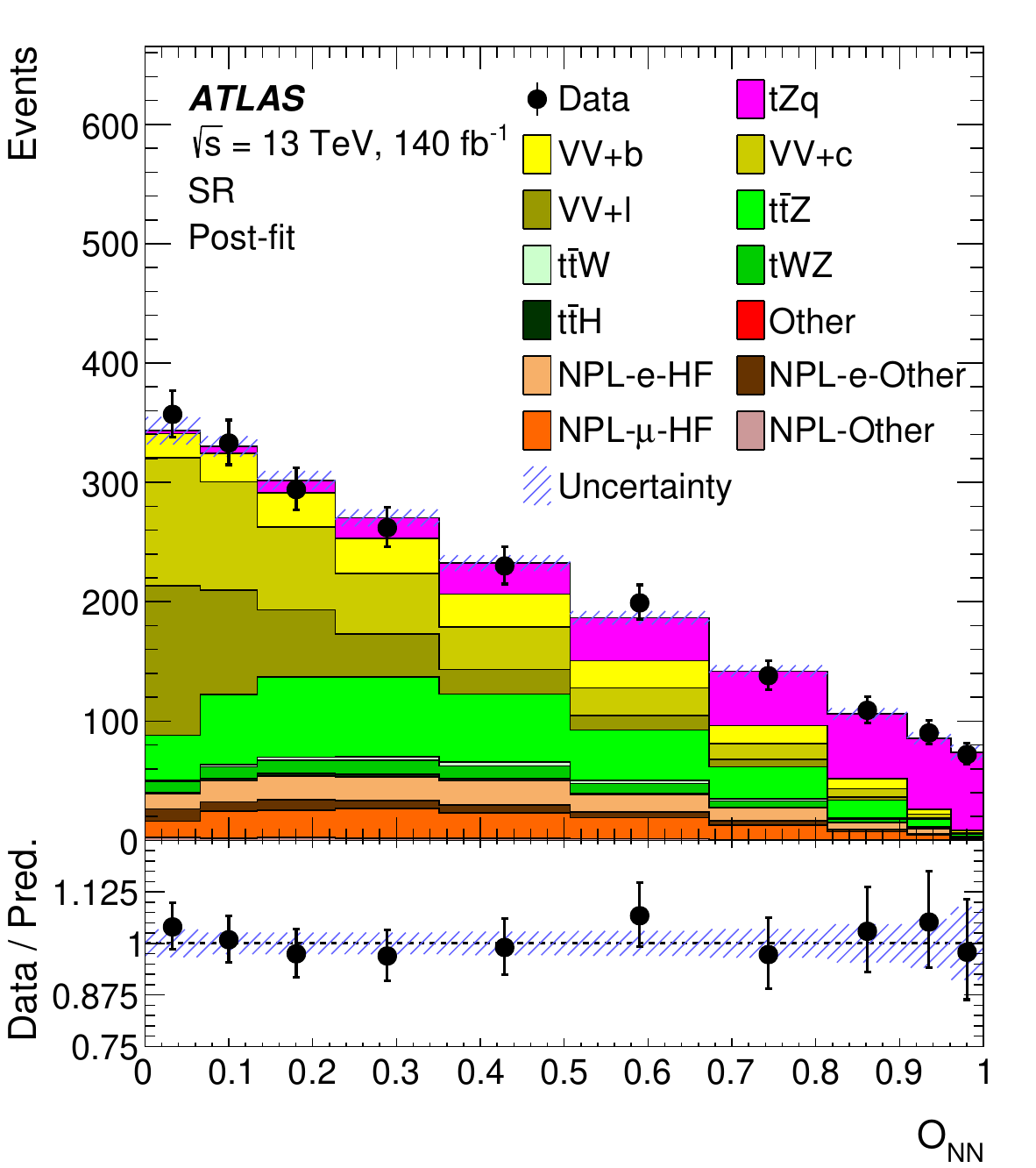}
\caption{}
\end{subfigure}
\end{tabular}
\caption{
(a) Pre- and (b) post-fit distributions of the NN discriminant in the SR.
The lower panel shows the ratio of the data to the total pre- or post-fit prediction (\enquote{Pred.}).
The hatched band represents the total uncertainty in the prediction, including statistical and systematic contributions.
}
\label{fig:inclusive:sr}
\end{figure}

\begin{figure}[!htbp]
\centering
\begin{tabular}{cc}
\begin{subfigure}[b]{.35\textwidth}
\includegraphics[width=\textwidth]{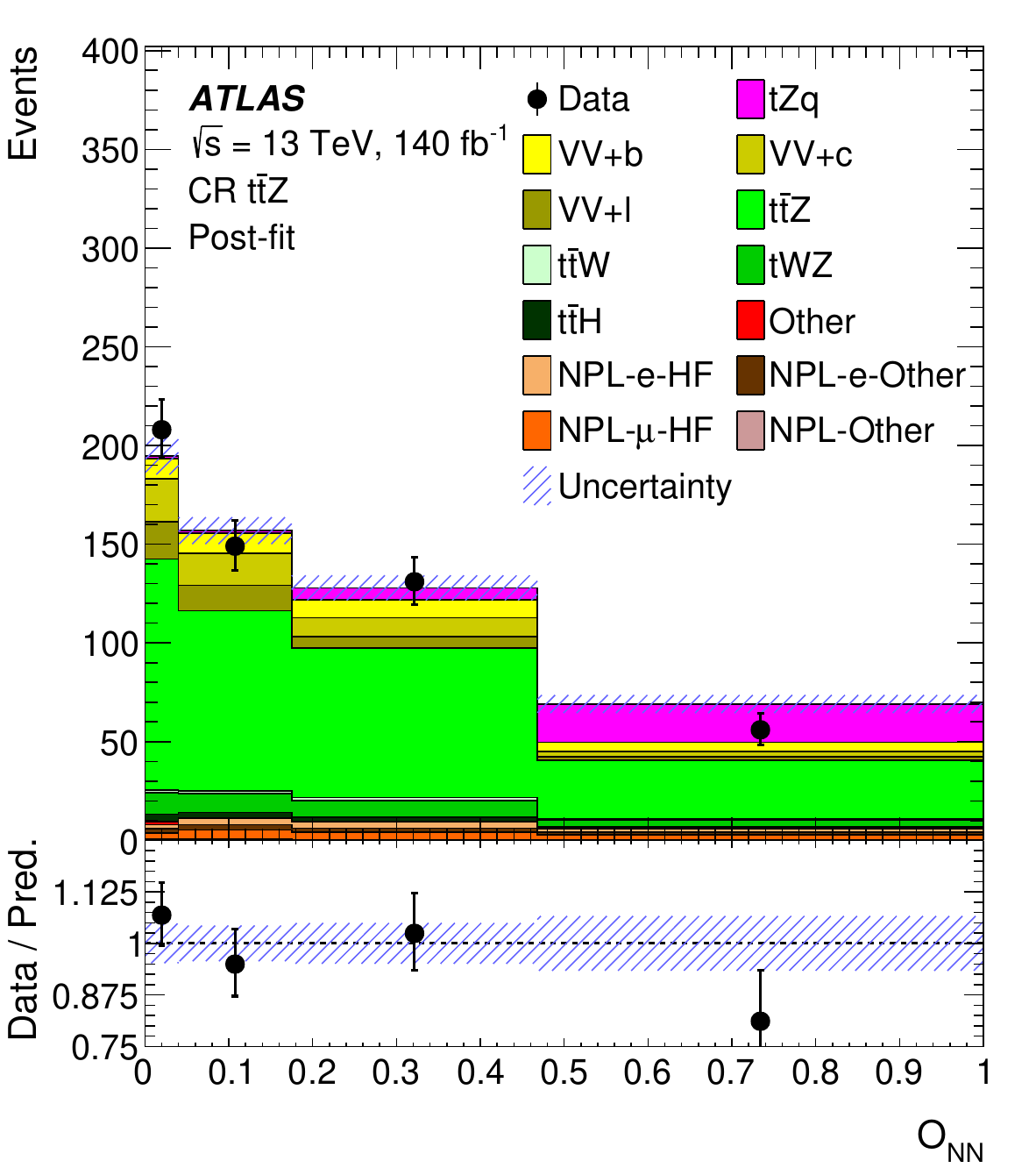}
\caption{}
\end{subfigure}
&
\begin{subfigure}[b]{.35\textwidth}
\includegraphics[width=\textwidth]{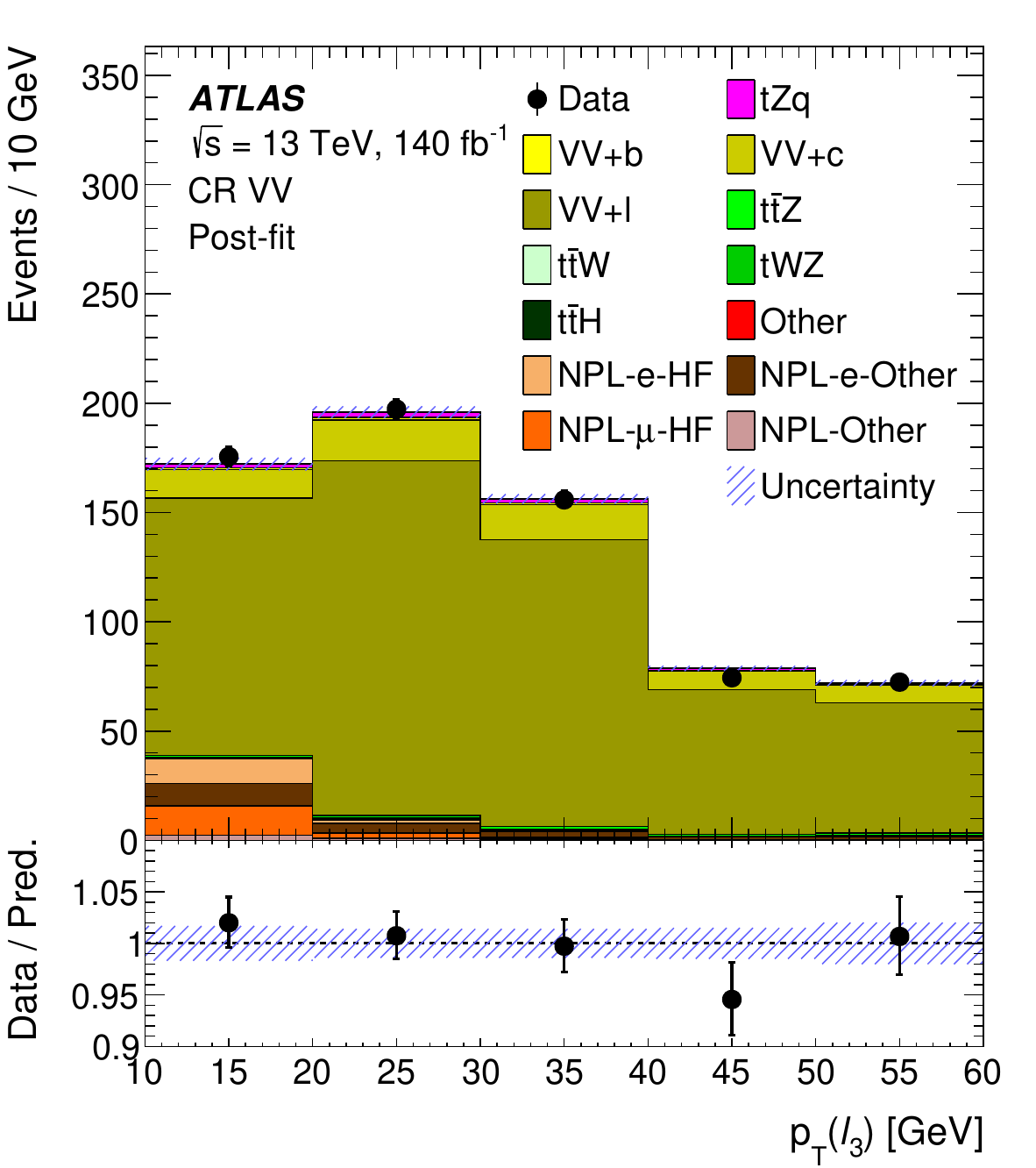}
\caption{}
\end{subfigure} \\

\begin{subfigure}[b]{.35\textwidth}
\includegraphics[width=\textwidth]{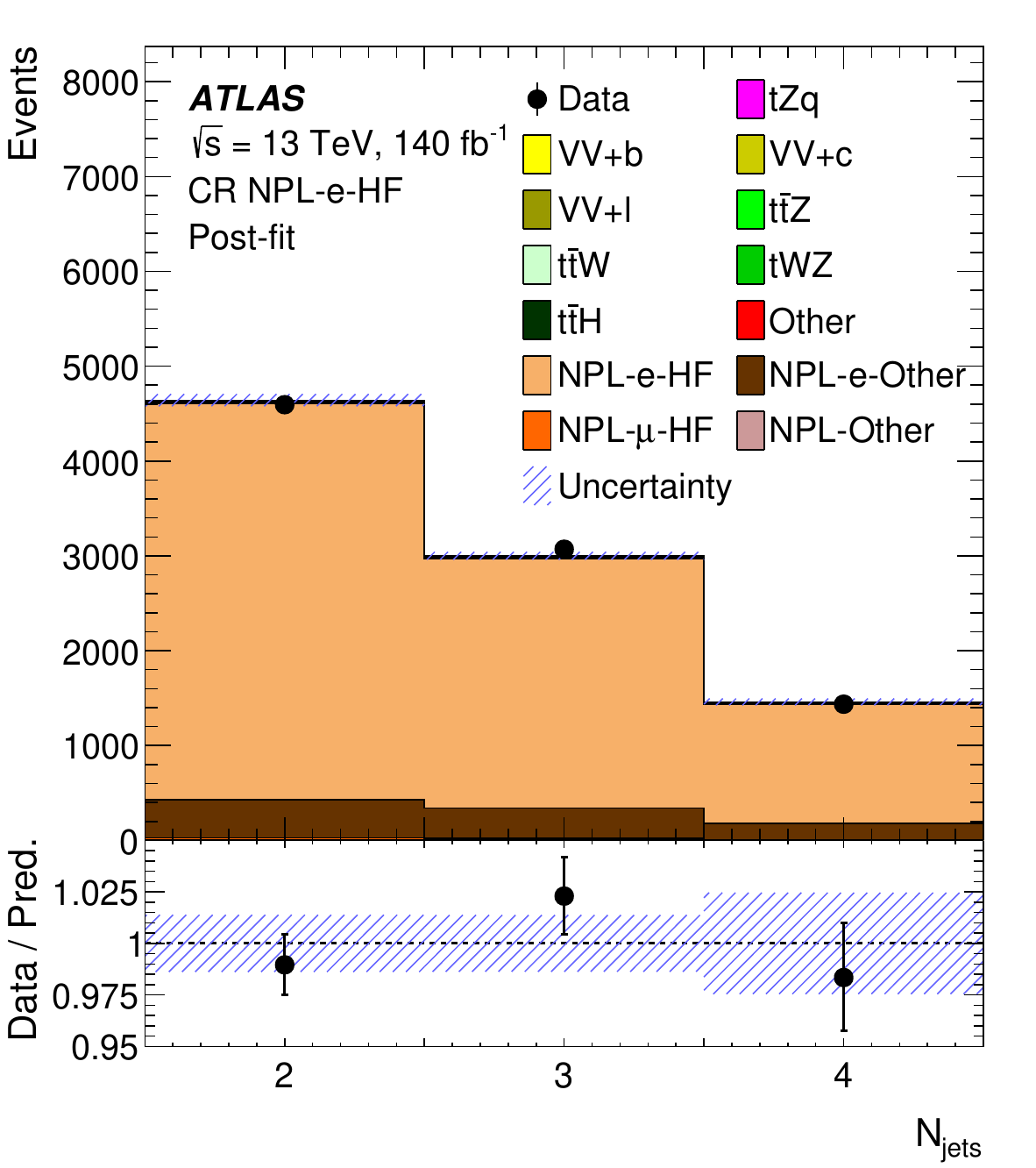}
\caption{}
\end{subfigure}
&
\begin{subfigure}[b]{.35\textwidth}
\includegraphics[width=\textwidth]{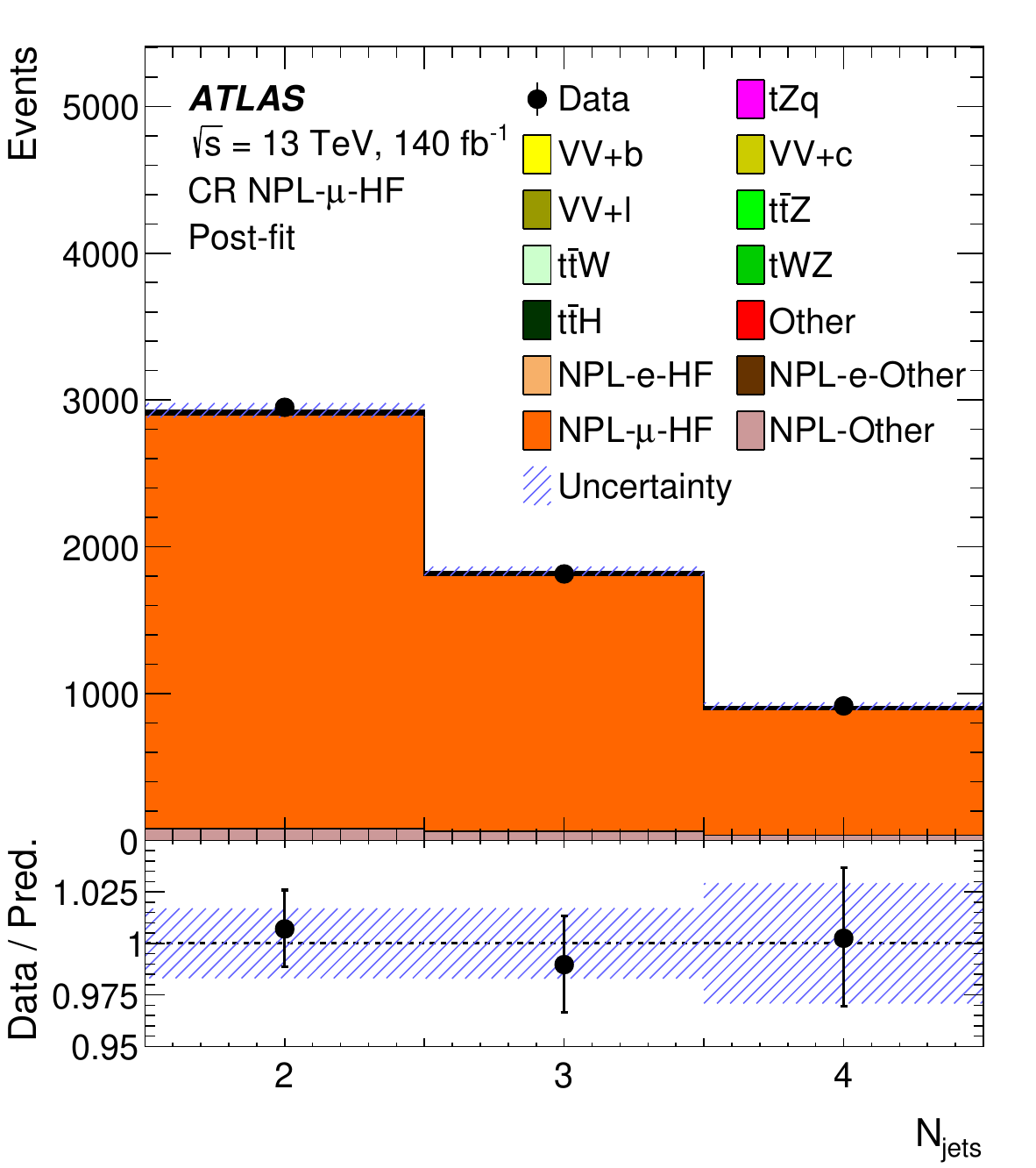}
\caption{}
\end{subfigure} \\

\begin{subfigure}[b]{.35\textwidth}
\includegraphics[width=\textwidth]{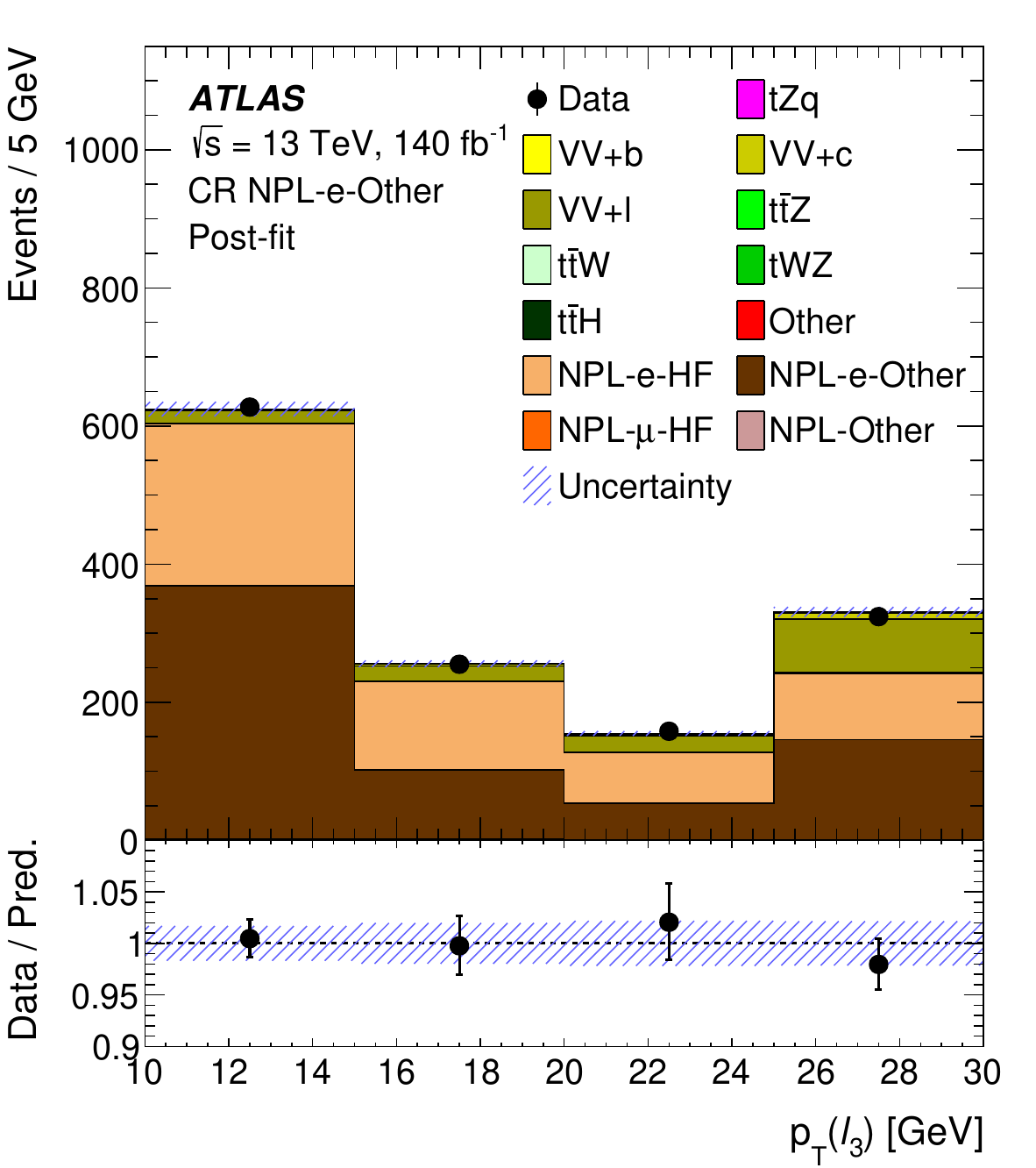}
\caption{}
\end{subfigure}
&
\begin{subfigure}[b]{.35\textwidth}
\includegraphics[width=\textwidth]{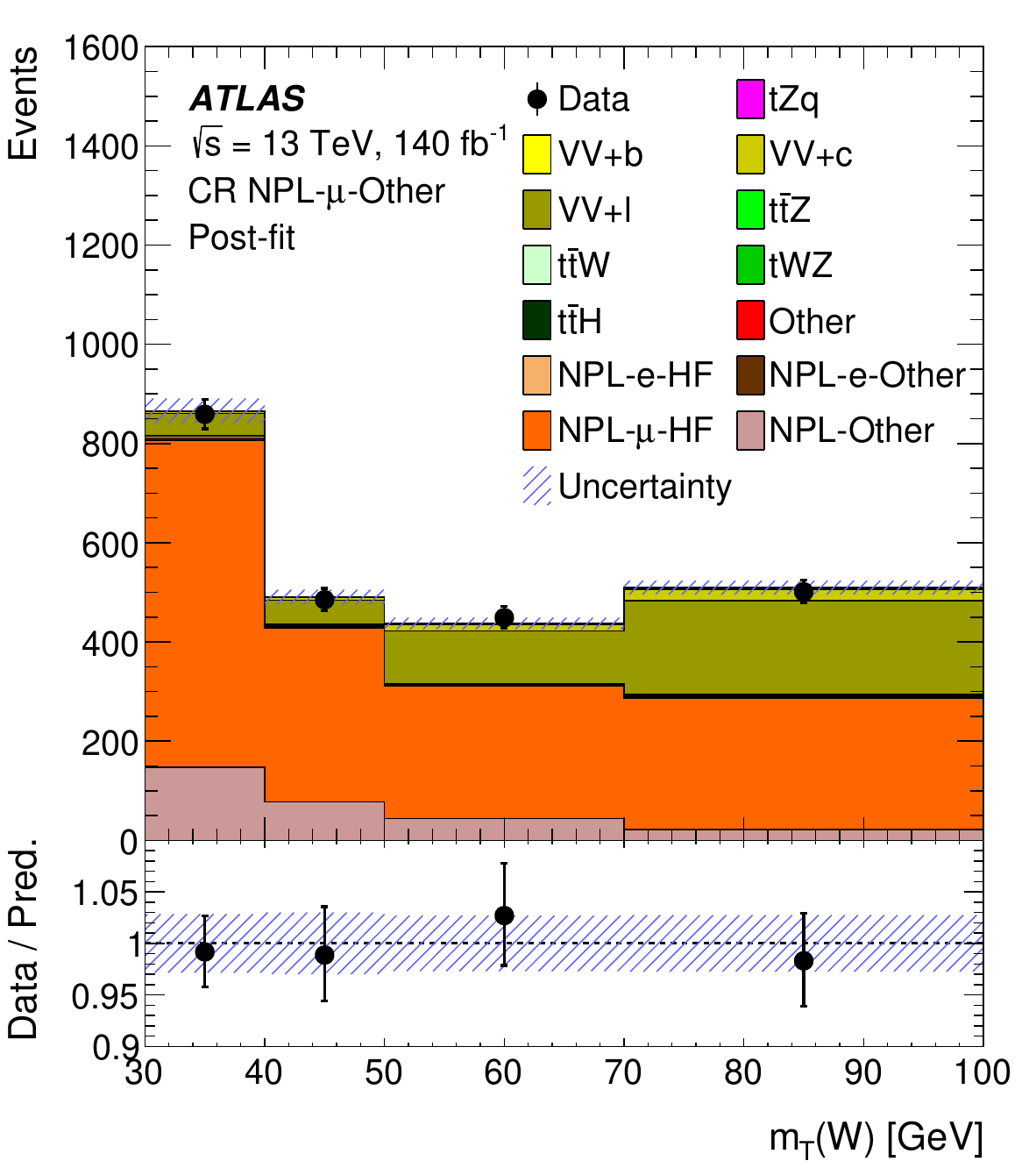}
\caption{}
\end{subfigure}
\end{tabular}
\caption{
Post-fit distributions of the discriminant in the control regions:
(a) \CRttZ,
(b) \CRVV,
(c) \CRelHF,
(d) \CRmuHF,
(e) \CRelOther, and
(f) \CRmuOther.
The lower panels show the ratio of the data to the prediction (\enquote{Pred.}).
The hatched bands represent the total uncertainty in the prediction, including both statistical and systematic contributions.
Where applicable, overflow events are included in the rightmost bin.
}
\label{fig:inclusive:crs}
\end{figure}

\Cref{fig:inclusive:sr} shows the comparison between data and pre- and post-fit background and signal distributions in the SR. \Cref{fig:inclusive:crs} shows the comparison between data and post-fit background and signal distributions in the CRs. The goodness-of-fit (GoF) test~\cite{cousins_saturated} is used to quantify the agreement between the data and the model under consideration. This test accounts for the penalties arising from shifts in the nuisance parameters during the fit. For this fit, the GoF test returns a value of 0.9.

The numbers of fitted signal and background events compared with the data are shown in \cref{tab:inclusive:yields:srcr,tab:inclusive:yields:npl}.
The background normalisation factors obtained from the fit are found to be compatible with unity within uncertainties, except for the \NPLelOther component, for which the fitted value is $0.59 \pm 0.06$.
This value reflects the pre-fit overestimation observed in the dedicated \NPLelOther control region, shown in \cref{fig:app_inclusive:crs}.
A sizeable correction is not unexpected for backgrounds containing non-prompt or misidentified leptons, whose rates and kinematic properties are difficult to model accurately in simulation. The use of a dedicated control region allows the normalisation of this component to be constrained directly from data rather than relying on the absolute simulation prediction.

\begin{table}[!htbp]
\sisetup{
round-mode = places,
round-precision = 1
}
\tablesetup
\caption{Post-fit event yields in the signal region and the \CRttZ and \CRVV control regions. The signal and background predictions are shown after the fit to data. The quoted uncertainties include the statistical and systematic uncertainties of the yields, computed taking into account correlations among nuisance parameters and among processes.}
\label{tab:inclusive:yields:srcr}
\centering
\begin{tabular}{l
S[table-format=4.1]@{$\,\pm\,$}S[table-format=2.1]
S[table-format=4.1]@{$\,\pm\,$}S[table-format=2.1]
S[table-format=4.1]@{$\,\pm\,$}S[table-format=2.1]
}
\toprule
Process
& \multicolumn{2}{c}{SR}
& \multicolumn{2}{c}{\CRttZ}
& \multicolumn{2}{c}{\CRVV} \\
\midrule
\tZq  & {\numRP[0]{322.063}} & {\numRP[0]{29.7047}}  & 27.8971 & 8.3157   & 74.7693 & 8.16807 \\
\VVBF & 181.75  & 5.81892  & 34.0928 & 3.64716  & 35.2264 & 3.59582 \\
\VVCF & {\numRP[0]{401.457}} & {\numRP[0]{24.147}}   & 50.188  & 6.42917  & {\numRP[0]{640.916}} & {\numRP[0]{31.539}} \\
\VVLF & {\numRP[0]{347.499}} & {\numRP[0]{32.5269}}  & 39.5197 & 6.69305  & {\numRP[0]{5370.16}} & {\numRP[0]{88.075}} \\
\ttZ  & {\numRP[0]{381.222}} & {\numRP[0]{55.051}}   & {\numRP[0]{313.68}}  & {\numRP[0]{28.93}}    & {\pho\numRP[0]{52.7266}} & {\numRP[0]{10.3286}} \\
\ttW  & 17.0744 & 0.381203 & 4.55483 & 0.325573 & 1.98182 & 0.183057 \\
\tWZ  & 70.0539 & 7.93476  & 32.6527 & 9.22749  & 18.9869 & 0.998581 \\
\ttH  & 10.9472 & 0.960111 & 9.58015 & 0.880346 & 1.23202 & 0.132682 \\
Other & 1.00815 & 0.49694  & 1.97351 & 0.969311 & 3.44389 & 1.69361 \\
\NPLelHF & {\numRP[0]{126.407}} & {\numRP[0]{12.6152}} & 9.96487 & 2.30638 & {\numRP[0]{136.886}} & {\numRP[0]{11.8945}} \\
\NPLelOther & 51.5204 & 8.9482 & 8.35418 & 1.77371 & {\numRP[0]{199.656}} & {\numRP[0]{35.9669}} \\
\NPLmuHF & {\numRP[0]{147.188}} & {\numRP[0]{10.7345}} & 14.1304 & 1.54346 & {\numRP[0]{169.522}} & {\numRP[0]{23.8815}} \\
\NPLOther & 12.902 & 5.15479 & 1.9126 & 0.826261 & {\pho\numRP[0]{42.7765}} & {\numRP[0]{19.0141}} \\
\midrule
Total & {\numRP[0]{2071.09}} & {\numRP[0]{42.1149}} & {\numRP[0]{548.501}} & {\numRP[0]{22.3376}} & {\numRP[0]{6748.29}} & {\numRP[0]{80.8683}} \\
\midrule
Data & \multicolumn{2}{c}{2084} & \multicolumn{2}{c}{544} & \multicolumn{2}{c}{6755} \\
\bottomrule
\end{tabular}
\end{table}

\begin{table}[!htbp]
\sisetup{
round-mode = places,
round-precision = 1
}
\tablesetup
\caption{Post-fit event yields in the NPL control regions. The signal and background predictions are shown after the fit to data. The quoted uncertainties include the statistical and systematic uncertainties of the yields, computed taking into account correlations among nuisance parameters and among processes.}
\label{tab:inclusive:yields:npl}
\centering
\begin{tabular}{l
S[table-format=4.1]@{$\,\pm\,$}S[table-number-alignment=left, table-format=3.1]
S[table-format=4.1]@{$\,\pm\,$}S[table-number-alignment=left, table-format=3.1]
S[table-format=4.1]@{$\,\pm\,$}S[table-number-alignment=left, table-format=3.1]
S[table-format=4.1]@{$\,\pm\,$}S[table-number-alignment=left, table-format=3.1]
}
\toprule
Process
& \multicolumn{2}{c}{\CRelHF}
& \multicolumn{2}{c}{\CRmuHF}
& \multicolumn{2}{c}{\CRelOther}
& \multicolumn{2}{c}{\CRmuOther} \\
\midrule
\tZq & 8.31052 & 2.25945 & 5.89302 & 1.53854 & 11.0424 & 1.25572 & 5.74321 & 0.759788 \\
\VVBF & 7.11296 & 0.429395 & 4.8637 & 0.247773 & 4.93671 & 0.60537 & 2.99916 & 0.356032 \\
\VVCF & 12.6111 & 0.882222 & 8.70843 & 0.738362 & 80.2871 & 4.02655 & 44.0803 & 2.24495 \\
\VVLF & 12.5352 & 1.2722 & 9.48322 & 1.22274 & {\numRP[0]{707.382}} & {\numRP[0]{19.1446}} & {\numRP[0]{389.037}} & {\numRP[0]{12.7469}} \\
\ttZ & 27.759 & 12.3883 & 14.0678 & 1.76821 & 11.7638 & 3.66988 & 5.21209 & 0.981586 \\
\ttW & 15.1577 & 0.329351 & 8.01675 & 0.189102 & 0.382519 & 0.0354567 & 0.146089 & 0.0217173 \\
\tWZ & 2.05449 & 0.1517 & 1.45545 & 0.293787 & 3.13264 & 0.199178 & 1.65281 & 0.0895392 \\
\ttH & 10.9073 & 0.95226 & 6.28297 & 0.545693 & 0.327803 & 0.0383592 & 0.168719 & 0.0209855 \\
Other & 0.420853 & 0.208905 & 0.105603 & 0.053333 & 0.989415 & 0.486399 & \multicolumn{2}{c}{< 0.1}  \\
\NPLelHF & {\numRF[3]{8054.55}\pho[.]} & {\numRF[2]{377.009}} & 16.0156 & 4.38099 & {\numRF[3]{2654}} & {\numRF[2]{178.961}} & 8.83139 & 1.85096 \\
\NPLelOther & {\numRF[2]{899.323}} & {\numRF[2]{328.013}} & 11.9704 & 3.57882 & {\numRF[3]{3329.4}} & {\numRF[2]{199.272}} & 13.2442 & 2.40724 \\
\NPLmuHF & 39.6465 & 5.86711 & {\numRF[3]{5413.33}} & {\numRF[2]{108.15}} & 16.8962 & 3.76774 & {\numRF[3]{1541.27}} & {\numRF[2]{116.469}} \\
\NPLOther & 13.3631 & 5.27855 & {\numRP[0]{176.936}} & {\numRP[0]{69.8699}} & 7.94065 & 3.70448 & {\numRF[2]{291.198}} & {\numRF[2]{110.41}} \\
\midrule
Total & {\numRP[0]{9103.75}} & {\numRP[0]{101.436}} & {\numRP[0]{5677.12}} & {\numRP[0]{76.4404}} & {\numRP[0]{6828.48}} & {\numRP[0]{82.1147}} & {\numRP[0]{2303.59}} & {\numRP[0]{44.2055}} \\
\midrule
Data & \multicolumn{2}{c}{9100} & \multicolumn{2}{c}{5681} & \multicolumn{2}{c}{6822} & \multicolumn{2}{c}{2294} \\
\bottomrule
\end{tabular}
\end{table}

The results of the fit yield a \tZq production cross-section, including non-resonant dilepton pairs with $m_{\ell^+\ell^-} > \SI{30}{\GeV}$, of
\begin{equation}
\sigma(\tZq) = \tZqxsec,
\end{equation}
corresponding to a total uncertainty of $\pm$10\%. The statistical uncertainty is dominant for this result.
This result is in good agreement with the 5FS NLO QCD SM prediction of \qty{84.0}{}$^{+4.7}_{-1.0}$\qty{}{\fb} from Ref.~\cite{tZq-Pagani}, evaluated for the same dilepton phase space.

The impact of the various sources of systematic uncertainties on the \tZq cross-section, broken down into major categories, is summarised in \cref{tab:inclusive:impact}.
These category-level impacts are obtained from the impacts of the individual nuisance parameters entering the fit.
The contribution of each nuisance parameter to the uncertainty in the measured cross-section is obtained from the corresponding off-diagonal element of the post-fit covariance matrix~\cite{pinto2024uncertaintycomponentsprofilelikelihood}. In the case of nuisance parameters representing the MC statistical uncertainty in each bin, their impacts are evaluated by dividing their post-fit off-diagonal elements by their corresponding pre-fit uncertainties.
For each category, the uncertainty contribution is obtained by summing in quadrature the contributions from its nuisance parameters. The total systematic uncertainty is calculated similarly by combining all systematic uncertainty components in quadrature. The statistical uncertainty is calculated as the square root of the difference between the total variance and the total systematic variance.
The dominant source of systematic uncertainty arises from the modelling of the \tZq signal.

\begin{table}[!tbp]
\tablesetup
\caption{Contribution (in percent) from each group of systematic uncertainty to the total uncertainty on $\sigma(\tZq)$, ranked by impact.}
\label{tab:inclusive:impact}
\centering
\begin{tabular}{l S}
\toprule
Uncertainty source & \multicolumn{1}{c}{$\Delta\sigma(\tZq)/\sigma[\%]$} \\
\midrule
\tZq modelling & 4.3 \\
Prompt background modelling & 2.3 \\
Jets & 1.4 \\
Leptons & 1.3 \\
$b$-tagging & 1.2 \\
Luminosity & 0.8 \\
MC statistics & 0.8 \\
Pile-up reweighting & 0.7 \\
Non-prompt-lepton background modelling & 0.7 \\
\tZq PDF & 0.4 \\
Other background & 0.1 \\
\midrule
Total systematic uncertainty & 5.6 \\
\midrule
Statistical uncertainty & 8.5 \\
\midrule
Total uncertainty & 10.2 \\
\bottomrule
\end{tabular}
\end{table}

In addition to the inclusive measurement, the \tZq cross-section is measured separately for top-quark and top-antiquark production, as well as their ratio. The separation between top-quark and top-antiquark events is achieved using the charge of the lepton originating from the \PW boson, which is positive (negative) for top-(anti)quark production.
The effect of lepton charge misidentification, which results in migration between the top-quark and top-antiquark categories, is found to be negligible.

The fit strategy follows that of the inclusive measurement, with the SR further divided according to the lepton charge, while the CRs remain unchanged. Independent signal-strength parameters are introduced for the top-quark and top-antiquark components, allowing their cross-sections to be extracted simultaneously.
A third fit is performed to determine the ratio of top-quark to top-antiquark production. The fit is parameterised in terms of the ratio of the corresponding signal strengths, which is converted to the physical cross-section ratio, \Rt, using the corresponding reference cross-sections. The GoF for these two fits is 0.4.

\begin{figure}
\centering
\begin{tabular}{cc}
\begin{subfigure}[b]{.45\textwidth}
\includegraphics[width=\textwidth]{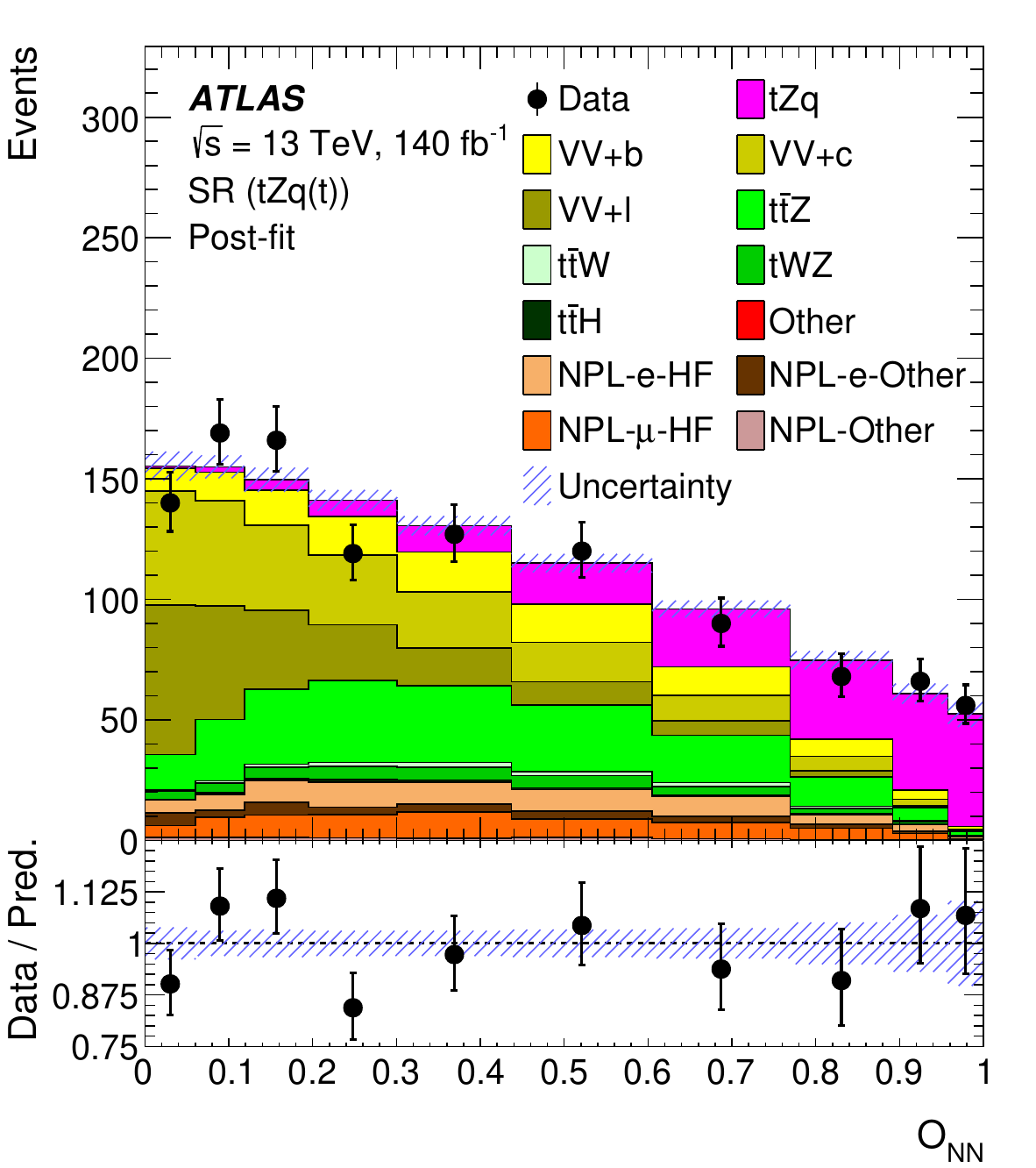}
\caption{}
\end{subfigure}
&
\begin{subfigure}[b]{.45\textwidth}
\includegraphics[width=\textwidth]{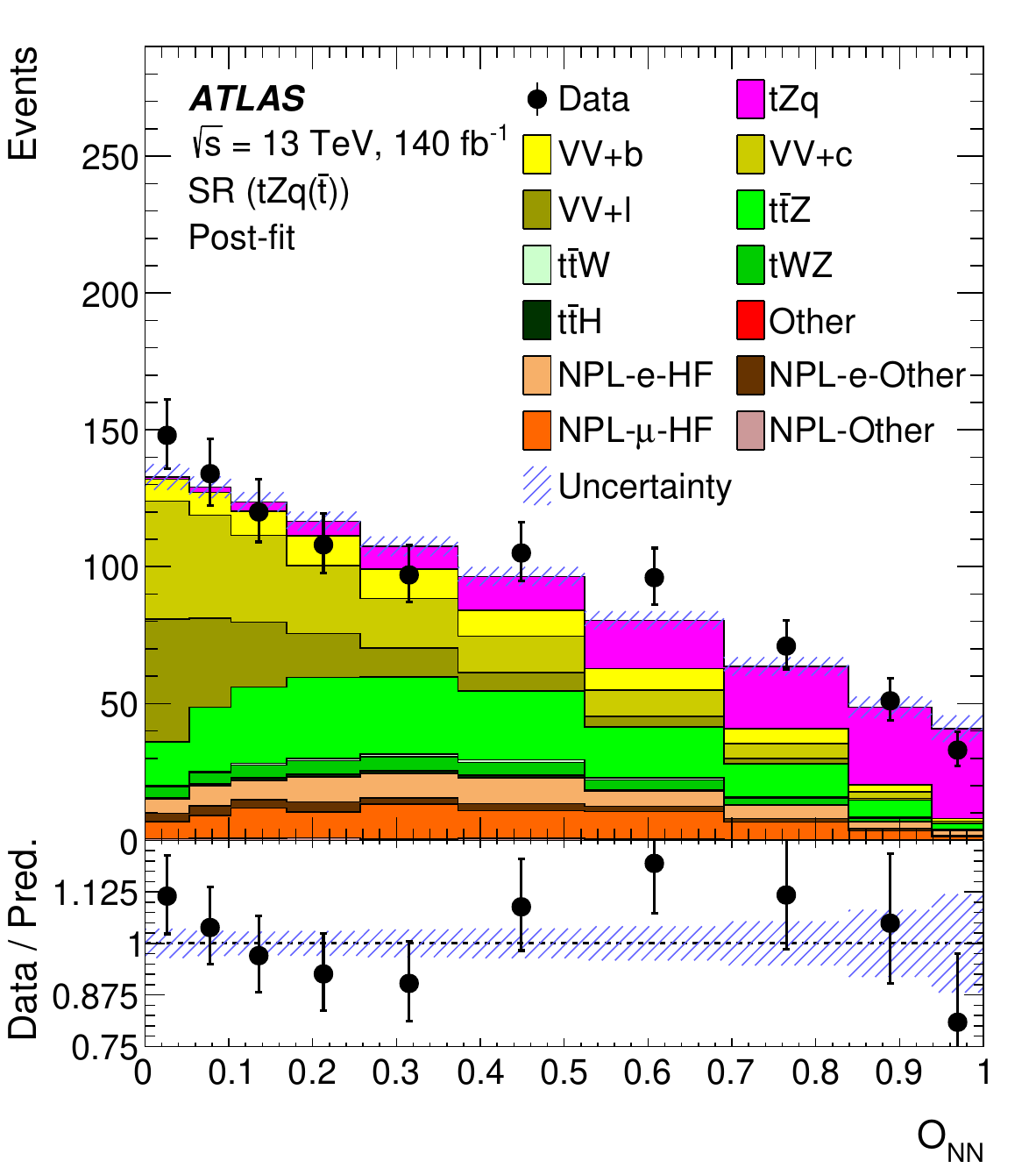}
\caption{}
\end{subfigure} \\
\end{tabular}
\caption{
Post-fit distributions of the NN discriminant in the split SR:
(a) SR(\tZqt),
(b) SR(\tZqtbar).
The lower panels show the ratio of the data to the prediction (\enquote{Pred.}).
The hatched bands represent the total uncertainty in the prediction, including both statistical and systematic contributions.
}
\label{fig:inclusive:srsplit}
\end{figure}

\Cref{fig:inclusive:srsplit} shows the comparison between data and post-fit background and signal distributions, in the split SRs.
The numbers of fitted signal and background events compared with the data in the split SRs are shown in \cref{tab:inclusive:yields:srsplit}.

\begin{table}[!htbp]
\sisetup{
round-mode = places,
round-precision = 1
}
\tablesetup
\caption{Post-fit event yields in the split signal regions. The signal and background predictions are shown after the fit to data. The quoted uncertainties include the statistical and systematic uncertainties of the yields, computed taking into account correlations among nuisance parameters and among processes.}
\label{tab:inclusive:yields:srsplit}
\centering
\begin{tabular}{l
S[table-format=4.1]@{$\,\pm\,$}S[table-format=3.1]
S[table-format=4.1]@{$\,\pm\,$}S[table-format=3.1]
}
\toprule
Process
& \multicolumn{2}{c}{SR(\tZqt)}
& \multicolumn{2}{c}{SR(\tZqtbar)} \\
\midrule
\tZq   & {\numRP[0]{185.62}}  & {\numRP[0]{20.509}}  & {\numRP[0]{133.715}} & {\numRP[0]{20.3109}} \\
\VVBF  & 108.304 & 3.41841 & 72.9708 & 2.7963 \\
\VVCF  & {\numRP[0]{214.406}} & {\numRP[0]{12.9477}} & {\numRP[0]{186.996}} & {\numRP[0]{11.9524}} \\
\VVLF  & {\numRP[0]{199.629}} & {\numRP[0]{18.9746}} & {\numRP[0]{140.919}} & {\numRP[0]{13.2947}} \\
\ttZ   & {\numRP[0]{203.91}}  & {\numRP[0]{31.3518}} & {\numRP[0]{189.132}} & {\numRP[0]{25.8511}} \\
\ttW   & 10.9645 & 0.30528 & 6.11077 & 0.170918 \\
\tWZ   & 35.0042 & 3.87476 & 34.9171 & 4.07064 \\
\ttH   & 5.42283 & 0.47759 & 5.52345 & 0.488275 \\
Other  & 0.551744 & 0.272718 & 0.448218 & 0.221659 \\
\NPLelHF    & {\pho\numRP[0]{64.9131}} & {\numRP[0]{10.7987}} & 62.0038 & 4.48883 \\
\NPLelOther & 28.0734 & 6.22623 & 22.2568 & 4.20423 \\
\NPLmuHF    & {\pho\numRP[0]{65.6506}} & {\numRP[0]{10.1958}} & 79.3491 & 4.83941 \\
\NPLOther   & 8.0852  & 3.27682 & 5.07517 & 1.94057 \\
\midrule
Total  & {\numRP[0]{1130.53}} & {\numRP[0]{27.915}} & {\numRP[0]{939.416}} & {\numRP[0]{23.869}} \\
\midrule
Data   & \multicolumn{2}{c}{1121} & \multicolumn{2}{c}{963} \\
\bottomrule
\end{tabular}
\end{table}

The cross-sections for \tZq production are measured separately for top quarks and top antiquarks to be
\begin{align}
\sigma\bigl(\tZq(t)\bigr)     &= \tZqxsect, \\
\sigma\bigl(\tZq(\bar{t})\bigr) &= \tZqxsectbar,
\end{align}
with the corresponding ratio measured as
\begin{equation}
\Rt = \tZqxsecr.
\end{equation}
This result is consistent with the nominal 5FS MC prediction of 1.79, obtained using the \NNPDF[3.0nlo] PDF set in the same phase space.
In all cases, the statistical uncertainty is the dominant contribution.
The impact of the various sources of systematic uncertainties on the cross-sections and on \Rt, broken down into major categories, is summarised in \cref{tab:inclusive:impactsplit}. For the cross-sections, the dominant source of systematic uncertainty arises from the modelling of the \tZq signal, while for \Rt~it arises from the modelling of the non-prompt-lepton background.

\begin{table}[htbp]
\tablesetup
\caption{Contribution (in percent) from each group of systematic uncertainty to the total uncertainty on $\sigma\bigl(\tZqt\bigr)$, $\sigma\bigl(\tZqtbar\bigr)$ and \Rt, ranked by impact.}
\label{tab:inclusive:impactsplit}
\centering
\begin{tabular}{l SSS}
\toprule
Uncertainty source
& \multicolumn{1}{c}{$\Delta\sigma\bigl(\tZqt\bigr)/\sigma\ [\%]$}
& \multicolumn{1}{c}{$\Delta\sigma\bigl(\tZqtbar\bigr)/\sigma\ [\%]$}
& \multicolumn{1}{c}{$\Delta\Rt/\Rt\ [\%]$} \\
\midrule
\tZq modelling & 4.2 & 4.4 & 1.3 \\
Prompt background modelling & 1.9 & 3.7 & 1.9 \\
$b$-tagging & 1.5 & 1.8 & 0.7 \\
Leptons & 1.5 & 1.5 & 0.5 \\
Jets & 1.3 & 2.5 & 1.7 \\
Pile-up reweighting & 0.9 & 0.8 & 0.1 \\
Luminosity & 0.8 & 0.8 & <0.1 \\
MC statistics & 0.8 & 1.5 & 1.7 \\
Non-prompt-lepton background modelling & 1.5 & 2.2 & 3.3 \\
\tZq PDF & 0.4 & 0.4 & 0.8 \\
Other background & <0.1 & 0.1 & <0.1 \\
\midrule
Total systematic uncertainty & 5.6 & 7.1 & 4.5 \\
\midrule
Statistical uncertainty & 10.1 & 15.1 & 17.6 \\
\midrule
Total uncertainty & 11.7 & 16.8 & 18.3 \\
\bottomrule
\end{tabular}
\end{table}

The measured cross-sections are compared with MC predictions evaluated in the particle-level and parton-level phase spaces defined in \cref{sec:sel:fid}. Separate fits are performed for the inclusive and fiducial measurements. For the fiducial measurements, associated acceptance effects are removed using truth-level information, such that the extracted cross-sections correspond directly to the fiducial volume.
The theoretical predictions are obtained from \(\sigma = \sigma_\text{th}(\tZq) \cdot \epsilon\), where \(\sigma_\text{th}(\tZq)\) denotes the \tZq production cross-section in the $m_{\ell\ell}>\qty{5}{\GeV}$ phase space calculated in the 5FS, and \(\epsilon\) is the corresponding selection efficiency. The efficiencies are defined relative to the same $m_{\ell\ell}>\qty{5}{\GeV}$ phase space.
The efficiencies used to derive the fiducial predictions are summarised in \cref{tab:fiducial:eff}.
The efficiencies include the branching fractions of the leptonic \PW-boson decay modes.
The uncertainties on the predictions are obtained by propagating the relative uncertainties on \(\sigma(\tZq)\) and \(\epsilon\) in quadrature. The uncertainty on \(\sigma(\tZq)\) includes contributions from PDF variations and renormalisation and factorisation scale variations. The uncertainty on \(\epsilon\) accounts for modelling effects from renormalisation and factorisation scale variations, evaluated using a seven-point envelope, PDF variations, the \texttt{Var3cUp} and \texttt{Var3cDown} variations of the \PYTHIA A14 tune, and differences between \PYTHIA and \HERWIG parton-shower models. These contributions are added in quadrature.

\begin{table}[htbp]
\tablesetup[1.5]
\centering
\caption{Selection efficiencies, including the branching fractions of the leptonic \PW-boson decay modes, for the particle-level and parton-level phase spaces.}
\label{tab:fiducial:eff}
\begin{tabular}{llc}
\toprule
Phase space & Process & Efficiency [\%] \\
\midrule
& \tZqt    & $4.15^{+0.18}_{-0.21}$ \\
Particle level & \tZqtbar & $4.22^{+0.12}_{-0.18}$ \\
& \tZq     & $4.17^{+0.16}_{-0.19}$ \\
\midrule
& \tZqt    & $10.03 \pm 0.02$ \\
Parton level & \tZqtbar & $9.86\pm0.02$ \\
& \tZq     & $9.97 \pm 0.02$ \\
\bottomrule
\end{tabular}
\end{table}

The measured and predicted fiducial cross-sections, obtained as described above, are summarised in \cref{tab:fiducial:results}. The measured particle-level and parton-level cross-sections are compared to the corresponding predictions and are found to be consistent within the quoted uncertainties.

\begin{table}[htbp]
\tablesetup[1.5]
\centering
\caption{Measured and predicted fiducial cross-sections for \tZqt, \tZqtbar, and their combination
in the particle-level and parton-level phase spaces. The measured uncertainties are quoted separately as statistical and systematic uncertainties. The uncertainties on the MC predictions include contributions from the inclusive cross-section and the corresponding selection efficiency.}
\label{tab:fiducial:results}
\begin{tabular}{llcc}
\toprule
Phase space & Process & Measured [fb] & MC prediction [fb] \\
\midrule
\multirow{3}{*}{Particle level}
& \tZqt
& $3.09^{+0.35}_{-0.33}~(\mathrm{stat.})^{+0.14}_{-0.15}~(\mathrm{syst.})$
& $3.45^{+0.29}_{-0.20}$ \\
& \tZqtbar
& $2.21^{+0.33}_{-0.31}~(\mathrm{stat.})^{+0.15}_{-0.14}~(\mathrm{syst.})$
& $1.96^{+0.16}_{-0.10}$ \\
& \tZq
& $5.36^{+0.49}_{-0.47}~(\mathrm{stat.})^{+0.27}_{-0.28}~(\mathrm{syst.})$
& $5.41^{+0.44}_{-0.28}$ \\
\midrule
\multirow{3}{*}{Parton level}
& \tZqt
& $7.45^{+0.83}_{-0.76}~(\mathrm{stat.})^{+0.40}_{-0.39}~(\mathrm{syst.})$
& $8.34^{+0.60}_{-0.23}$ \\
& \tZqtbar
& $5.26^{+0.79}_{-0.71}~(\mathrm{stat.})^{+0.45}_{-0.43}~(\mathrm{syst.})$
& $4.58^{+0.35}_{-0.13}$ \\
& \tZq
& $12.60^{+1.15}_{-1.07}~(\mathrm{stat.})^{+0.65}_{-0.64}~(\mathrm{syst.})$
& $12.92^{+0.92}_{-0.32}$ \\
\bottomrule
\end{tabular}
\end{table}


%

%

%
\section{Differential cross-section measurement and spin asymmetry}
\label{sec:differential}

The differential cross-sections are measured at both parton and particle level using a PLU procedure, as described in Ref.~\cite{TOPQ-2020-20}. The parton-level results are defined in a phase space that is inclusive with respect to the kinematics of the final-state objects, to facilitate comparisons with fixed-order theoretical predictions, while the particle-level results are measured in a fiducial phase space closely matched to the detector-level selection, as detailed in \cref{sec:sel:fid}.

In the PLU approach, the detector-level distributions are modelled as the sum of contributions from each truth-level bin, which are treated as independent signal components in the fit. The parameters of interest correspond to the normalisation factors associated with the truth-level bins and are determined simultaneously in a binned profile-likelihood fit to the data in the signal and control regions.

\begin{figure}[htbp]
\centering
\begin{subfigure}{0.32\textwidth}
\centering
\includegraphics[width=0.95\textwidth]{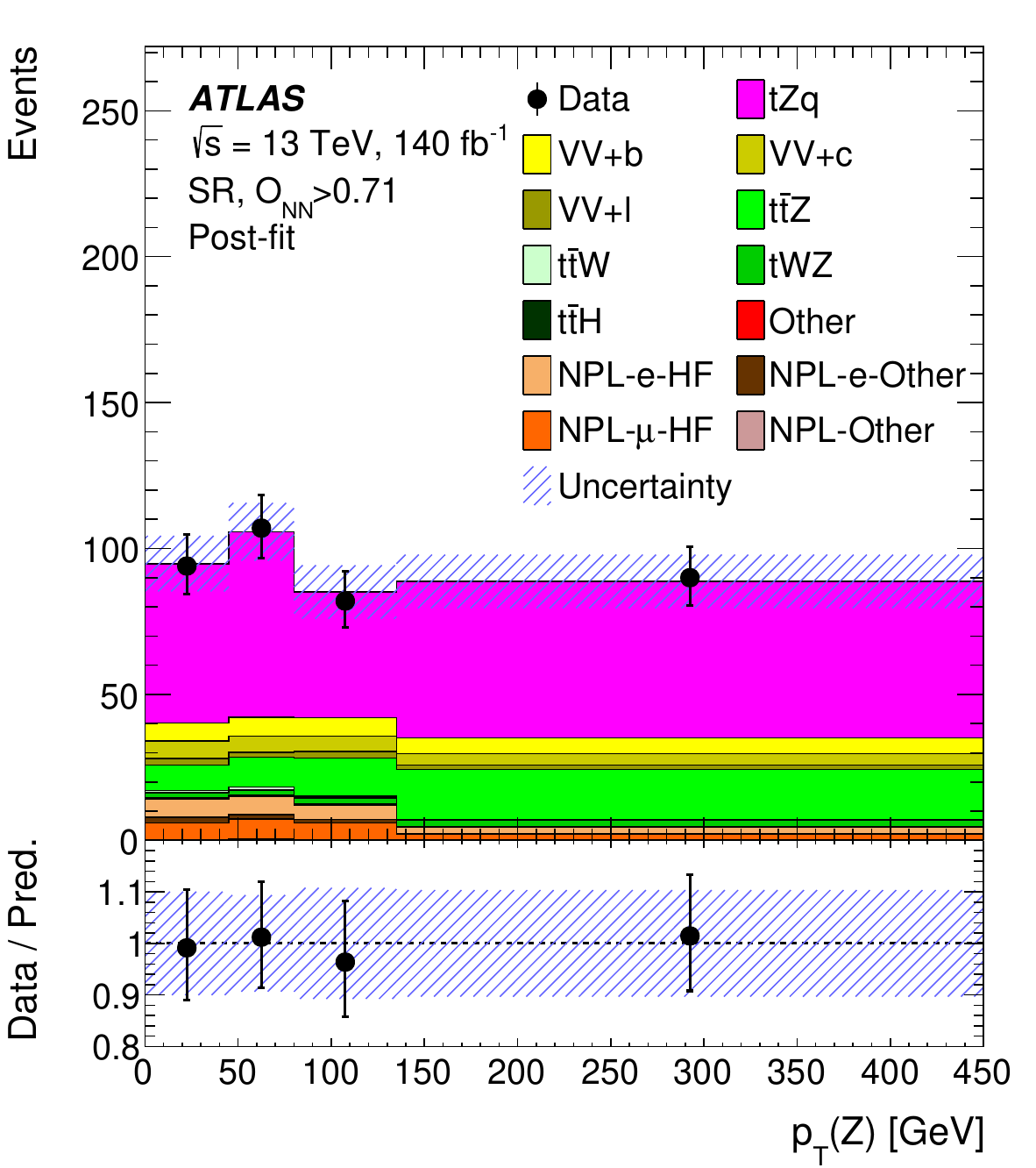}
\caption{}
\end{subfigure}
\begin{subfigure}{0.32\textwidth}
\centering
\includegraphics[width=0.95\textwidth]{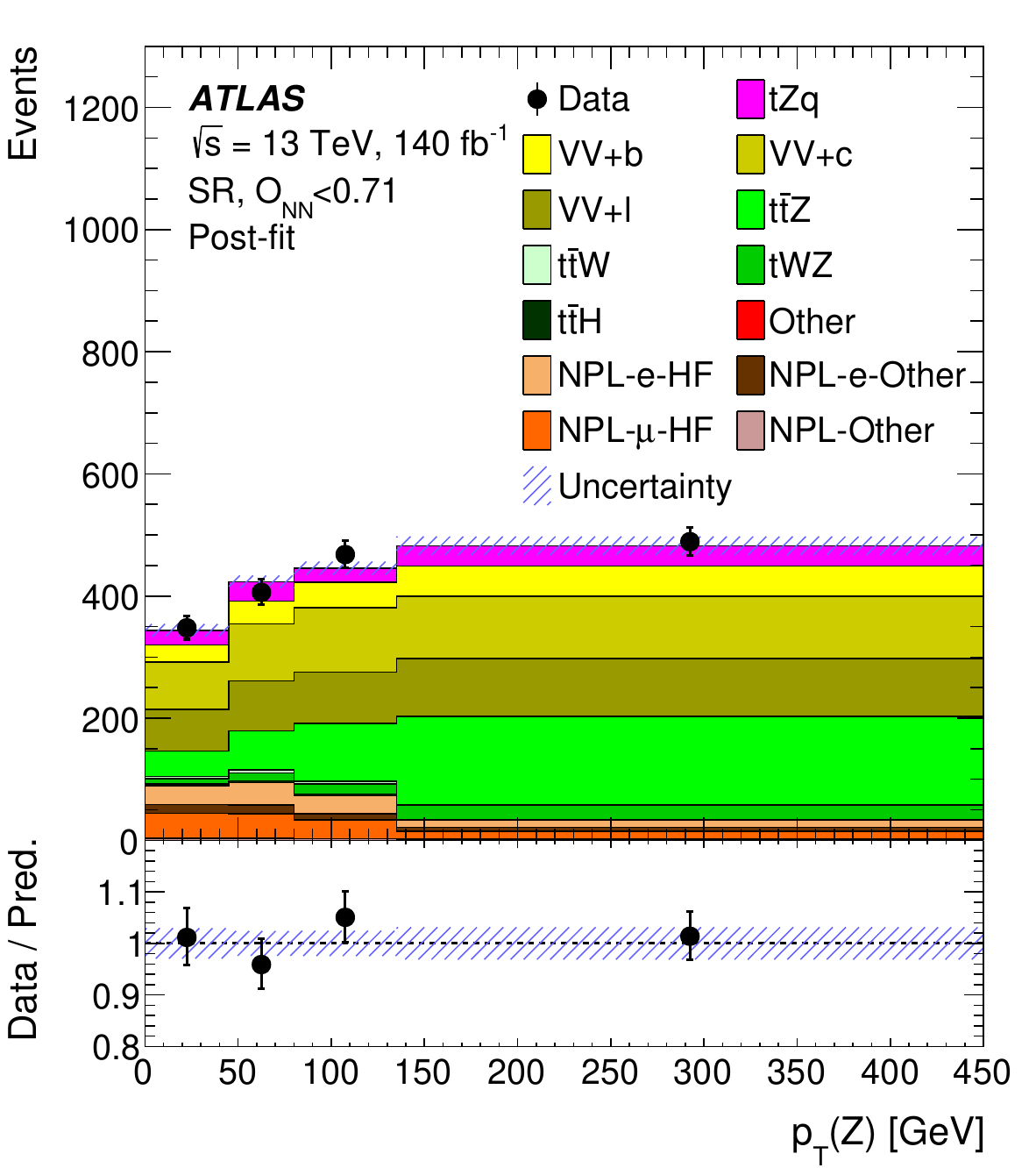}
\caption{}
\end{subfigure}
\begin{subfigure}{0.32\textwidth}
\centering
\includegraphics[width=0.95\textwidth]{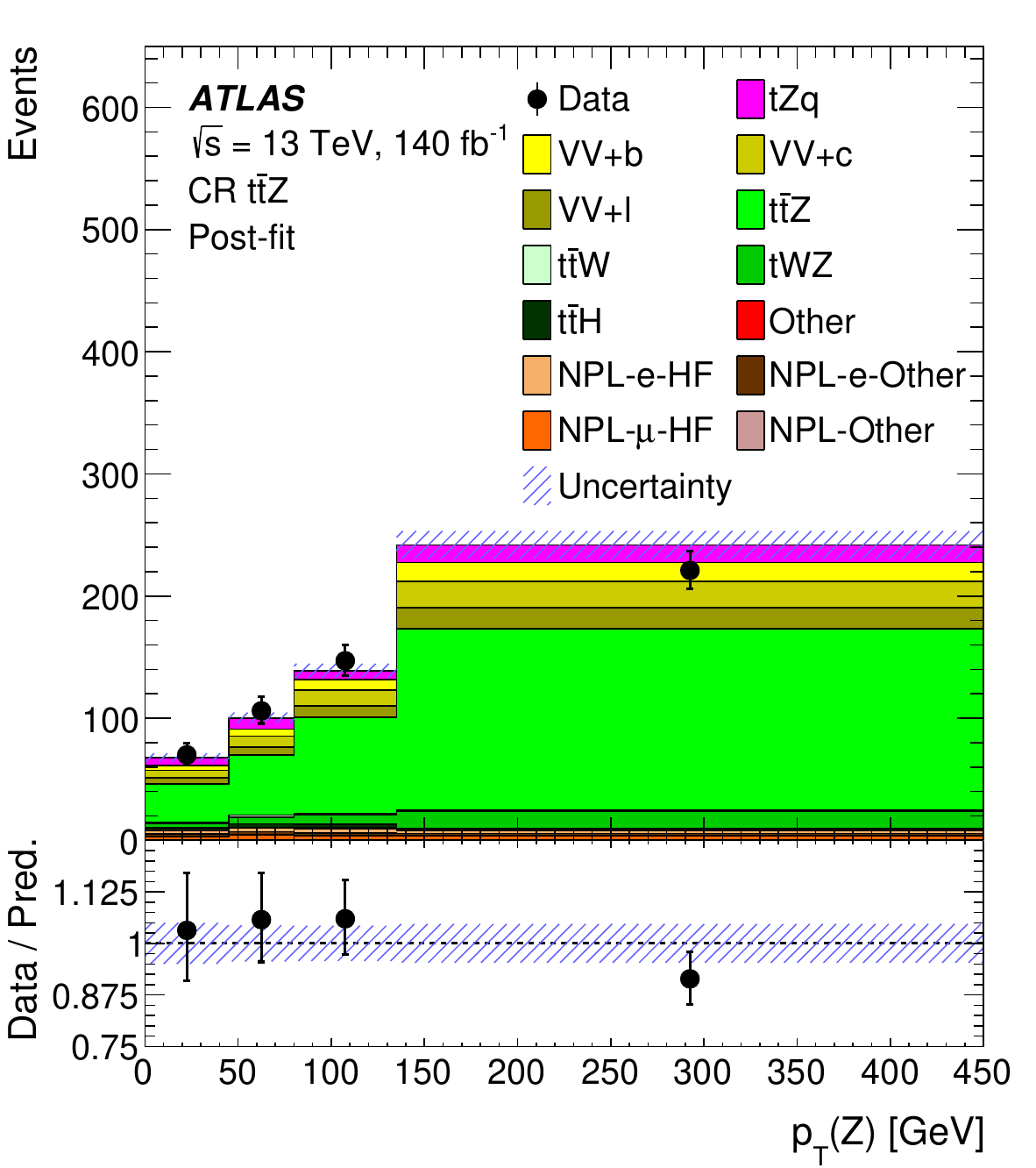}
\caption{}
\end{subfigure}
\begin{subfigure}{0.32\textwidth}
\centering
\includegraphics[width=0.95\textwidth]{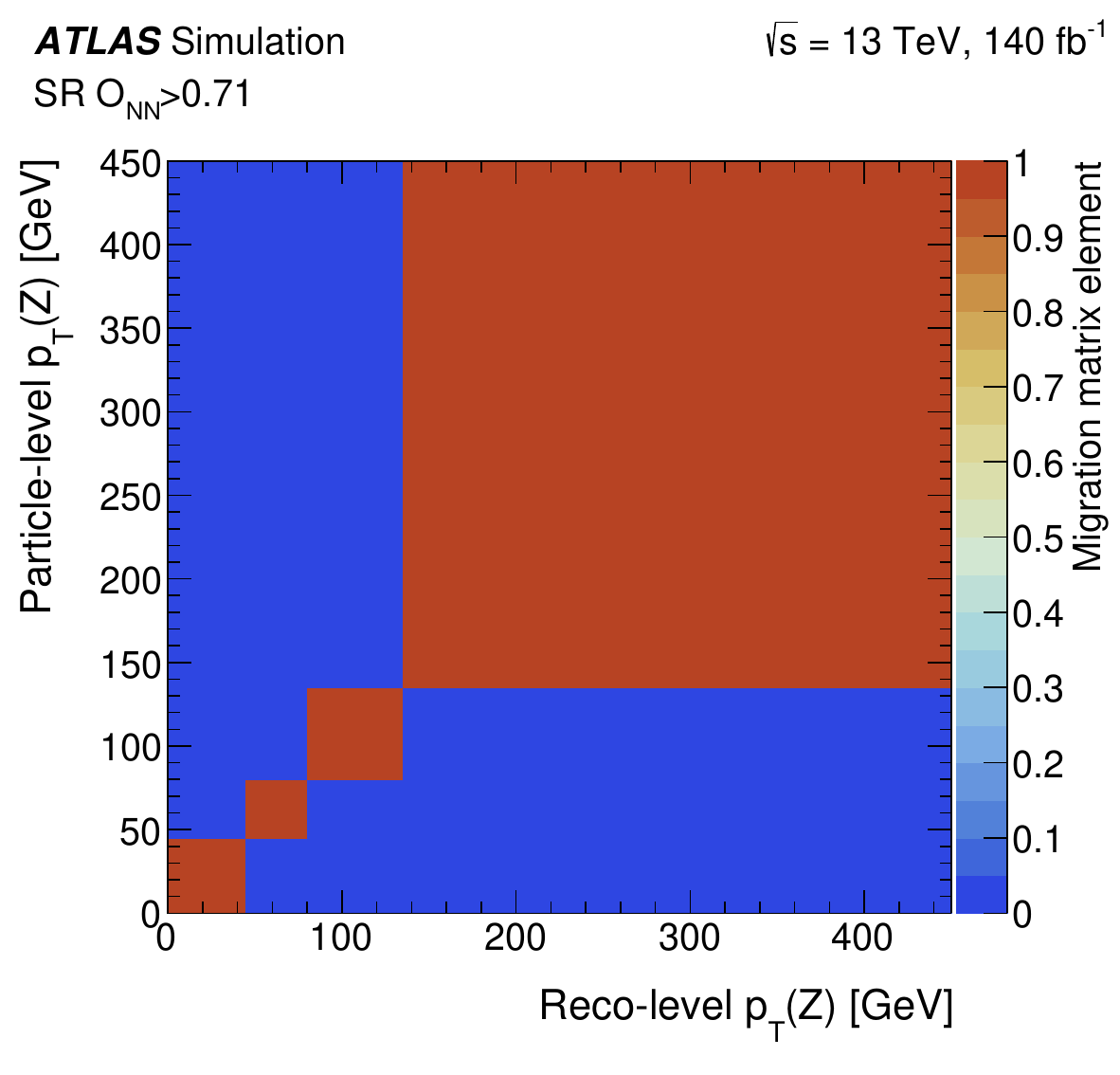}
\caption{}
\end{subfigure}
\begin{subfigure}{0.32\textwidth}
\centering
\includegraphics[width=0.95\textwidth]{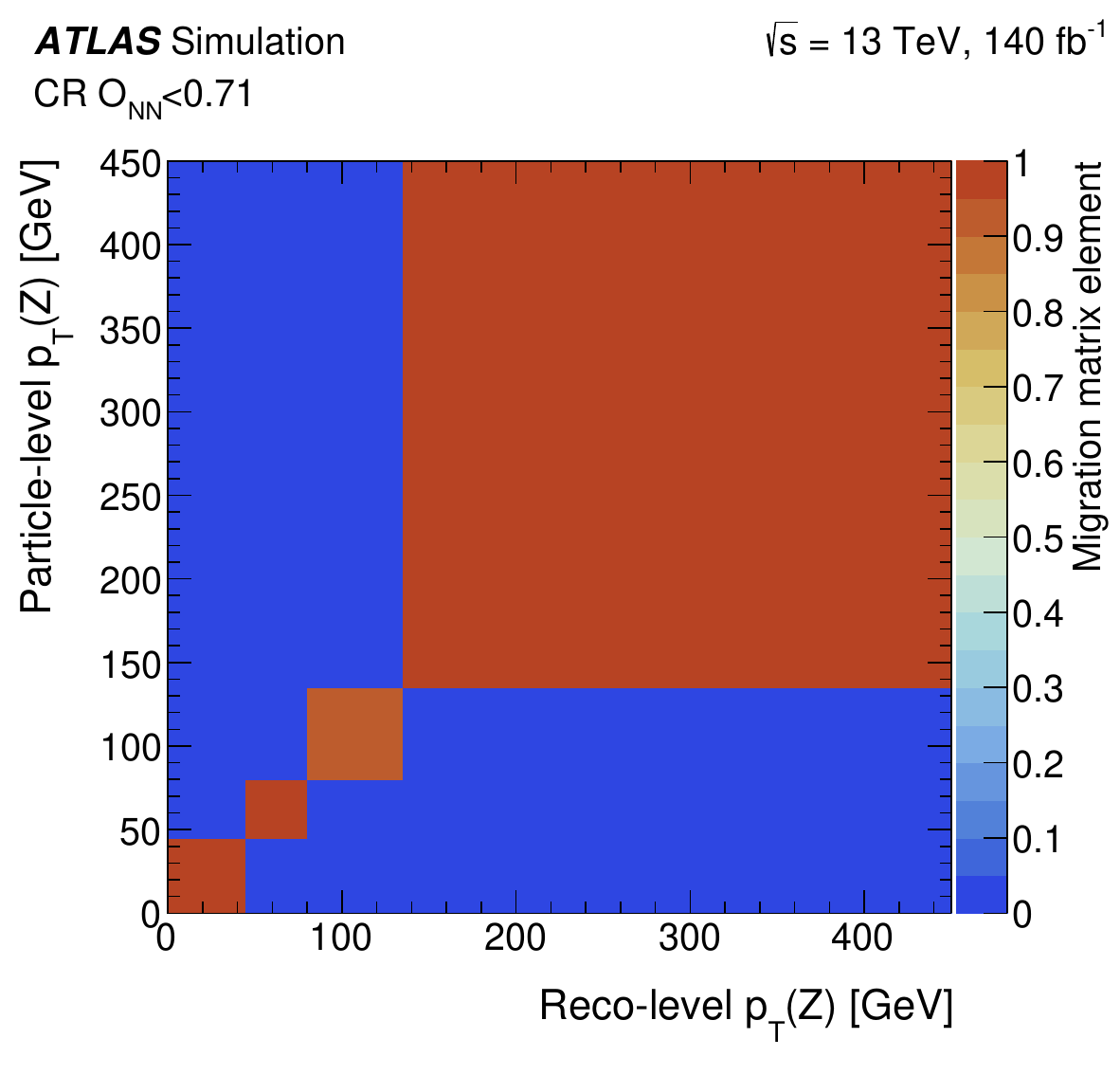}
\caption{}
\end{subfigure}
\begin{subfigure}{0.32\textwidth}
\centering
\includegraphics[width=0.95\textwidth]{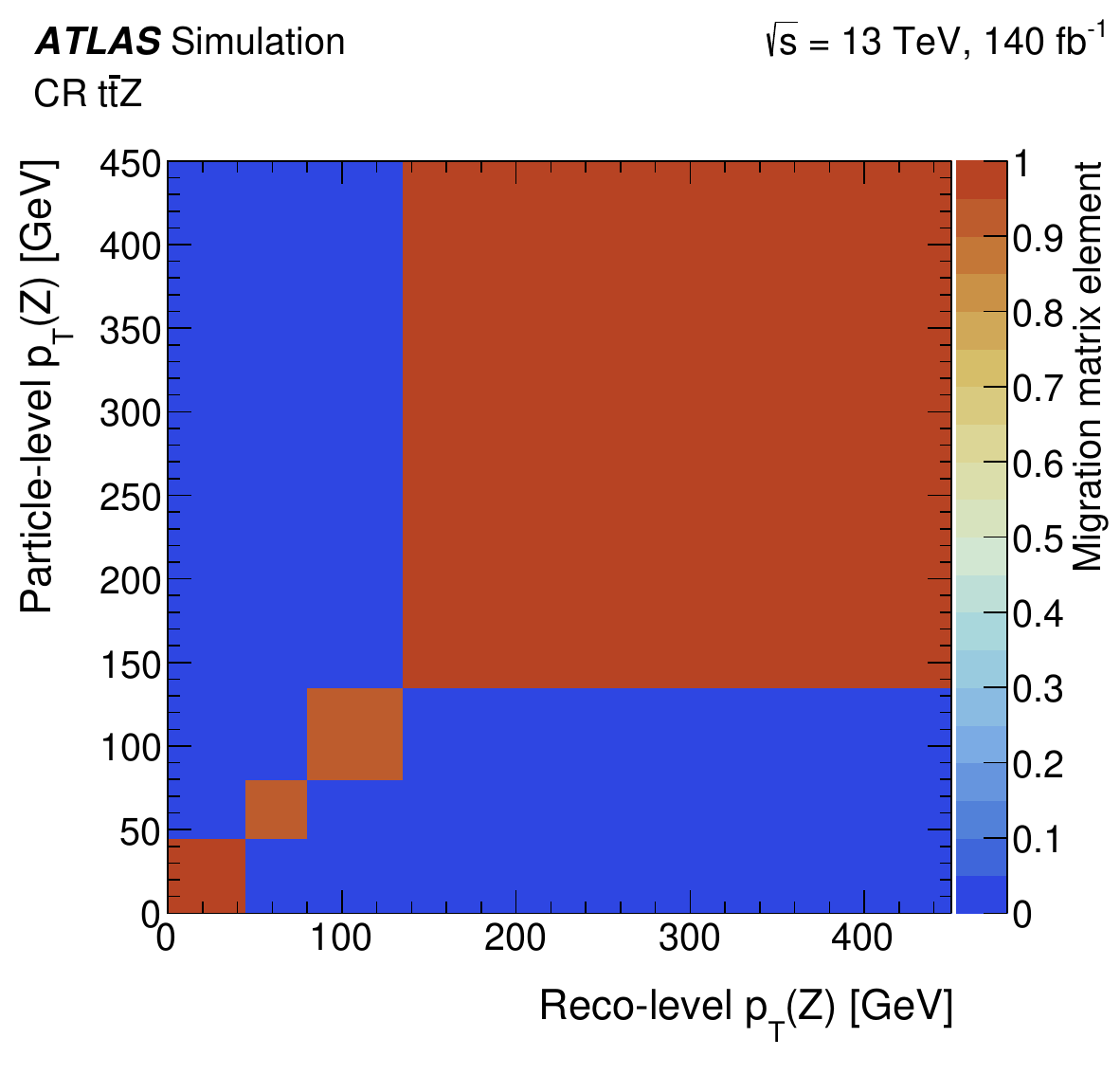}
\caption{}
\end{subfigure}
\begin{subfigure}{0.32\textwidth}
\centering
\includegraphics[width=0.95\textwidth]{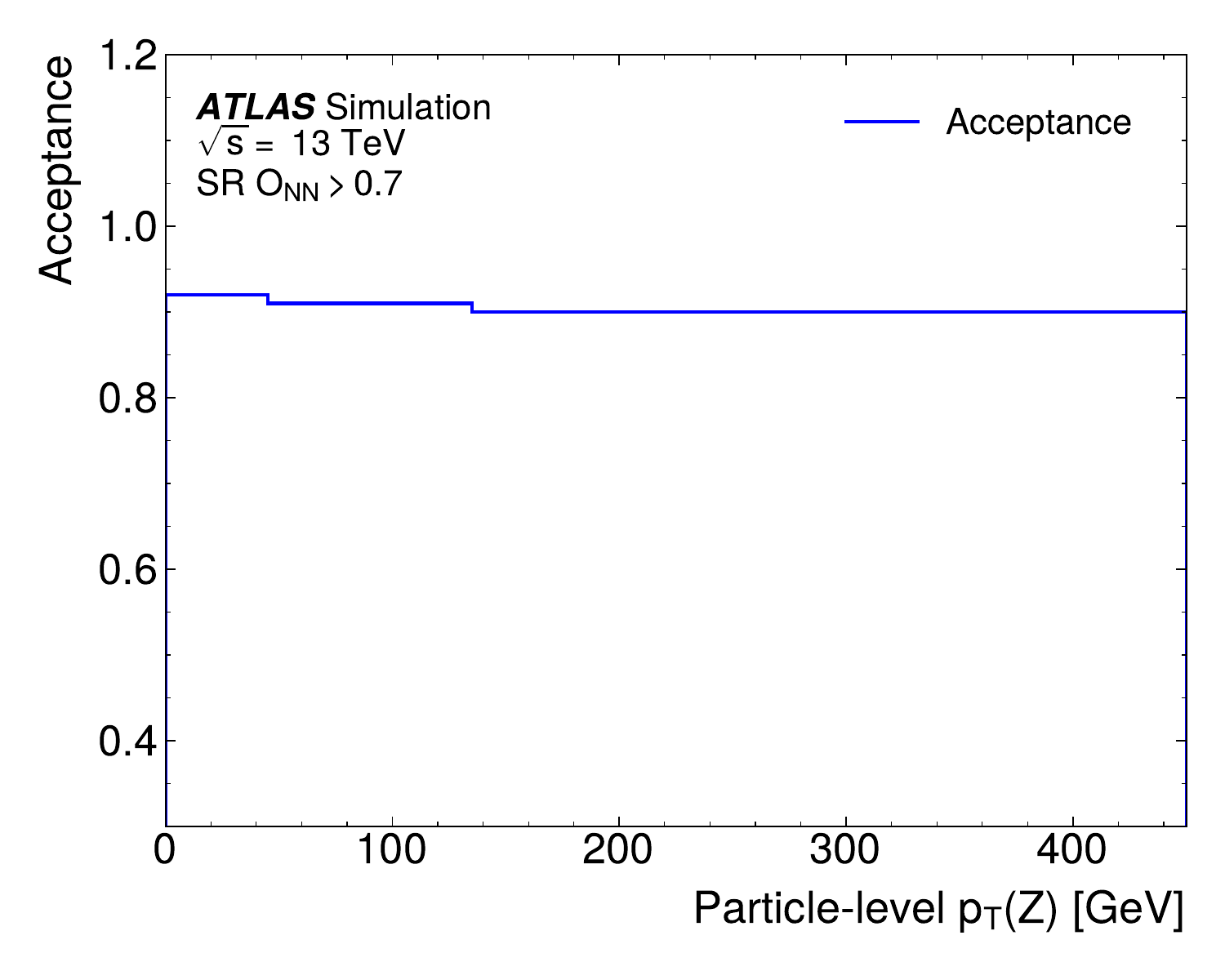}
\caption{}
\end{subfigure}
\begin{subfigure}{0.32\textwidth}
\centering
\includegraphics[width=0.95\textwidth]{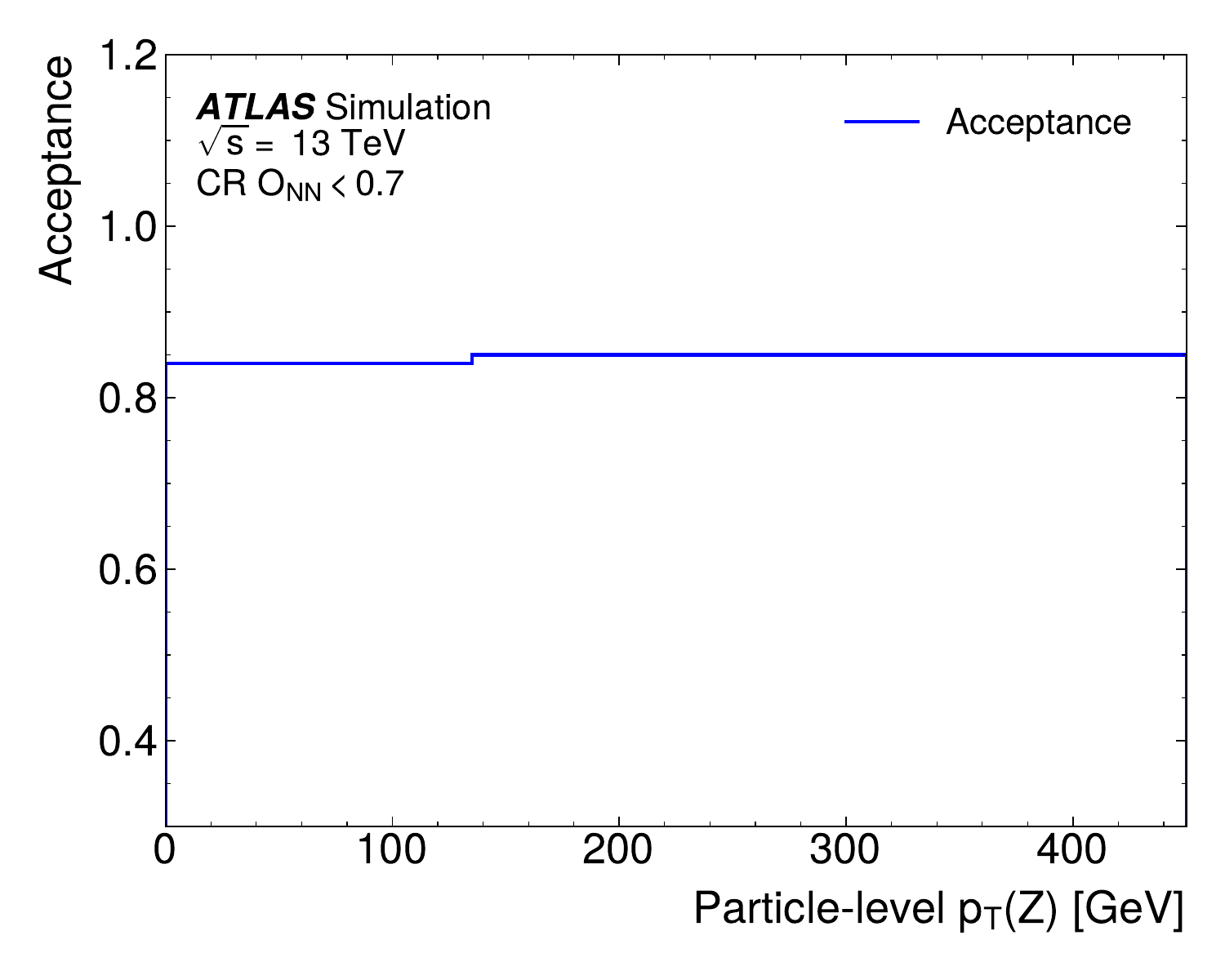}
\caption{}
\end{subfigure}
\begin{subfigure}{0.32\textwidth}
\centering
\includegraphics[width=0.95\textwidth]{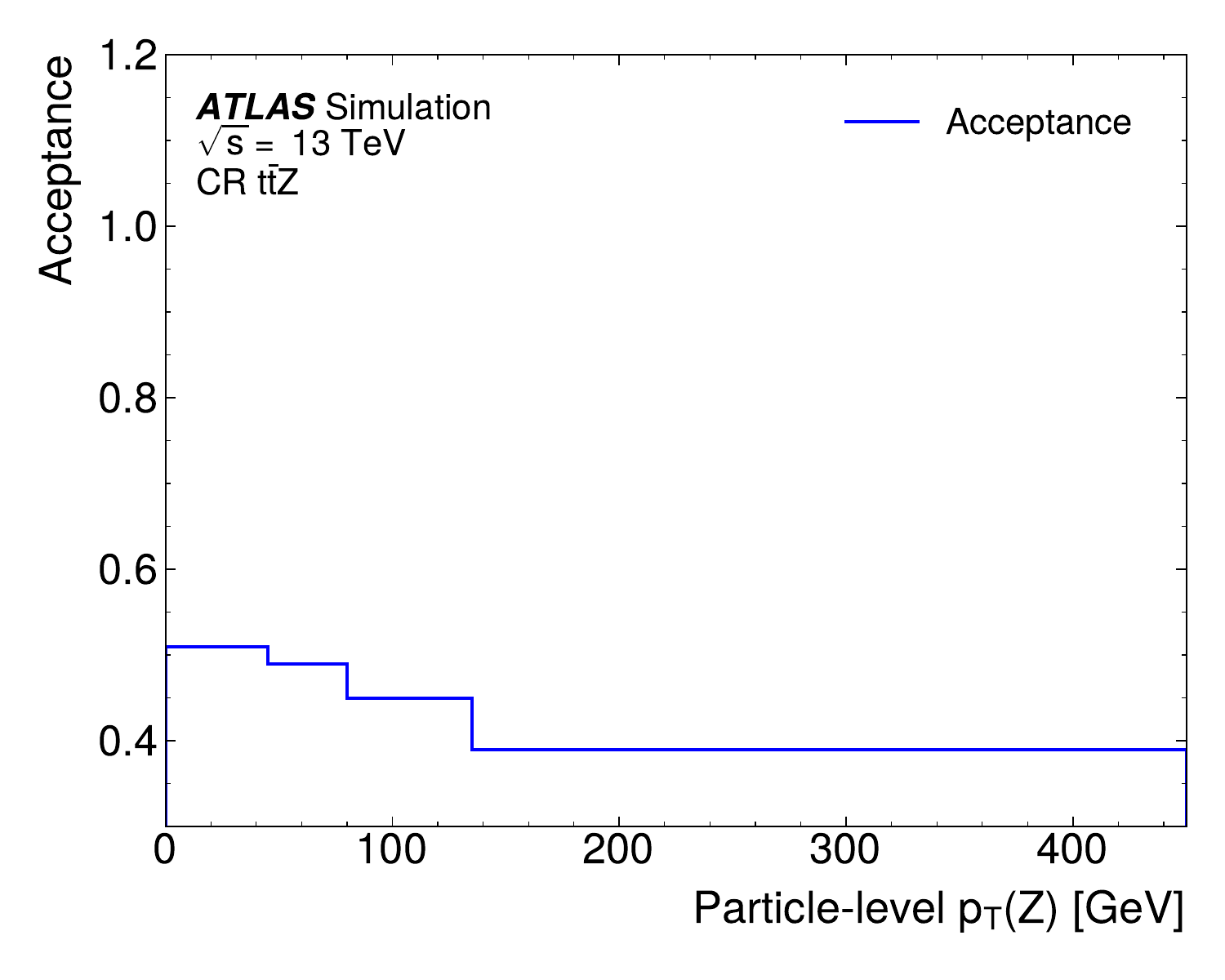}
\caption{}
\end{subfigure}
\begin{subfigure}{0.32\textwidth}
\centering
\includegraphics[width=0.95\textwidth]{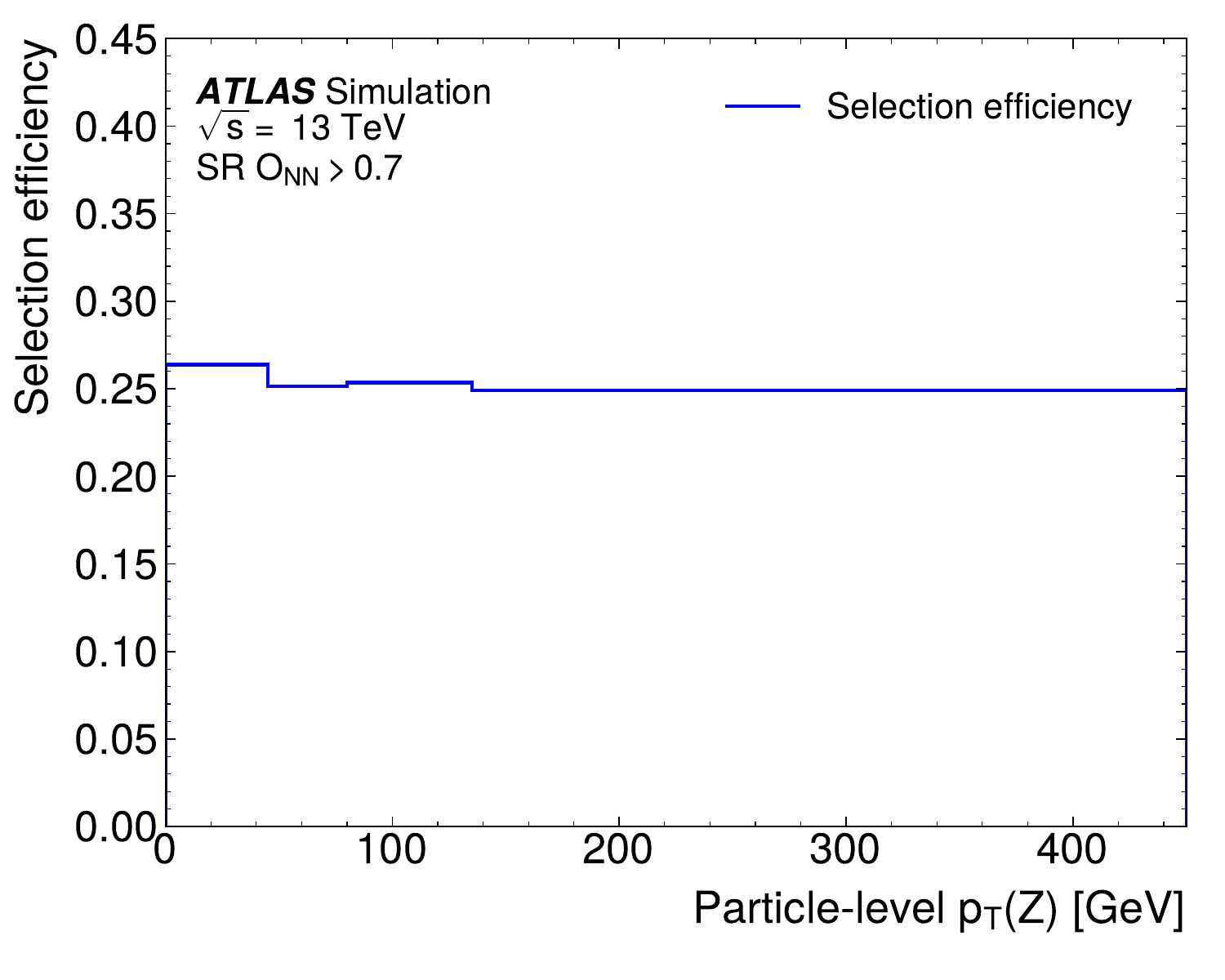}
\caption{}
\end{subfigure}
\begin{subfigure}{0.32\textwidth}
\centering
\includegraphics[width=0.95\textwidth]{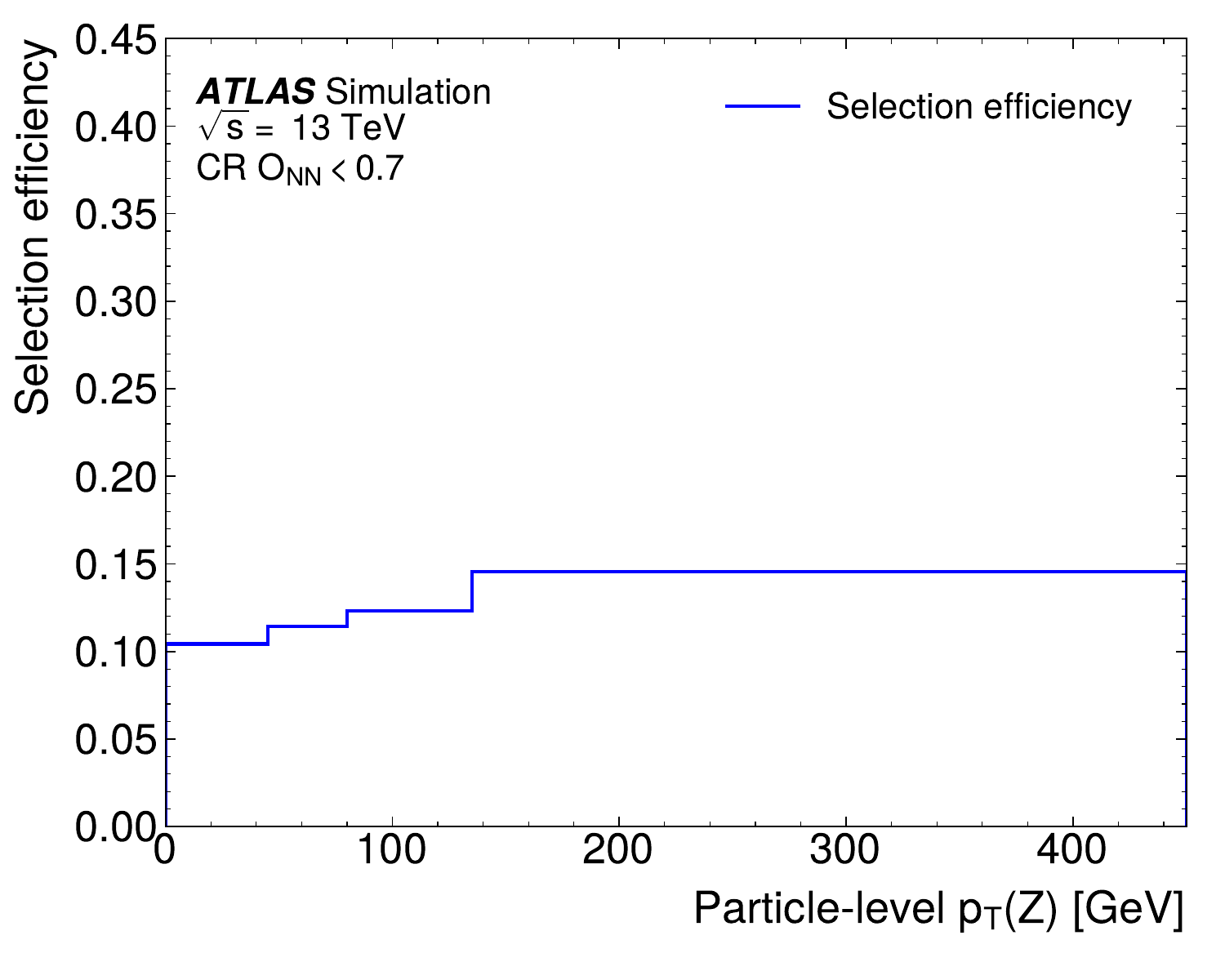}
\caption{}
\end{subfigure}
\begin{subfigure}{0.32\textwidth}
\centering
\includegraphics[width=0.95\textwidth]{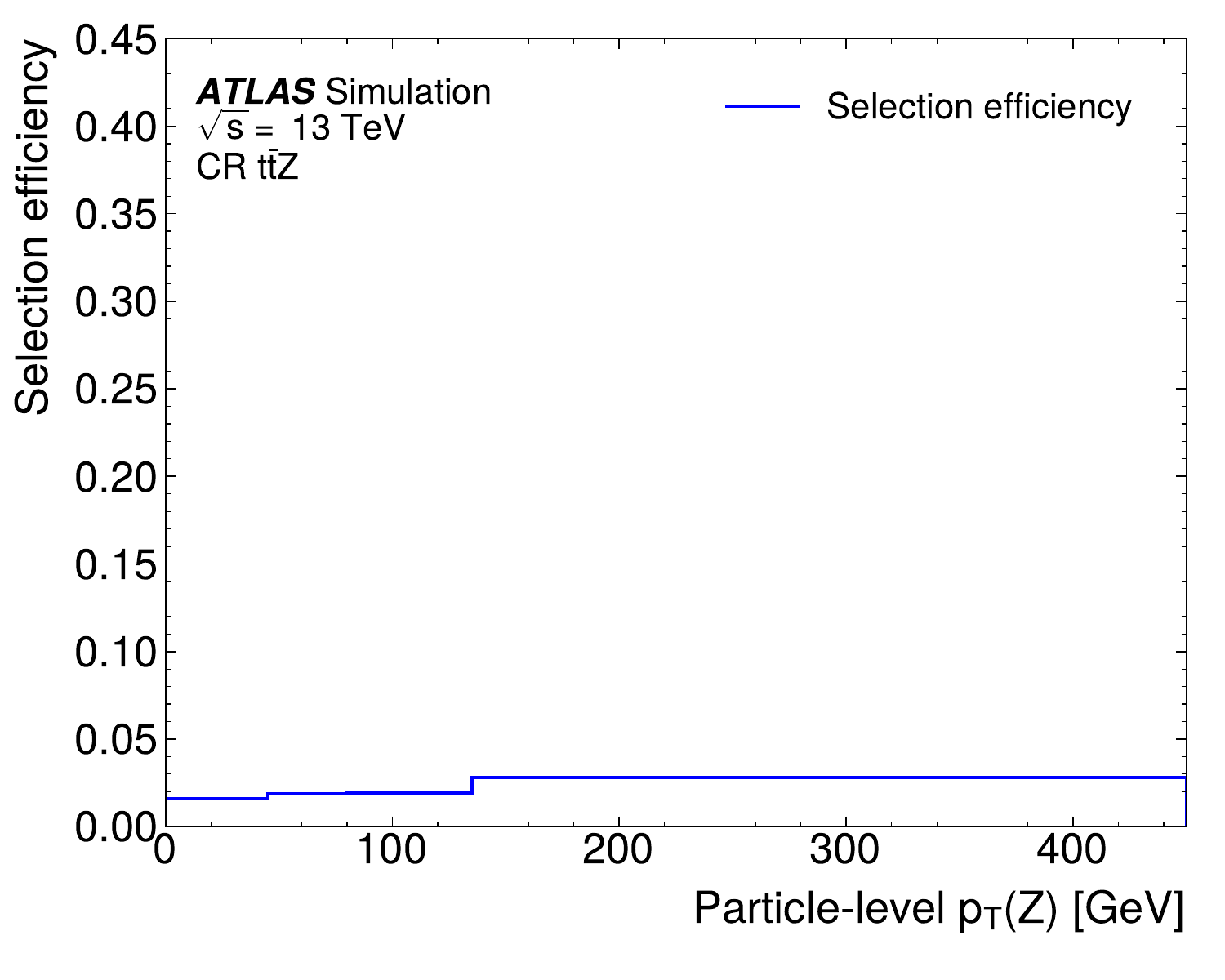}
\caption{}
\end{subfigure}
\caption{ (a, b, c) Detector-level distributions,
together with (d, e, f) migration matrices,
(g, h, i) acceptance, and (j, k, l) selection efficiency  histograms for the \pTZ observable
in the (a, d, g, j) high-\ONN score region of the SR ,
in the (b, e, h, k) low-\ONN score region of the SR, and
in the (c, f, i, l) \CRttZ.
Migration matrices and corrections apply to the particle level.}
\label{fig:diff:mig:Zpt:particle}
\end{figure}

The relation between truth-level and detector-level observables is described by response matrices derived from simulated events, separately for the parton-level and particle-level definitions. These matrices encode the effects of detector resolution and bin-to-bin migrations. Corrections for detector efficiency and acceptance are included through factors derived from simulation, accounting for the probability for a truth-level event to be reconstructed and for the fraction of reconstructed events originating from the fiducial phase space. Representative detector-level distributions, migration matrices, and the corresponding acceptance and selection efficiency corrections are shown in \cref{fig:diff:mig:Zpt:particle}.

Background contributions are included in the likelihood and constrained using dedicated control regions, while systematic uncertainties are incorporated as nuisance parameters and profiled in the fit.

The variables used for the differential measurements at parton level and particle level are listed in \cref{tab:differential:vars}. These observables probe the kinematics of the \tZq system and are sensitive to its production dynamics and top-quark polarisation.

\begin{table}[htbp]
\tablesetup[1.3]
\caption{List of unfolded variables.}
\centering
\begin{tabular}{l l p{8cm}}
\toprule
Variable & Parton/particle/both &  Definition \\
\midrule
\pTZ & both & Transverse momentum of the reconstructed \PZ boson \\
\etaZ & both & Pseudorapidity of the reconstructed \PZ boson \\
\pTt & both & Transverse momentum of the reconstructed top quark \\
\etat & both & Pseudorapidity of the reconstructed top quark \\
\pTsjet & particle & Transverse momentum of the spectator jet \\
\etasjet & particle & Pseudorapidity of the spectator jet \\
\dRtZ & both & Angular distance between the top quark and the \PZ boson \\
\dRtsjet & particle & Angular distance between the top quark and the spectator jet \\
\dphiZl & both & Azimuthal angular distance between the two leptons originating from the \PZ boson \\
\cosTheta & both & Cosine of the top-quark polarisation angle, defined as the angle between the spectator quark or jet and the lepton from the top-quark decay in the top-quark rest frame \\
\mlll & both & Invariant mass of the trilepton system \\
\bottomrule
\end{tabular}
\label{tab:differential:vars}
\end{table}

The profile-likelihood fit is performed simultaneously in two signal regions and multiple control regions. The \ONN is used to define regions of different signal purity in the SR. Events are categorised according to the NN score into a signal-enriched region (with $\ONN > 0.71$) and a signal-depleted region (with $\ONN < 0.71$). The threshold of 0.71 was chosen by maximising the figure of merit $s/\sqrt{s+b}$, where $s$ and $b$ denote the expected signal and background yields in the signal-enriched region. The signal-enriched region provides enhanced sensitivity to the signal, while the signal-depleted region contributes to the constraint of the background normalisations. In addition, the signal contribution in the \CRttZ is unfolded simultaneously using the same observable as in the signal regions, while the remaining control regions are used to constrain the background contributions, in particular those from non-prompt leptons and diboson processes. An additional normalisation uncertainty is assigned to the \tZq contribution in the NPL and diboson CRs, corresponding to the uncertainty on the inclusive cross-section measurement. The impact of this uncertainty on the unfolded results is negligible, reflecting both the small \tZq contribution in these regions and the fact that they are not used as unfolding regions.

The binning is optimised to balance statistical precision and bin-to-bin migration effects, to ensure a similar sensitivity across bins and a minimum of \qty{60}{\%} in the diagonal elements of the migration matrix,  preserving the shape of the distributions.
The final binning is summarised in \cref{tab:differential:bins}.

\begin{table}[htbp]
\tablesetup[1.3]
\caption{Optimised binning for the unfolded observables. For \pTZ, \pTt, \pTsjet and \mlll, the last bin includes overflow events.}
\centering
\begin{tabular}{l c l}
\toprule
Variable & Number of bins & Binning \\
\midrule
\pTZ      & 4 & [0, 45, 80, 135, 450] $\si{\giga\electronvolt}$ \\
\etaZ     & 4 & [0, 0.70, 1.40, 2.10, 4.00] \\
\pTt      & 3 & [0, 95, 190, 450] $\si{\giga\electronvolt}$ \\
\etat     & 3 & [0, 0.70, 1.70, 4.00] \\
\pTsjet   & 4 & [0, 65, 100, 170, 400] $\si{\giga\electronvolt}$ \\
\etasjet  & 4 & [0, 2.00, 3.00, 3.50, 4.50] \\
\dRtZ     & 3 & [0, 2.60, 3.50, 7.00] \\
\dRtsjet  & 4 & [0, 3.40, 4.20, 5.20, 7.00] \\
\dphiZl   & 4 & [0, 0.40, 0.60, 0.80, 1.00] \\
\cosTheta & 3 & [-1.00, 0.00, 0.58, 1.00] \\
\mlll     & 4 & [0, 150, 210, 390, 600] $\si{\giga\electronvolt}$ \\
\bottomrule
\end{tabular}
\label{tab:differential:bins}
\end{table}

\begin{figure}[htbp]
\centering
\begin{subfigure}{0.45\textwidth}
\centering
\includegraphics[width=1\textwidth]{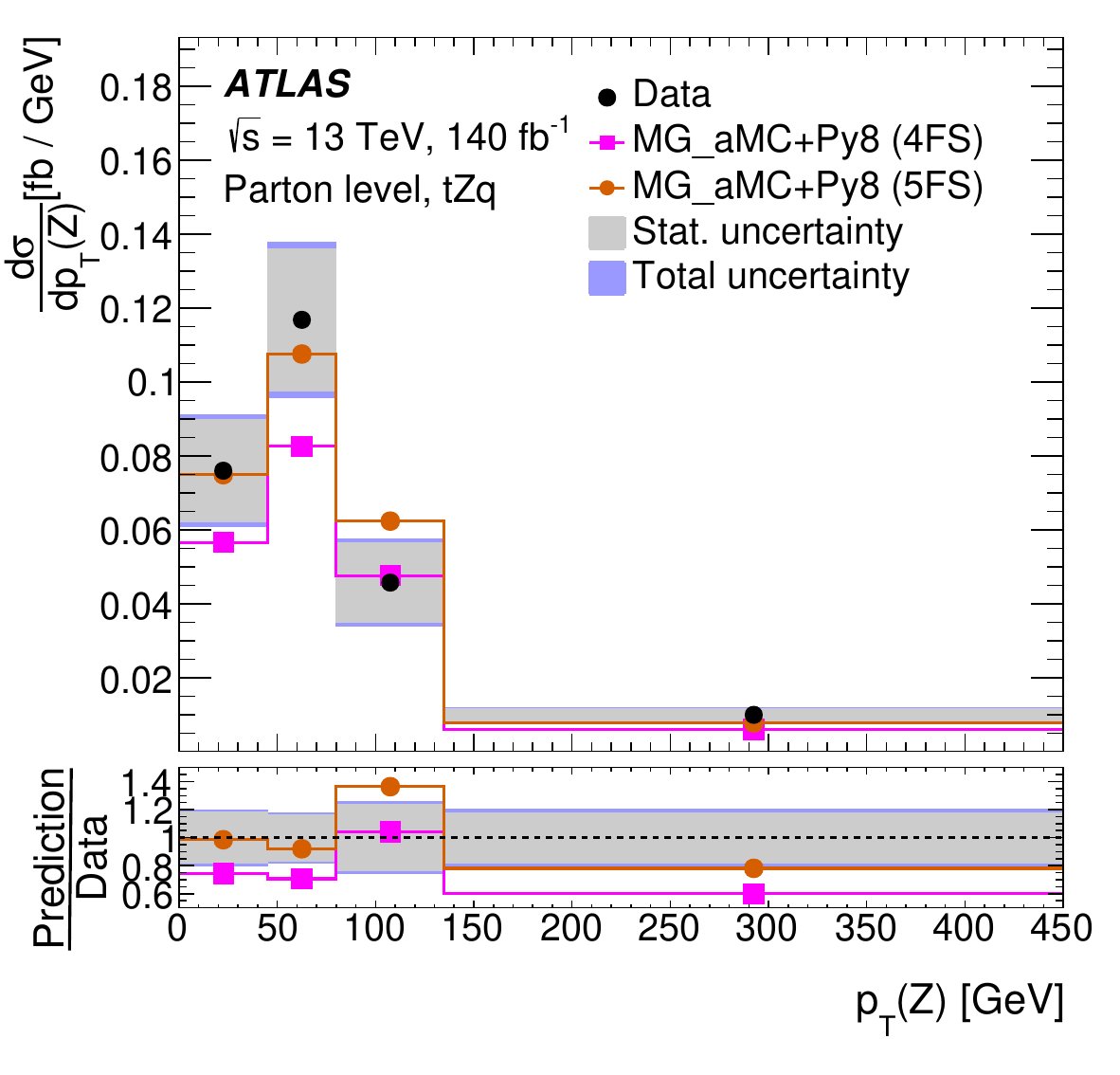}
\caption{}
\end{subfigure}
\begin{subfigure}{0.45\textwidth}
\centering
\includegraphics[width=1\textwidth]{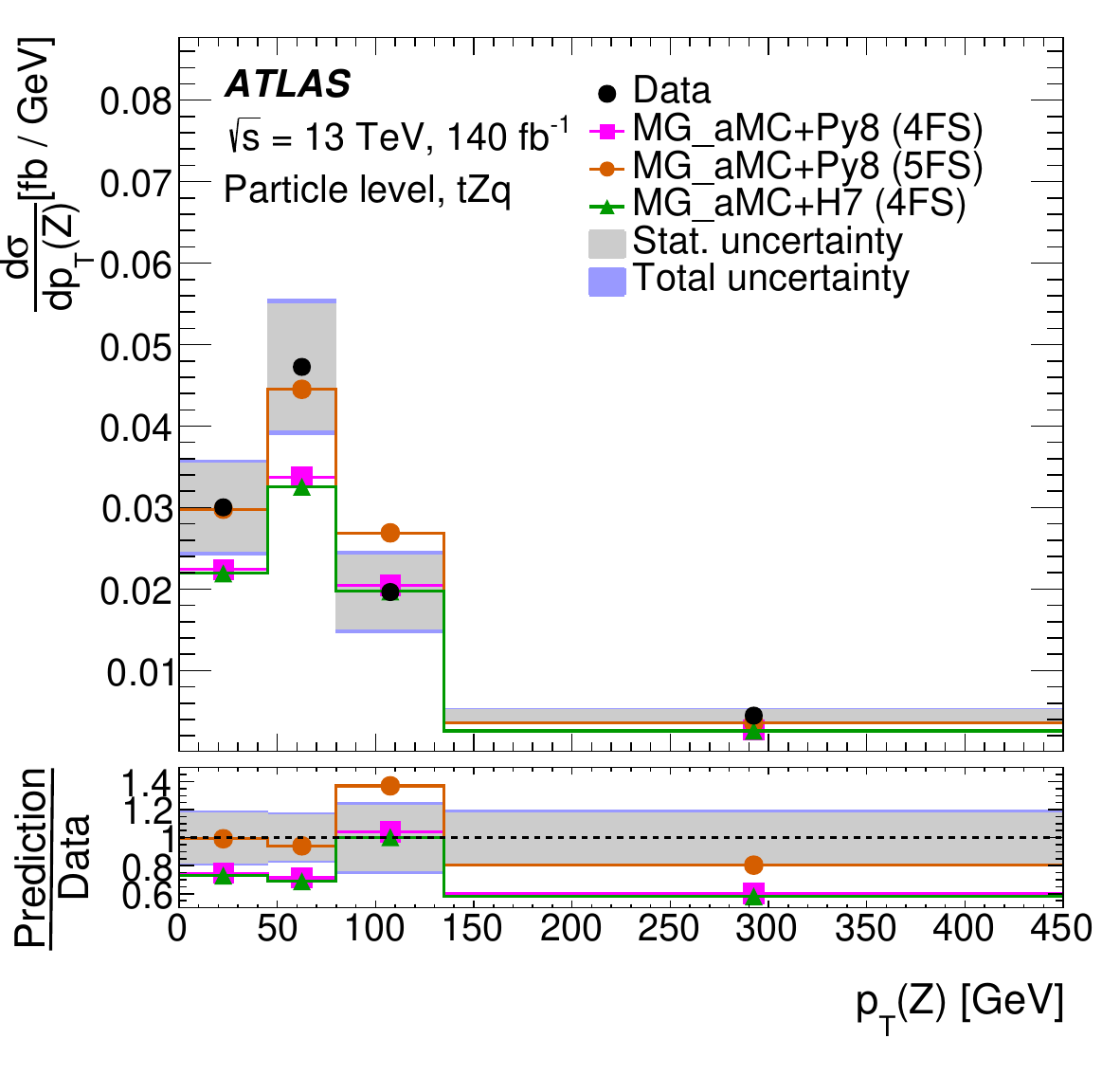}
\caption{}
\end{subfigure}
\begin{subfigure}{0.45\textwidth}
\centering
\includegraphics[width=1\textwidth]{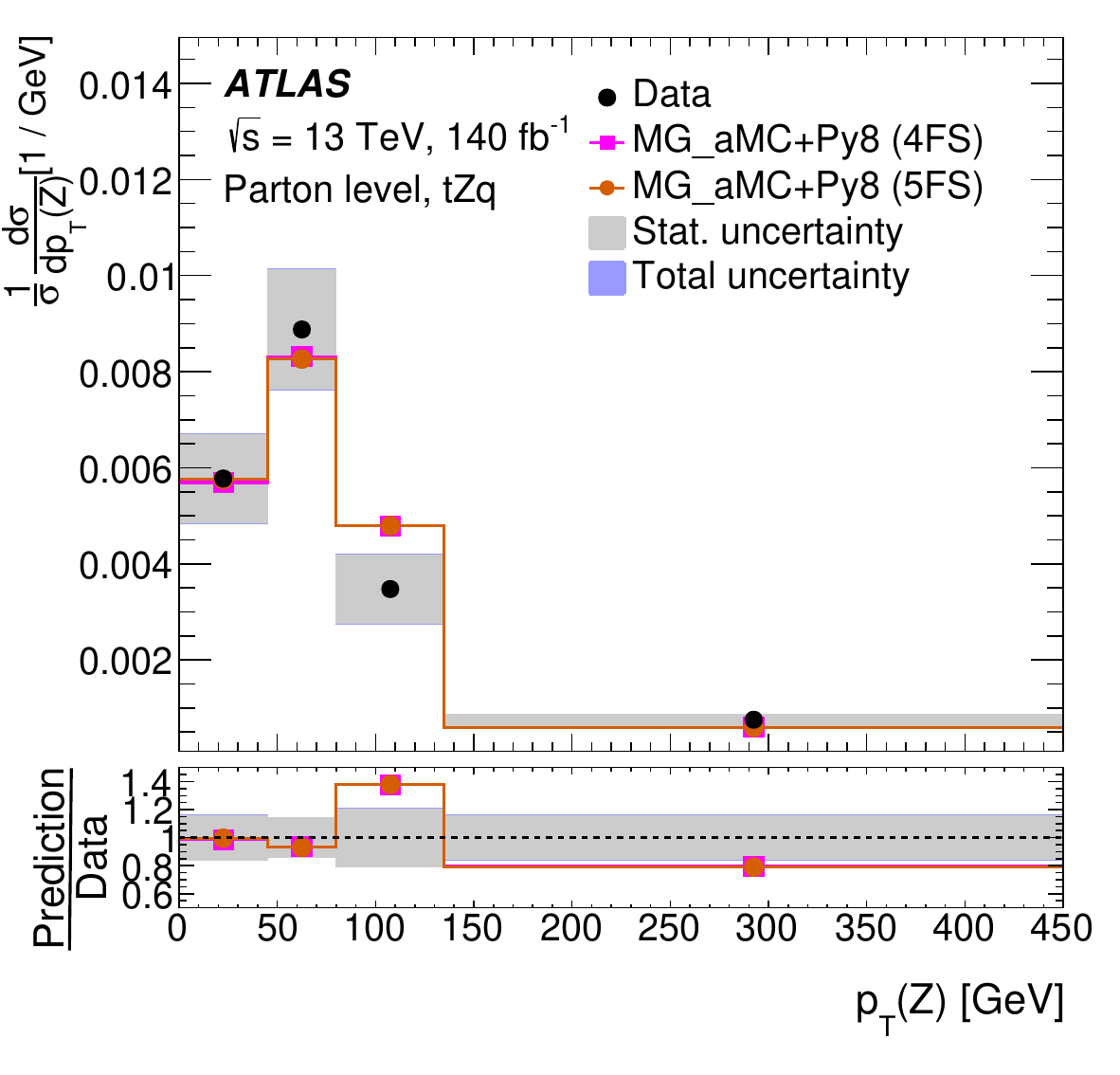}
\caption{}
\end{subfigure}
\begin{subfigure}{0.45\textwidth}
\centering
\includegraphics[width=1\textwidth]{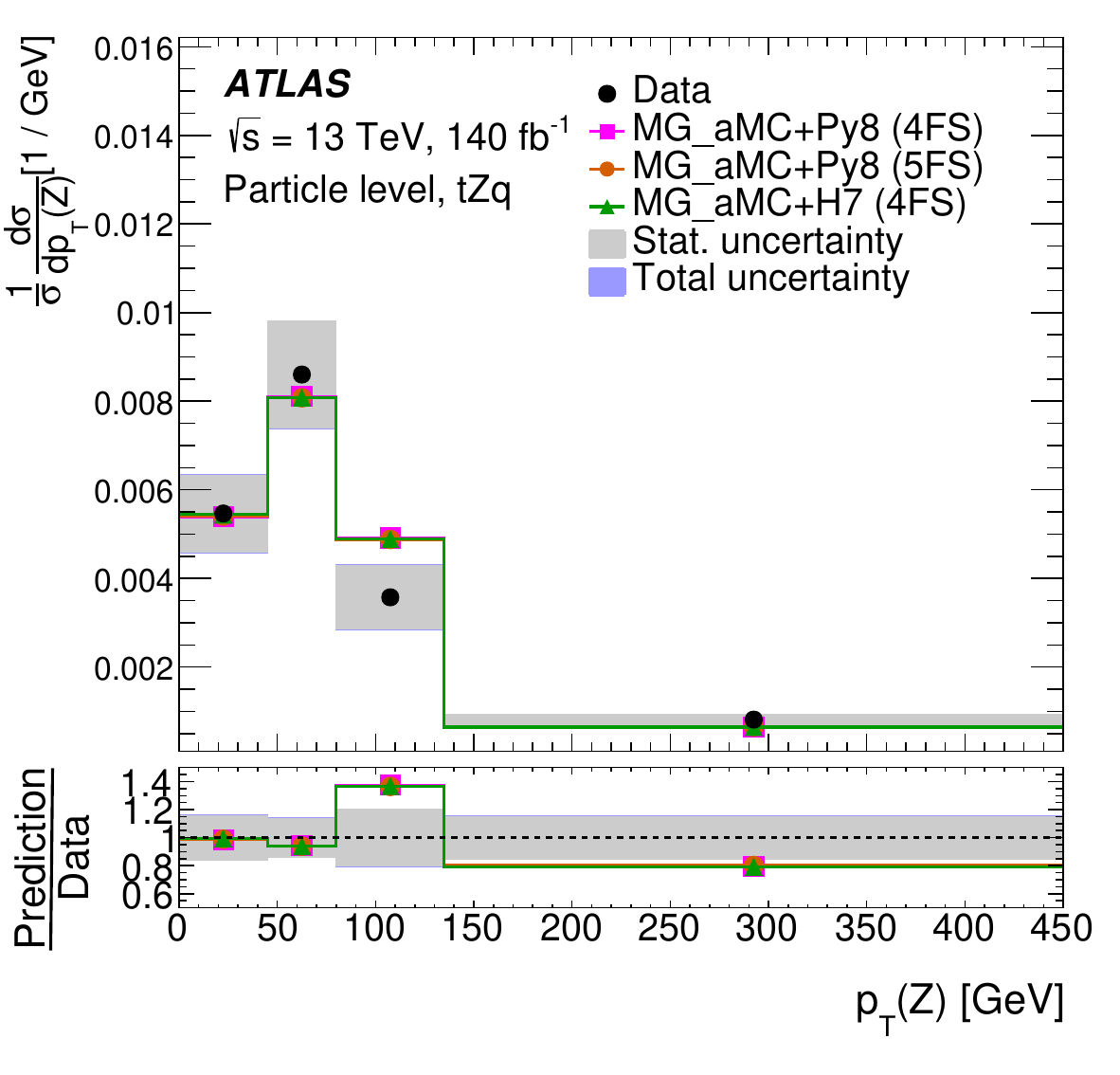}
\caption{}
\end{subfigure}
\caption{Differential cross-section measurements as a function of \pTZ observable in the SR, (a, b) absolute and (c, d) normalised, unfolded to (a, c) parton level and (b, d) particle level.
The light blue band corresponds to the total uncertainty of the measurement; in some cases, it is almost fully covered by the light grey band, representing the dominant statistical uncertainty.
The measurements are compared with the generator predictions indicated in the legend and described in the text.
The lower panels show the ratios of the predictions to the (a, b) measured cross-sections and the (c, d) measured normalised differential cross-sections.
}
\label{fig:diff:Zpt}
\end{figure}

\begin{figure}[htbp]
\centering
\begin{subfigure}{0.45\textwidth}
\centering
\includegraphics[width=1\textwidth]{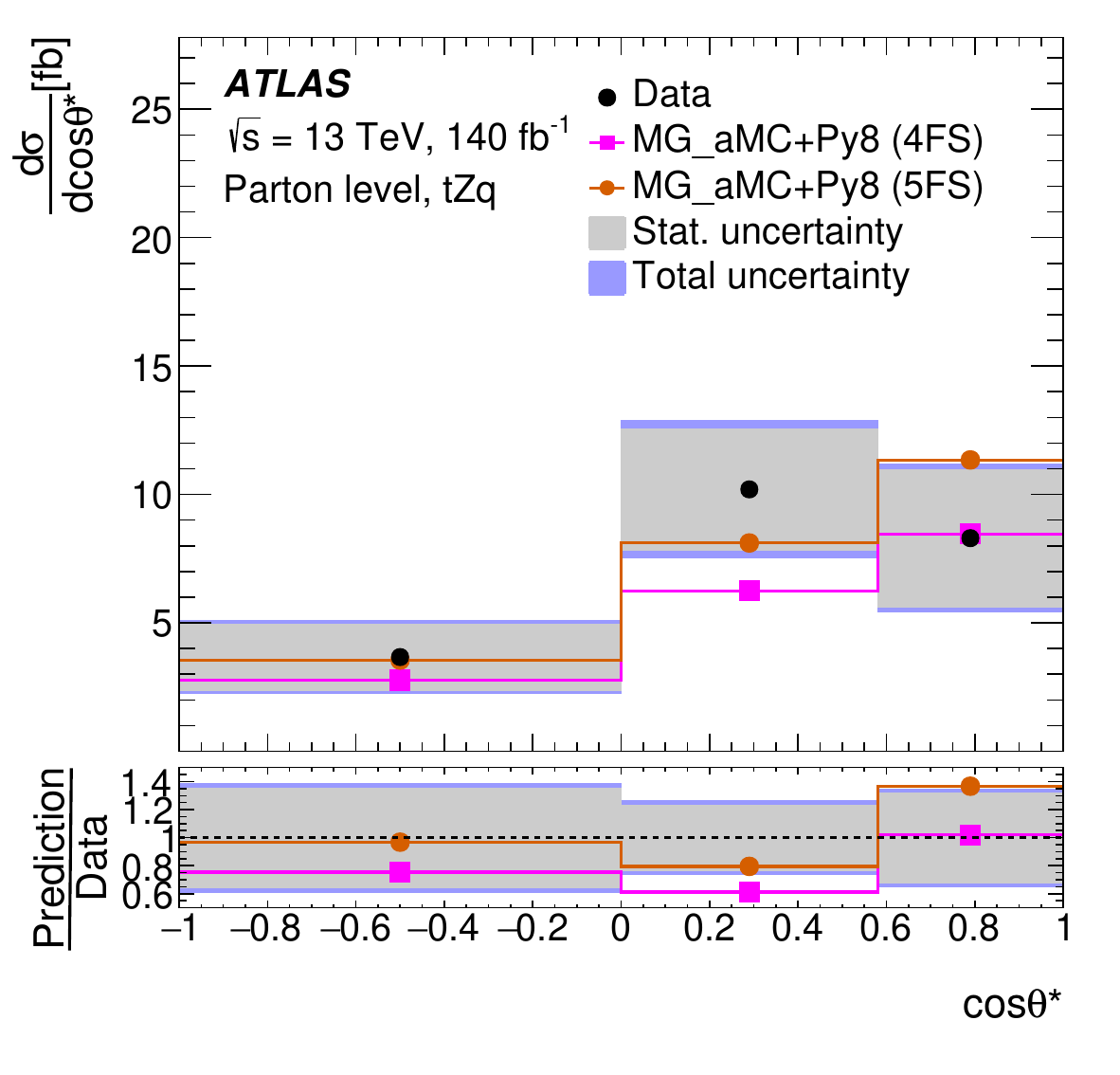}
\caption{}
\end{subfigure}
\begin{subfigure}{0.45\textwidth}
\centering
\includegraphics[width=1\textwidth]{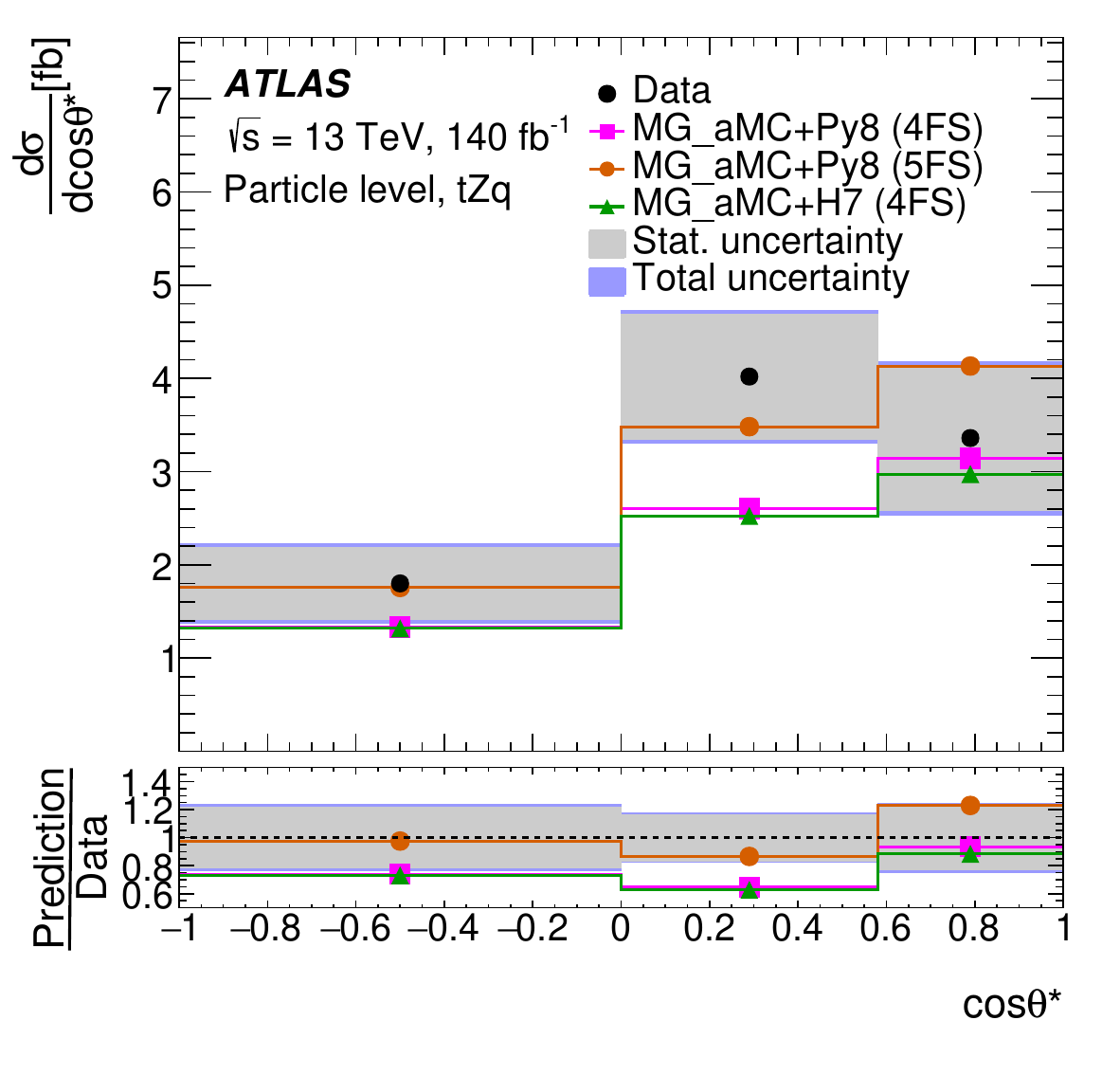}
\caption{}
\end{subfigure}
\begin{subfigure}{0.45\textwidth}
\centering
\includegraphics[width=1\textwidth]{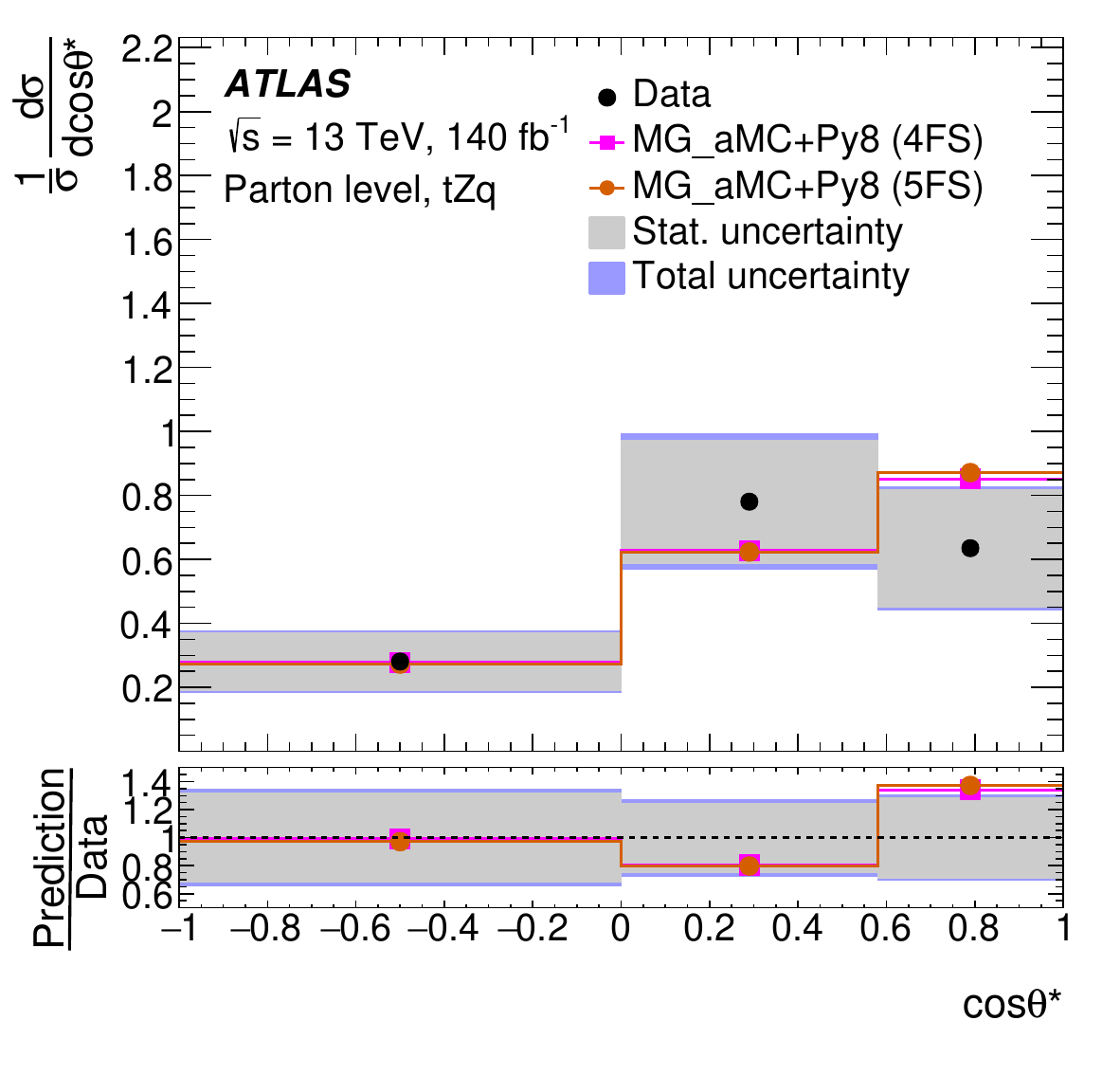}
\caption{}
\end{subfigure}
\begin{subfigure}{0.45\textwidth}
\centering
\includegraphics[width=1\textwidth]{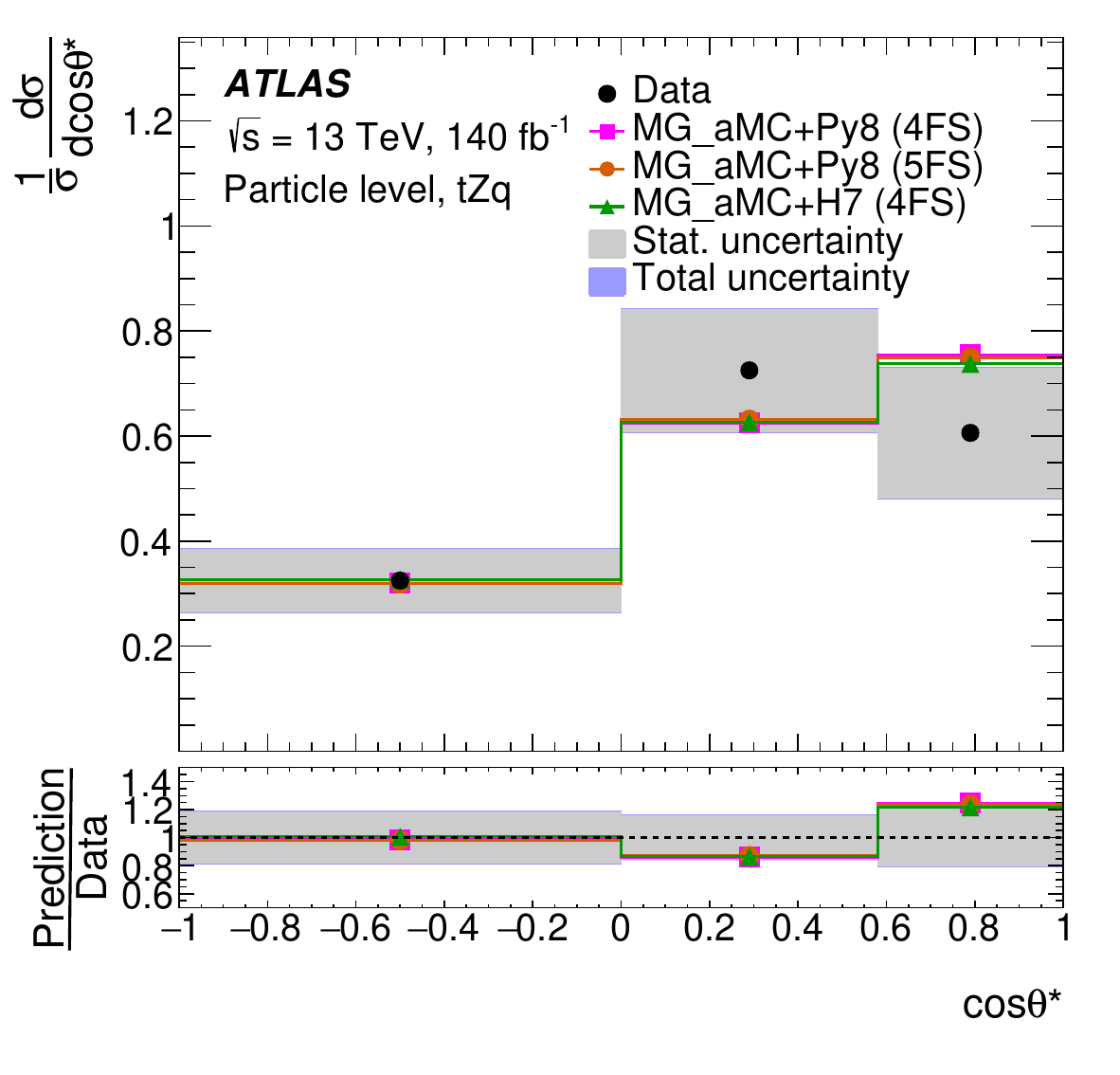}
\caption{}
\end{subfigure}
\caption{Differential cross-section measurements as a function of \cosTheta observable in the SR,
(a, b) absolute and (c, d) normalised, unfolded to (a, c) parton level and (b, d) particle level.
The light blue band corresponds to the total uncertainty of the measurement; in some cases, it is almost fully covered by the light grey band, representing the dominant statistical uncertainty.
The measurements are compared with the generator predictions indicated in the legend and described in the text.
The lower panels show the ratios of the predictions to the (a, b) measured cross-sections and the (c, d) measured normalised differential cross-sections.
}
\label{fig:diff:cosTheta}
\end{figure}

\begin{figure}[htbp]
\centering
\begin{subfigure}{0.45\textwidth}
\centering
\includegraphics[width=1\textwidth]{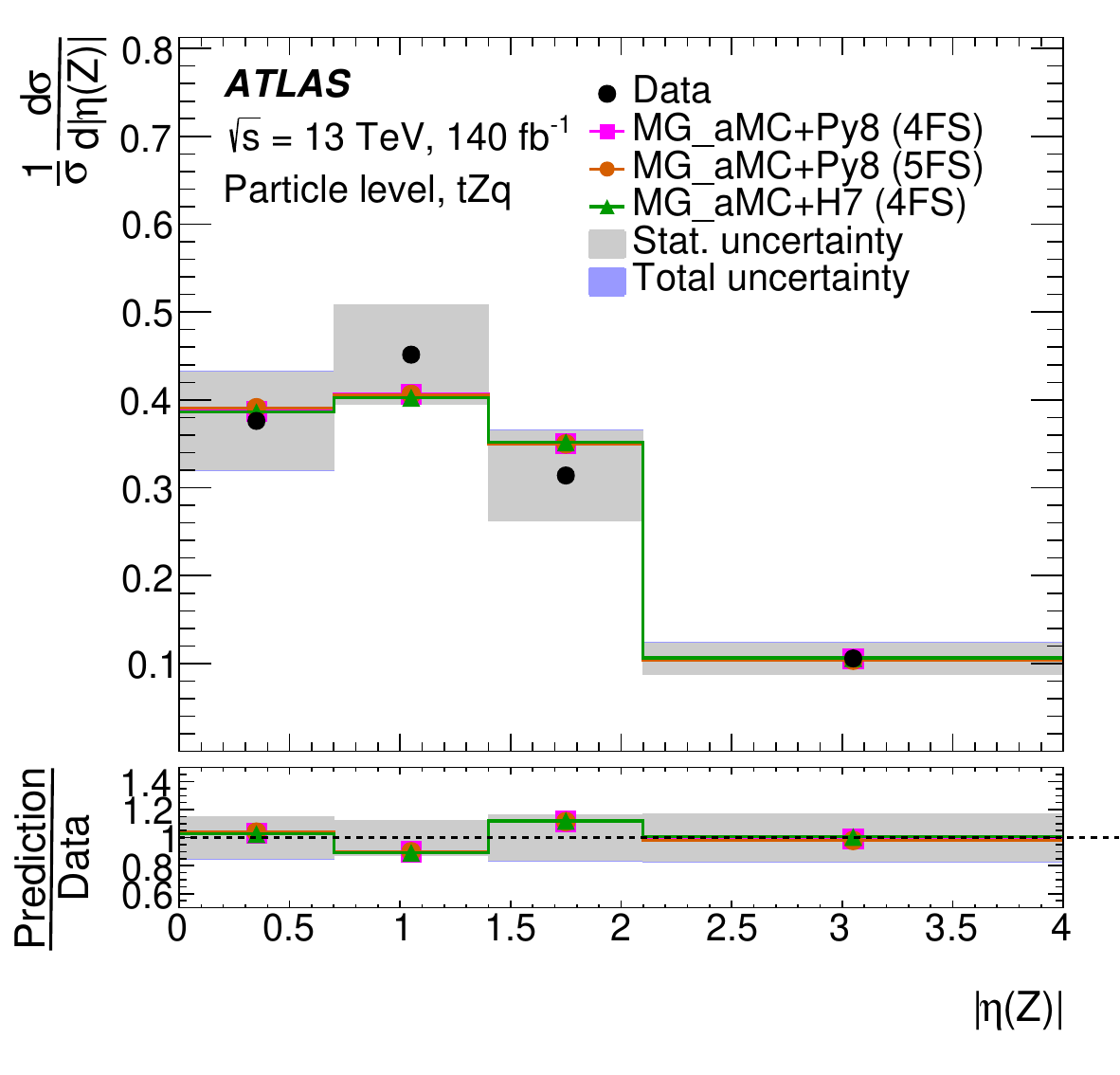}
\caption{}
\end{subfigure} \\
\begin{subfigure}{0.45\textwidth}
\centering
\includegraphics[width=1\textwidth]{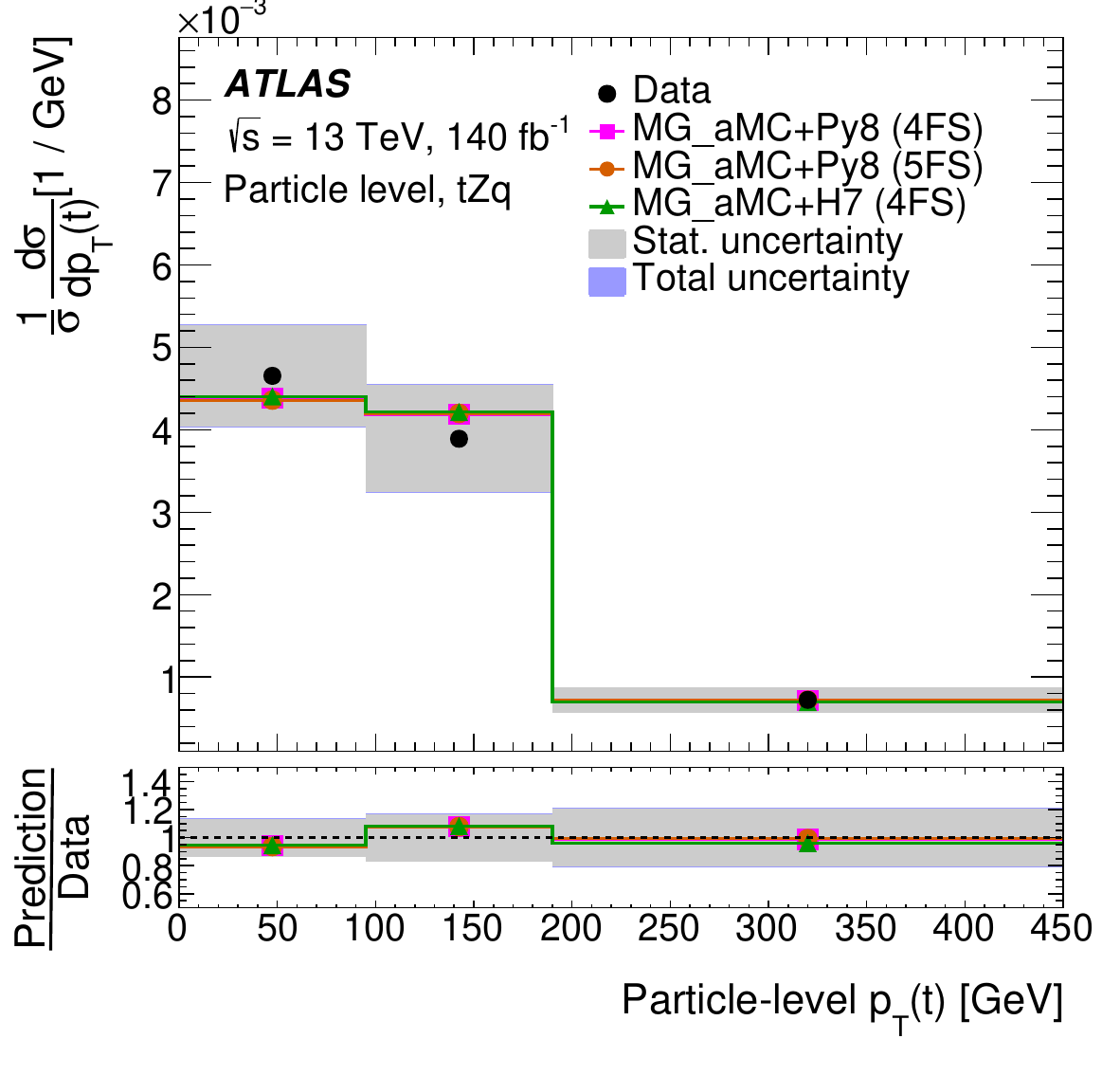}
\caption{}
\end{subfigure}
\begin{subfigure}{0.45\textwidth}
\centering
\includegraphics[width=1\textwidth]{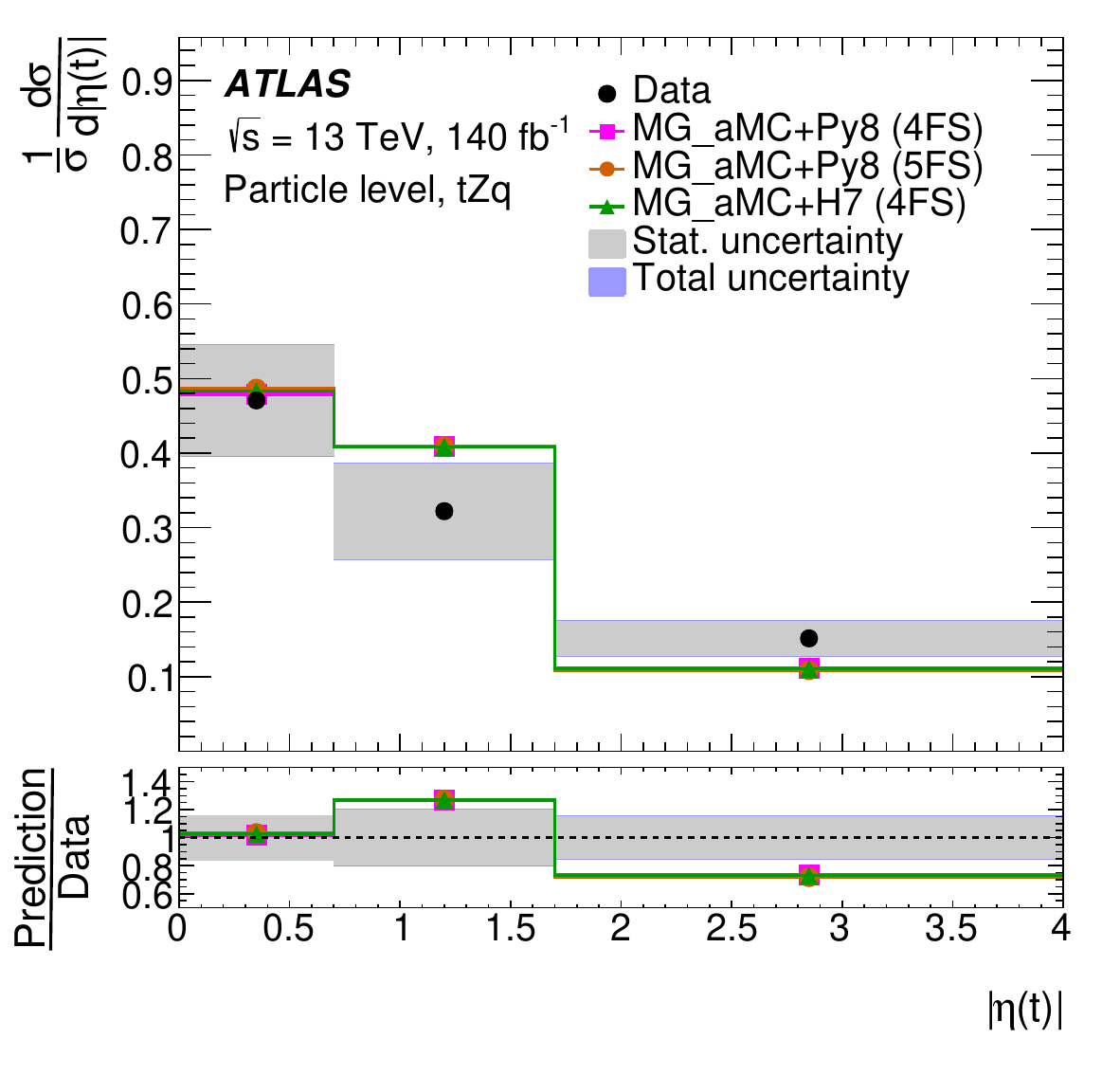}
\caption{}
\end{subfigure}
\caption{Normalised differential cross-section measurements as a function of
(a) \etaZ, (b) \pTt, (c) \etat
observable in the SR, unfolded to particle level.
The light blue band corresponds to the total uncertainty of the measurement; in some cases, it is almost fully covered by the light grey band, representing the dominant statistical uncertainty.
The measurements are compared with the generator predictions indicated in the legend and described in the text.
The lower panels show the ratios of the predictions to the measured normalised differential cross-sections.
}
\label{fig:diff:others1}
\end{figure}

\begin{figure}[htbp]
\centering
\begin{subfigure}{0.45\textwidth}
\centering
\includegraphics[width=1\textwidth]{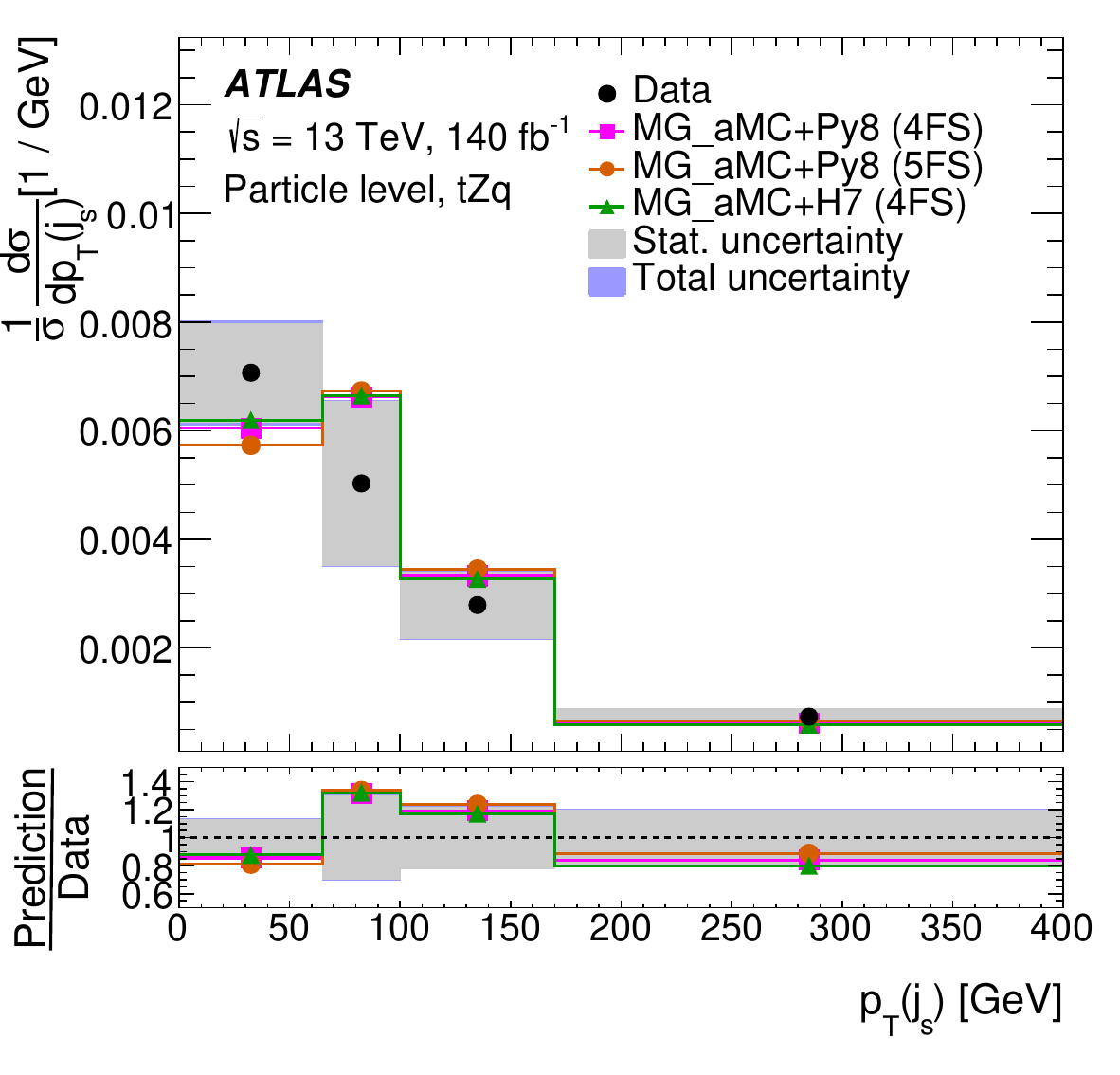}
\caption{}
\end{subfigure}
\begin{subfigure}{0.45\textwidth}
\centering
\includegraphics[width=1\textwidth]{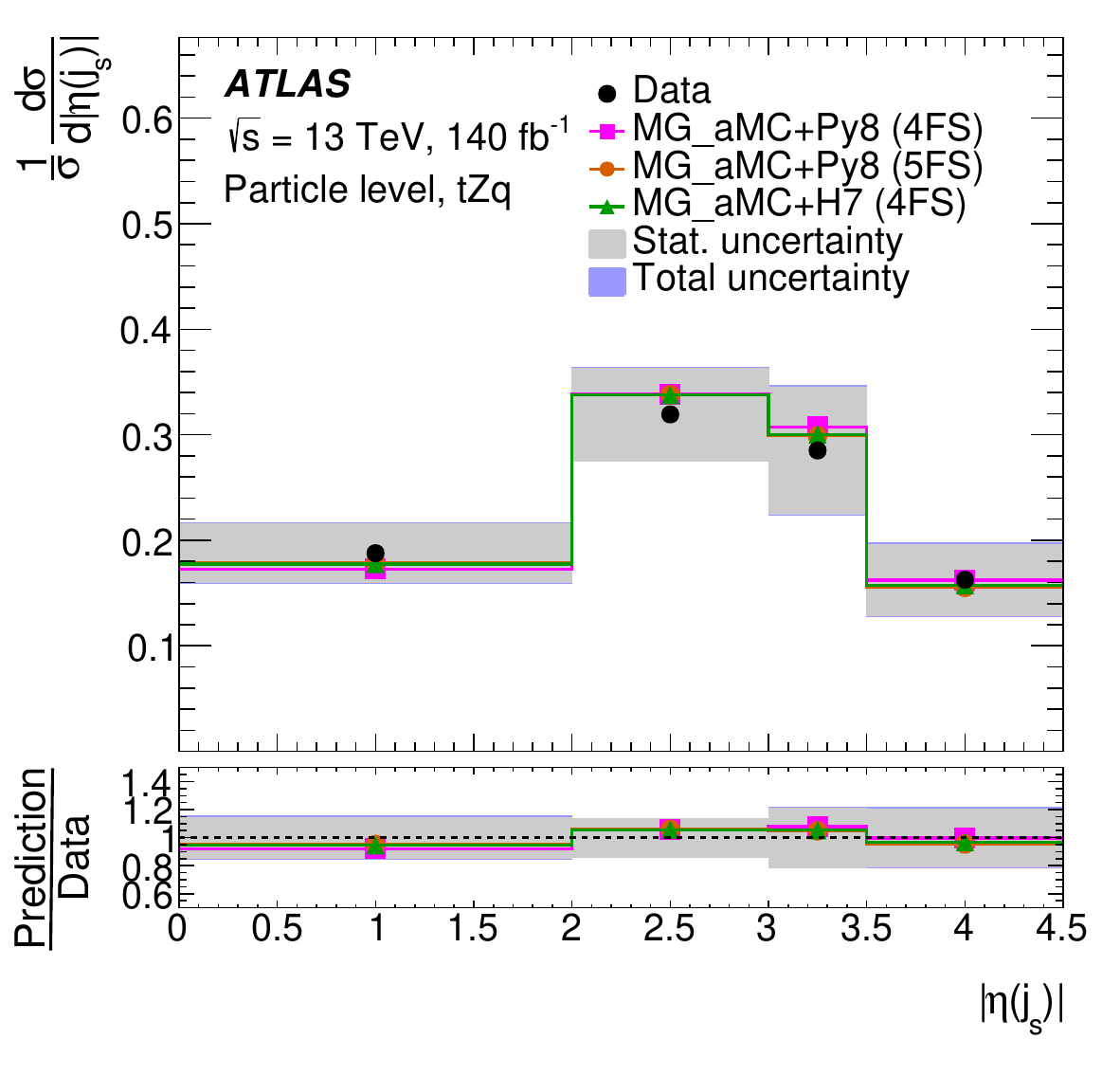}
\caption{}
\end{subfigure}
\begin{subfigure}{0.45\textwidth}
\centering
\includegraphics[width=1\textwidth]{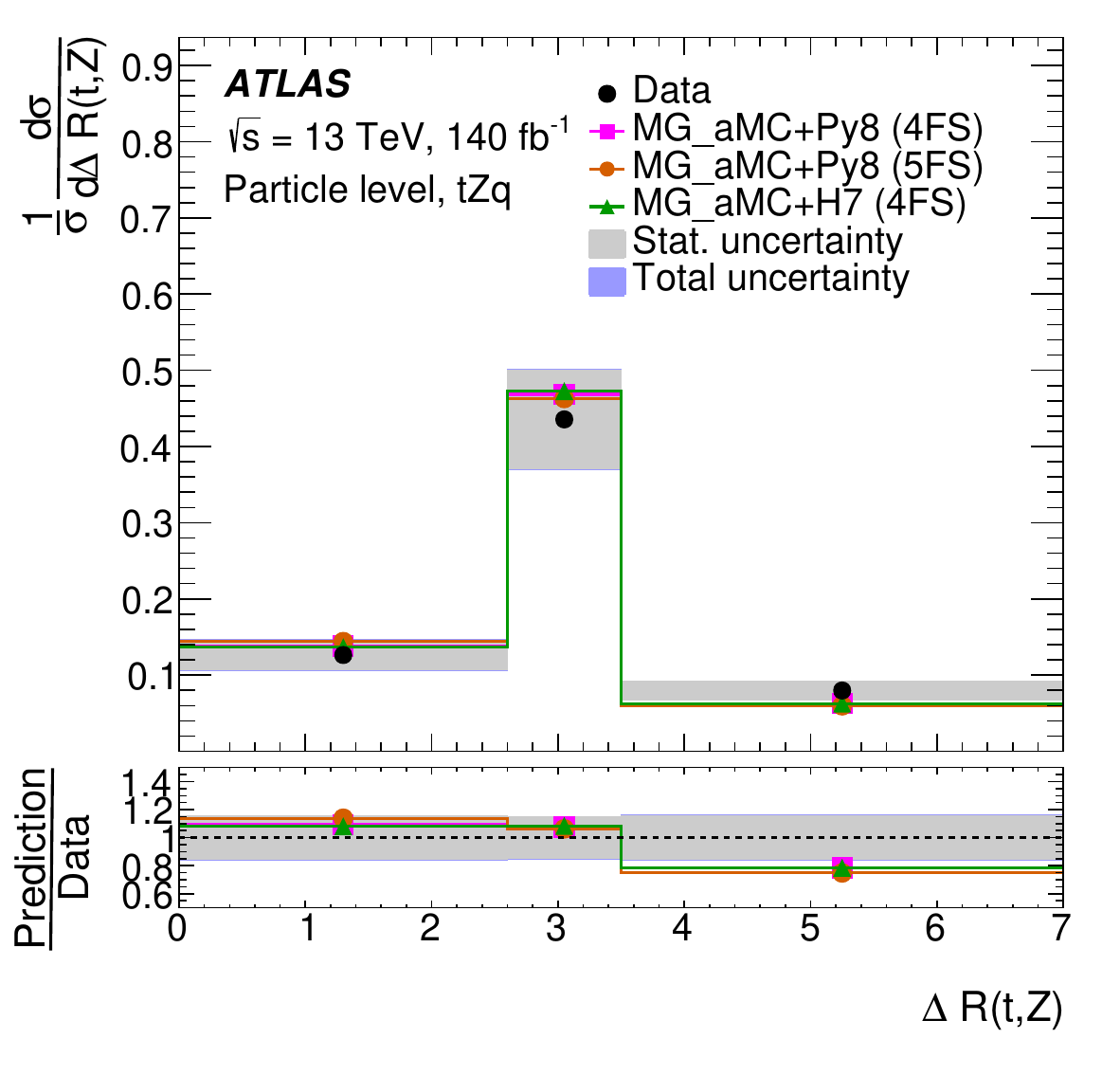}
\caption{}
\end{subfigure}
\begin{subfigure}{0.45\textwidth}
\centering
\includegraphics[width=1\textwidth]{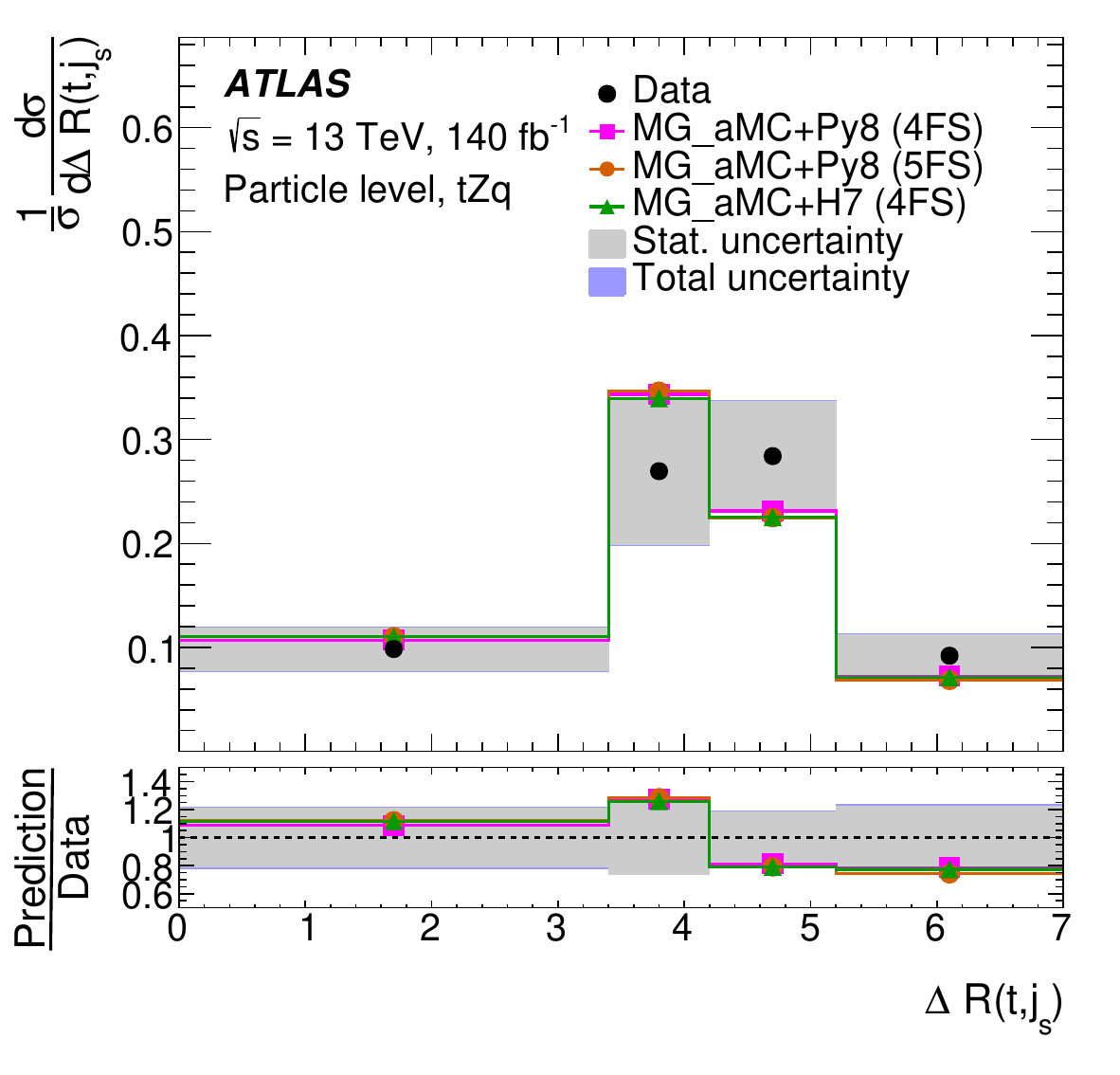}
\caption{}
\end{subfigure}
\caption{Normalised differential cross-section measurements as a function of
(a) \pTsjet, (b) \etasjet, (c) \dRtZ, (d) \dRtsjet
observable in the SR, unfolded to particle level.
The light blue band corresponds to the total uncertainty of the measurement; in some cases, it is almost fully covered by the light grey band, representing the dominant statistical uncertainty.
The measurements are compared with the generator predictions indicated in the legend and described in the text.
The lower panels show the ratios of the predictions to the measured normalised differential cross-sections.
}
\label{fig:diff:others2}
\end{figure}

\begin{figure}[tbp]
\centering
\begin{subfigure}{0.45\textwidth}
\centering
\includegraphics[width=1\textwidth]{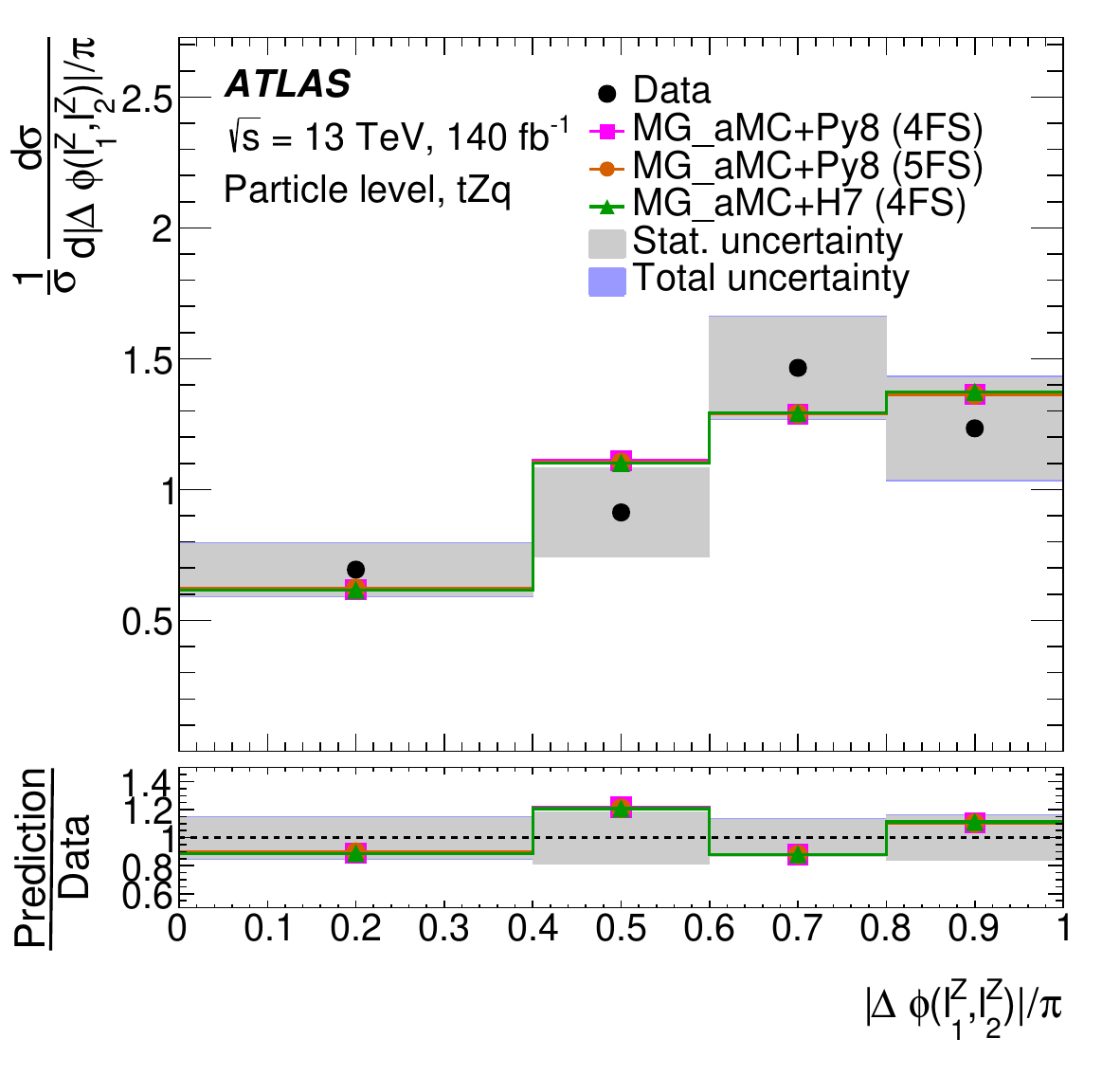}
\caption{}
\end{subfigure}
\begin{subfigure}{0.45\textwidth}
\centering
\includegraphics[width=1\textwidth]{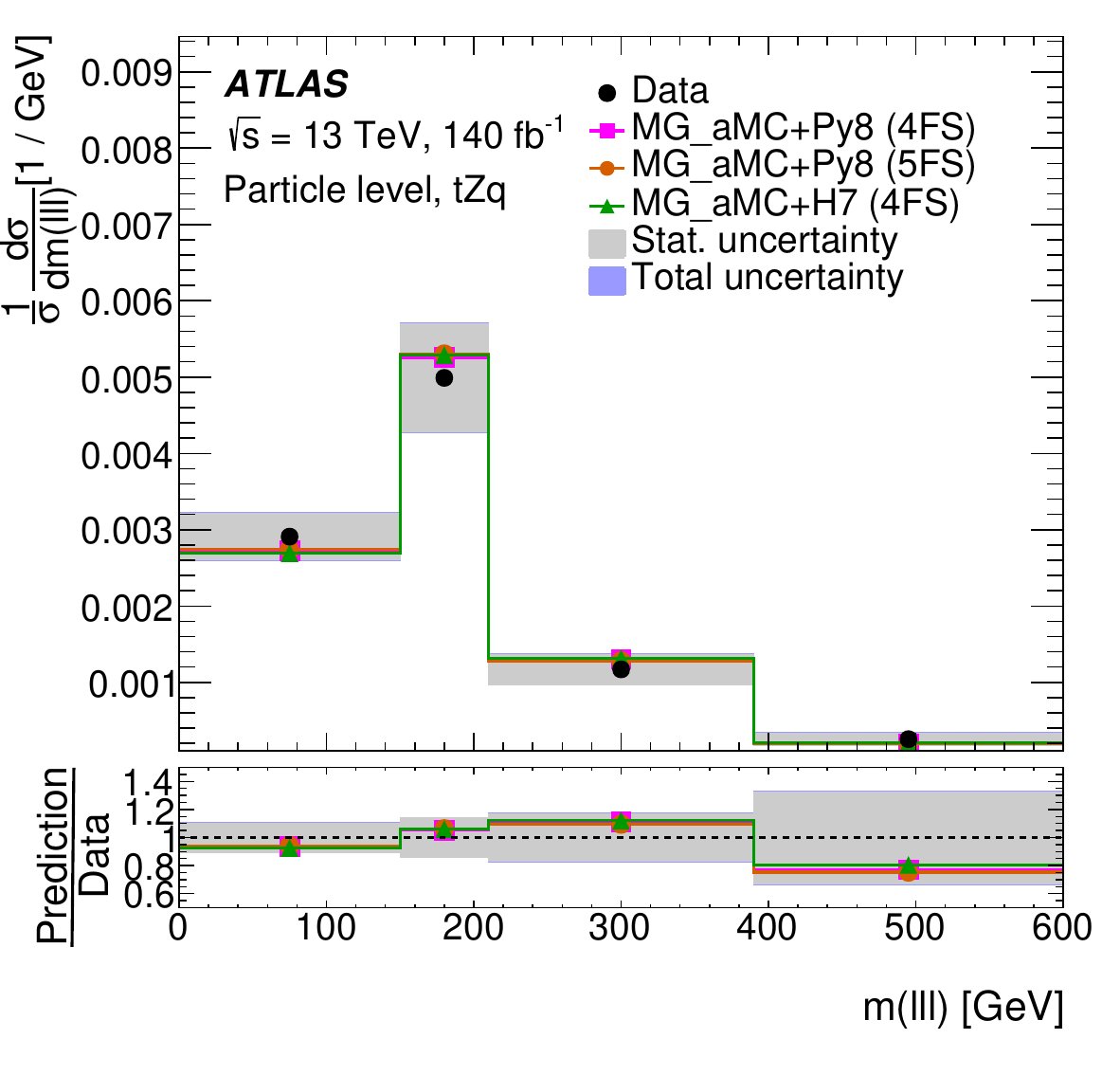}
\caption{}
\end{subfigure}
\caption{Normalised differential cross-section measurements as a function of
(a) \dphiZl and (b) \mlll
in the SR, unfolded to particle level.
The light blue band corresponds to the total uncertainty of the measurement; in some cases, it is almost fully covered by the light grey band, representing the dominant statistical uncertainty.
The measurements are compared with the generator predictions indicated in the legend and described in the text.
The lower panels show the ratios of the predictions to the measured normalised differential cross-sections.
}
\label{fig:diff:others3}
\end{figure}

The unfolded differential cross-sections are shown in \cref{fig:diff:Zpt,fig:diff:cosTheta} for the \pTZ and \cosTheta observables, respectively. Both absolute and normalised cross-sections are presented, unfolded to parton and particle level. For the remaining observables, only the particle-level unfolded normalised differential cross-sections are shown in \cref{fig:diff:others1,fig:diff:others2,fig:diff:others3}. The particle- and parton-level absolute differential cross-sections, together with the normalised parton-level differential cross-sections, where available, are presented in the Appendix in \cref{fig:diff:Zeta,fig:diff:toppt,fig:diff:topeta,fig:diff:sjetpt,fig:diff:sjeteta,fig:diff:dRtopZ,fig:diff:dRtopsjet,fig:diff:dPhilZ1lZ2,fig:diff:mlll}.
The measurements are compared with predictions obtained using \MGNLO+\PYTHIA[8] in both the 4FS and 5FS. At particle level, an additional prediction obtained with \MGNLO+\HERWIG[7] in the 4FS is also considered to assess the impact of the parton-shower model.
The 4FS predictions are presented with their original cross-section normalisation, which is 30\% below the 5FS prediction overall.
The measured spectra are well described by the 5FS predictions across the full phase space. In particular, the overall shapes of the distributions, as well as the behaviour in the highest-\pt and highest-mass regions, are reproduced within the experimental uncertainties. The differences between the predictions are generally small compared with the experimental uncertainties, which are dominated by the statistical component in all bins. The 4FS predictions show good shape agreement, while differences in the absolute cross-sections are largely associated with the different overall normalisation, which results in predicted spectra being below the measurement.
The level of agreement is quantified by the $p$-values reported in \cref{tab:diff:chi2:abs} and \cref{tab:diff:chi2:norm} for the absolute and normalised differential cross-sections, respectively. All 5FS comparisons
yield $p$-values larger than 0.19, indicating no significant discrepancy between the measured distributions and the predictions. The lowest $p$-values are observed for the \etat distribution, although the agreement remains statistically acceptable. Similar levels of agreement are observed for the particle-level and parton-level measurements, and for the absolute and normalised cross-sections for the 5FS predictions.
The compatibility tests use the full covariance matrix of the unfolded measurement, obtained by inverting the Hessian matrix evaluated at the likelihood minimum and including the systematic uncertainties propagated through the unfolding fit. No theoretical uncertainties are assigned to the generator predictions entering the compatibility tests.

\begin{table}[htbp]
\tablesetup[1.3]
\caption{Summary of the compatibility tests, expressed as $p$-values, between the unfolded absolute differential spectra and the predictions at particle level and parton level.}
\label{tab:diff:chi2:abs}
\centering
\begin{tabular}{l | cc | ccc}
\toprule
& \multicolumn{2}{c|}{Parton level} & \multicolumn{3}{c}{Particle level} \\
\midrule
& MG\_aMC+Py8 & MG\_aMC+Py8 & MG\_aMC+Py8 & MGa\_MC+Py8 & MG\_aMC+H7 \\
& (4FS)     & (5FS)     & (4FS)     & (5FS)     & (4FS) \\
\midrule
Variable & $p$-value & $p$-value & $p$-value & $p$-value & $p$-value \\
\midrule
\pTZ      & 0.14 & 0.39 & 0.12 & 0.41 & 0.09 \\
\etaZ     & 0.13 & 0.88 & 0.18 & 0.92 & 0.12 \\
\pTt      & 0.18 & 0.96 & 0.15 & 0.96 & 0.10 \\
\etat     & 0.07 & 0.38 & 0.05 & 0.30 & 0.03 \\
\pTsjet   & --   & --   & 0.20 & 0.47 & 0.14  \\
\etasjet  & --   & --   & 0.39 & 0.99 & 0.27 \\
\dRtZ     & 0.07 & 0.54 & 0.06 & 0.41 & 0.04 \\
\dRtsjet  & --   & --   & 0.11 & 0.40 & 0.06 \\
\dphiZl   & 0.22 & 0.65 & 0.19 & 0.67 & 0.14 \\
\cosTheta & 0.14 & 0.66 & 0.11 & 0.70 & 0.08 \\
\mlll     & 0.26 & 0.89 & 0.19 & 0.86 & 0.13 \\
\bottomrule
\end{tabular}
\end{table}

\begin{table}[htbp]
\tablesetup[1.3]
\caption{Summary of the compatibility tests, expressed as $p$-values, between the unfolded normalised differential spectra and the predictions at particle level and parton level.}
\label{tab:diff:chi2:norm}
\centering
\begin{tabular}{l | cc | ccc}
\toprule
& \multicolumn{2}{c|}{Parton level} & \multicolumn{3}{c}{Particle level} \\
\midrule
& MG\_aMC+Py8 & MG\_aMC+Py8 & MG\_aMC+Py8 & MG\_aMC+Py8 & MG\_aMC+H7 \\
& (4FS)     & (5FS)     & (4FS)     & (5FS)     & (4FS) \\
\midrule
Variable & $p$-value & $p$-value & $p$-value & $p$-value & $p$-value \\
\midrule
\pTZ      & 0.26 & 0.26 & 0.26 & 0.28 & 0.27 \\
\etaZ     & 0.82 & 0.82 & 0.84 & 0.83 & 0.81\\
\pTt      & 0.90 & 0.90 & 0.89 & 0.88 & 0.88 \\
\etat     & 0.19 & 0.18 & 0.22 & 0.19 & 0.20 \\
\pTsjet   & --   & --   & 0.41 & 0.32 & 0.39 \\
\etasjet  & --   & --   & 0.95 & 0.96 & 0.96 \\
\dRtZ     & 0.40 & 0.37 & 0.40 & 0.28 & 0.41 \\
\dRtsjet  & --   & --   & 0.43 & 0.31 & 0.39 \\
\dphiZl   & 0.49 & 0.48 & 0.48 & 0.51 & 0.50 \\
\cosTheta & 0.52 & 0.46 & 0.47 & 0.50 & 0.54 \\
\mlll     & 0.78 & 0.78 & 0.78 & 0.78 & 0.78 \\
\bottomrule
\end{tabular}
\end{table}

\begin{figure}[tbp]
\centering
\begin{subfigure}{0.45\textwidth}
\centering
\includegraphics[width=\textwidth]{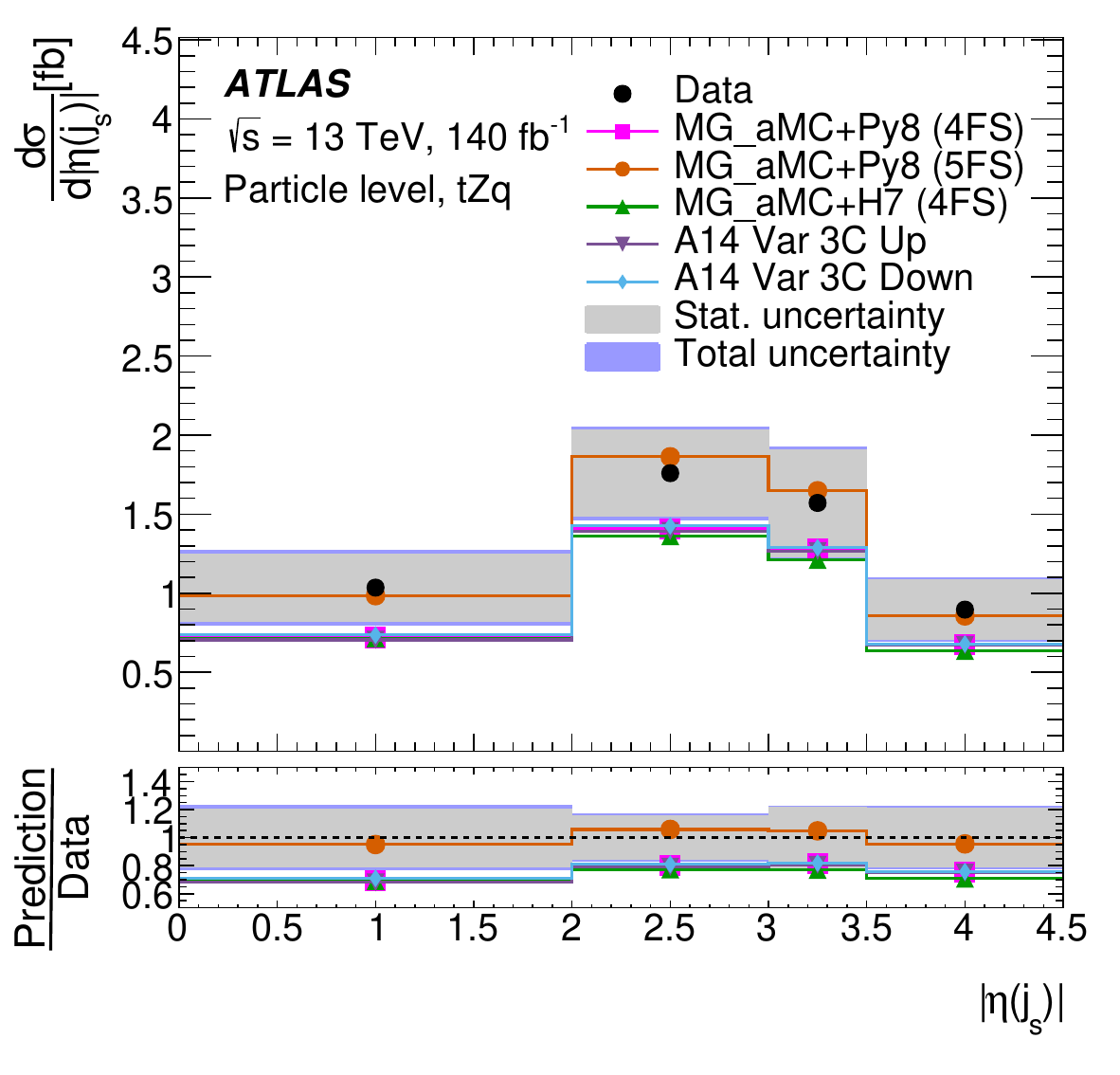}
\caption{}
\end{subfigure}
\begin{subfigure}{0.45\textwidth}
\centering
\includegraphics[width=\textwidth]{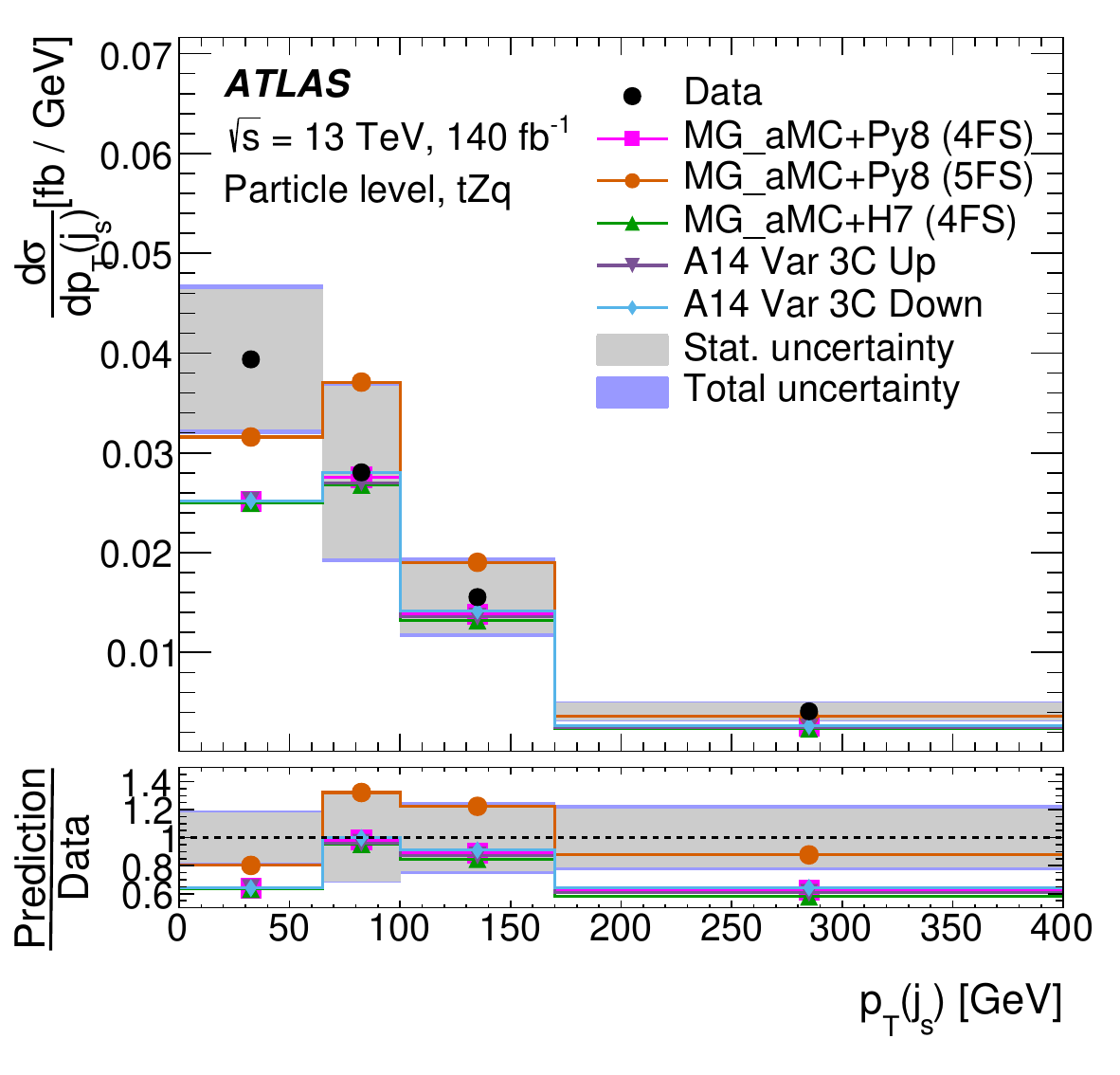}
\caption{}
\end{subfigure}
\vspace{0.5cm}
\begin{subfigure}{0.45\textwidth}
\centering
\includegraphics[width=\textwidth]{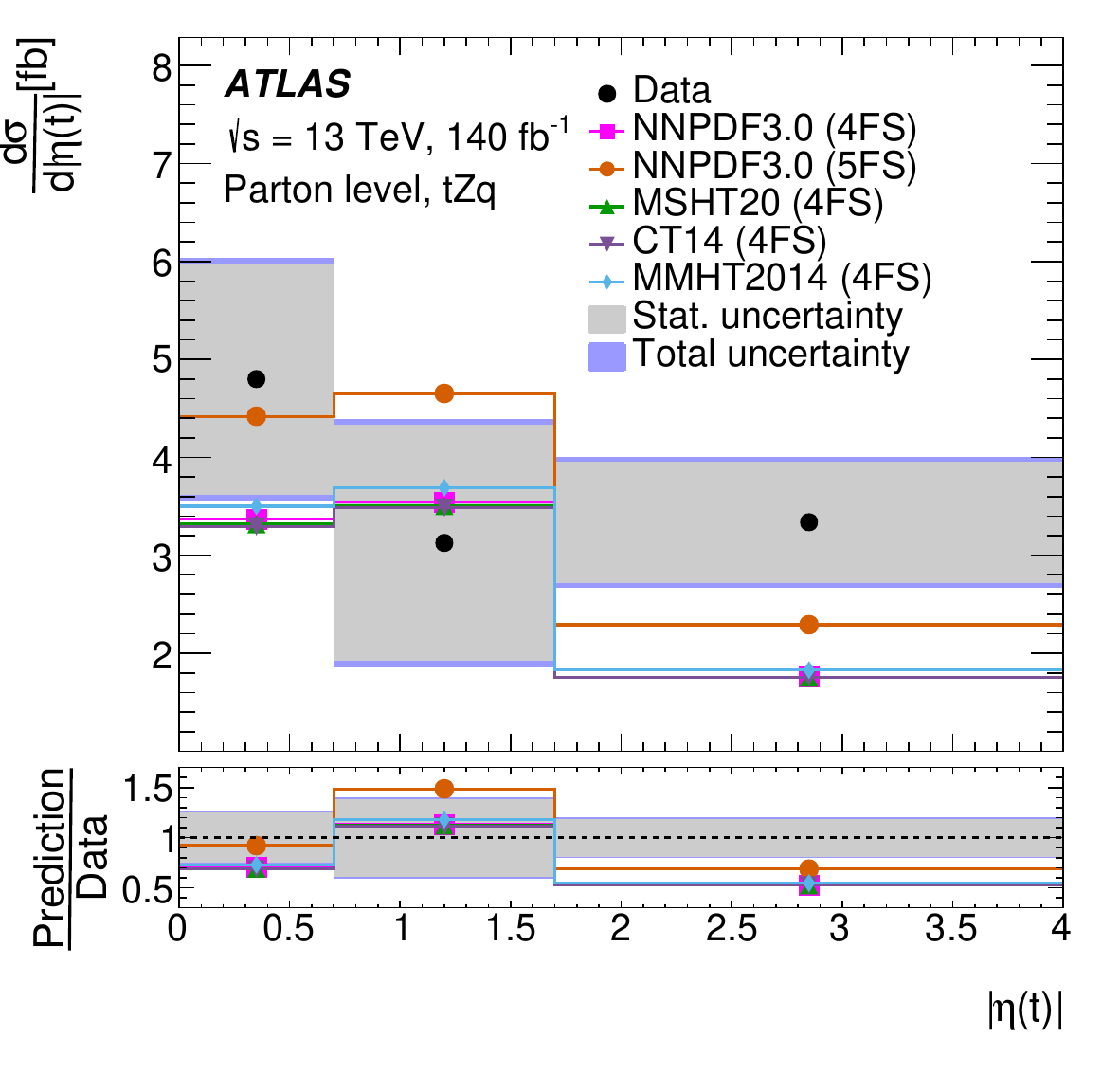}
\caption{}
\end{subfigure}
\begin{subfigure}{0.45\textwidth}
\centering
\includegraphics[width=\textwidth]{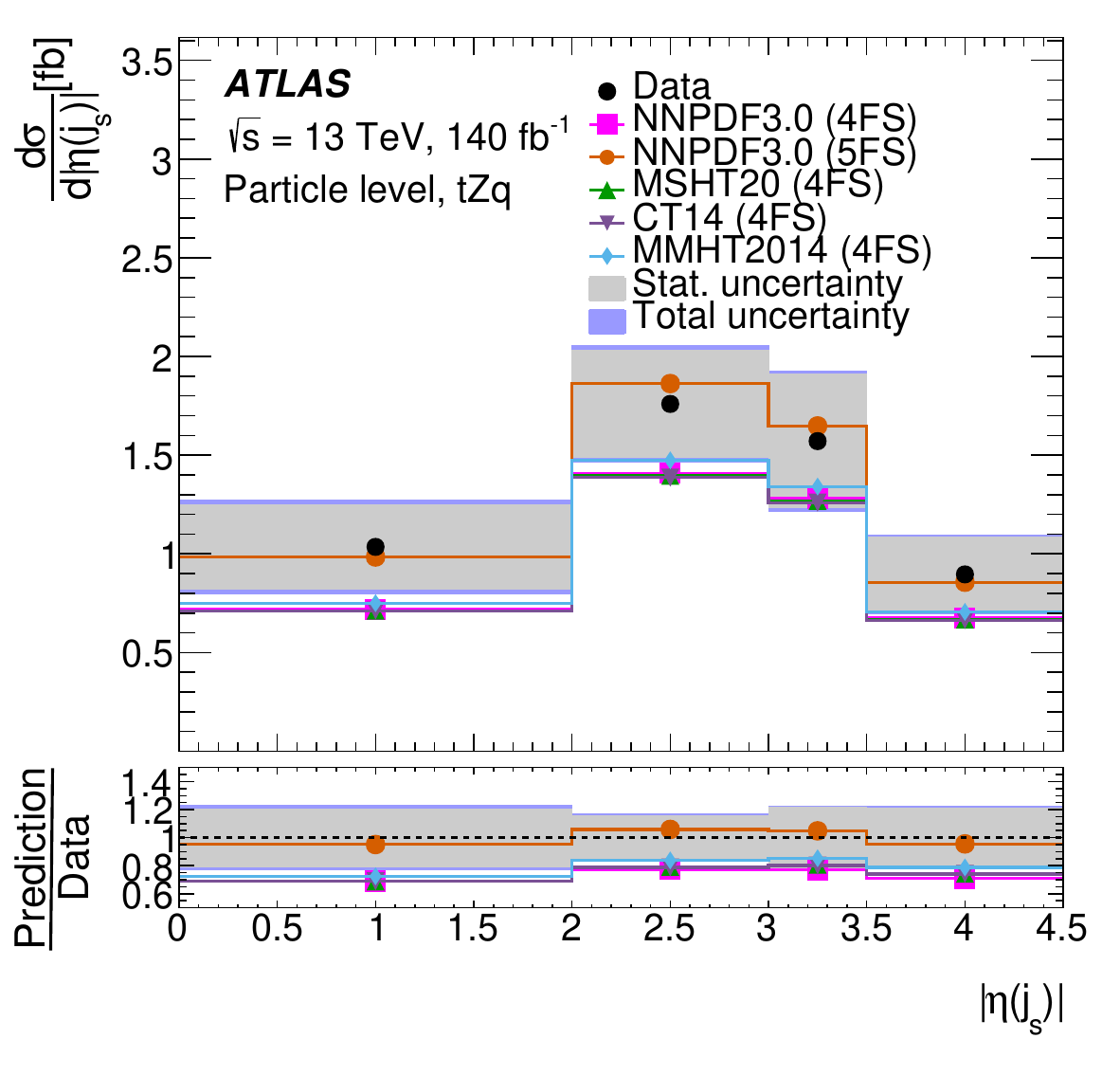}
\caption{}
\end{subfigure}
\caption{Differential cross-section measurements illustrating the impact of alternative theoretical predictions in the SR. The parton-shower comparison is shown for (a) \etasjet and (b) \pTsjet at particle level, while the PDF comparison is shown for (c) \etat at parton level and (d) \etasjet at particle level. The light blue band corresponds to the total uncertainty of the measurement; in some cases, it is almost fully covered by the light grey band, representing the dominant statistical uncertainty.
The lower panels show the ratios of the predictions to the measured cross-sections.
}
\label{fig:diff:theory}
\end{figure}

Additional comparisons with alternative theoretical predictions are shown in \cref{fig:diff:theory}. The impact of the parton-shower modelling is illustrated for the particle-level \etasjet and \pTsjet observables by comparing the nominal \MGNLO+\PYTHIA[8] prediction with alternative shower configurations. The impact of the PDF choice is shown for the parton-level \etat observable and the particle-level \etasjet observable by comparing the nominal \texttt{NNPDF3.0} prediction with alternative PDF sets. In all cases, the differences between the theoretical predictions are small compared with the experimental uncertainties and remain well within the uncertainty bands of the measurements. For the spectator-jet observables considered, no clear preference for a particular parton-shower configuration is observed, despite the potential sensitivity of these observables to the modelling of QCD radiation and the parton shower. Similarly, no clear preference for a particular PDF set is observed within the current experimental precision.

The spin asymmetry is extracted from the parton-level differential distribution of \cosTheta shown in \cref{fig:diff:cosTheta}, assuming the differential cross-section dependence~\cite{Aguilar_Saavedra_2014}
\begin{equation}
\frac{d\sigma}{d\cosTheta} = \sigma \left(\frac{1}{2} + A \cosTheta \right),
\label{eq:Asymmetry}
\end{equation}
where $\sigma$ denotes the fiducial cross section in the parton-level phase space, $A$ is the spin asymmetry, and $\cos\theta^\star$ is defined as in \cref{tab:differential:vars}.

The spin asymmetry is measured by fitting the unfolded differential cross-section distribution of \cosTheta at parton level, as per \cref{eq:Asymmetry}. Since this distribution combines top-quark and top-antiquark events, a single asymmetry parameter $A$ is extracted for the combined sample. Both the fiducial cross-section $\sigma$ and the asymmetry $A$ are treated as free-floating parameters in the fit. The fit is performed by minimising a $\chi^2$ function (\cref{eq:fit_asymm}) constructed from the unfolded cross-section measurements $\mathbf{d}$ and the full covariance matrix $\mathbf{V}$. In Eq.~(\ref{eq:fit_asymm}), $\mathbf{m}$ is the prediction of the fit model evaluated at the bin centres. The covariance matrix is built considering both statistical and systematic uncertainties. For the statistical-only uncertainty, the covariance matrix built considering only the statistical uncertainty from the unfolding fit is used.
\begin{equation}
\chi^2(\mathbf{m}) =
(\mathbf{d}-\mathbf{m})^{\mathrm{T}}
\mathbf{V}^{-1}
(\mathbf{d}-\mathbf{m}),
\label{eq:fit_asymm}
\end{equation}

\begin{figure}
\centering
\includegraphics[width=0.8\textwidth]{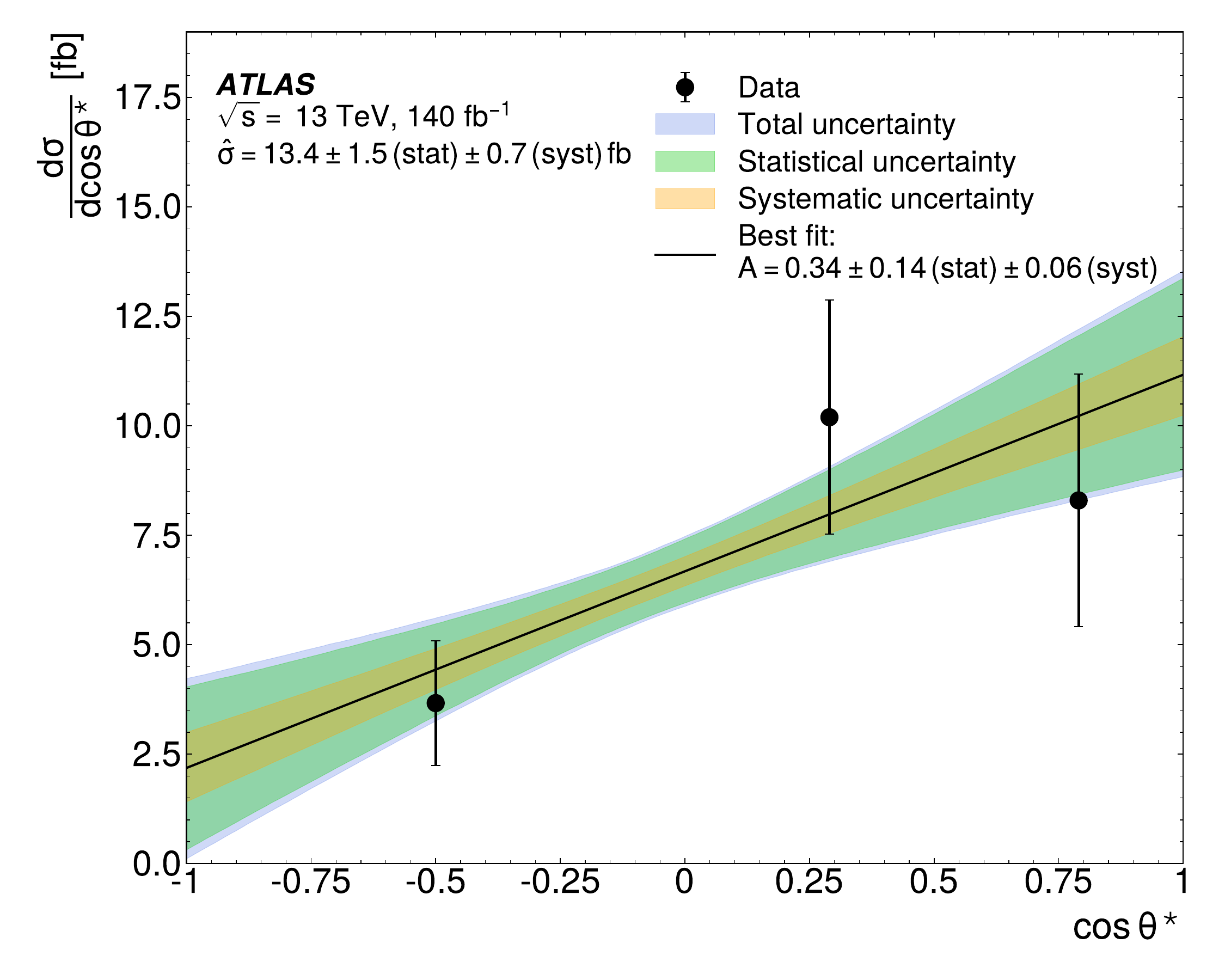}
\caption{$\chi^2$ fit to the unfolded differential values of \cosTheta performed to measure the spin asymmetry, as per \cref{eq:fit_asymm}. The fiducial cross-section $\hat{\sigma}$ and the asymmetry $A$ are left as free-floating parameters in the fit. Their fitted values, together with their statistical and systematic uncertainties, are quoted directly on the plot. The full covariance matrix $\mathbf{V}$, including both statistical and systematic uncertainties, is used in the fit.}
\label{fig:diff:spinasym}
\end{figure}

\Cref{fig:diff:spinasym} reports the results of the fit. The asymmetry is measured to be
\begin{equation}
A = 0.34 \pm 0.14 \ \text{(stat.)} \pm 0.06 \ \text{(syst.)}.
\end{equation}
The simultaneously fitted fiducial cross section is $\hat{\sigma} = 13.4 \pm 1.5 \ \text{(stat.)} \pm 0.7 \ \text{(syst.)}\,\text{fb}$.
This value is consistent with the parton-level fiducial cross section reported in \cref{sec:inclusive}, and with the value obtained by integrating the unfolded \cosTheta distribution.
The measured asymmetry is in good agreement with the CMS measurement of $A = 0.54 \pm 0.16 \ \text{(stat.)} \pm 0.06 \ \text{(syst.)}$ \cite{CMS-TOP-20-010} and the SM prediction obtained from \MGNLO[3.5.1] of $A_\ell^\text{4FS}=0.444^{+0.006}_{-0.003}(\text{Scale})$, the latter estimated by varying the renormalisation and factorisation scales by 0.5 and 2.

%
%
%
%
%
%
%
%
%
%
%
%
%
%
%
%
%
%
%
%
%
%
%
%
%
%
%

%
%
%
%
%
%
%
%
%
%
%
%
%
%
%
%
%
%
%
%
%
%
%
%
%
%
%

%
%
%
%
%
%
%
%
%
%
%
%
%
%
%
%
%
%
%
%
%
%
%
%
%
%
%

%
%
%
%
%
%
%
%
%
%
%
%
%
%
%
%
%

%
%
%
%
%
%
%
%
%
%
%
%
%
%
%
%
%

%
%
%
%
%
%
%
%
%
%
%
%
%
%
%
%
%
%
%
%
%
%
%
%
%
%
%

%
%
%
%
%
%
%
%
%
%
%
%
%
%
%
%
%

%
%
%
%
%
%
%
%
%
%
%
%
%
%
%
%
%
%
%
%
%
%
%
%
%
%

%
%
%
%
%
%
%
%
%
%
%
%
%
%
%
%
%
%
%
%
%
%
%
%
%
%

%
%
%
%
%
%
%
%
%
%
%
%
%
%
%
%
%
%
%
%
%
%
%
%
%
%


%

%

%
\section{SMEFT interpretation}
\label{sec:smeft}

The SMEFT~\cite{Brivio:2017vri,Brivio:2020onw} provides a phenomenological extension of the SM beyond the dimension-four terms of the Lagrangian.
Using the degrees of freedom and gauge symmetries of the SM, the SMEFT constructs an infinite series of higher-dimensional operators ordered by their canonical dimension.
New-physics effects arising from theories characterised by a mass scale larger than the energies probed at the LHC can then be described in terms of these operators and their associated Wilson coefficients (WCs), without specifying a particular underlying model.
Agreement of the data with the SM can therefore be translated into constraints on the corresponding WCs.

The SMEFT Lagrangian can be written as
\begin{equation}
\mathcal{L}_\mathrm{SMEFT}
=
\mathcal{L}_\mathrm{SM}
+
\sum_{d>4}\mathcal{L}^{(d)},
\qquad
\mathcal{L}^{(d)}
=
\sum_i
\frac{C_i^{(d)}}{\Lambda^{d-4}}
\mathcal{O}_i^{(d)},
\label{eq:smeft:lagrangian}
\end{equation}
where \(\Lambda\) denotes the scale of new physics, \(\mathcal{O}_i^{(d)}\) are operators of mass dimension \(d\), and \(C_i^{(d)}\) are the corresponding WCs.
The conventional choice $\Lambda = \qty{1}{\TeV}$ is used.
The interpretation is restricted to dimension-six operators described by the Warsaw basis~\cite{Grzadkowski:2010es}, which provide the leading SMEFT contributions here.

The \tZq process is sensitive to SMEFT effects involving top-quark electroweak interactions, since its production involves the coupling of a top quark to electroweak bosons.
The EFT interpretation therefore targets a set of dimension-six WCs affecting \tZq production, including operators involving top-boson and four-quark interactions, with the corresponding effects evaluated at detector level.
The following seven dimension-six SMEFT operators are considered,
\begin{equation}
\centering
\mathcal{O}_{\Pqt \PW}, \mathcal{O}_{\Pqt \PB}, \mathcal{O}_{\PH Q}^{(1)}, \mathcal{O}_{\PH Q}^{(3)}, \mathcal{O}_{\PH \Pqt}, \mathcal{O}_{Q q}^{(3,1)}, \mathcal{O}_{Q q}^{(3,8)} \,.
\label{eq:smeft:WCs}
\end{equation}
Resulting in nine WCs, due to the inclusion of the real and imaginary parts of the $\mathcal{O}_{\Pqt \PW}$ and  $\mathcal{O}_{\Pqt \PB}$ operators.

Using EFT Monte Carlo samples produced with the \texttt{SMEFTsim 3.0} UFO model~\cite{Brivio:2020onw}, the dependence of an observable $\mathscr{O}$, evaluated at detector level, on the Wilson coefficients can be parametrised as
\begin{equation}
\mathscr{O}
=
\mathscr{O}_{\mathrm{SM}}
+
\sum_i C_i A_i
+
\sum_{i,j} C_i C_j B_{ij}.
\label{eq:smeft:parametrisation}
\end{equation}
$\mathscr{O}$ corresponds to the predicted yield in a bin of the reconstructed \PZ-boson transverse momentum distribution.
The linear terms $A_i$ describe the interference between the SM and SMEFT amplitudes, while the quadratic terms $B_{ij}$ describe pure SMEFT contributions, including both the squared contribution of a single operator and interference between different SMEFT operators.
The coefficients of \cref{eq:smeft:parametrisation} are determined from the EFT Monte Carlo samples and are used to parameterise the dependence of the predicted yields on the Wilson coefficients in the likelihood fit.
The effect of the EFT operators is parameterised separately for the \tZq and \ttZ processes on a bin-by-bin basis using a second-order polynomial function. This parameterisation describes the relative change in the predicted yield arising from the interference between the SMEFT and SM amplitudes at $\mathcal{O}(\Lambda^{-2})$ (linear term) and from pure SMEFT contributions at $\mathcal{O}(\Lambda^{-4})$ (quadratic term). Higher-order QCD effects are approximated by scaling the EFT prediction, generated at LO in QCD, to the corresponding SM prediction at NLO in QCD on a bin-by-bin basis. Once the EFT parameterisation is obtained, individual profile-likelihood fits are performed, in which the Wilson coefficient under consideration is treated as a freely floating parameter while all other Wilson coefficients are fixed to zero.
To perform the measurement, a methodology consistent with previous ATLAS SMEFT analyses~\cite{TOPQ-2019-18,TOPQ-2023-02} is adopted.

The EFT interpretation includes contributions from both the \tZq and \ttZ processes, as the latter improves the overall sensitivity to several of the considered Wilson coefficients.
The detector-level $\pT(Z)$ distribution is used as the fit observable.
The distribution is binned in four bins ($0,100,170,240,1000\ \si{\giga\electronvolt}$), and the SR is split into the same two regions used for the differential fits, by applying a cut on the NN score at 0.71. All the remaining CRs are kept the same as in the inclusive fit.

\begin{figure}[htbp]
\centering
\includegraphics[width=0.7\textwidth]{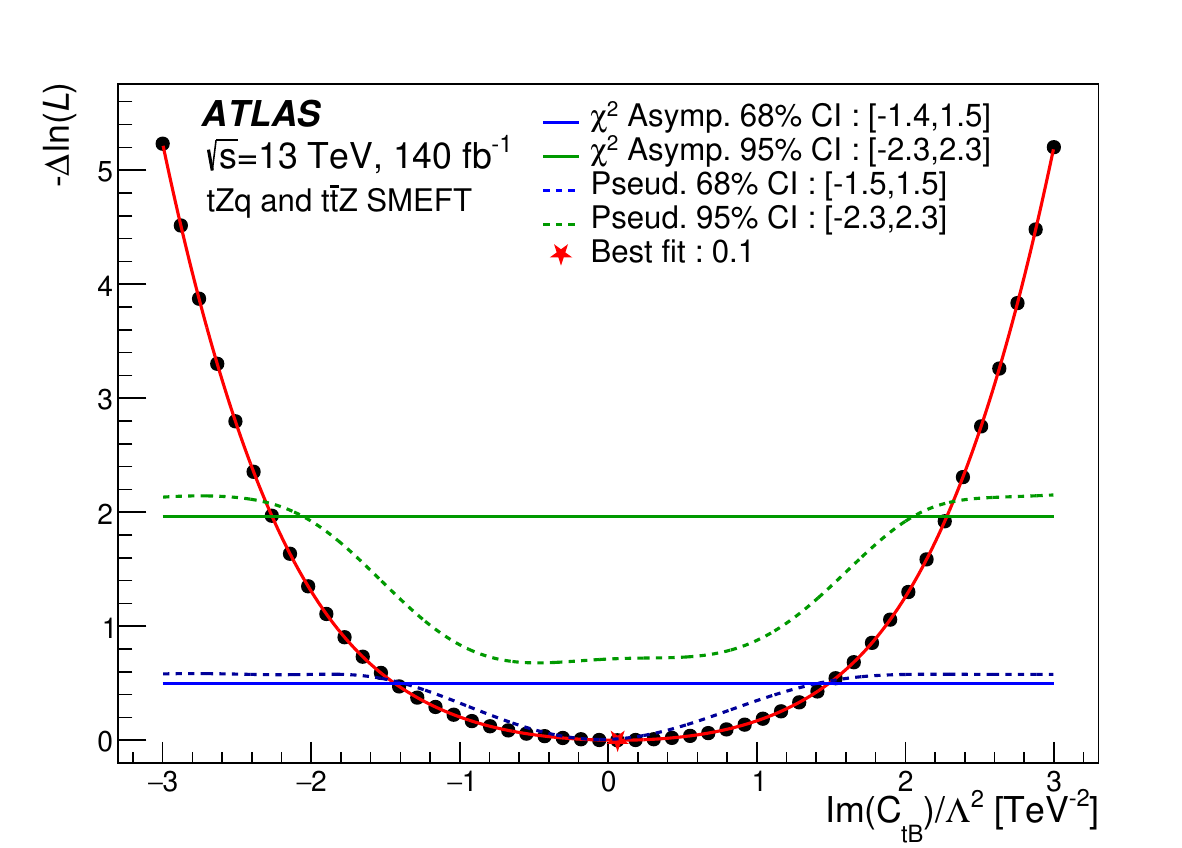}
\caption{Scans of the negative log-likelihood value as a function of the WC value for Im($C_{tB}$). The solid lines show the asymptotic $\chi^{2}$ 68\% and 95\% thresholds for calculating confidence intervals. The dashed lines show the pseudodata coverage-adjusted 68\% and 95\% confidence interval thresholds.}
\label{fig:EFTctBImLHScan}
\end{figure}

\begin{figure}[htbp]
\centering
\includegraphics[width=0.8\textwidth]{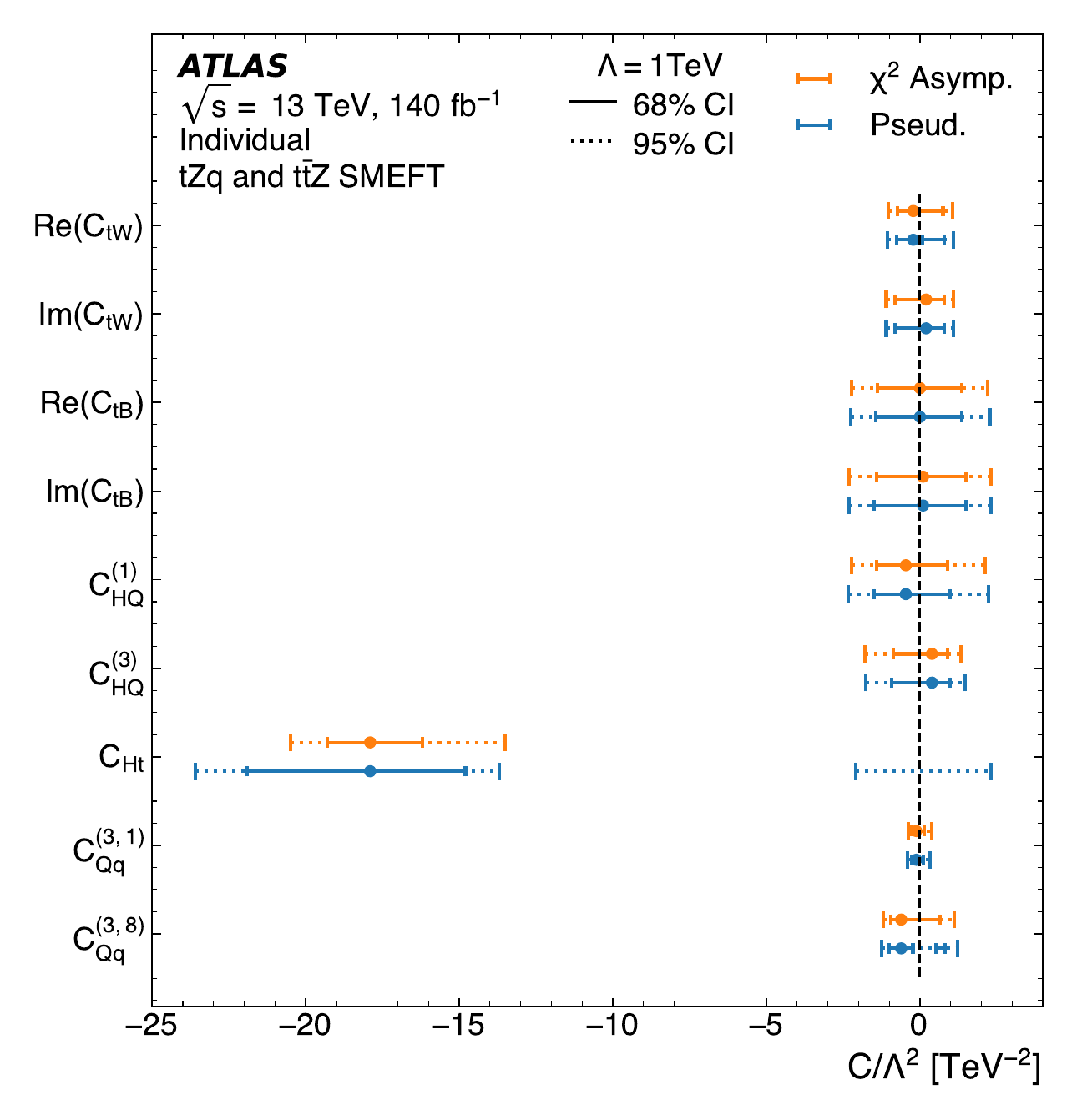}
\caption{Results of the fit to data for each WC fitted individually considering linear+quadratic effects. For each WC the upper bar corresponds to the asymptotic $\chi^{2}$ 68\% (solid) and 95\% (dashed) thresholds for calculating confidence intervals, while the lower bar corresponds to the pseudodata coverage-adjusted 68\% (solid) and 95\% (dashed) confidence interval thresholds.
The vertical dashed line indicates the Standard Model expectation, $C_i/\Lambda^2=0$.}
\label{fig:EFTresults}
\end{figure}

For operators where the (quadratic) pure-EFT $\mathcal{O}(\Lambda^{-4})$ dominate a breakdown of a key axiom of Wilks' theorem~\cite{Wilks:1938dza} can occur, meaning the profile-likelihood ratio (PLR) may not follow the expected $\chi^{2}$-distribution~\cite{Bernlochner:2022oiw}. This can lead to the coverage in the test statistic differing from what is expected in the asymptotic case. The true coverage can be determined by generating toy pseudodata and taking the 68\% and 95\% percentiles of the fitted pseudodata test statistics. In the following the 68\% and 95\% Confidence Intervals for both the non-coverage adjusted ($\chi^{2}$ Asymptotic) and the coverage-adjusted (Pseudodata) results are presented. Scans of the fitted negative log-likelihood value as a function of the WC value are shown in~\cref{fig:EFTctBImLHScan} for Im($C_{tB}$). The solid lines show the asymptotic $\chi^{2}$ 68\% and 95\% thresholds for calculating confidence intervals. The dashed lines show the pseudodata coverage-adjusted 68\% and 95\% confidence interval thresholds. The confidence intervals calculated with the coverage from pseudodata are quoted from intersections of the curves with the likelihood scan.
\cref{fig:EFTresults} presents the observed limits for the nine WCs, with a summary reported in \cref{tab:EFTresults}.
For $C_{Ht}$, the best-fit value and the 68\% Confidence Level (CL) interval are displaced from zero. However, the profile-likelihood scan exhibits two minima, one at a negative value of $C_{Ht}$ and another close to the SM value, and $C_{Ht}=0$ remains within the 95\% CL interval. This result therefore does not constitute a significant deviation from the SM.
The constraints on $C_{HQ}^{(1)}$, $C_{HQ}^{(3)}$, and $C_{Qq}^{(3,1)}$ are comparable to the best existing ATLAS limits. Other operators do not show an improvement over the existing limits~\cite{TOPQ-2020-17,TOPQ-2023-44,TOPQ-2018-11,TOPQ-2018-24,TOPQ-2020-20} owing to the limited statistical sensitivity of the analysis.

\begin{table}[htbp]
\tablesetup[1.5]
\caption{Results of the fit to data for each WC fitted individually considering linear+quadratic effects. The best fit value, 68\% and 95\% confidence intervals for the $\chi^{2}$ asymptotic assumption and from the coverage adjusted pseudodata are shown.}
\centering

%
\begin{tabular}{l c cc cc}
\toprule
Wilson & Best fit & \multicolumn{2}{c}{$\chi^{2}$ Asymptotic} & \multicolumn{2}{c}{Pseudodata} \\
Coefficient & & 68\% CI & 95\% CI & 68\% CI & 95\% CI \\
\midrule
$\text{Re}(C_{tW})$ & $-0.2$ & $[-0.7,0.8]$ & $[-1.0,1.1]$ & $[-0.8,0.0]\bigcup[0.1,0.8]$ & $[-1.1,1.1]$\\
$\text{Im}(C_{tW})$ & $0.2$ & $[-0.8,0.8]$ & $[-1.1,1.1]$ & $[-0.8,0.8]$ & $[-1.1,1.1]$\\
$\text{Re}(C_{tB})$ & $0.0$ & $[-1.4,1.4]$ & $[-2.2,2.2]$ & $[-1.4,1.4]$ & $[-2.3,2.3]$\\
$\text{Im}(C_{tB})$ & $0.1$ & $[-1.4,1.5]$ & $[-2.3,2.3]$ & $[-1.5,1.5]$ & $[-2.3,2.3]$\\
$C_{HQ}^{(1)}$ & $-0.5$ & $[-1.4,0.9]$ & $[-2.2,2.1]$ & $[-1.5,1.0]$ & $[-2.3,2.2]$\\
$C_{HQ}^{(3)}$ & $0.4$ & $[-0.9,0.9]$ & $[-1.8,1.3]$ & $[-0.9,1.0]$ & $[-1.8,1.5]$\\
$C_{Ht}$ & $-18$ & $[-19,-16]$ & $[-20,-14]$ & $[-22,-15]$ & $[-24,-14]\bigcup[-2,2]$\\
$C_{Qq}^{(3,1)}$ & $-0.12$ & $[-0.26,0.13]$ & $[-0.37,0.39]$ & $[-0.27,0.11]$ & $[-0.40,0.33]$\\
$C_{Qq}^{(3,8)}$ & $-0.6$ & $[-0.9,0.7]$ & $[-1.2,1.1]$ & $[-1.0,-0.2]\bigcup[0.5,0.8]$ & $[-1.2,1.2]$\\
\bottomrule
\end{tabular}

%
%
%
%
%
%
%
%
%
%
%
%
%
%
%
%


\label{tab:EFTresults}
\end{table}

\begin{figure}[tbp]
\centering
\begin{subfigure}{0.45\textwidth}
\centering
\includegraphics[width=\textwidth]{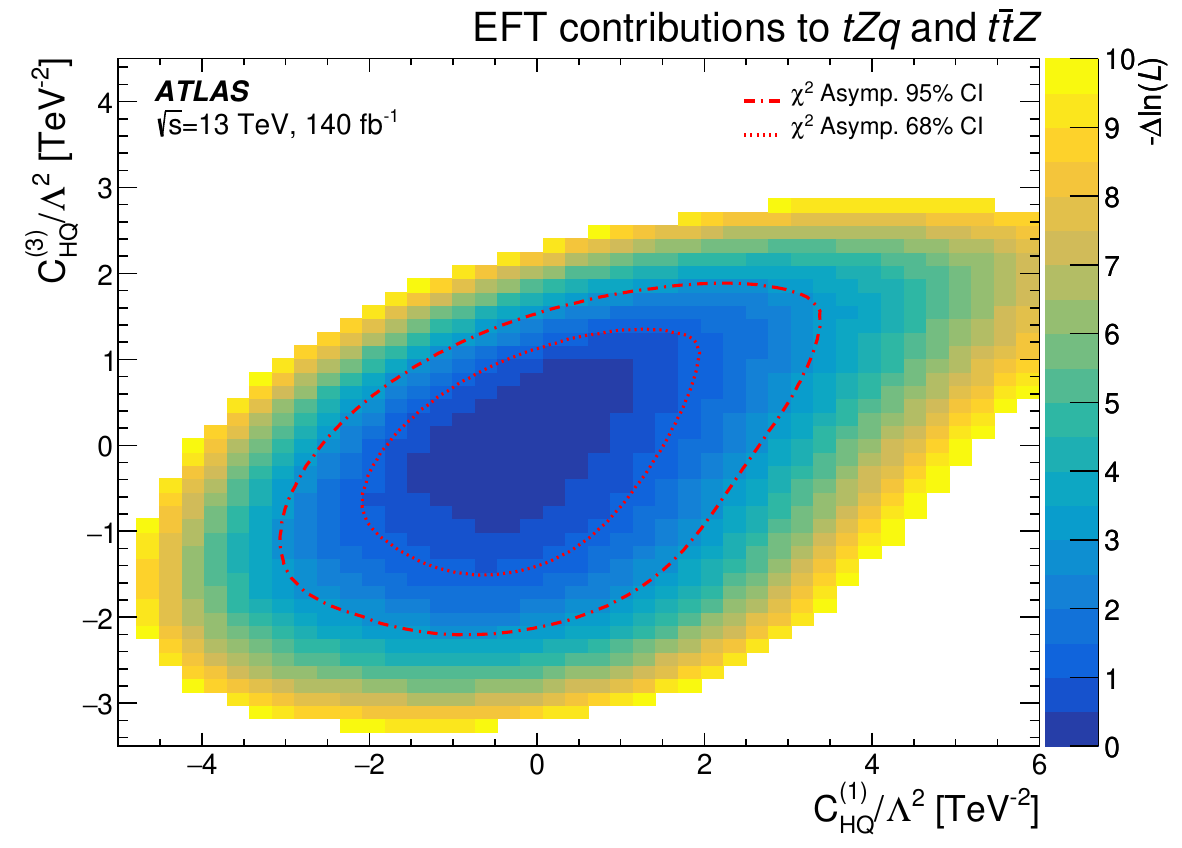}
\caption{}
\label{fig:eft:cHQ1:cHQ3:combined}
\end{subfigure} \\
\hfill
\begin{subfigure}{0.45\textwidth}
\centering
\includegraphics[width=\textwidth]{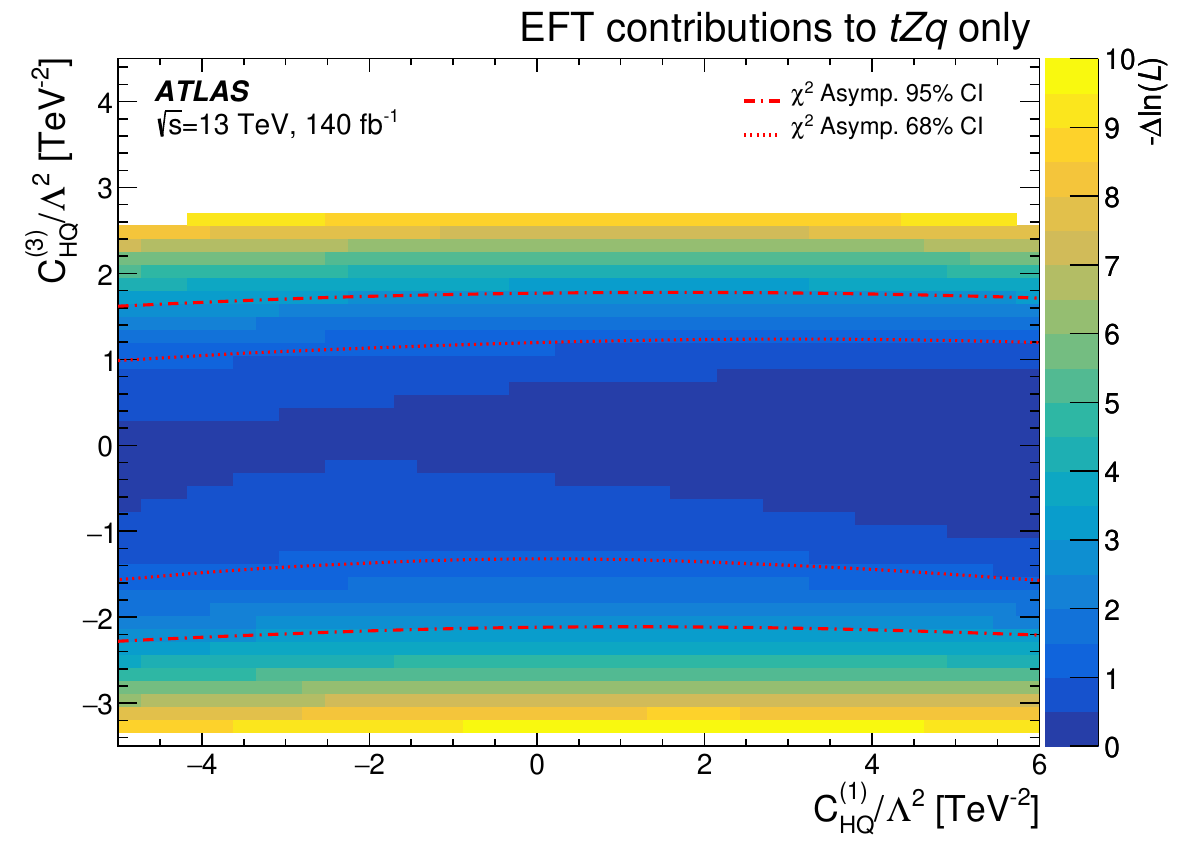}
\caption{}
\label{fig:eft:cHQ1:cHQ3:tZq}
\end{subfigure}
\hfill
\begin{subfigure}{0.45\textwidth}
\centering
\includegraphics[width=\textwidth]{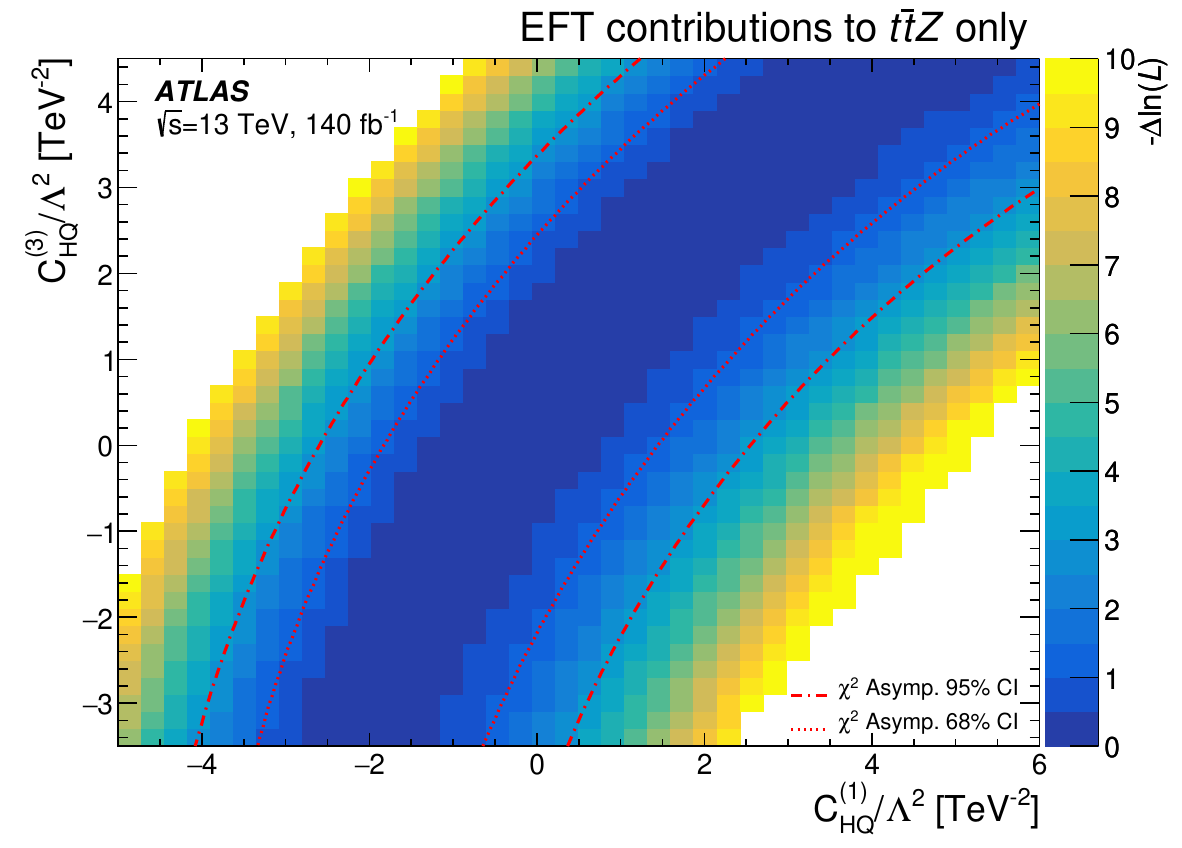}
\caption{}
\label{fig:eft:cHQ1:cHQ3:ttZ}
\end{subfigure}
\caption{Two-dimensional likelihood scans in the $C_{HQ}^{(1)}$--$C_{HQ}^{(3)}$ plane, obtained using (a) the combination of the \tZq and \ttZ signals, (b) only the \tZq signal, and (c) only the \ttZ signal. The dashed and dotted contours correspond to the asymptotic 95\% and 68\% confidence intervals, respectively.}
\label{fig:eft:cHQ1:cHQ3}
\end{figure}

The two-dimensional constraints on the Wilson coefficients $C_{HQ}^{(1)}$ and $C_{HQ}^{(3)}$ are shown in \cref{fig:eft:cHQ1:cHQ3}. The likelihood scans are performed using the \tZq and \ttZ signals separately, as well as their combination. The two processes exhibit different dependences on the two operators. The \tZq-only fit shows a stronger sensitivity to $C_{HQ}^{(3)}$, with a comparatively weaker dependence on $C_{HQ}^{(1)}$, resulting in contours that are mostly elongated along the $C_{HQ}^{(1)}$ direction. In contrast, the \ttZ-only fit constrains a different linear combination of the two coefficients, leading to a markedly different orientation of the allowed region. The combination of the two processes therefore exploits their complementarity and reduces the degeneracy between $C_{HQ}^{(1)}$ and $C_{HQ}^{(3)}$, yielding a more constrained allowed region than either process alone.

%
\FloatBarrier

%

%
\section{Conclusions}
\label{sec:conclusions}

Measurements of the production of a single top quark in association with a \PZ boson are presented using the full Run~2 dataset of proton--proton collisions collected with the ATLAS detector at $\sqrt{s}=\qty{13}{\TeV}$, corresponding to an integrated luminosity of \lumi. The analysis is performed in the trilepton final state and exploits a neural-network discriminant to separate the signal from the dominant backgrounds.

The inclusive \tZq production cross-section is measured to be \tZqxsec for dilepton invariant masses $m_{\ell\ell}>\qty{30}{\GeV}$, in agreement with the Standard Model prediction. Separate measurements of the \tZqt and \tZqtbar production cross-sections and their ratio are also performed and found to be consistent with theoretical expectations.

Differential cross-section measurements are presented at both particle level and parton level for a variety of observables sensitive to the kinematic properties of the \tZq system. The unfolded distributions are found to be well described by the Standard Model predictions within the experimental and theoretical uncertainties.
Integrated fiducial cross-sections are also measured in the particle-level and parton-level phase spaces and are consistent with the corresponding Standard Model predictions.

The top-quark spin asymmetry is measured using the unfolded distribution of the top-quark polarisation angle and is found to be $A = 0.34 \pm 0.14~(\mathrm{stat.}) \pm 0.06~(\mathrm{syst.})$,
in agreement with the Standard Model prediction.

The results are further interpreted in the framework of the Standard Model Effective Field Theory. Constraints are set on nine Wilson coefficients corresponding to seven dimension-six operators of the Warsaw basis. No significant deviation from the Standard Model expectation is observed.

These measurements provide the most precise ATLAS study of \tZq production to date and constitute a comprehensive test of the electroweak interactions of the top quark. They also provide valuable input for future global SMEFT interpretations and for precision studies of rare top-quark processes.


%
%
%
%
\section*{Acknowledgements}

%
%

%
%

We thank CERN for the very successful operation of the LHC and its injectors, as well as the support staff at
CERN and at our institutions worldwide without whom ATLAS could not be operated efficiently.

The crucial computing support from all WLCG partners is acknowledged gratefully, in particular from CERN, the ATLAS Tier-1 facilities at TRIUMF/SFU (Canada), NDGF (Denmark, Norway, Sweden), CC-IN2P3 (France), KIT/GridKA (Germany), INFN-CNAF (Italy), NL-T1 (Netherlands), PIC (Spain), RAL (UK) and BNL (USA), the Tier-2 facilities worldwide and large non-WLCG resource providers. Major contributors of computing resources are listed in Ref.~\cite{ATL-SOFT-PUB-2026-001}.

We gratefully acknowledge the support of ANPCyT, Argentina; YerPhI, Armenia; ARC, Australia; BMWFW and FWF, Austria; ANAS, Azerbaijan; CNPq and FAPESP, Brazil; NSERC, NRC and CFI, Canada; CERN; ANID, Chile; CAS, MOST, NSFC and FRFCU, China; Minciencias, Colombia; MEYS CR, Czech Republic; DNRF and DNSRC, Denmark; IN2P3-CNRS and CEA-DRF/IRFU, France; SRNSFG, Georgia; BMFTR, HGF and MPG, Germany; GSRI, Greece; RGC and Hong Kong SAR, China; ICHEP and Academy of Sciences and Humanities, Israel; INFN, Italy; MEXT and JSPS, Japan; CNRST, Morocco; NWO, Netherlands; RCN, Norway; MNiSW, Poland; FCT, Portugal; MNE/IFA, Romania; MSTDI, Serbia; MSSR, Slovakia; ARIS and MVZI, Slovenia; DSI/NRF, South Africa; MICIU/AEI, Spain; SRC and Wallenberg Foundation, Sweden; SERI, SNSF and Cantons of Bern and Geneva, Switzerland; NSTC, Taipei; TENMAK, T\"urkiye; STFC/UKRI, United Kingdom; DOE and NSF, United States of America.

Individual groups and members have received support from BCKDF, CANARIE, CRC and DRAC, Canada; CERN-CZ, FORTE and PRIMUS, Czech Republic; COST, ERC, ERDF, Horizon 2020 and Marie Sk{\l}odowska-Curie Actions, European Union; Investissements d'Avenir Labex, Investissements d'Avenir Idex and ANR, France; DFG and AvH Foundation, Germany; Herakleitos, Thales and Aristeia programmes co-financed by EU-ESF and the Greek NSRF, Greece; BSF-NSF and MINERVA, Israel; NCN and NAWA, Poland; La Caixa Banking Foundation, CERCA and AGAUR programs from Generalitat de Catalunya and PROMETEO and GenT Programmes Generalitat Valenciana, Spain; G\"{o}ran Gustafssons Stiftelse, Sweden; The Royal Society and Leverhulme Trust, United Kingdom; Eric and Wendy Schmidt Fund for Strategic Innovation, United States of America.

In addition, individual members wish to acknowledge support from Chile: Agencia Nacional de Investigaci\'on y Desarrollo (ANID FONDECYT reg. 1230987, FONDECYT 1230812, FONDECYT 1240864, Fondecyt 3240661, Fondecyt Regular 1240721); China: Fundamental Research Funds for the Central Universities (010-63263105), Chinese Ministry of Science and Technology (MOST-2023YFA1605700, MOST-2023YFA1609300), National Natural Science Foundation of China (NSFC 12275265, NSFC-W2543005); Czech Republic: Czech Science Foundation (GACR - 24-11373S), Ministry of Education Youth and Sports (ERC-CZ-LL2327, FORTE CZ.02.01.01/00/22\_008/0004632); EU: H2020 European Research Council (ERC - 101002463); European Union: European Research Council (BARD No. 101116429, ERC - 101219398, ERC 101089007), European Regional Development Fund (HE COFUND GA No.101081355, ERDF), Marie Sklodowska-Curie Actions (GAP-101168829); France: Agence Nationale de la Recherche (ANR-22-EDIR-0002, ANR-24-CE31-0504-01); Germany: Deutsche Forschungsgemeinschaft (DFG - 469666862); China: Research Grants Council (GRF); Italy: Istituto Nazionale di Fisica Nucleare (LHC-MIUR - 28003/2025), Ministero dell'Università e della Ricerca (NextGenEU I53D23000820006 M4C2.1.1, SOE2024\_0000023); Japan: Japan Society for the Promotion of Science (JSPS KAKENHI  JP25H0063, JSPS KAKENHI JP22H04944, JSPS KAKENHI JP24K23939, JSPS KAKENHI JP24KK0251, JSPS KAKENHI JP25H00650, JSPS KAKENHI JP25H01291, JSPS KAKENHI JP25K01011, JSPS KAKENHI JP25K01023, JSPS KAKENHI JP25KK0047); Poland: Polish National Science Centre (NCN 2021/42/E/ST2/00350, NCN OPUS 2023/51/B/ST2/02507, NCN OPUS nr 2022/47/B/ST2/03059, NCN UMO-2019/34/E/ST2/00393, UMO-2022/47/O/ST2/00148, UMO-2023/49/B/ST2/04085, UMO-2023/51/B/ST2/00920, UMO-2024/53/N/ST2/00869); Spain: Agència de Gestió d'Ajuts Universitaris i de Recerca. (AGAUR - 2023 BP 00141), Ministry of Science and Innovation (RYC2019-028510-I, RYC2020-030254-I, RYC2021-031273-I, RYC2022-038164-I), Ministerio de Ciencia, Innovación y Universidades/Agencia Estatal de Investigaci\'on (EU NextGenerationEU (PRTR-C17.I1), PID2022-142604OB-C22); Sweden: Carl Trygger Foundation (Carl Trygger Foundation CTS 22:2312), Swedish Research Council (Swedish Research Council 2023-04654, VR 2021-03651, VR 2022-03845, VR 2022-04683, VR 2023-03403, VR 2024-05451, VR 2025-05940), Knut and Alice Wallenberg Foundation (KAW 2023.0366); United Kingdom: The Binks Trust, Royal Society (NIF-R1-231091); United States of America: U.S. Department of Energy (ECA DE-AC02-76SF00515), John Templeton Foundation (John Templeton Foundation 63206), Neubauer Family Foundation.

%
%


%

%
%
%
%
%
%
%
%
%
%

%
\clearpage
\appendix
\part*{Appendix}
\addcontentsline{toc}{part}{Appendix}

%

This appendix provides additional information about the inclusive cross-section measurement, as well as the full set of differential cross-sections (absolute and normalised, unfolded to particle level and parton level).

\begin{figure}[!htbp]
\centering
\begin{tabular}{cc}
\begin{subfigure}[b]{.35\textwidth}
\includegraphics[width=\textwidth]{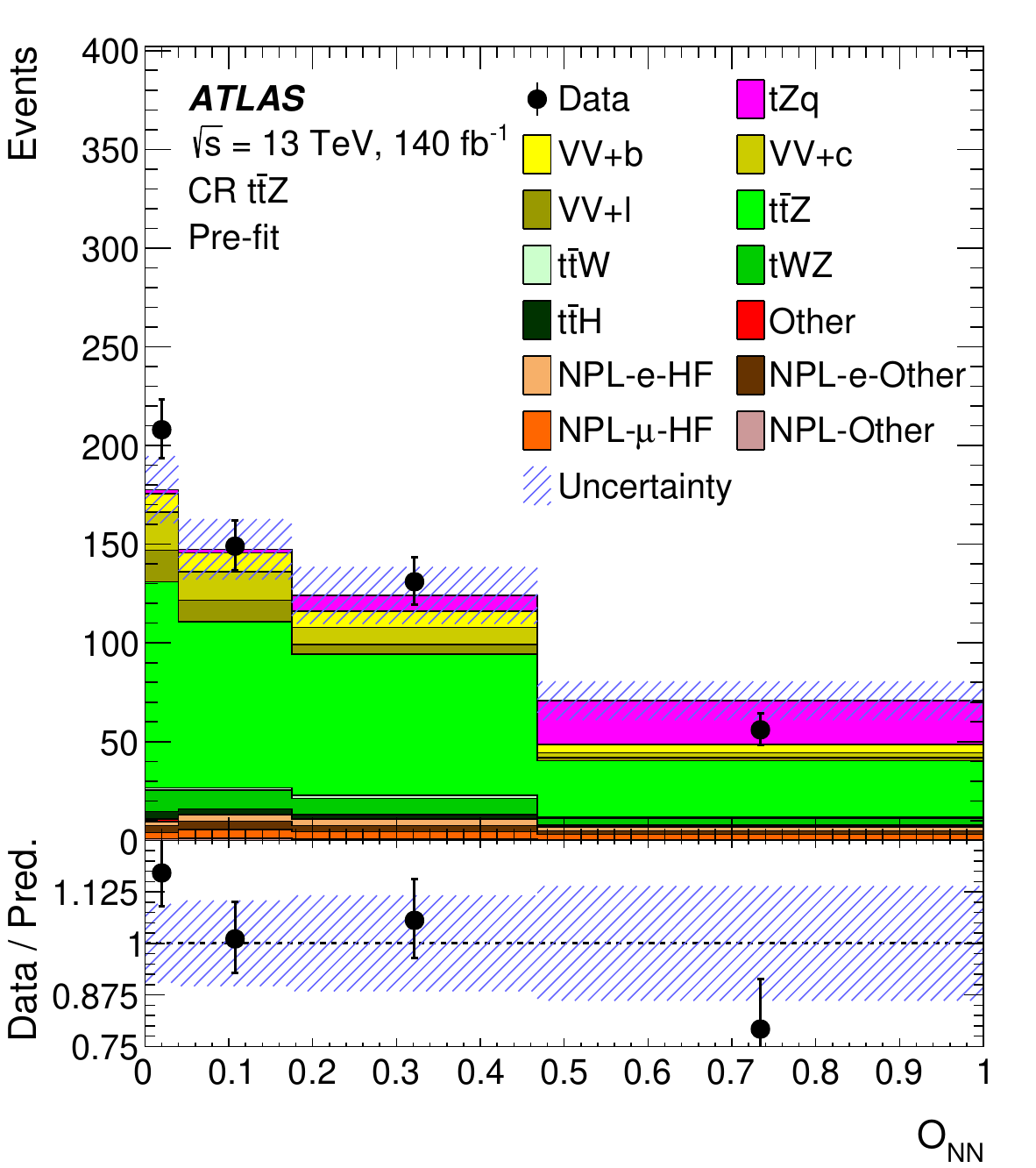}
\caption{}
\end{subfigure}
&
\begin{subfigure}[b]{.35\textwidth}
\includegraphics[width=\textwidth]{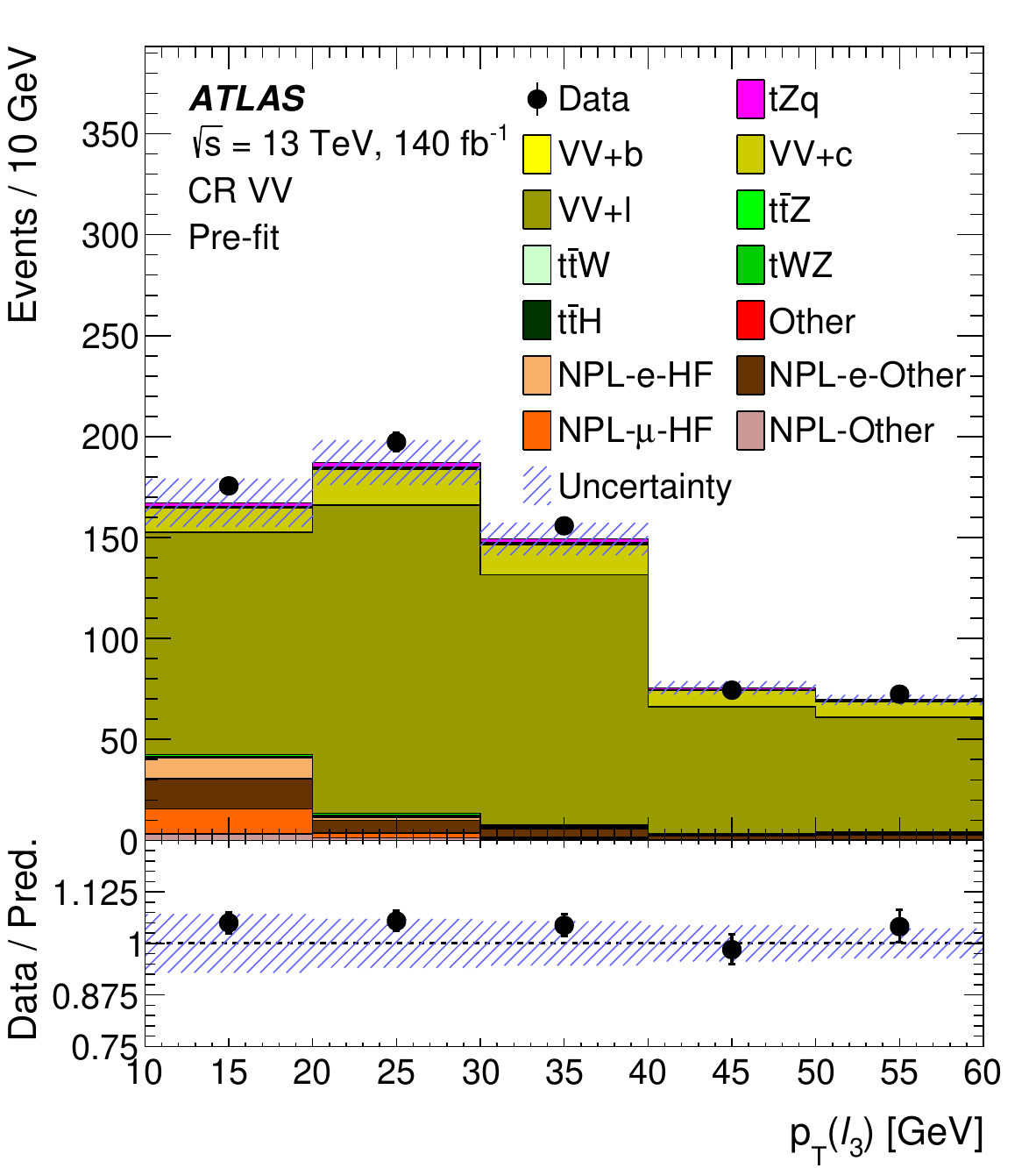}
\caption{}
\end{subfigure} \\

\begin{subfigure}[b]{.35\textwidth}
\includegraphics[width=\textwidth]{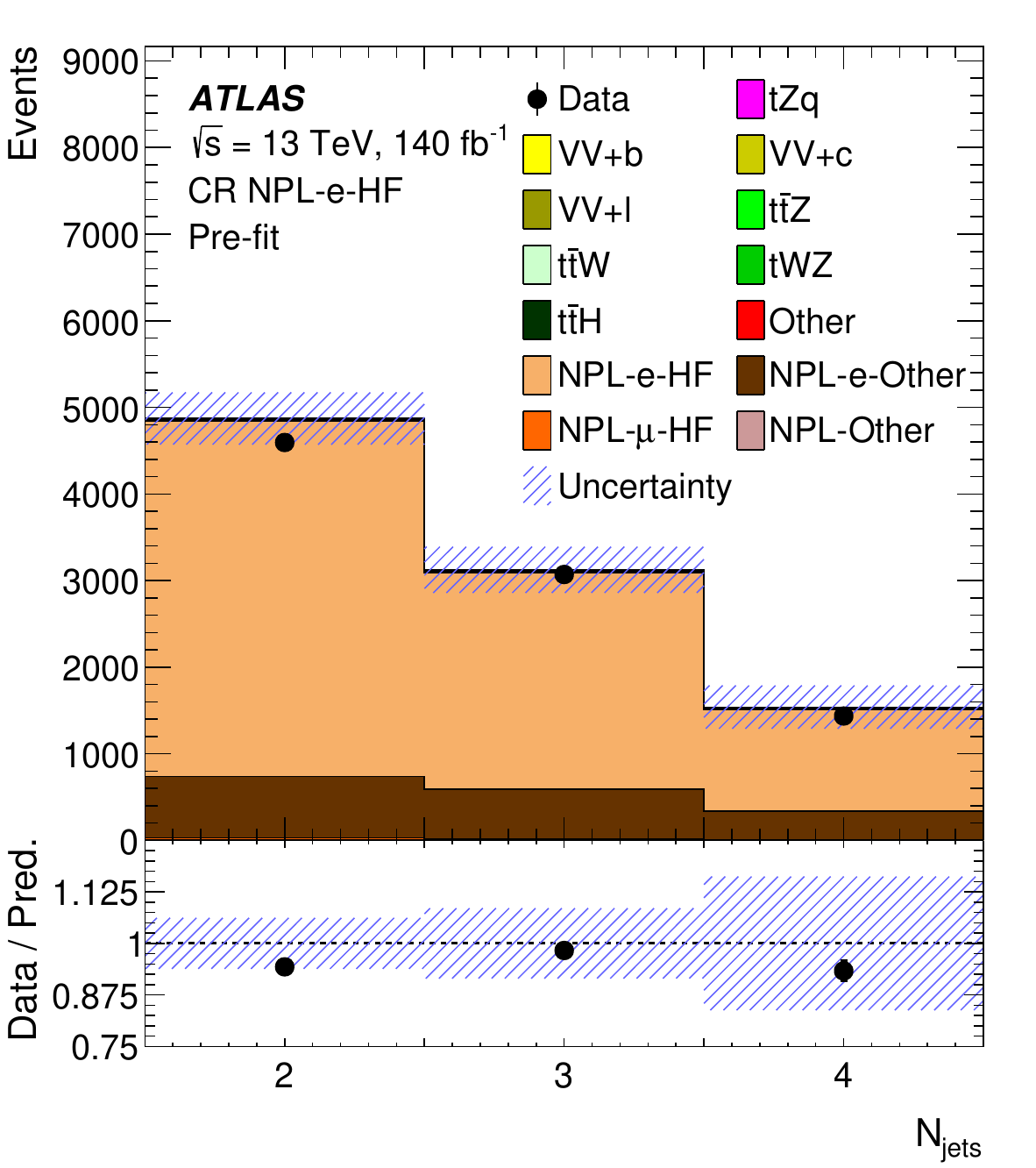}
\caption{}
\end{subfigure}
&
\begin{subfigure}[b]{.35\textwidth}
\includegraphics[width=\textwidth]{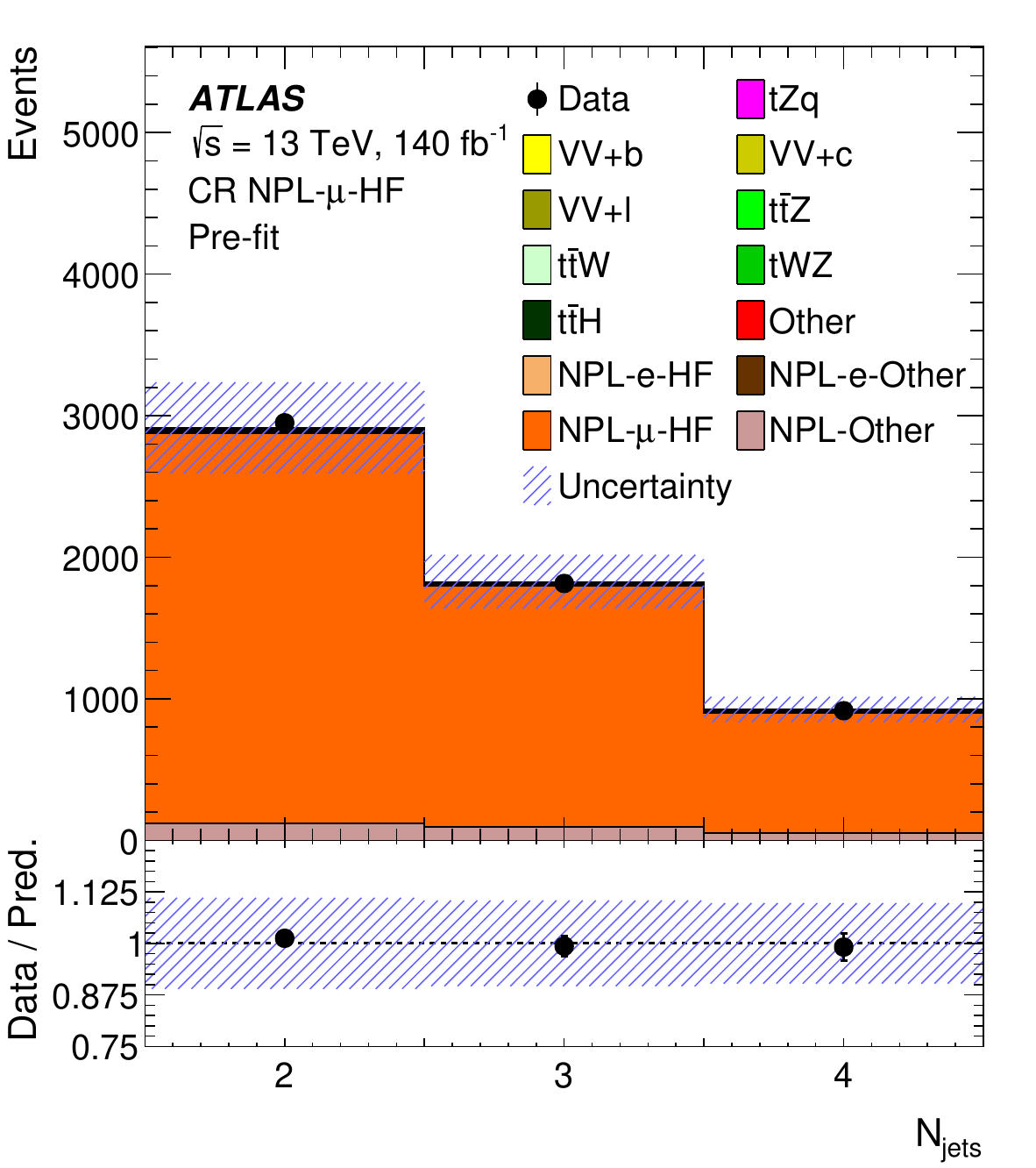}
\caption{}
\end{subfigure} \\
\begin{subfigure}[b]{.35\textwidth}
\includegraphics[width=\textwidth]{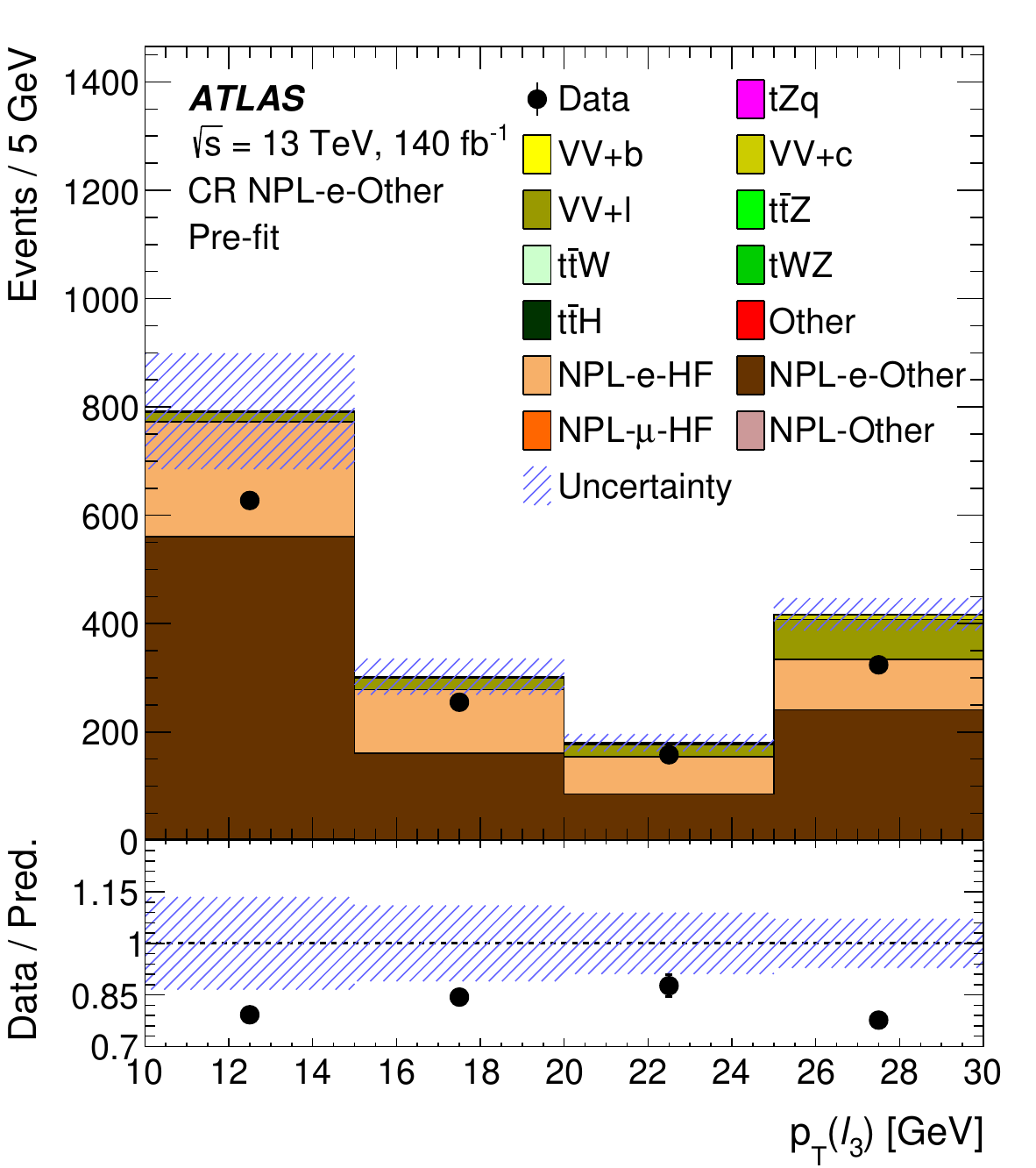}
\caption{}
\end{subfigure}
&
\begin{subfigure}[b]{.35\textwidth}
\includegraphics[width=\textwidth]{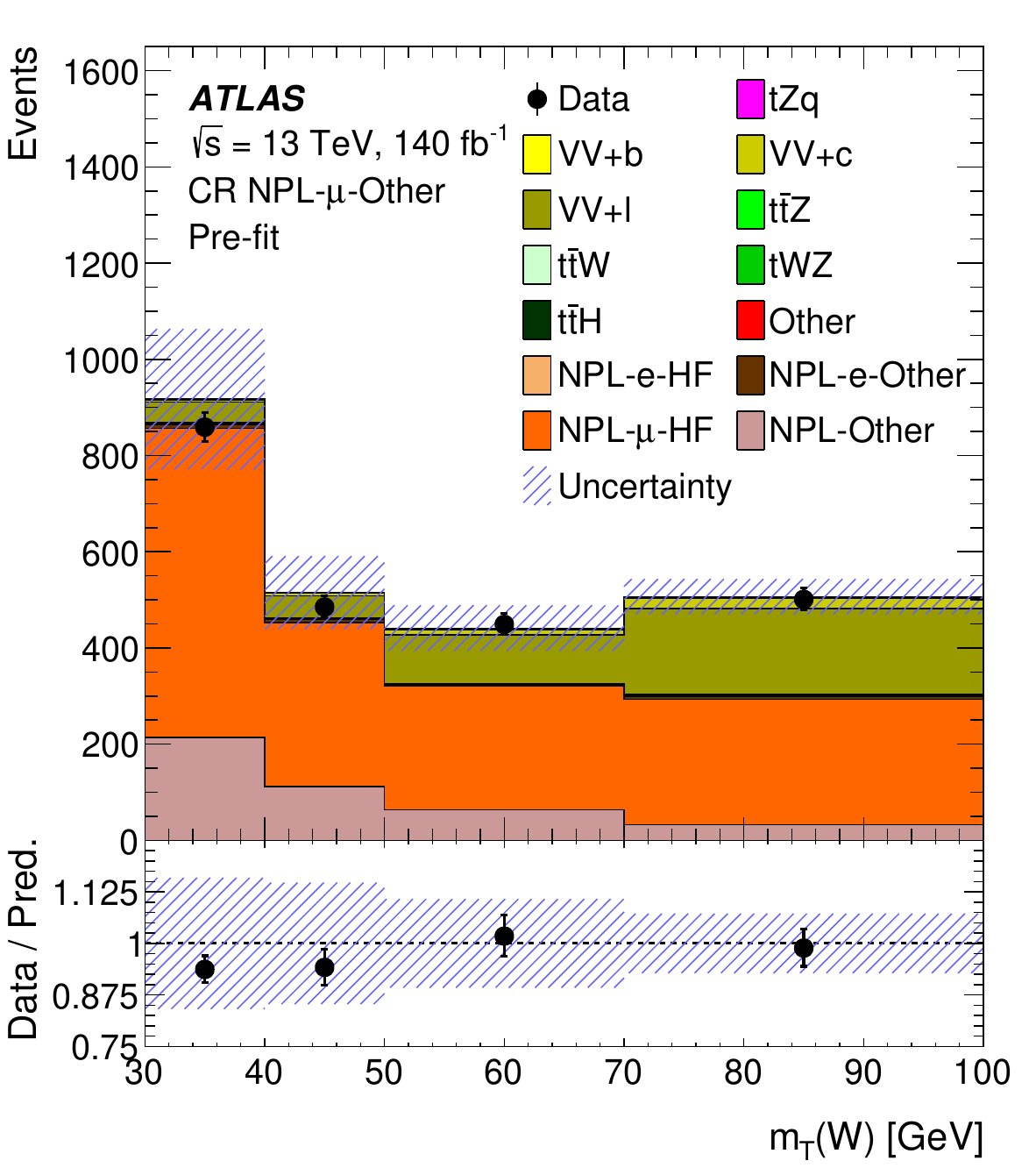}
\caption{}
\end{subfigure}
\end{tabular}
\caption{
Pre-fit distributions of the discriminant in the control regions:
(a) \CRttZ,
(b) \CRVV,
(c) \CRelHF,
(d) \CRmuHF,
(e) \CRelOther, and
(f) \CRmuOther.
The lower panels show the ratio of the data to the prediction (\enquote{Pred.}).
The hatched bands represent the total uncertainty in the prediction, including both statistical and systematic contributions.
Where applicable, overflow events are included in the rightmost bin.
}
\label{fig:app_inclusive:crs}
\end{figure}

\Cref{fig:app_inclusive:crs} shows the pre-fit distributions in the CRs used for the inclusive fit. Among these, only the \NPLelOther CR exhibits poor pre-fit data/MC agreement. This region is mainly composed of \NPLelHF and \NPLelOther contributions. Since \NPLelHF is the dominant contribution in the \NPLelHF CR, where the pre-fit data/MC agreement is close to unity, the fit is expected to reduce the \NPLelOther contribution in order to improve the agreement with data. This behaviour is reflected in the post-fit normalisation factors: \NPLelOther is the only component whose fitted normalisation factor is significantly below unity.

\begin{figure}
\centering
\begin{tabular}{cc}
\begin{subfigure}[b]{.45\textwidth}
\includegraphics[width=\textwidth]{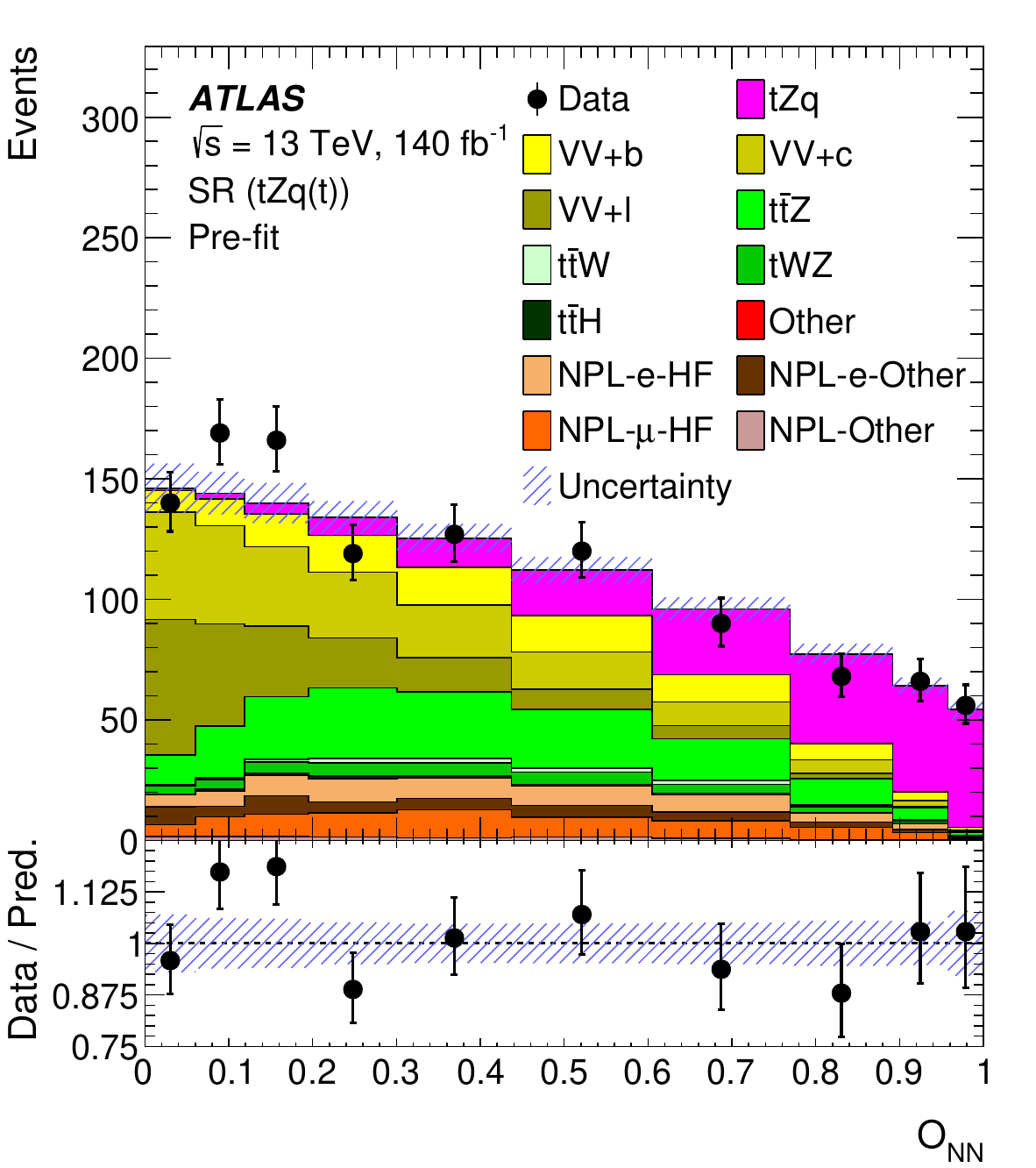}
\caption{}
\end{subfigure}
&
\begin{subfigure}[b]{.45\textwidth}
\includegraphics[width=\textwidth]{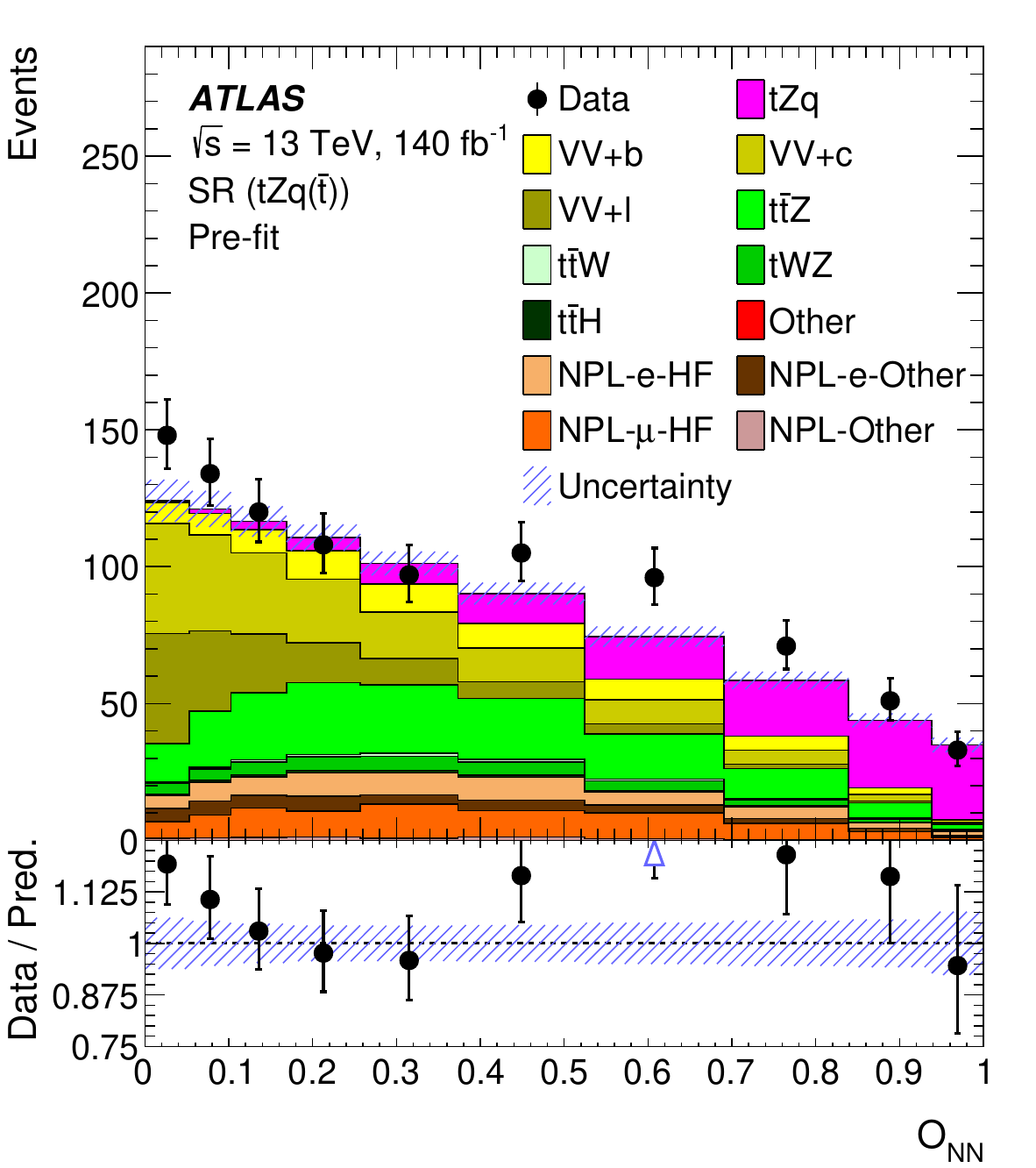}
\caption{}
\end{subfigure} \\
\end{tabular}
\caption{
Pre-fit distributions of the NN discriminant in the split SR:
(a) SR(\tZqt),
(b) SR(\tZqtbar).
The lower panels show the ratio of the data to the prediction (\enquote{Pred.}).
The hatched bands represent the total uncertainty in the prediction, including both statistical and systematic contributions.
}
\label{fig:app_inclusive:srsplit}
\end{figure}

\Cref{fig:app_inclusive:srsplit} compares the data with the pre-fit background and signal distributions in the SRs, split according to the charge of the lepton associated with the \PW boson.


%

%

\begin{figure}[htbp]
\centering
\begin{subfigure}{0.45\textwidth}
\centering
\includegraphics[width=1\textwidth]{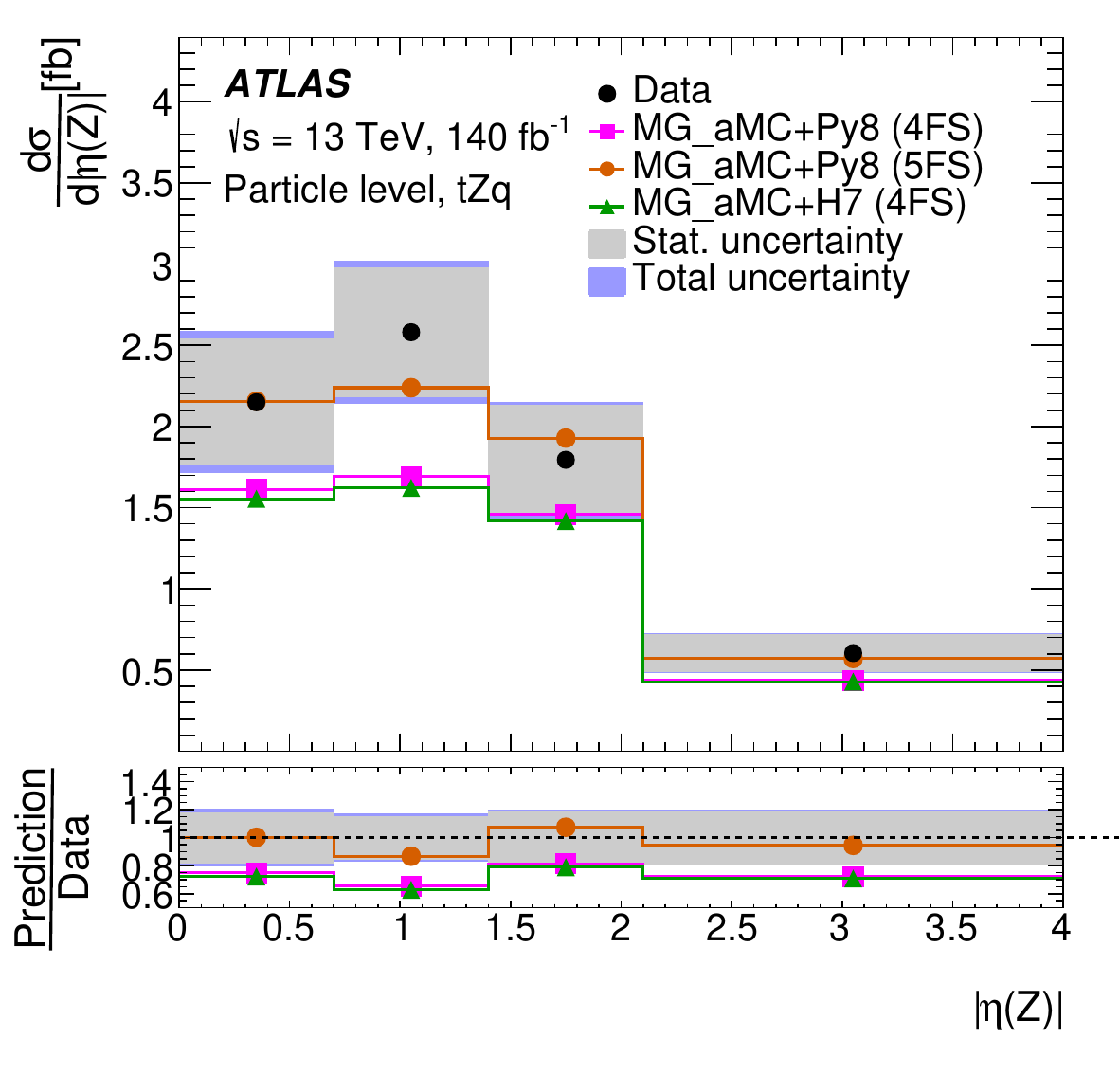}
\caption{}
\end{subfigure} \\
\begin{subfigure}{0.45\textwidth}
\centering
\includegraphics[width=1\textwidth]{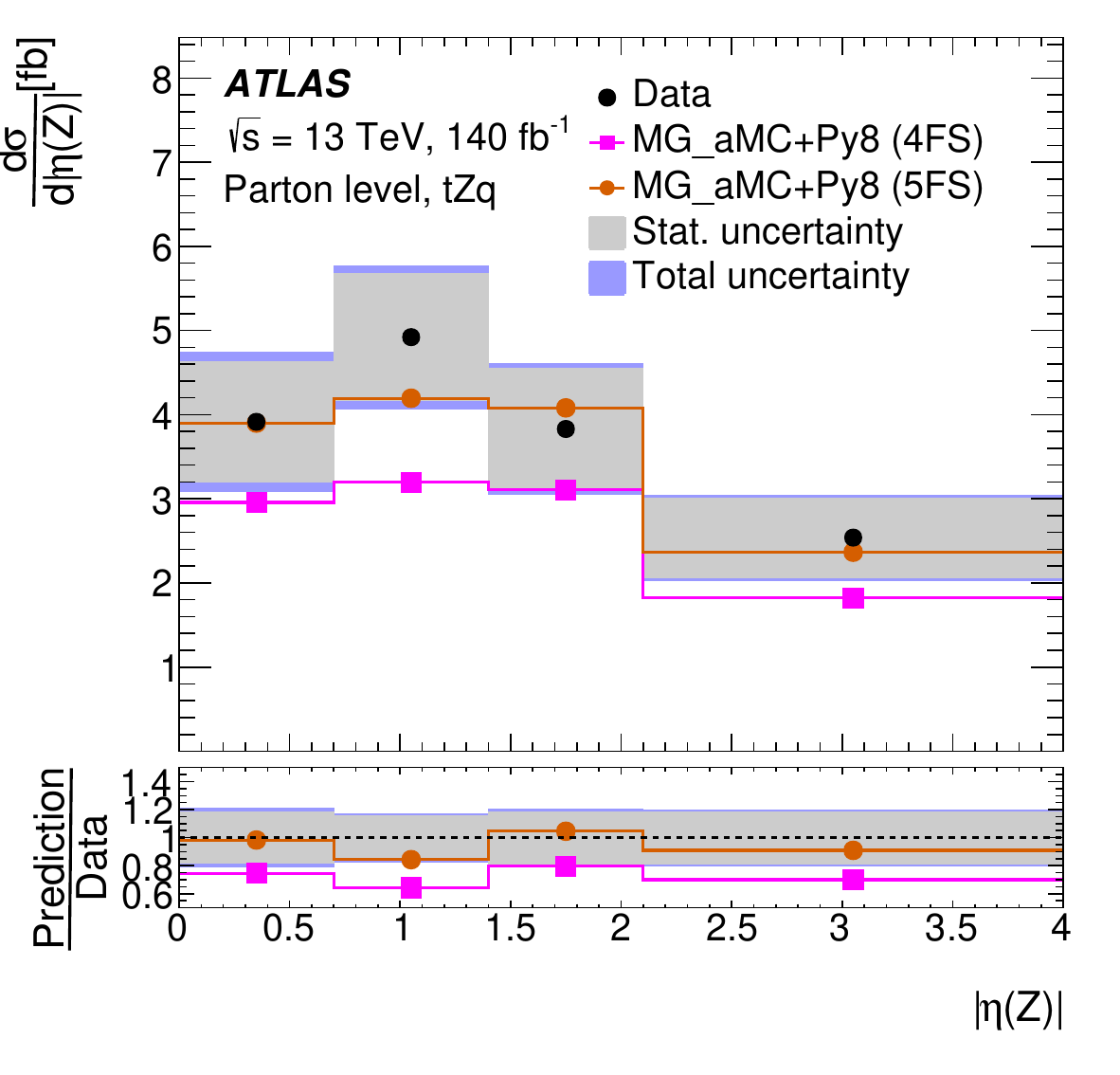}
\caption{}
\end{subfigure}
\begin{subfigure}{0.45\textwidth}
\centering
\includegraphics[width=1\textwidth]{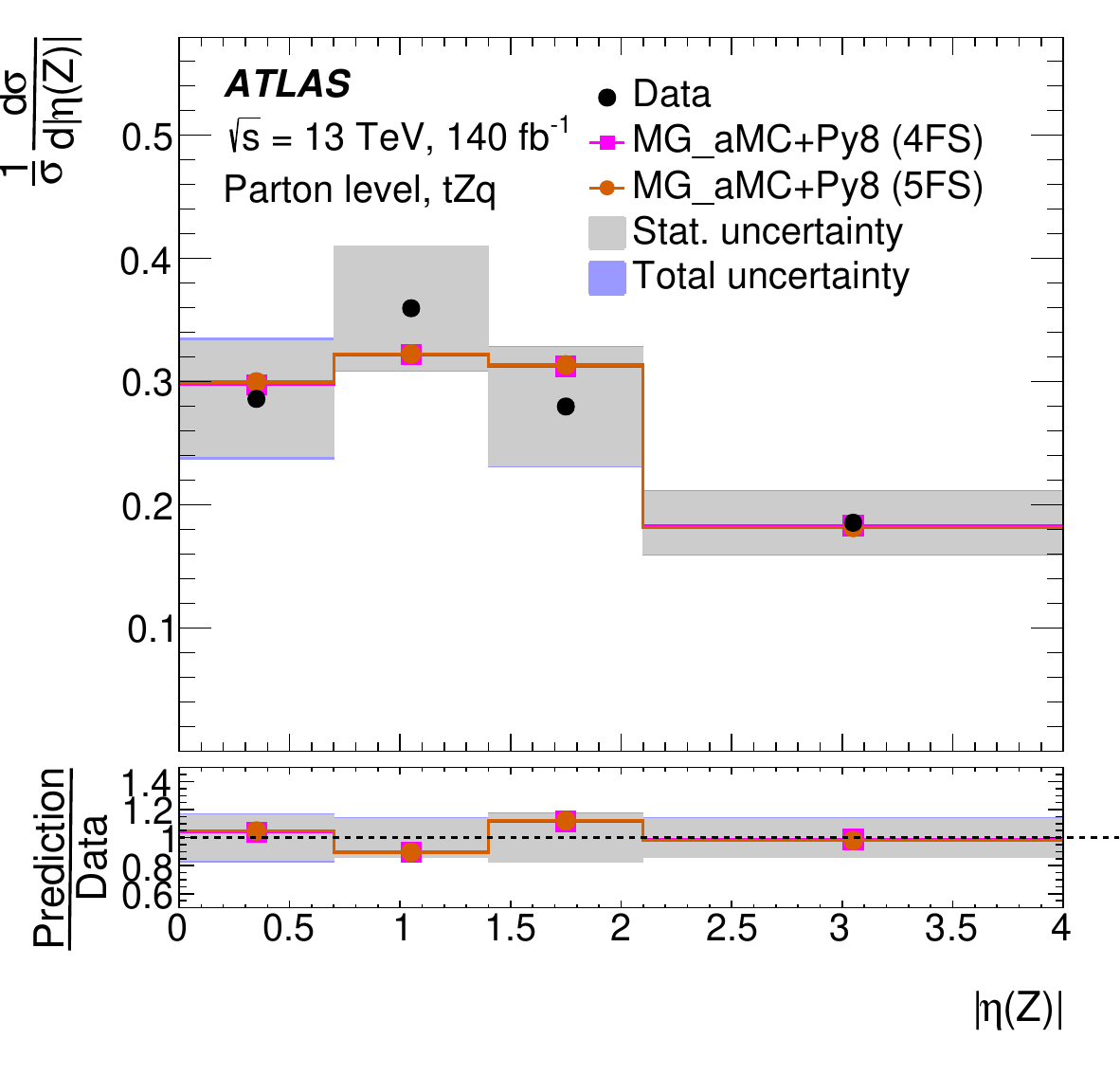}
\caption{}
\end{subfigure}
\caption{Differential cross-section measurements as a function of \etaZ observable in the SR,
(a, b) absolute and (c) normalised, unfolded to (a) particle level and (b, c) parton level.
The light blue band corresponds to the total uncertainty of the measurement; in some cases, it is almost fully covered by the light grey band, representing the dominant statistical uncertainty.
The measurements are compared with the generator predictions indicated in the legend and described in the text.
The lower panels show the ratios of the predictions to the (a, b) measured cross-sections and the (c) measured normalised differential cross-sections.
}
\label{fig:diff:Zeta}
\end{figure}

\begin{figure}[htbp]
\centering
\begin{subfigure}{0.45\textwidth}
\centering
\includegraphics[width=1\textwidth]{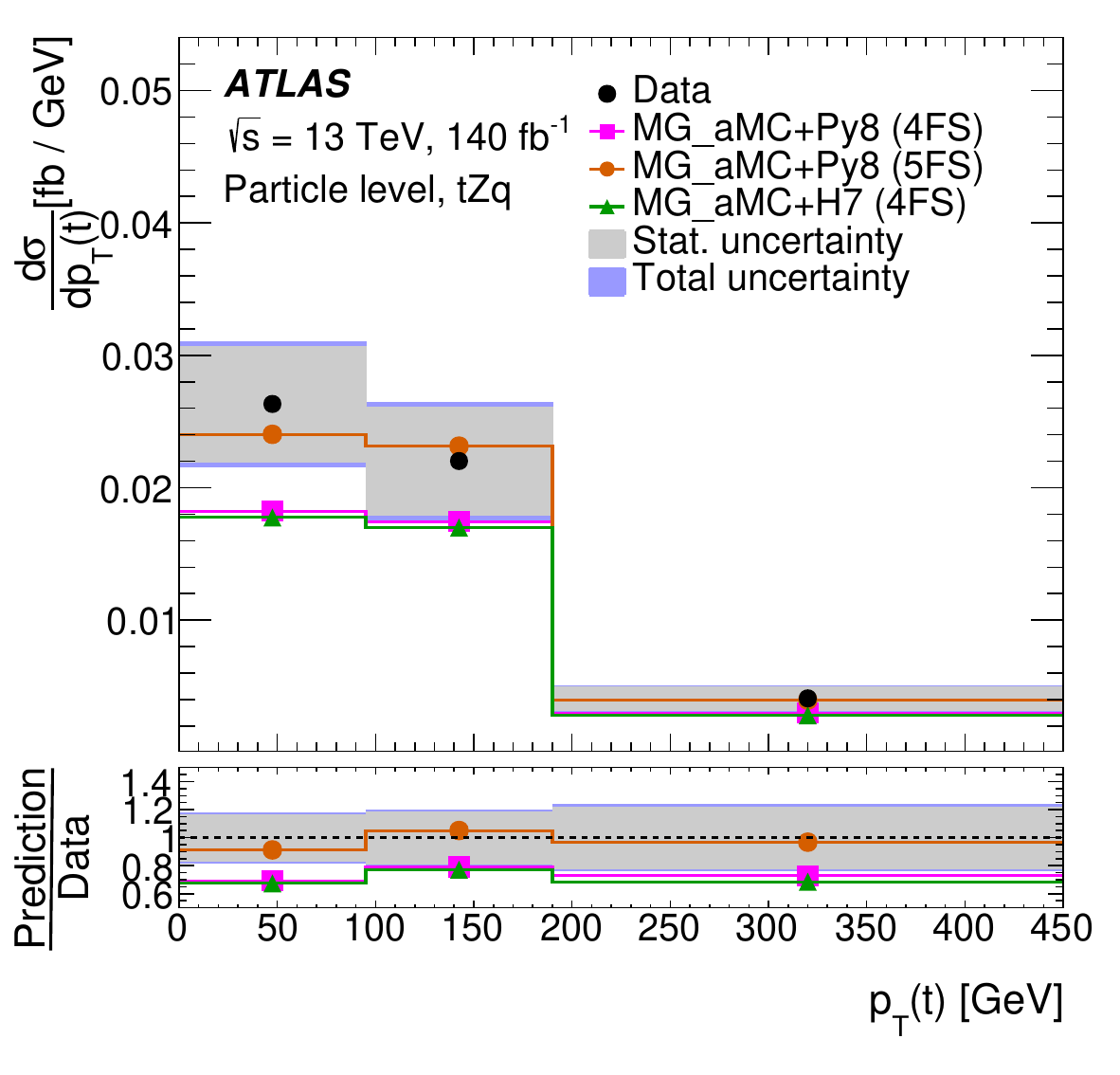}
\caption{}
\end{subfigure} \\
\begin{subfigure}{0.45\textwidth}
\centering
\includegraphics[width=1\textwidth]{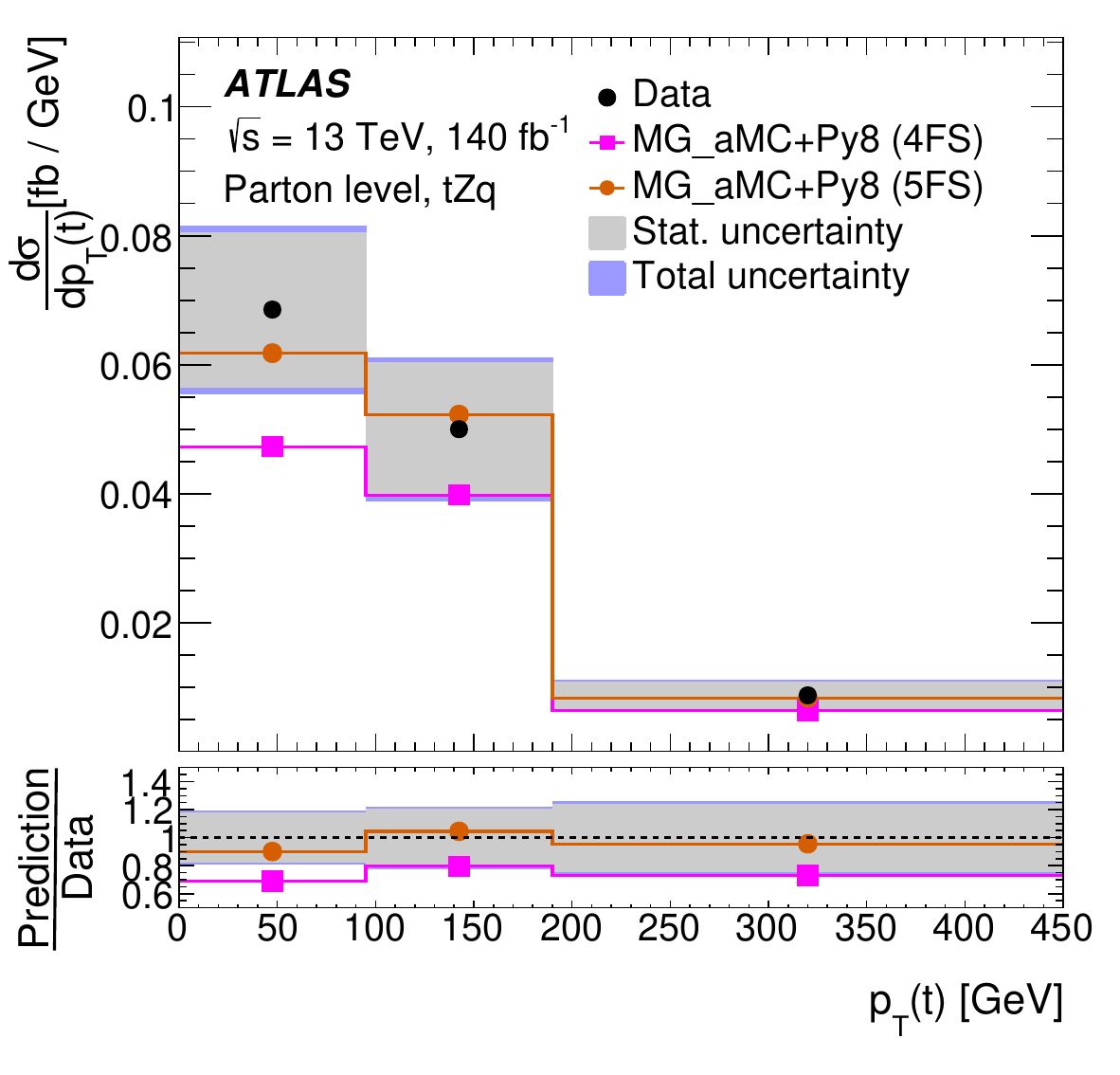}
\caption{}
\end{subfigure}
\begin{subfigure}{0.45\textwidth}
\centering
\includegraphics[width=1\textwidth]{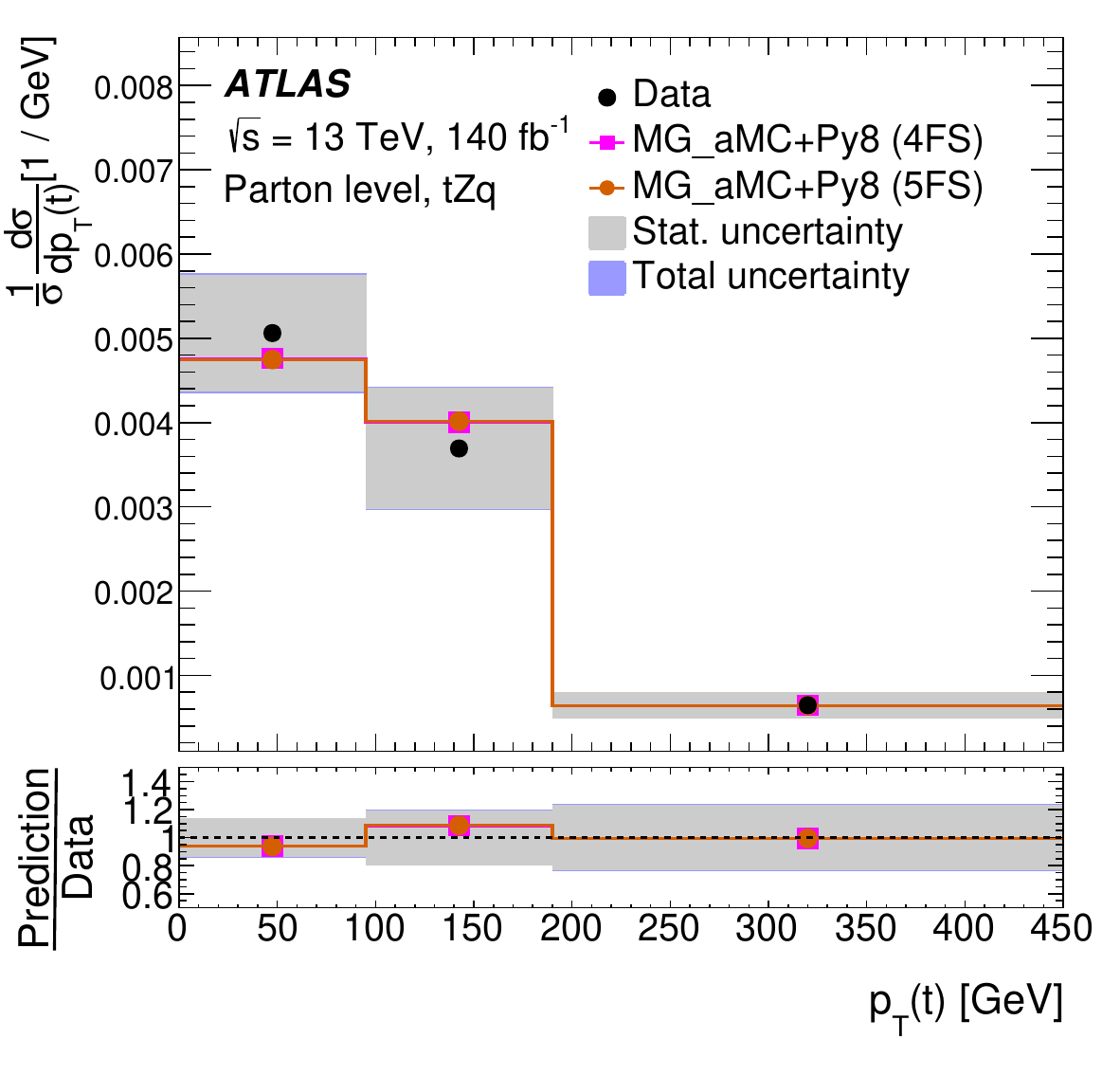}
\caption{}
\end{subfigure}
\caption{Differential cross-section measurements as a function of \pTt observable in the SR,
(a, b) absolute and (c) normalised, unfolded to (a) particle level and (b, c) parton level.
The light blue band corresponds to the total uncertainty of the measurement; in some cases, it is almost fully covered by the light grey band, representing the dominant statistical uncertainty.
The measurements are compared with the generator predictions indicated in the legend and described in the text.
The lower panels show the ratios of the predictions to the (a, b) measured cross-sections and the (c) measured normalised differential cross-sections.
}
\label{fig:diff:toppt}
\end{figure}

\begin{figure}[htbp]
\centering
\begin{subfigure}{0.45\textwidth}
\centering
\includegraphics[width=1\textwidth]{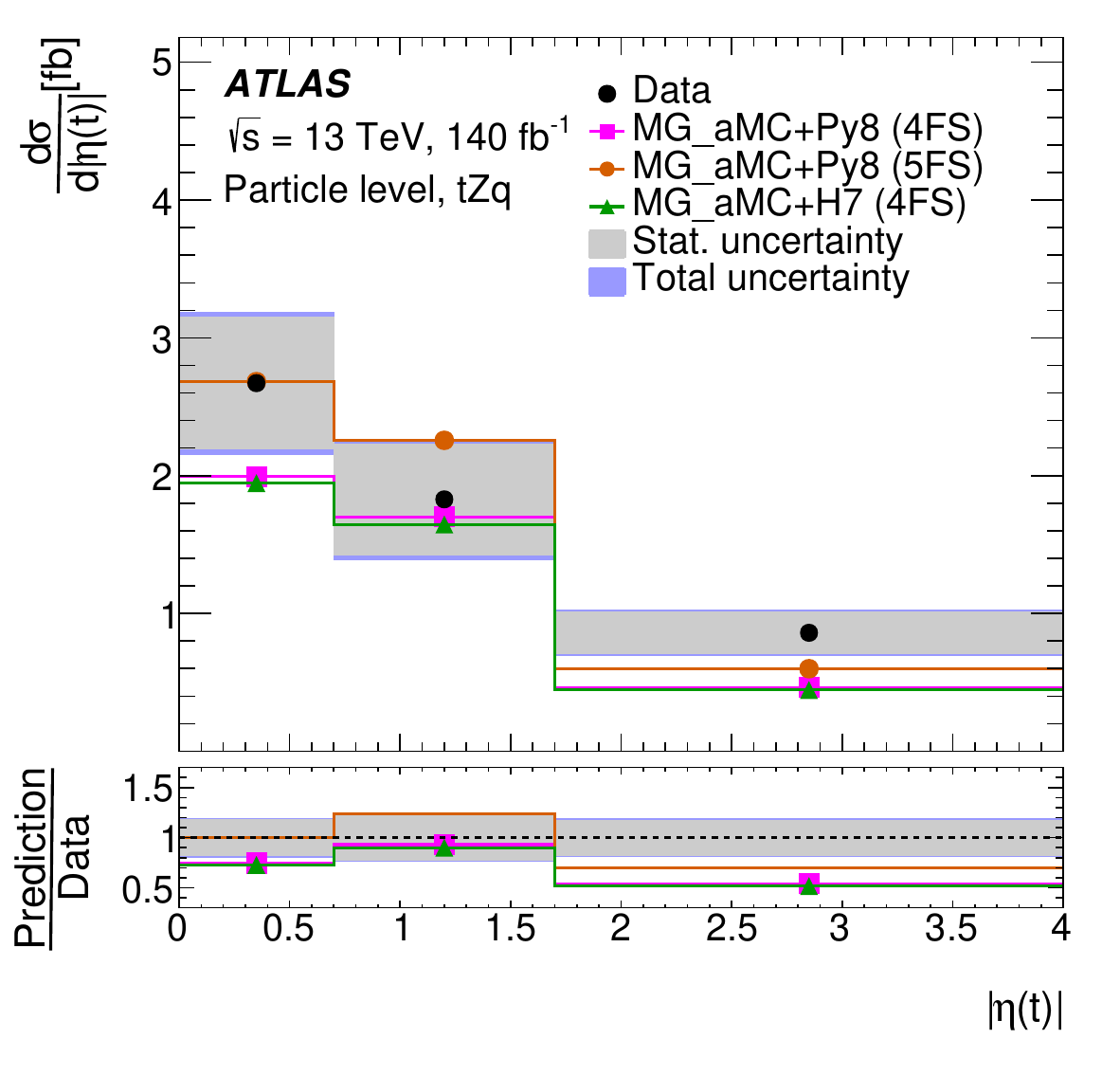}
\caption{}
\end{subfigure} \\
\begin{subfigure}{0.45\textwidth}
\centering
\includegraphics[width=1\textwidth]{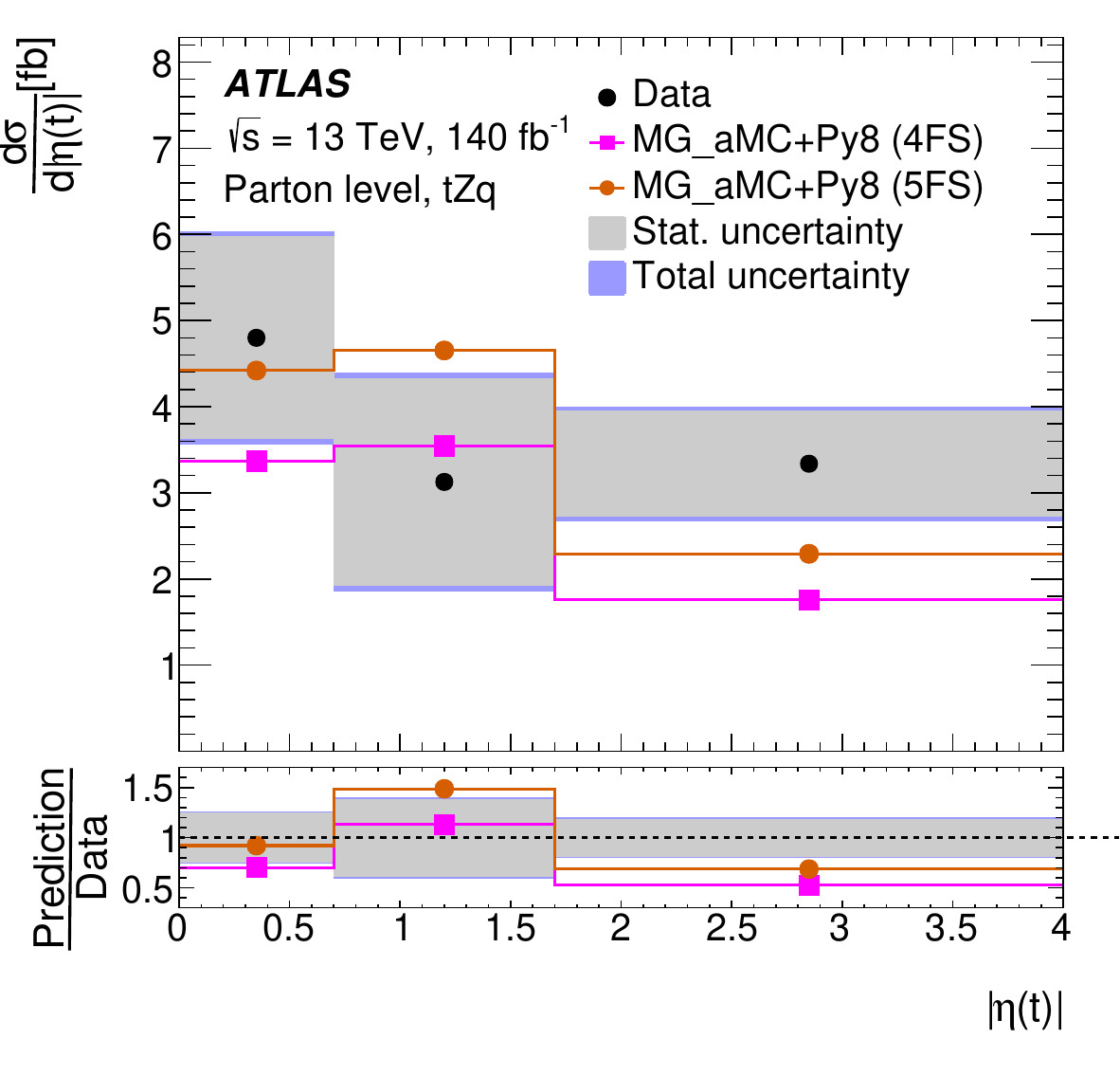}
\caption{}
\end{subfigure}
\begin{subfigure}{0.45\textwidth}
\centering
\includegraphics[width=1\textwidth]{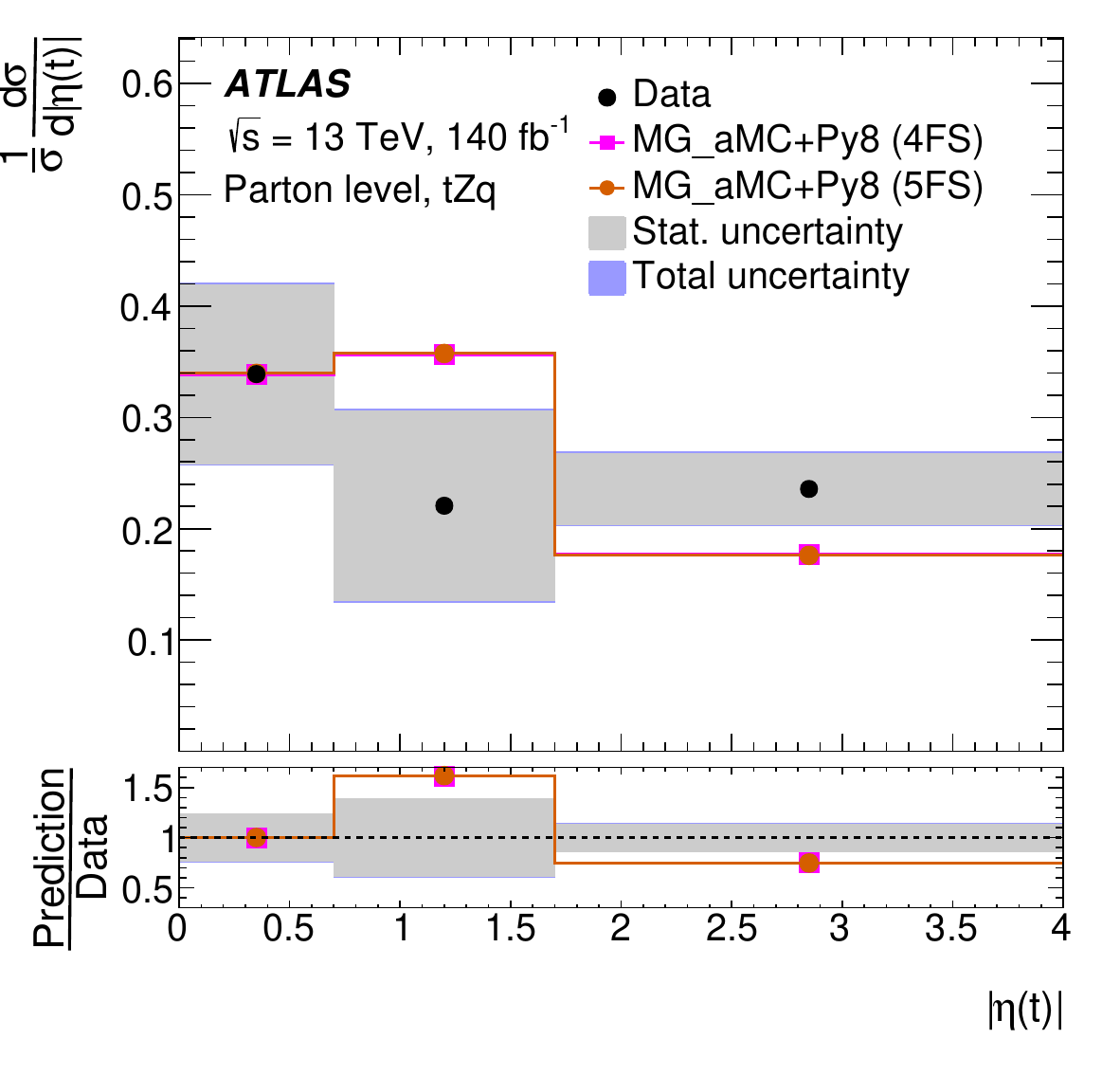}
\caption{}
\end{subfigure}
\caption{Differential cross-section measurements as a function of \etat observable in the SR,
(a, b) absolute and (c) normalised, unfolded to (a) particle level and (b, c) parton level.
The light blue band corresponds to the total uncertainty of the measurement; in some cases, it is almost fully covered by the light grey band, representing the dominant statistical uncertainty.
The measurements are compared with the generator predictions indicated in the legend and described in the text.
The lower panels show the ratios of the predictions to the (a, b) measured cross-sections and the (c) measured normalised differential cross-sections.
}
\label{fig:diff:topeta}
\end{figure}

\begin{figure}[htbp]
\centering
\begin{subfigure}{0.45\textwidth}
\centering
\includegraphics[width=1\textwidth]{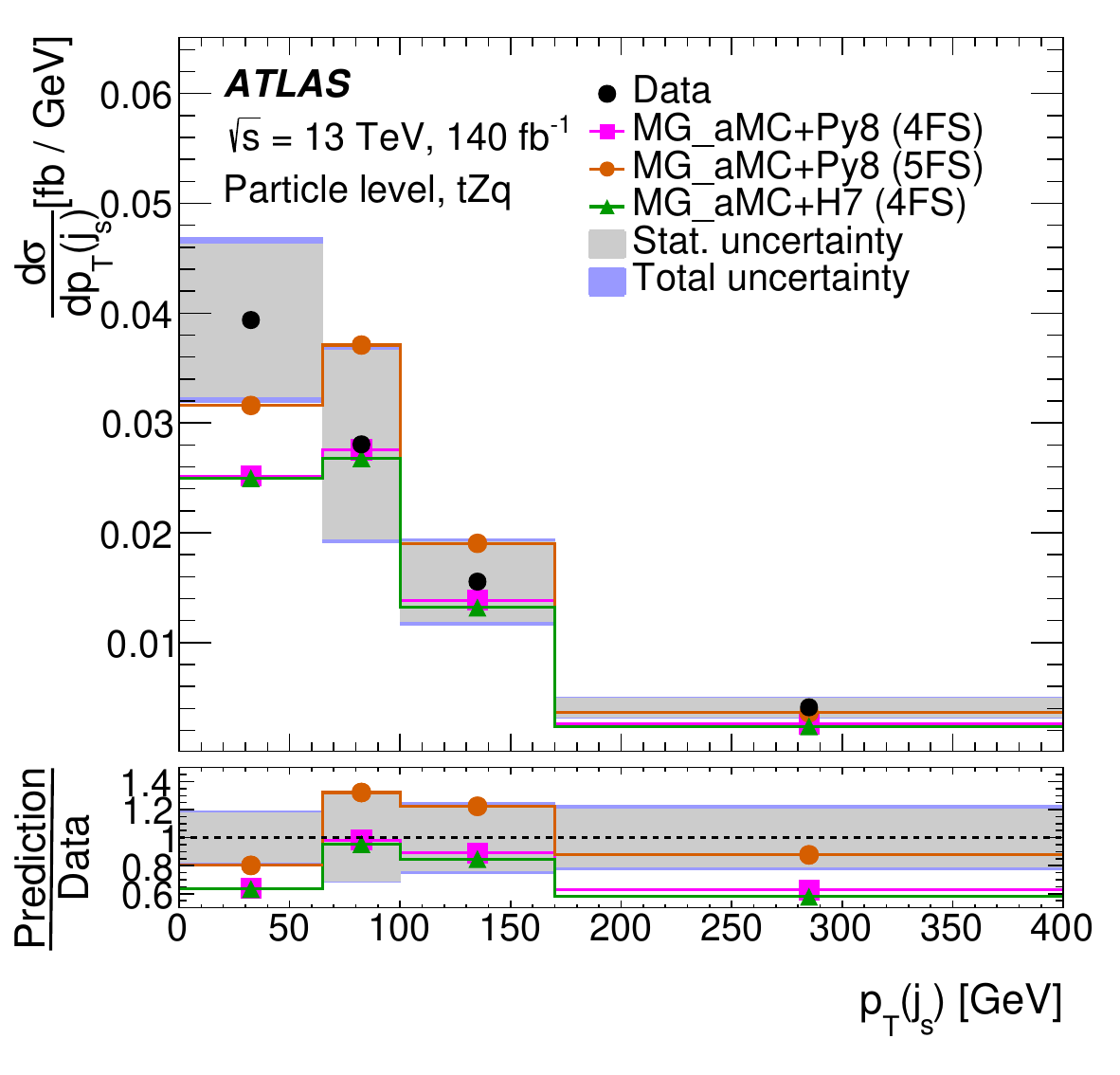}
\caption{}
\end{subfigure}
\caption{Absolute differential cross-section measurement as a function of \pTsjet observable in the SR, unfolded to particle level.
The light blue band corresponds to the total uncertainty of the measurement; in some cases, it is almost fully covered by the light grey band, representing the dominant statistical uncertainty.
The measurements are compared with the generator predictions indicated in the legend and described in the text.
The lower panel shows the ratios of the predictions to the measured cross-sections.
}
\label{fig:diff:sjetpt}
\end{figure}

\begin{figure}[htbp]
\centering
\begin{subfigure}{0.45\textwidth}
\centering
\includegraphics[width=1\textwidth]{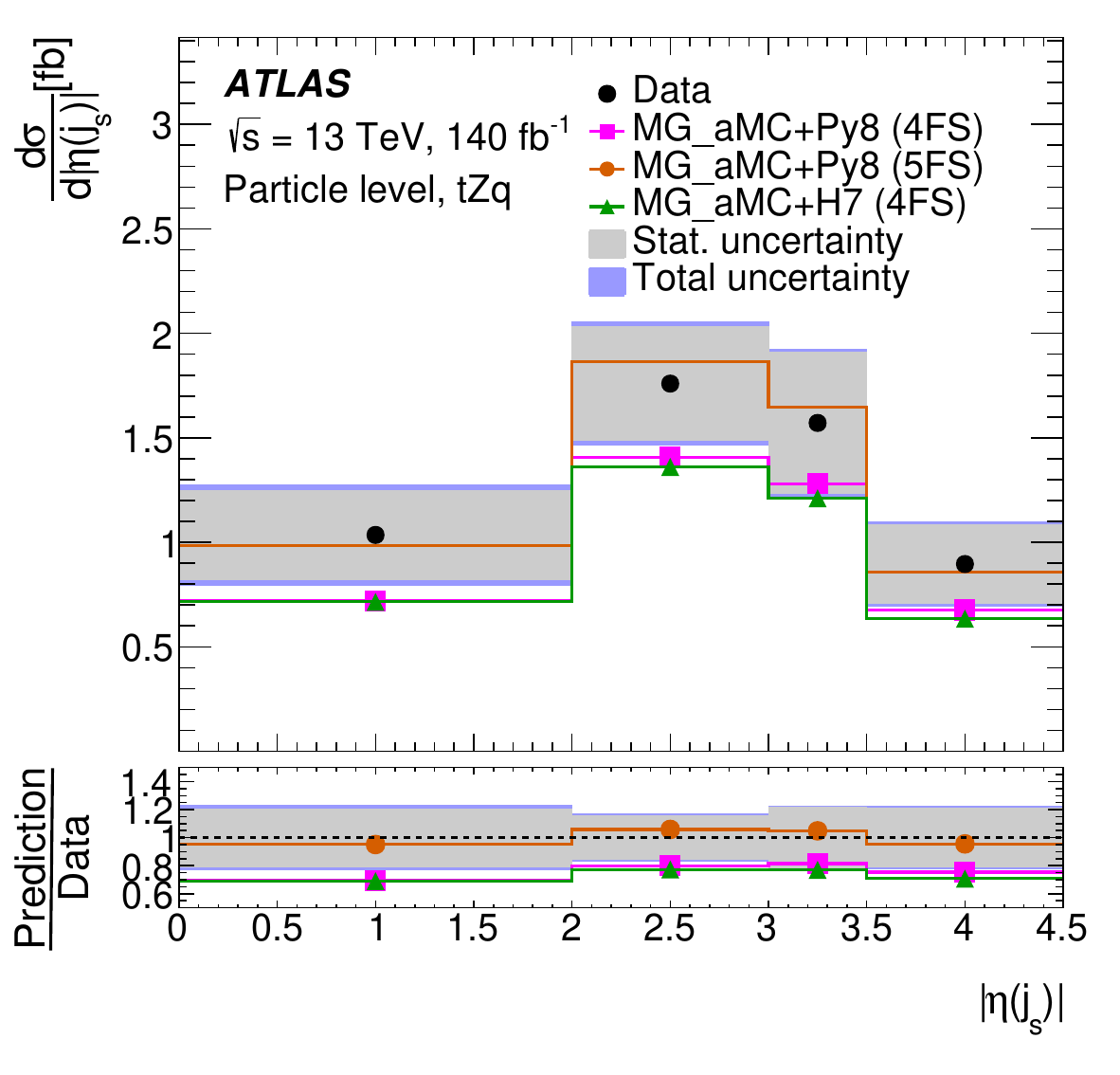}
\caption{}
\end{subfigure}
\caption{Absolute differential cross-section measurement as a function of \etasjet observable in the SR, unfolded to particle level.
The light blue band corresponds to the total uncertainty of the measurement; in some cases, it is almost fully covered by the light grey band, representing the dominant statistical uncertainty.
The measurements are compared with the generator predictions indicated in the legend and described in the text.
The lower panel shows the ratios of the predictions to the measured cross-sections.
}
\label{fig:diff:sjeteta}
\end{figure}

\begin{figure}[htbp]
\centering
\begin{subfigure}{0.45\textwidth}
\centering
\includegraphics[width=1\textwidth]{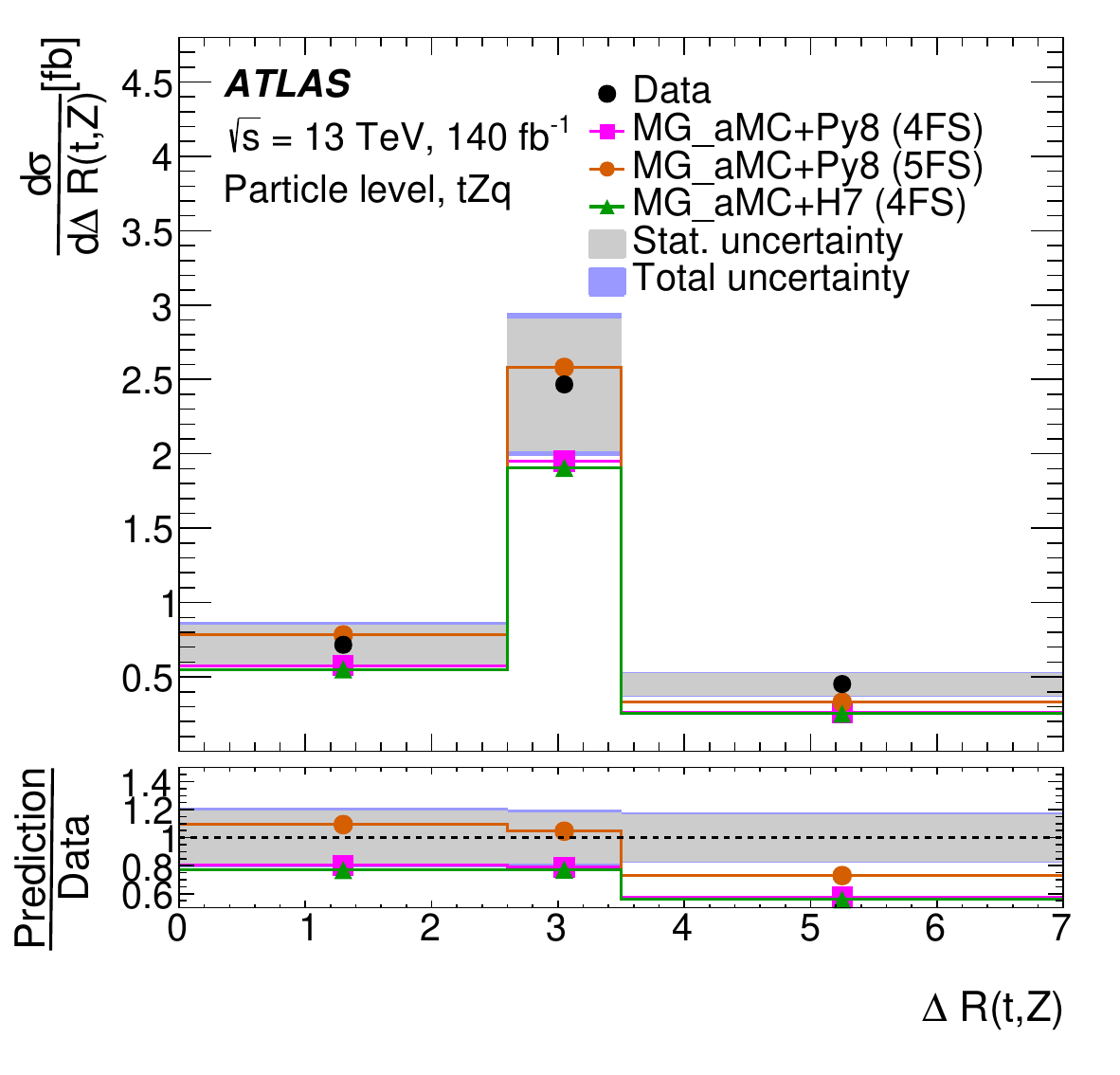}
\caption{}
\end{subfigure} \\
\begin{subfigure}{0.45\textwidth}
\centering
\includegraphics[width=1\textwidth]{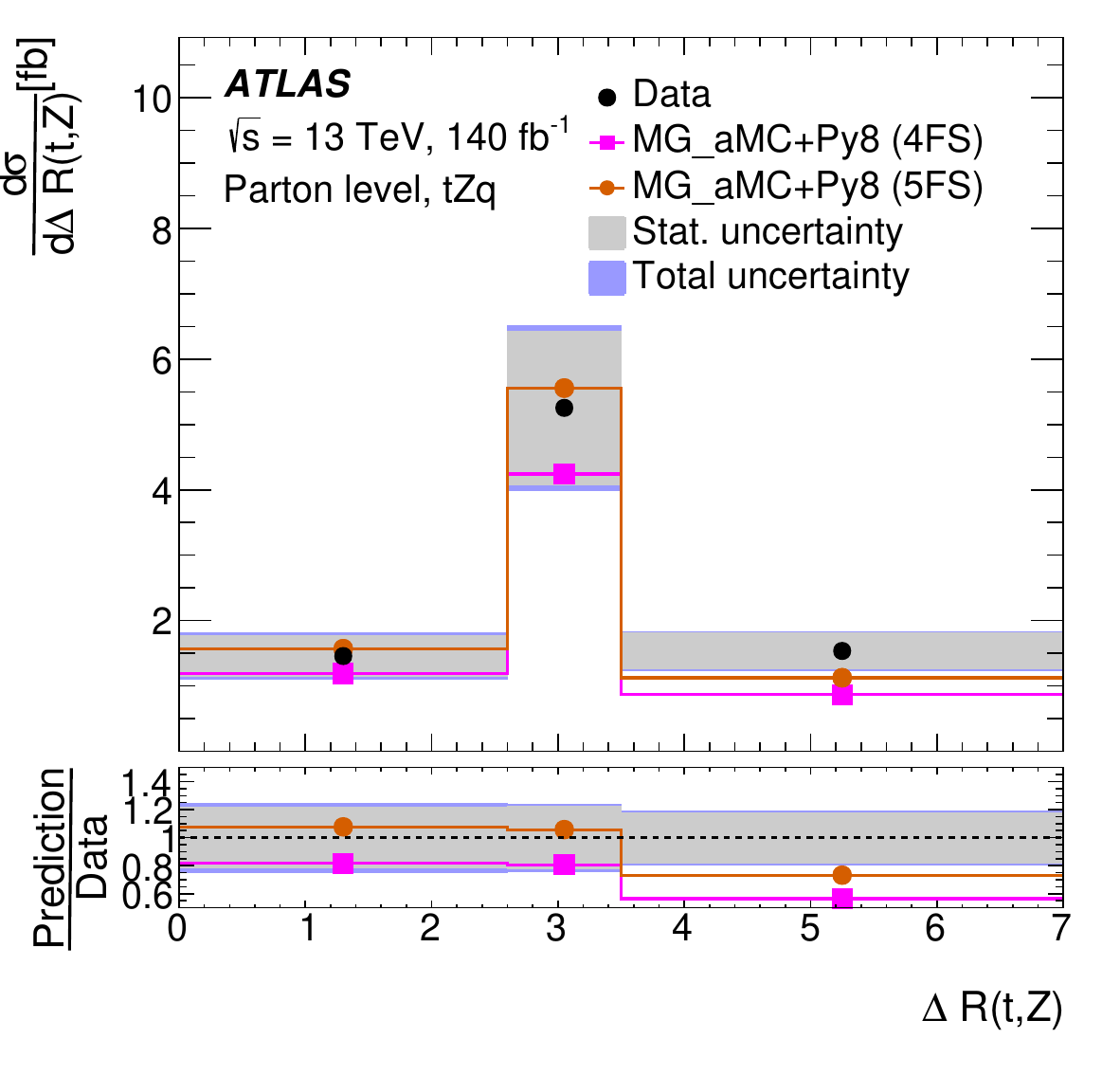}
\caption{}
\end{subfigure}
\begin{subfigure}{0.45\textwidth}
\centering
\includegraphics[width=1\textwidth]{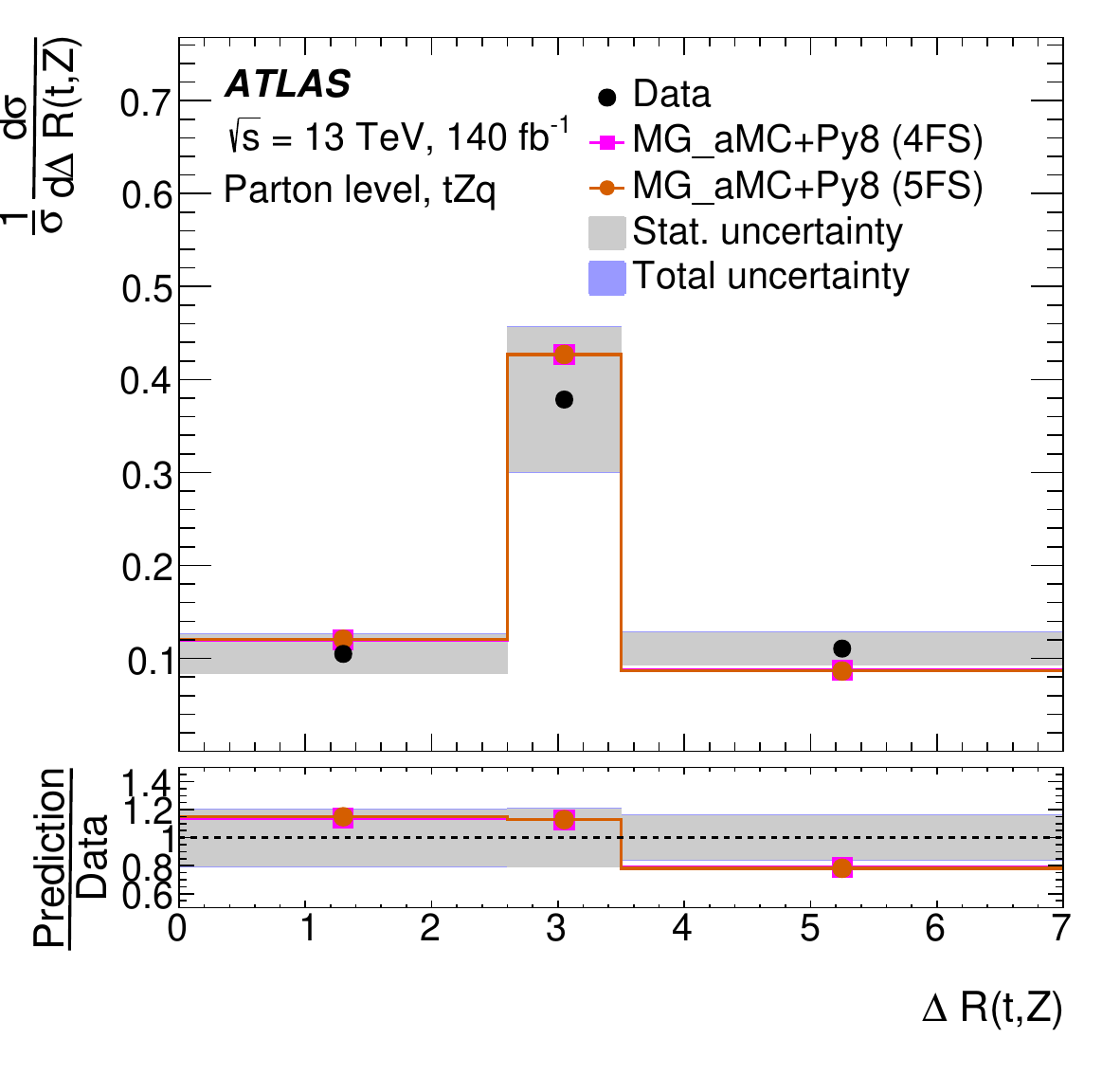}
\caption{}
\end{subfigure}
\caption{Differential cross-section measurements as a function of \dRtZ observable in the SR,
(a, b) absolute and (c) normalised, unfolded to (a) particle level and (b, c) parton level.
The light blue band corresponds to the total uncertainty of the measurement; in some cases, it is almost fully covered by the light grey band, representing the dominant statistical uncertainty.
The measurements are compared with the generator predictions indicated in the legend and described in the text.
The lower panels show the ratios of the predictions to the (a, b) measured cross-sections and the (c) measured normalised differential cross-sections.
}
\label{fig:diff:dRtopZ}
\end{figure}

\begin{figure}[htbp]
\centering
\begin{subfigure}{0.45\textwidth}
\centering
\includegraphics[width=1\textwidth]{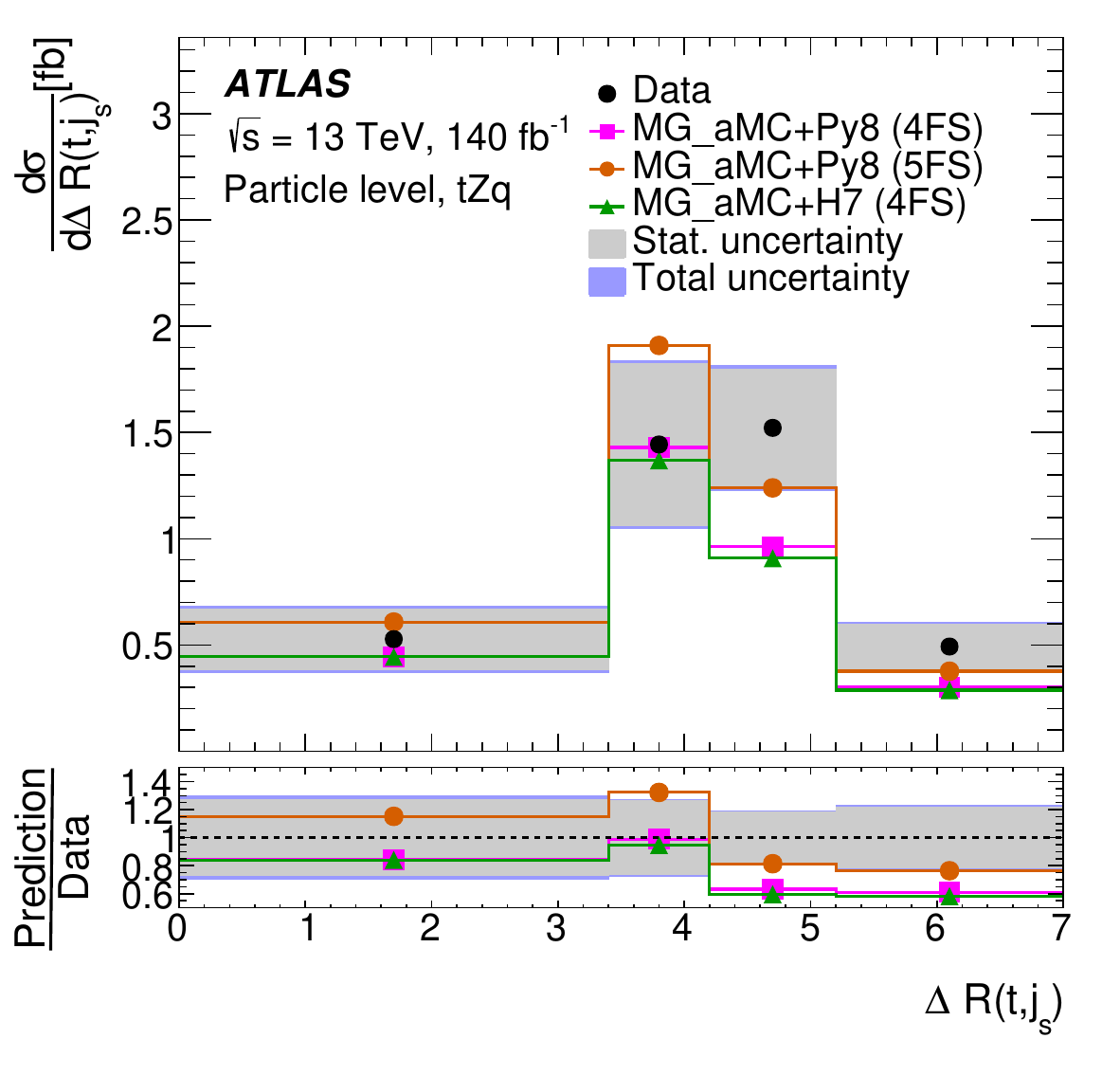}
\caption{}
\end{subfigure}
\caption{Absolute differential cross-section measurement as a function of \dRtsjet observable in the SR, unfolded to particle level.
The light blue band corresponds to the total uncertainty of the measurement; in some cases, it is almost fully covered by the light grey band, representing the dominant statistical uncertainty.
The measurements are compared with the generator predictions indicated in the legend and described in the text.
The lower panel shows the ratios of the predictions to the measured cross-sections.
}
\label{fig:diff:dRtopsjet}
\end{figure}

\begin{figure}[htbp]
\centering
\begin{subfigure}{0.45\textwidth}
\centering
\includegraphics[width=1\textwidth]{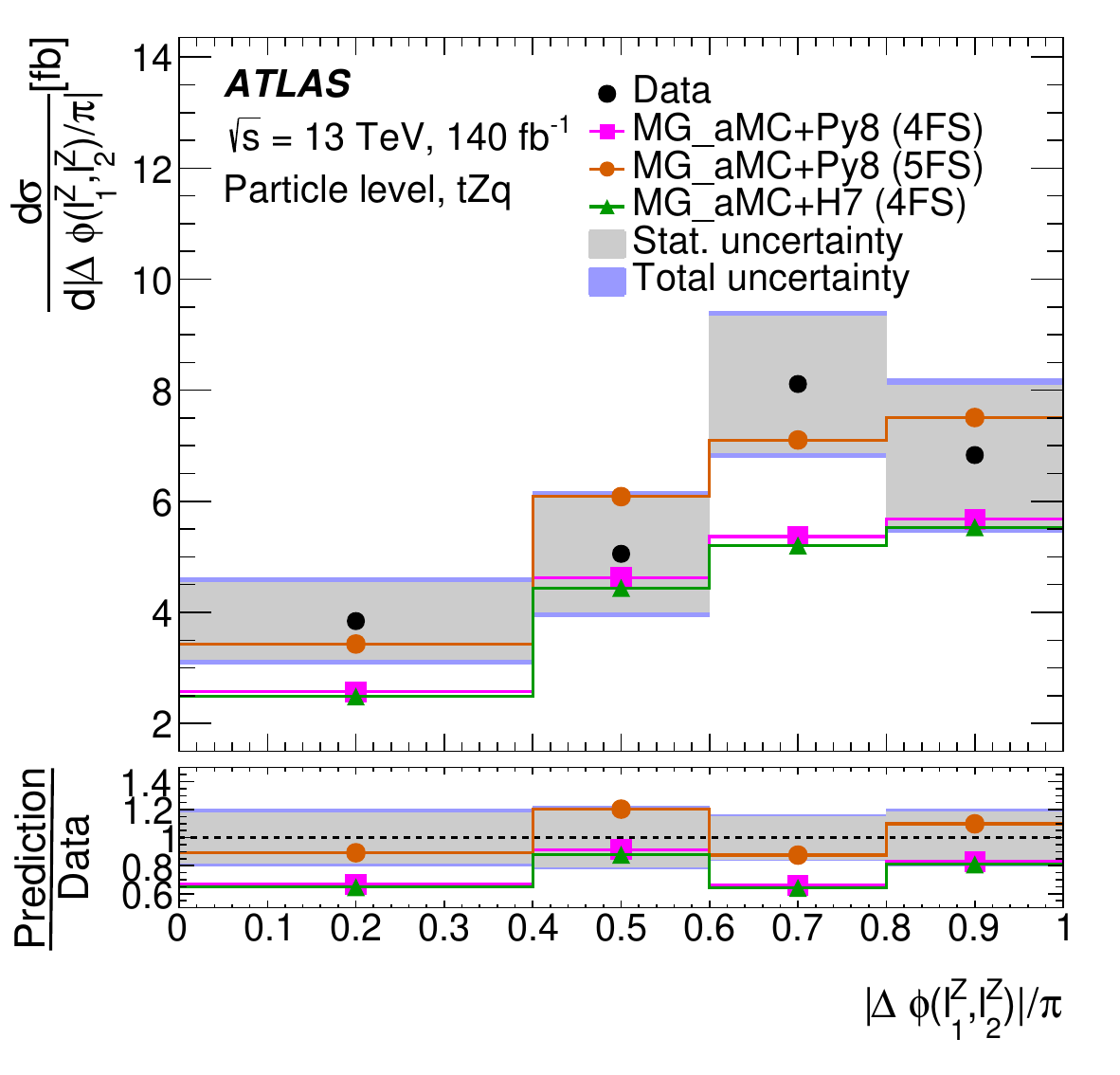}
\caption{}
\end{subfigure} \\
\begin{subfigure}{0.45\textwidth}
\centering
\includegraphics[width=1\textwidth]{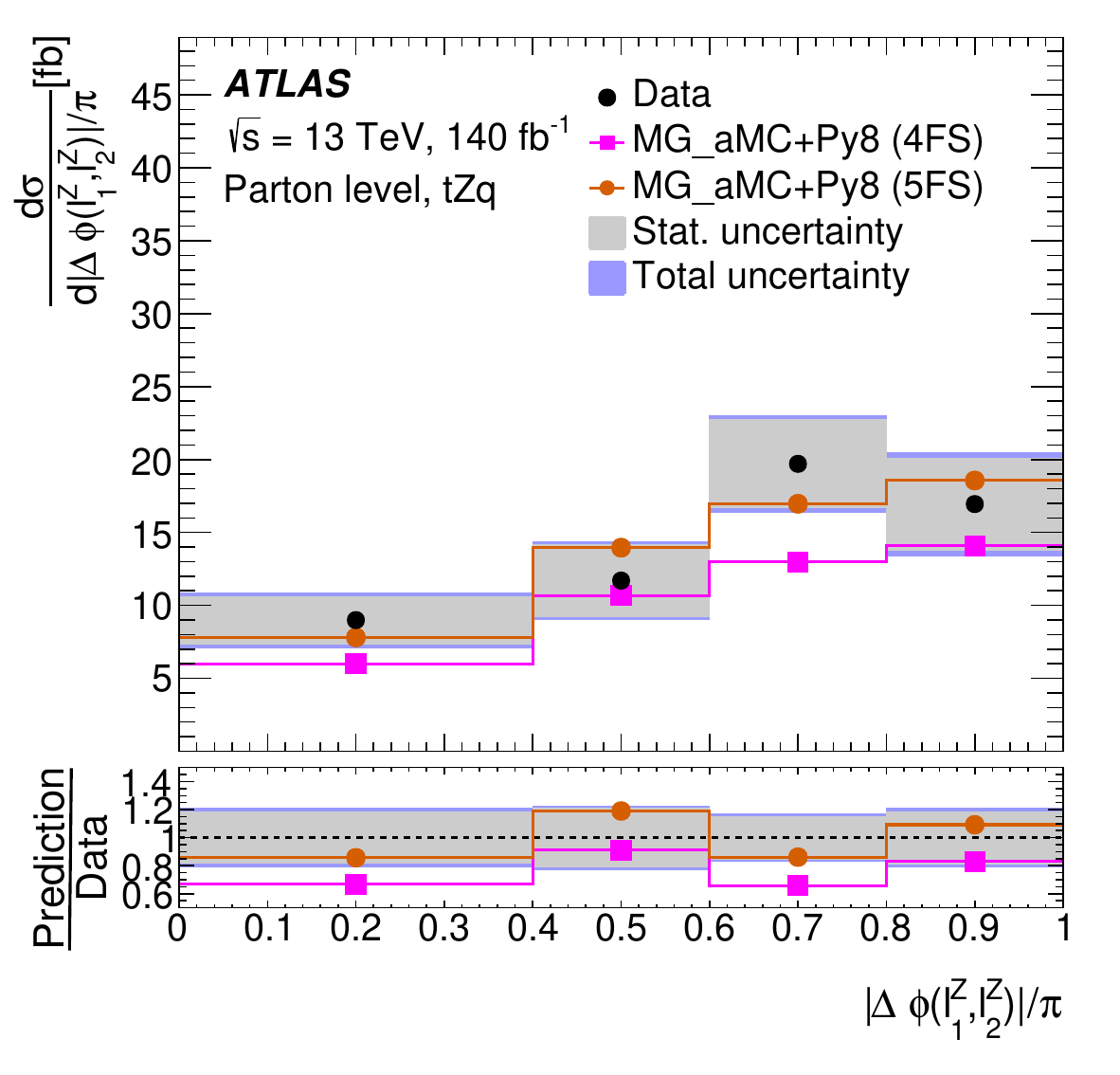}
\caption{}
\end{subfigure}
\begin{subfigure}{0.45\textwidth}
\centering
\includegraphics[width=1\textwidth]{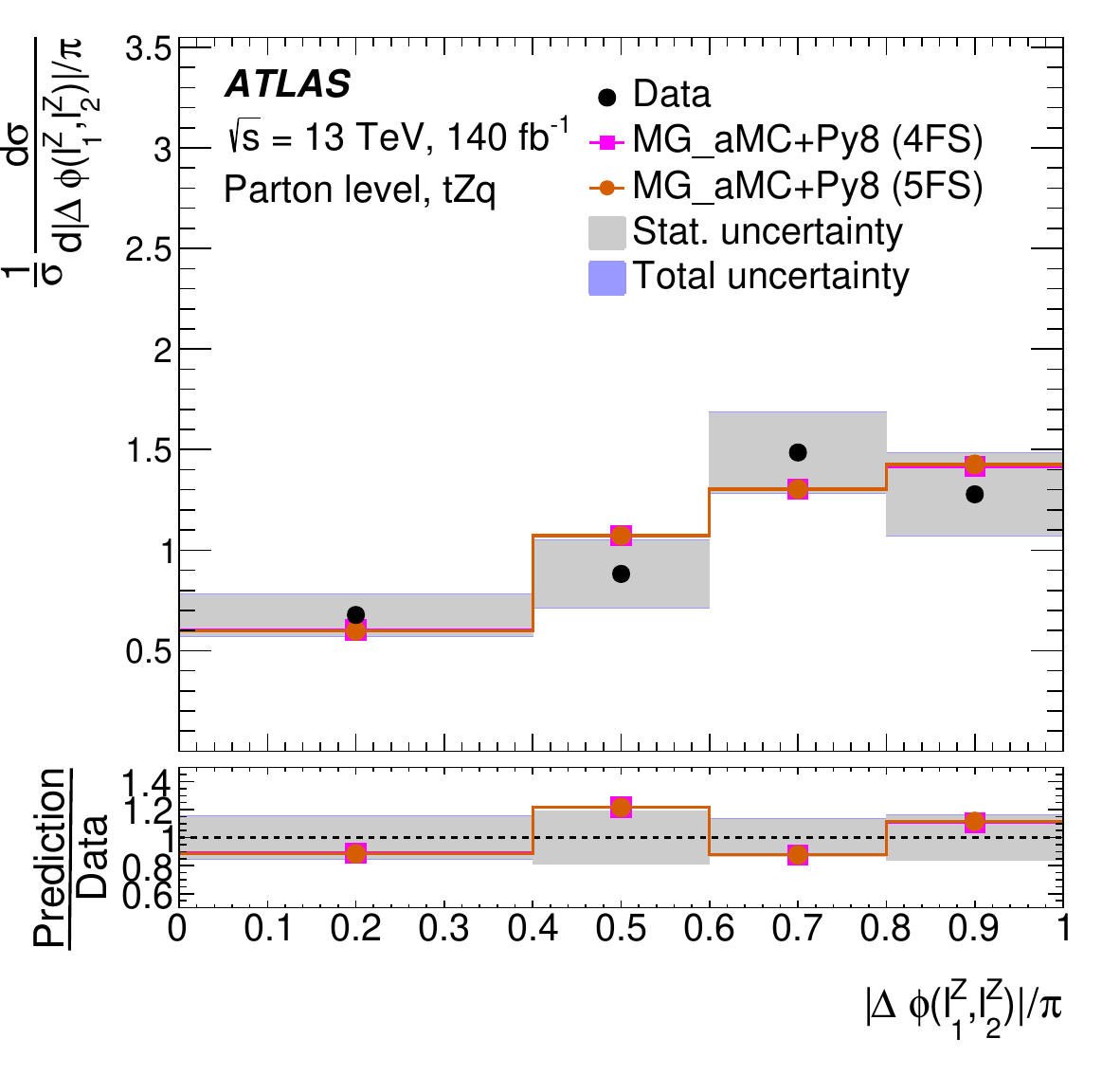}
\caption{}
\end{subfigure}
\caption{Differential cross-section measurements as a function of \dphiZl observable in the SR,
(a, b) absolute and (c) normalised, unfolded to (a) particle level and (b, c) parton level.
The light blue band corresponds to the total uncertainty of the measurement; in some cases, it is almost fully covered by the light grey band, representing the dominant statistical uncertainty.
The measurements are compared with the generator predictions indicated in the legend and described in the text.
The lower panels show the ratios of the predictions to the (a, b) measured cross-sections and the (c) measured normalised differential cross-sections.
}
\label{fig:diff:dPhilZ1lZ2}
\end{figure}

\begin{figure}[htbp]
\centering
\begin{subfigure}{0.45\textwidth}
\centering
\includegraphics[width=1\textwidth]{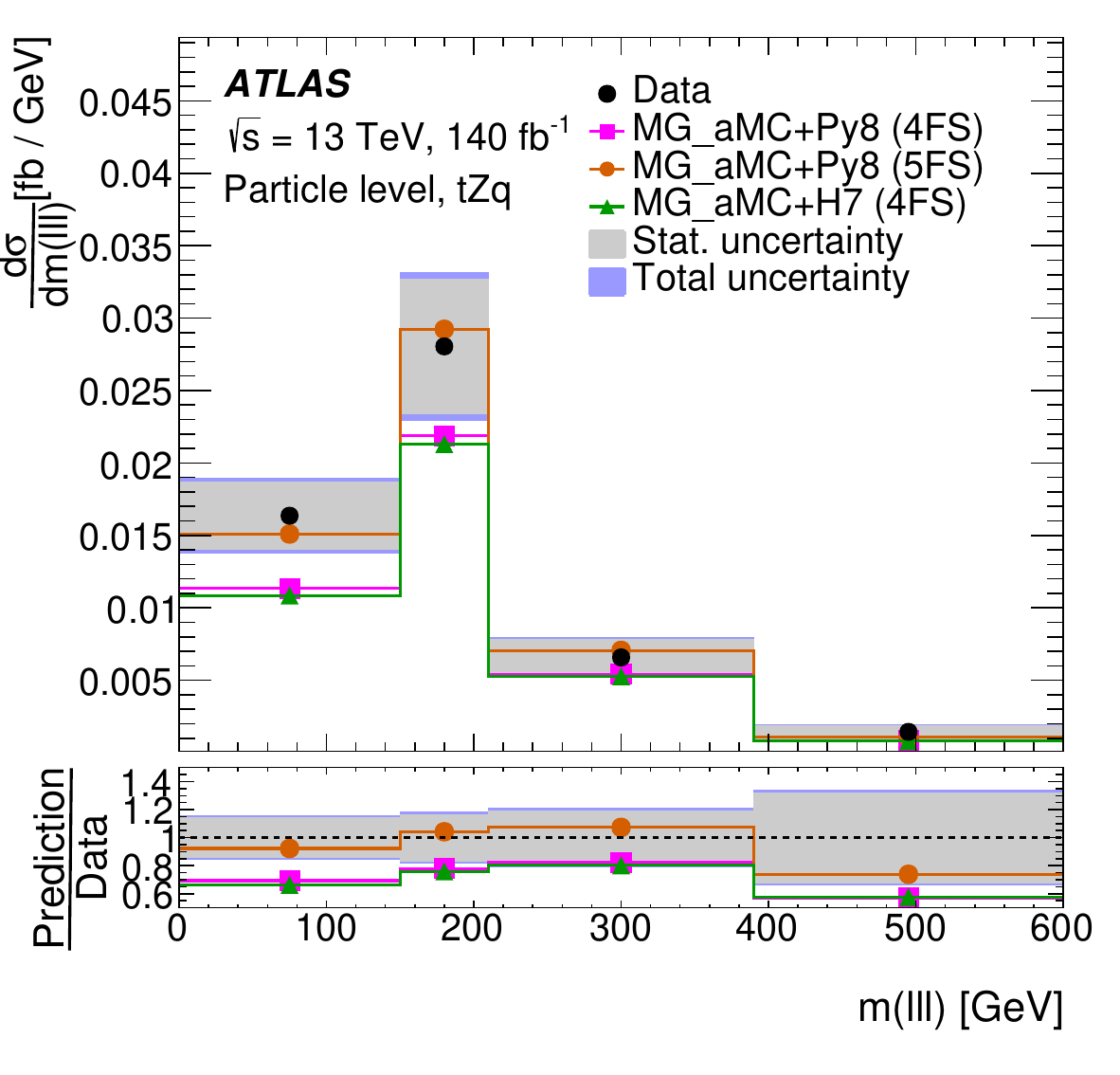}
\caption{}
\end{subfigure} \\
\begin{subfigure}{0.45\textwidth}
\centering
\includegraphics[width=1\textwidth]{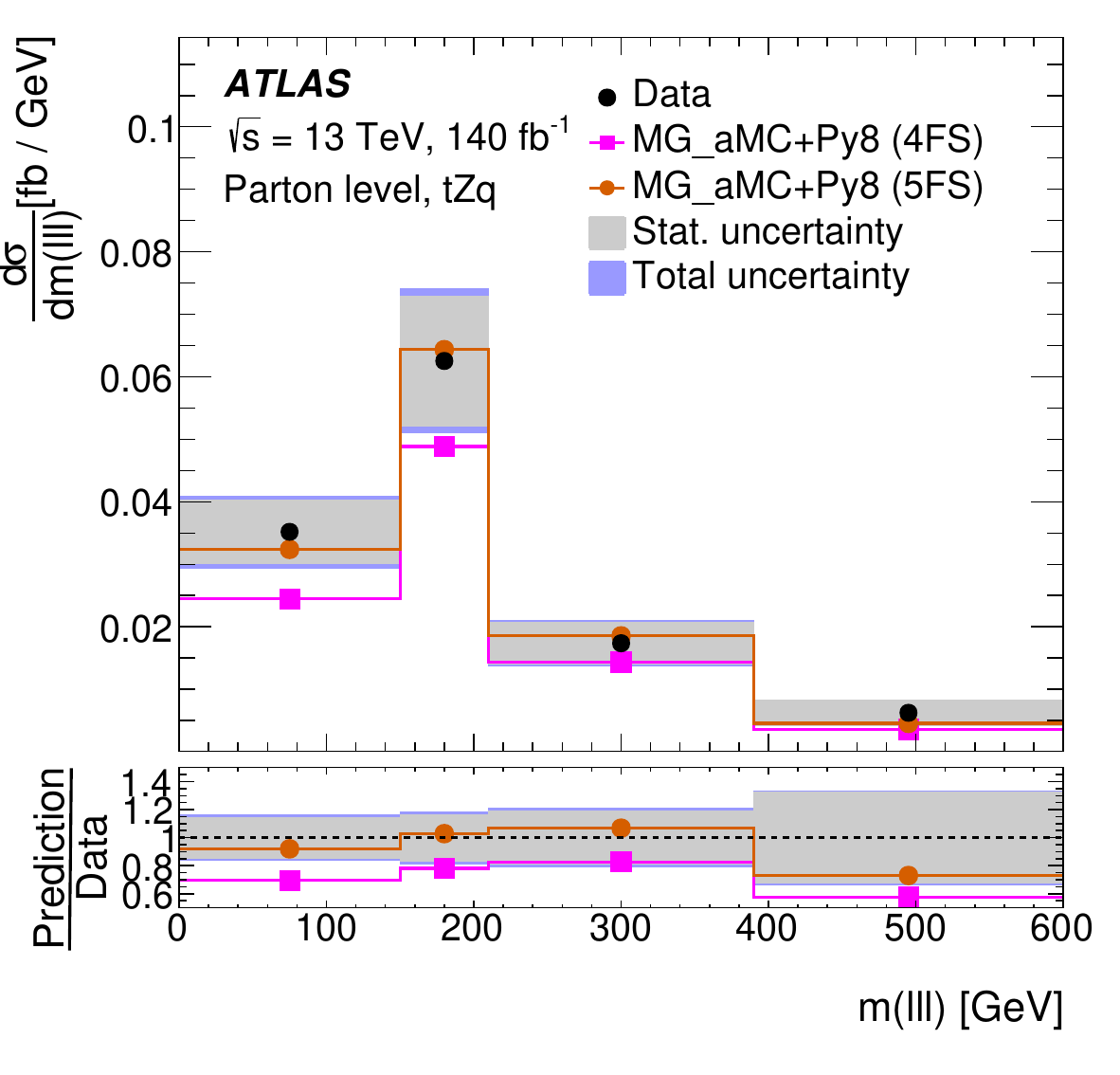}
\caption{}
\end{subfigure}
\begin{subfigure}{0.45\textwidth}
\centering
\includegraphics[width=1\textwidth]{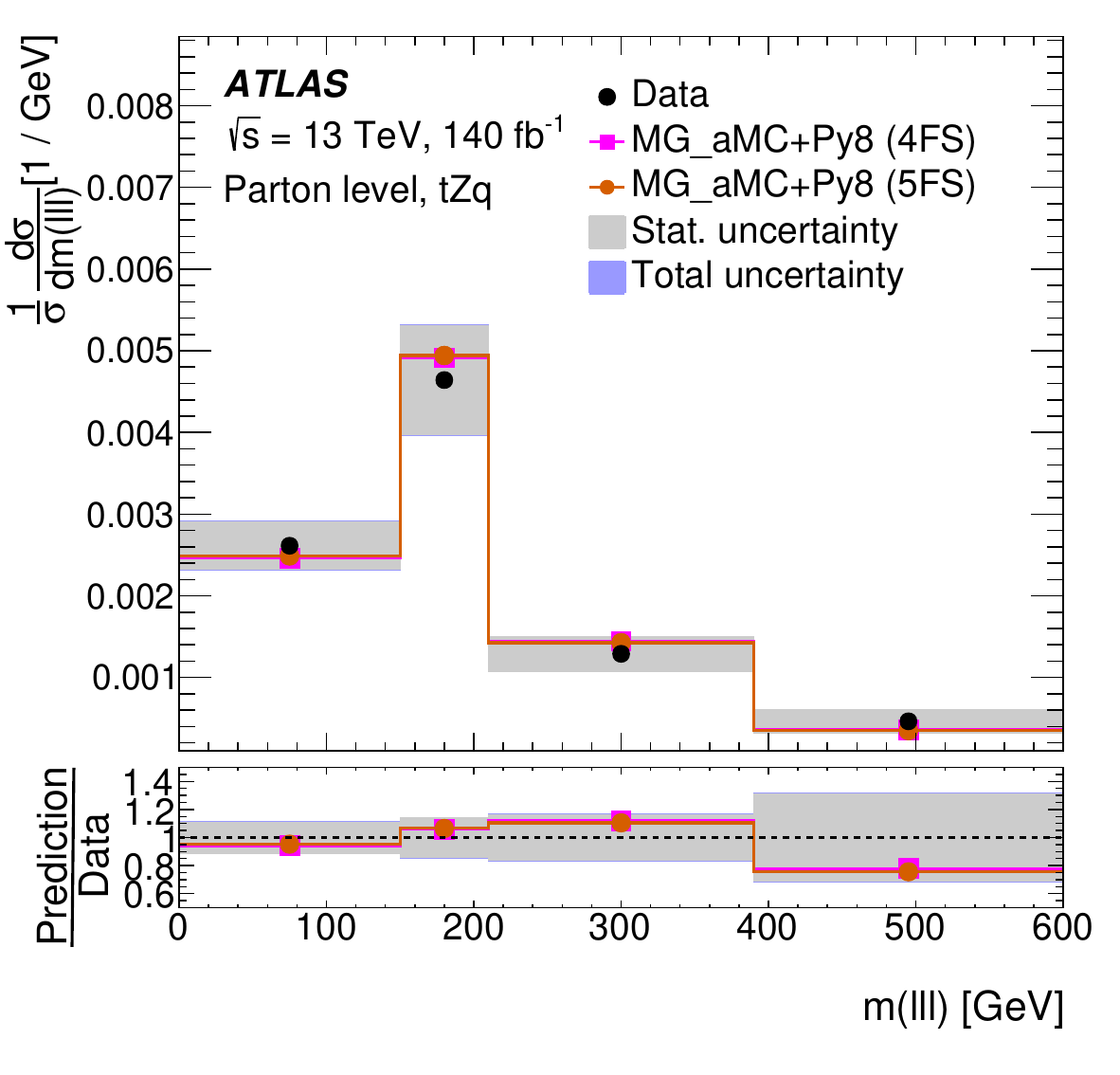}
\caption{}
\end{subfigure}
\caption{Differential cross-section measurements as a function of \mlll observable in the SR,
(a, b) absolute and (c) normalised, unfolded to (a) particle level and (b, c) parton level.
The light blue band corresponds to the total uncertainty of the measurement; in some cases, it is almost fully covered by the light grey band, representing the dominant statistical uncertainty.
The measurements are compared with the generator predictions indicated in the legend and described in the text.
The lower panels show the ratios of the predictions to the (a, b) measured cross-sections and the (c) measured normalised differential cross-sections.
}
\label{fig:diff:mlll}
\end{figure}


\clearpage
\printbibliography
\clearpage
 
\begin{flushleft}
\hypersetup{urlcolor=black}
{\Large The ATLAS Collaboration}

\bigskip

\AtlasOrcid[0000-0002-6665-4934]{G.~Aad}$^\textrm{\scriptsize 102}$,
\AtlasOrcid[0000-0001-7616-1554]{E.~Aakvaag}$^\textrm{\scriptsize 17}$,
\AtlasOrcid[0000-0002-5888-2734]{B.~Abbott}$^\textrm{\scriptsize 121}$,
\AtlasOrcid[0000-0002-0287-5869]{S.~Abdelhameed}$^\textrm{\scriptsize 83b}$,
\AtlasOrcid[0000-0002-1002-1652]{K.~Abeling}$^\textrm{\scriptsize 54}$,
\AtlasOrcid[0000-0001-5763-2760]{N.J.~Abicht}$^\textrm{\scriptsize 48}$,
\AtlasOrcid[0000-0002-8496-9294]{S.H.~Abidi}$^\textrm{\scriptsize 30}$,
\AtlasOrcid[0009-0003-6578-220X]{M.~Aboelela}$^\textrm{\scriptsize 44}$,
\AtlasOrcid[0000-0002-9987-2292]{A.~Aboulhorma}$^\textrm{\scriptsize 36e}$,
\AtlasOrcid[0000-0001-5329-6640]{H.~Abramowicz}$^\textrm{\scriptsize 154}$
% ...
\bigskip
\\

% ...
$^{17}$Department for Physics and Technology, University of Bergen, Bergen; Norway.\\
% ...
$^{30}$Physics Department, Brookhaven National Laboratory, Upton NY; United States of America.\\
% ...
$^{36}$$^{(a)}$Facult\'e des Sciences Ain Chock, Universit\'e Hassan II de Casablanca;$^{(b)}$Facult\'{e} des Sciences, Universit\'{e} Ibn-Tofail, K\'{e}nitra;$^{(c)}$Facult\'e des Sciences Semlalia, Universit\'e Cadi Ayyad, LPHEA-Marrakech;$^{(d)}$LPMR, Facult\'e des Sciences, Universit\'e Mohamed Premier, Oujda;$^{(e)}$Facult\'e des sciences, Universit\'e Mohammed V, Rabat;$^{(f)}$Institute of Applied Physics, Mohammed VI Polytechnic University, Ben Guerir; Morocco.\\
% ...
$^{44}$Physics Department, Southern Methodist University, Dallas TX; United States of America.\\
% ...
$^{48}$Fakult\"{a}t Physik, Technische Universit{\"a}t Dortmund, Dortmund; Germany.\\
% ...
$^{54}$II. Physikalisches Institut, Georg-August-Universit\"{a}t G\"ottingen, G\"ottingen; Germany.\\
% ...
$^{83}$$^{(a)}$Khalifa University of Science and Technology, Abu Dhabi;$^{(b)}$New York University Abu Dhabi, Abu Dhabi;$^{(c)}$United Arab Emirates University, Al Ain;$^{(d)}$University of Sharjah, Sharjah; United Arab Emirates.\\
% ...
$^{102}$CPPM, Aix-Marseille Universit\'e, CNRS/IN2P3, Marseille; France.\\
% ...
$^{121}$Homer L. Dodge Department of Physics and Astronomy, University of Oklahoma, Norman OK; United States of America.\\
% ...
$^{154}$Raymond and Beverly Sackler School of Physics and Astronomy, Tel Aviv University, Tel Aviv; Israel.\\
% ...
$^{a}$ Also at Affiliated with an institute formerly covered by a cooperation agreement with CERN.\\
$^{b}$ Also at An-Najah National University, Nablus; Palestine.\\
$^{c}$ Also at Borough of Manhattan Community College, City University of New York, New York NY; United States of America.\\
$^{d}$ Also at Center for Interdisciplinary Research and Innovation (CIRI-AUTH), Thessaloniki; Greece.\\
$^{e}$ Also at Centre of Physics of the Universities of Minho and Porto (CF-UM-UP); Portugal.\\
$^{f}$ Also at CERN, Geneva; Switzerland.\\
% ...

\end{flushleft}


%

%
%

%

%
%
%
%

\end{document}